\documentclass[a4paper,12pt,%
  normalheadings,%
  headsepline,%
  headinclude,%
  footexclude,%
  bibtotoc,%
  BCOR10mm,%
  openany,%
  pointlessnumbers%
]{scrbook}

\usepackage{natbib}
\bibpunct{[}{]}{;}{a}{}{,}
\let\cite\citep

\usepackage{graphicx,color}

\usepackage{hyperref}
\hypersetup{%
  bookmarksopen=true, % 
  bookmarksnumbered=true, % 
  pdftitle={Functional Renormalization Group Analysis of Luttinger liquids %
    with impurities},
  pdfsubject={Doctoral dissertation},
  pdfauthor={Sabine Andergassen},
  pdfkeywords={correlated electrons}
}

\usepackage{feynmp}
\ifpdfoutput{\DeclareGraphicsRule{*}{mps}{*}{}}{} % 
\let\comma=, % 

\usepackage[fleqn]{amsmath}
\allowdisplaybreaks % 
\ifx\undefined\gtrsim % 
  \usepackage{amssymb} % 
\fi

\providecommand{\abb}[1]{{{#1}}}

\def\fRG{f\kern1pt\abb{RG}}

\def\ie{i.\kern.5pt e.}
\def\zB{z.\kern1pt B.}
\def\PhD{Ph.\kern1pt D.}
\DeclareMathOperator{\Tr}{Tr}

\providecommand{\inner}[2]{\left(#1,#2\right)}

\newcommand{\ket}[1]{\lvert#1\rangle}
\newcommand{\bra}[1]{\langle#1\rvert}

\renewcommand{\Re}{\mathrm{Re}}
\renewcommand{\Im}{\mathrm{Im}}
\newcommand{\at}{\Big\vert}
\newcommand{\dmux}[3]{d\mu_{#1}[#2,#3]}
\newcommand{\dmu}[2]{\dmux{#1}{#2}{\bar #2}}

\newcommand{\Laplace}[1]{\Delta_{#1}}
\newcommand{\ddd}[1]{\frac{\delta}{\delta #1}}

\newcommand{\G}{\mathcal{G}}

\def\b0{{\bf 0}}

\def\cG{{\cal G}}
\def\cO{{\cal O}}

\def\cV{{\cal V}}
\def\cZ{{\cal Z}}
\def\cW{{\cal W}}
\def\cY{{\cal Y}}

\def\tG{\tilde G}

\def\Re{{\rm Re}}
\def\Im{{\rm Im}}
\def\bra{\langle}
\def\ket{\rangle}
\def\up{\uparrow}
\def\down{\downarrow}
\def\lra{\leftrightarrow}
\def\alf{\alpha}
\def\eps{\epsilon}
\def\gam{\gamma}
\def\Gam{\Gamma}
\def\lam{\lambda}
\def\Lam{\Lambda}
\def\om{\omega}

\def\sg{\sigma}
\def\Sg{\Sigma}
\def\etab{\bar{\eta}}
\def\chib{\bar{\chi}}
\def\psib{\bar{\psi}}
\def\phib{\bar{\phi}}
\def\Psib{\bar{\Psi}}
\def\Phib{\bar{\Phi}}

\def\lra{\leftrightarrow}

\newcommand{\fig}[4][clip]{\begin{figure}[ht!]%
    \centerline{\includegraphics*[#1]{#2}}
    \vskip 4mm 
    \setcapindent{0mm}\setcapwidth[c]{\textwidth}\caption{\small\label{#3}#4}\end{figure}
}

\newcommand{\setsize}[1]{\linespread{#1}\normalsize}

\originalTeX

\begin{document}
\hyphenation{pa-ra-me-tri-za-tion bo-so-ni-za-tion trans-for-ma-tion
  back-scat-ter-ing qua-drat-ic re-nor-ma-li-za-tion}

\setkomafont{caption}{\itshape}
\setkomafont{pagehead}{  \sffamily}

\mainmatter
\pagestyle{empty}

\begin{flushleft}
  \large\sffamily
  {\LARGE\bfseries
     Functional Renormalization-Group Analysis \\[1ex]
     of Luttinger Liquids with Impurities} \\[5ex]
  {\Large\bfseries Sabine Andergassen} \\[4ex]
  Dissertation accepted by the University of Stuttgart%
  \footnote{\raggedright\sffamily Version without the German summary; full
    version available online at \\
    \href{http://elib.uni-stuttgart.de/opus/volltexte/2006/2534/}
    {\ttfamily http://elib.uni-stuttgart.de/opus/volltexte/2006/2534/}} \\
  for the degree of Doctor of Natural Sciences \\[2ex]
  Examiner: Prof.\ Dr.\ Walter Metzner \\
  Co-examiner: Prof.\ Dr.\ Ulrich Weiss \\[2ex]
  Max-Planck-Institut f\"ur Festk\"orperforschung \\
  Stuttgart, Germany 2006
\end{flushleft}
\vfill
\clearpage

%%%%%%%%%%%%%%%%%%%%%%%%%%%%%%%%%%%%%%%%%%%%%%%%%%%%%%%%%%%%%%%%%%%%%%%%

\setsize{1.2}
\pagestyle{plain}
% Thesis abstract

\chapter*{Abstract}

In one-dimensional quantum wires the interplay of electron correlations 
and impurities strongly influences the low-energy physics.
The diversity of energy scales and the competition of correlations in 
interacting Fermi systems can be treated very efficiently with the
functional renormalization group (fRG), 
describing the gradual evolution from a microscopic model Hamiltonian 
to the effective low-energy action as a function 
of a continuously decreasing energy cutoff. 
The fRG provides the universal low-energy asymptotics as well as
nonuniversal properties, 
and in particular an answer to the important question at what scale the
ultimate asymptotics sets in.

The lowest order truncation of the fRG hierarchy of flow equations
considered previously for spinless fermions
is generalized to spin-$\frac{1}{2}$ systems and extended 
including renormalization of the two-particle interaction, in addition 
to renormalization of the impurity potential.
The underlying approximations are devised for weak interactions 
and arbitrary impurity strengths.
A comparison with numerical density-matrix renormalization results
for systems with up to $1000$ sites shows that the fRG 
is remarkably accurate even for intermediate interaction strengths.

We investigate the influence of impurities on spectral and transport 
properties of fermionic lattice models with short-range interactions. 
The results capture relevant energy scales and crossover phenomena,
in addition to the universal low-energy asymptotics.
For weak and intermediate impurity strengths 
the asymptotic behavior 
is approached only at rather low energy scales,
accessible only for very large systems.
For spin-$\frac{1}{2}$ systems
two-particle backscattering leads to striking effects,
which are not captured if the bulk system is approximated by
its low-energy fixed point, the Luttinger model.
In particular, the expected decrease of spectral weight near the 
impurity and of the conductance at low energy scales is often 
preceded by a pronounced increase, and the asymptotic power 
laws are modified by logarithmic corrections.

%%% Local Variables: 
%%% mode: latex
%%% TeX-master: "thesis"
%%% End: 

% Thesis acknowledgments

\chapter*{Acknowledgments}
\label{sec:ack}

Several people contributed to the realization of the present thesis, and I
would like to express my gratitude for their support during the last years.

First of all, I would like to thank my supervisor Prof.\ Walter Metzner,
who taught me to appreciate the beauty of simple ideas. 
I would like to thank him for his heedful supervision,
for guiding my first steps in scientific research,
for always having time for discussions, the numerous valuable suggestions,
for his constant encouragement and support,
and for the opportunity to attend several international 
workshops and conferences. 
It has been a great pleasure to work in his theory group, and
I wish to thank for the opportunity to join 
his collaboration with Prof.\ Kurt Sch\"onhammer and Prof.\ Volker 
Meden in G\"ottingen, and Prof.\ Uli Schollw\"ock in Aachen. 
It has been a particularly enriching experience and fruitful collabaration.
I would like to thank Prof.\ Kurt Sch\"onhammer and
Prof.\ Volker Meden for their support, the numerous discussions and
encouragement, the stimulating correspondence, the 
careful proofreading of the manuscript, and the warm hospitality.
To Prof.\ Uli Schollw\"ock I would like to express my thanks
for his precious suggestions.

I would like to thank Prof.\ Ulrich Weiss for his willingness to
co-report on the thesis.

I wish to thank Prof.\ Manfred Salmhofer for the
excellent lectures on the functional renormalization-group technique, 
and Prof.\ Carsten Honerkamp for his support and suggestions 
of further extensions and applications.

A special thank goes to my officemate Dr.\ Tilman Enss, for the numerous
discussions, his helpfulness and patient assistence in computer
issues. It was a great pleasure to share these years. I would also like to
thank him for a critical reading of the manuscript.

Let me also commemorate Xavier Barnab\'e-Th\'eriault from G\"ottingen, 
with whom I started, together with Dr.\ Tilman Enss,
to implement a small shared library, 
and who unfortunately died in a tragic accident on August 15, 2004.
He conveyed the curiosity and enthusiasm in approaching open problems.

I wish to thank Dr.\ Daniel Rohe for always finding time for my questions, 
his encouragement and a critical proofreading, Julius Reiss 
for his assistence and his sociocultural activities at the institute, 
Roland Gersch for a careful reading of the manuscript, 
Dr.\ habil.\ Karsten Held and Dr.\ habil.\ Dirk Manske 
for their help and encouragement,
as well as all other members of the Department Metzner, 
for the numerous discussions and
interesting conversations, their support and the friendly atmosphere.
I would also like to thank our secretary Mrs.\ Ingrid Knapp for her help
in all organizational matters, and the Computer Department of the 
Max Planck Institute for their support.

I owe particular thanks to Prof.\ Carlo Di Castro, 
Prof.\ Claudio Castellani,
and Dr.\ Massimo Capone in Rome for advices and valuable discussions.

%%% Local Variables: 
%%% mode: latex
%%% TeX-master: "thesis"
%%% End: 

\setsize{1.1}
\tableofcontents
\setsize{1.2}

%%%%%%%%%%%%%%%%%%%%%%%%%%%%%%%%%%%%%%%%%%%%%%%%%%%%%%%%%%%%%%%%%%%%%%%% 
\pagestyle{headings}
% Thesis introduction

\chapter{Introduction}
\label{sec:intro}

In one dimension metallic electron systems are strongly 
affected by interactions. 
Differently from the conventional Fermi-liquid behavior, 
the generic low-energy physics 
is described by the Luttinger-liquid phenomenology \cite{Gia03}. 
For various correlation functions
Luttinger-liquid theory predicts anomalous power laws;
for spin-rotation invariant systems the exponents
can be expressed in terms of a single interaction-dependent parameter
$K_{\rho}$.
An important aspect concerns the peculiar effects due to the 
interplay of impurities and interactions.
For Luttinger liquids with repulsive interactions 
already a single static impurity has a strong effect
\cite{LP74,Mat74,AR82,GS88}.
At low energy scales even a weak impurity effectively ``cuts''
the system into two parts with open boundary conditions at 
the end points, and physical observables are controlled by the 
open chain fixed point \cite{KF92b,KF92a}.
In particular, the impurity potential becomes
dressed by long-range oscillations 
leading to a characteristic power-law suppression 
of the local density of states near the impurity and the 
conductance through the impurity down to zero in the low-energy limit.
The asymptotic behavior is universal in the sense that the exponents
depend only on the properties of the bulk system via $K_{\rho}$,
while they do not depend on the impurity strength or shape.
These power laws are generally modified by logarithmic corrections
in the presence of two-particle backscattering.

The {\em asymptotic} low-energy properties of Luttinger liquids with a
single impurity are rather well understood. Universal power laws
and scaling functions have been obtained by bosonization, conformal
field theory and exact solutions for the low-energy asymptotics in 
special integrable cases \cite{Gia03}.
Numerical methods as exact diagonalization and the
density-matrix renormalization group (DMRG)
confirm the field-theoretical predictions and the validity
of the underlying assumptions for microscopic fermionic
systems with Luttinger-liquid behavior. 
The limited system size accessible to numerical solutions is however 
a serious constraint for a systematic analysis beyond the
perturbatively accessible weak and strong-impurity regimes.
The important question arises at what {\em scale} the
ultimate asymptotics sets in and 
asymptotic power laws are actually valid.
That scale can indeed be surprisingly
low, and the properties above it very different from the asymptotic
behavior.
Recently a functional renormalization group (fRG) method
has been introduced for a direct treatment of \emph{microscopic} models
of interacting fermions,
which does not only capture correctly the universal low-energy
asymptotics, but allows to compute observables on all energy
scales, providing thus also {\em nonuniversal} properties, 
and a possible key to the understanding of the behavior at intermediate
scales accessible in experiments.
Some of the nonuniversal properties can be computed numerically 
by the DMRG, 
but this method is limited to lattice systems with about $1000$ sites, 
and only a restricted set of observables can be evaluated with 
affordable computational effort.

The fRG provides a powerful computational 
tool to study interacting Fermi systems, 
especially low-dimensional systems with competing instabilities 
and entangled infrared singularities.
Starting point is an exact hierarchy of differential flow equations 
for the Green or vertex functions of the system, 
describing the gradual evolution from the microscopic model Hamiltonian 
to the effective action as a function 
of a continuously decreasing energy cutoff
introduced in the free propagator \cite{Salmhofer:9706188}. 
Approximations are then constructed by truncating the hierarchy and 
parametrizing the vertex functions with a manageable set of variables 
or functions.
The fRG captures the expected universal power laws at low energy,
as well as relevant energy scales and nonuniversal crossover phenomena 
at intermediate scales, as for the temperature dependence of the
conductance through a double barrier \cite{EMABMS04,Med04}.
The direct application to microscopic models
allows for a flexible modeling of different geometries, as
mesoscopic rings threaded by a magnetic flux \cite{MS03b,MS03a}
and Y junctions \cite{BSMS04,BSMS05}.

In previous applications to spinless Luttinger liquids with 
impurities \cite{MMSS02a,MMSS02b}
the fRG hierarchy of flow equations was truncated at first order,
where the renormalized vertex is approximated by the bare interaction.
Despite the simplicity of this scheme 
the effects of a single static impurity are captured qualitatively, 
and for spinless fermions in the weak coupling limit also quantitatively. 
It turned out that the asymptotic behavior typically holds 
only at very low energy scales and for very large systems, except for
very strong bare impurities.

In the present work we further develop and extend the fRG approach for 
Luttinger liquids with impurities to spin-$\frac{1}{2}$ fermions and
include two-particle vertex renormalization,
in addition to the renormalization of the impurity potential.
The underlying approximations are devised for weak interactions and 
arbitrary impurity strength.
A comparison with exact numerical DMRG results
for systems with up to 1000 sites shows however that the fRG with the
inclusion of vertex renormalization is remarkably accurate even
for intermediate interaction strengths. 
For spinless fermions this extension improves considerably the
quantitative accuracy of the results in
particular at intermediate interaction strengths, whereas
for spin-$\frac{1}{2}$ systems vertex renormalization is necessary to take 
into account that backscattering of particles with opposite spins at 
opposite Fermi points scales to zero in the low-energy limit.
Explicit flow equations are derived for various lattice fermion models
supplemented by different types of impurity potentials. 
We present results for spectral properties of single-particle excitations,
the oscillations in the density profile induced by impurities
or boundaries and the linear conductance 
for chains with up to $10^6$ lattice sites.
Two-particle backscattering leads to peculiar effects,
which are not captured if the bulk system is approximated by
its low-energy fixed point, the Luttinger model.
In particular, the expected decrease of spectral weight near the 
impurity and of the conductance at low energy scales is often 
preceded by a pronounced increase, and the asymptotic power 
laws are modified by logarithmic corrections.

The outline of the thesis is as follows.

\begin{itemize}
\item In Chapter~\ref{sec:imp} we give a short overview on general
aspects of Luttinger liquids with impurities.

\item The fRG formalism is developed in Chapter~\ref{sec:rg}.
We briefly review the fRG for interacting Fermi systems, and derive 
the hierarchy of differential flow equations for the 
one-particle irreducible (1PI) vertex functions.

\item In Chapter~\ref{sec:frglutt}
we describe the implementation of the fRG technique
for various one-dimensional microscopic lattice models with 
impurities, providing details on the parametrization of the 
two-particle vertex and different truncation schemes.
Parts of this chapter are published in

S.~Andergassen, T.~Enss, V.~Meden, W.~Metzner, U.~Schollw{\"o}ck, 
and K.~Sch{\"o}n-hammer, 
\textit{Functional renormalization group for Luttinger liquids with
  impurities},
\href{http://dx.doi.org/10.1103/PhysRevB.70.075102}{Phys.\ Rev.~B
  \textbf{70}, 075102 (2004)},
\href{http://arXiv.org/abs/cond-mat/0403517}
{\texttt{cond-mat/0403517}};

S.~Andergassen, T.~Enss, V.~Meden, W.~Metzner, U.~Schollw{\"o}ck, 
and K.~Sch{\"o}n-hammer,
\textit{Renormalization group analysis of the one-dimensional
  extended Hubbard model with a single impurity},
\href{http://arXiv.org/abs/cond-mat/0509021}
{\texttt{cond-mat/0509021}}.

\item In Chapter~\ref{sec:results}
we present results for spectral properties of 
single-particle excitations near an impurity or boundary, the 
density profile,
and 
transport properties in the presence of
a single and a double barrier. In the first part we focus on spinless 
fermions; the modifications due to the spin degree of freedom are 
addressed in the second part.
Parts of this chapter are presented in the above publications, 
and for the conductance in

V.~Meden, S.~Andergassen, W.~Metzner, U.~Schollw{\"o}ck, and
K.~Sch{\"o}nhammer,
\textit{Scaling of the conductance in a quantum wire},
\href{http://dx.doi.org/10.1209/epl/i2003-00624-x}{Europhys.~Lett. 
  \textbf{64}, 769 (2003)},
\href{http://arXiv.org/abs/cond-mat/0303460}
{\texttt{cond-mat/0303460}};

V.~Meden, T.~Enss, S.~Andergassen, W.~Metzner, and K.~Sch{\"o}nhammer,
\textit{Correlation effects on resonant tunneling in one-dimensional
  quantum wires},
\href{http://dx.doi.org/10.1103/PhysRevB.71.041302}
{Phys.\ Rev.~B \textbf{71}, 041302(R) (2005)},
\href{http://arXiv.org/abs/cond-mat/0403655}
{\texttt{cond-mat/0403655}};

T.~Enss, V.~Meden, S.~Andergassen, X.~Barnab\'e-Th\'eriault,
W.~Metzner, and K.~Sch{\"o}nhammer, 
\textit{Impurity and correlation effects on transport in
  one-dimen-sional quantum wires}, 
\href{http:/dx.doi.org/10.1103/PhysRevB.71.155401}
{Phys.\ Rev.~B \textbf{71}, 155401 (2005)},
\href{http://arXiv.org/abs/cond-mat/0411310}
{\texttt{cond-mat/0411310}};

S.~Andergassen, T.~Enss, and V.~Meden,
\textit{Kondo physics in transport through a quantum dot with 
  Luttinger liquid leads},
\href{http://arXiv.org/abs/cond-mat/0509576}
{\texttt{cond-mat/0509576}}.

\item We conclude in Chapter ~\ref{sec:concl} with a summary 
and an outlook on further applications and extensions of the present 
work.
\end{itemize}

%%% Local Variables: 
%%% mode: latex
%%% TeX-master: "thesis"
%%% End: 

% Thesis chapter on impurities in LL

\chapter{Impurities in Luttinger liquids}
\label{sec:imp}
\vspace{2cm}
\textit{The exactly soluble Luttinger model provides 
  a generic scenario 
  for one-dimensional Fermi systems with repulsive interactions,
  denoted as ``Luttinger liquid''. 
  The low-energy physics is completely determined by a
  few interaction-dependent characteristic parameters 
  describing the  power-law exponents of the correlation
  functions.
  Already a single static impurity leads to 
  peculiar modifications of the electronic properties of Luttinger 
  liquids.
  Even for a weak impurity potential, physical observables behave as 
  if the system is split into two parts in the low-energy limit.
  The local density of states near the
  impurity and the conductance through the impurity vanish as power 
  laws.}
\vspace{0.75cm}

%%%%%%%%%%%%%%%%%%%%%%%%%%%%%%%%%%%%%%%%%%%%%%%%%%%%%%%%%%%%%%%%%%%%%%%%

\section{Luttinger liquids}
\label{sec:imp:luttinger}

In one-dimensional interacting Fermi systems Fermi-liquid theory is not 
valid. 
The breakdown of Fermi-liquid theory is indicated already in second 
order perturbation theory, where the reduction of the quasi-particle 
weight at the Fermi surface due to interactions
diverges logarithmically. 
These divergencies can be treated by a 
weak-coupling renormalization-group method applied to an effective
low-energy theory known as $g$-ology model \cite{Sol79}. 
Depending on the values of the bare couplings, the renormalized 
couplings flow either to strong coupling, and hence out of the 
perturbatively controlled regime, or to a fixed-point Hamiltonian, 
the exactly soluble Luttinger model \cite{Tom50,Lut63,ML65}.
The term ``Luttinger liquid'' has been introduced for the latter 
systems, in analogy with the mapping of low-energy states of 
interacting electron systems onto the Fermi gas in higher dimensions 
for Fermi liquids \cite{Hal81a}.
The normal gapless metallic phase is characterized by 
{\it i)} a continuous momentum distribution with a power-law singularity at
the Fermi surface, described by a nonuniversal exponent $\alpha$;
{\it ii)} a single-particle density of states which vanishes as 
$|\omega|^{\alpha}$ near the Fermi energy, implying the absence of 
fermionic quasi-particles;
{\it iii)} finite charge and spin-density responses for long wavelengths 
and the existence of collective bosonic charge and spin-density modes;
{\it iv)} power-law singularities in various correlation functions with 
interaction-dependent exponents;
{\it v)} separation of spin and charge degrees of freedom.
There are several good reviews on one-dimensional Fermi systems, 
recent reviews are presented in Refs.~\cite{Voi95,Gia03}.
In the following we will summarize the most important results.

Theoretical work on interacting fermions in one dimension has 
progressed along different lines. Besides the perturbative 
investigation of the weak-coupling limit \cite{Sol79}, 
Luttinger-liquid theory is usually formulated using the bosonization 
technique \cite{ML65,Hal80,Hal81b,Hal81a,LP74,Mat74}.
A different approach is based on the Bethe-ansatz method for special 
integrable models \cite{Gia03}. The computation of correlation 
functions is however very difficult from the complicated
expressions for the eigenfunctions.

As proposed in a seminal work by Haldane \cite{Hal81a},
the low-energy physics of the Luttinger model is 
generic for interacting fermions in one dimension
with repulsive interactions. In the language of
the renormalization group the Luttinger model Hamiltonian is the
fixed-point Hamiltonian for a large class of one-dimensional 
fermions with repulsive interactions. 
The Luttinger model can be solved exactly at any interaction; 
it is characterized by a linear dispersion 
relation, and the electron-electron interaction is limited to forward 
scattering only \cite{Gia03}.
Umklapp and backscattering processes, as well as additional terms for 
more general models arising from band curvature
are irrelevant and vanish in the low-energy limit \cite{Hal81a}.
As in the Landau Fermi liquid a few parameters
completely determine the low-energy physics.
The charge degrees of freedom of Luttinger liquids are described by a
sound velocity $v_{\rho}$ and the dimensionless parameter $K_{\rho}$,
and the spin degrees of freedom are characterized by a spin-wave 
velocity $v_{\sigma}$ and $K_{\sigma}$.
All correlation functions are uniquely parametrized by $K_{\nu}$ and
the velocities of the collective modes 
$v_{\nu}$, with $\nu = \rho,\sigma$, in the low-energy limit;
the corresponding exponents are determined by $K_{\nu}$.
For noninteracting particles $K_{\rho} = K_{\sigma} = 1$. 
In the absence of a magnetic field, the ground state is 
spin-rotationally invariant and $K_{\sigma} = 1$, while
$K_{\rho} <1$ ($>1$) for repulsive (attractive) forces.

For $K_{\sigma} = 1$ the momentum distribution function 
exhibits a power-law singularity at the Fermi level with exponent
$\alpha = (K_{\rho}^{\phantom{-1}}\!\!\! + K_{\rho}^{-1} - 2)/4$
for any nonvanishing interaction \cite{Gia03}. 
For $\alpha <1$ the momentum distribution function 
near $k_F$ obeys a power law
\begin{equation}
  |n(k)-n(k_F)| \sim |k-k_F|^{\alpha} \; .
\end{equation}
The spectral function has the form
\begin{equation}
  N(\om)\sim|\om|^{\alpha}
\end{equation} 
in the low-energy limit.
Landau quasi-particle excitations are absent in the Luttinger liquid.
The power laws hold also for nonsoluble generalizations of the model 
with a nonlinear dispersion \cite{Hal81a}.

In the Luttinger model the charge and spin density modes are
exact undamped eigenstates, and any excited state of the 
model is a superposition of these elementary excitations.
This becomes particularly explicit in the bosonized
form of the Luttinger model \cite{ML65}. 
The Luttinger model Hamiltonian conserves charge and the
$z$ component of spin separately on each Fermi point.
Charge and spin excitations are completely independent, as
the respective terms in the Hamiltonian commute. This phenomenon 
is called ``spin-charge separation'' \cite{Gia03}, 
charge and spin propagate with different velocities.

Concerning the leading low-energy long-wavelength response 
functions there is no difference between Fermi and Luttinger liquids
\cite{Gia03}.
Thermodynamic properties as the compressibility and the 
susceptibility do not differ from the Fermi-liquid description 
and the modification due to the interaction leads to renormalized 
coefficients depending on $K_{\nu}$ and $v_{\nu}$.
Differences between Fermi and Luttinger-liquid behavior arise only
from the enhanced phase space for forward scattering in one
dimension.
Marked differences appear
in the single-particle propagator, which determines the momentum 
distribution function and the spectral density for single-particle
excitations. In a Fermi liquid residual interactions modify the
propagator only on a subleading level, leading for example to a small 
quasi-particle decay rate, while in a Luttinger liquid forward
scattering affects the leading low-energy behavior.
Another distinctive feature of Luttinger liquids is the singular
behavior of density correlations with momenta near $2k_F$ 
\cite{Gia03}.

Conservation laws play a crucial role in one-dimensional 
Fermi systems \cite{Metzner:9701012}.
In addition to the usual charge and spin conservation
the discrete structure of the Fermi surface in one dimension leads to 
an additional conservation law: separate charge 
conservation in low-energy scattering processes for particles near the
left and right Fermi points, respectively. 
Separate spin conservation is spoiled by the 
backscattering process generally present in  
models of spin-$\frac{1}{2}$ fermions. In most cases of interest,
in particular for the models considered in the present work, 
the backscattering amplitude scales to 
zero at low energies, and the separate spin 
conservation is restored asymptotically.
The velocities associated with the corresponding conserved currents 
provide a complete parametrization of the low-energy physics 
\cite{Hal81a,MC93}.

%%%%%%%%%%%%%%%%%%%%%%%%%%%%%%%%%%%%%%%%%%%%%%%%%%%%%%%%%%%%%%%%%%%%%%%%

\section{Impurity effects}
\label{sec:imp:impeff}

An important aspect of Luttinger-liquid behavior concerns the peculiar 
modification of the electronic properties in the presence of impurities.
For Luttinger liquids with repulsive interactions ($K_{\rho}<1$) 
already a single static impurity has a strong effect at low energy
scales, even if the impurity potential is relatively 
weak \cite{LP74,Mat74,AR82,GS88,KF92b,KF92a,FN93,FN93s,YGM94}.
In general the interplay of disorder and interactions is still a 
challenging issue, 
although the properties of noninteracting disordered
electronic systems are rather well understood.
In one-dimensional noninteracting systems disorder leads to localization 
of all electrons; the localization length
characterizing the exponential decay of the wave function is of the 
same order as the mean free path \cite{Gia03}.
On the other hand interactions strongly affect the properties of the 
pure system, leading to Luttinger-liquid behavior. Thus in one dimension
a particularly strong mutual influence of disorder and interactions 
is expected.

Relevant parameters in the description of disorder are
the strength of the individual impurity $V$ and the impurity density 
$n_{\rm imp}$. The variation of these two parameters leads to different 
physical effects. 
In the limiting case of very weak individual impurities with a dense 
distribution the effect of a single impurity is negligible and collective 
effects dominate;
the corresponding relevant length scale is $\sim 1/n_{\rm imp}$.
As a consequence of the central-limit theorem, 
for continuous systems the disorder can be 
described by a Gaussian distribution in the limit 
$n_{\rm imp} \to \infty$ and $V \to 0$ for constant $n_{\rm imp} V^2$ 
measuring the disorder strength \cite{Gia03}.
The main results for Gaussian disorder from a perturbative treatment 
can be summarized as follows \cite{Gia03}. 
Interactions are effectively renormalized by disorder, which is reversely 
affected by interactions.
Repulsive interactions generally enhance localization 
whereas attractive ones reduce this effect.
For spinless fermions attractive interactions 
enhance superconducting fluctuations, 
leading to an effective screening of the disorder. 
For spin-$\frac{1}{2}$ fermions
a competing effect arises. The tendency towards a uniform charge 
distribution inhibits the coupling to disorder, leading to
an increase in the localization length for strong interactions
in the pure Hubbard model.
For the extended Hubbard model with a local as well as nearest-neighbor 
interaction this effect is reduced.
The opposite limit examined in the present work corresponds to  
strong and dilute impurities. In this case
collective effects do not play any role and the problem essentially 
reduces to a single isolated impurity.
An interesting unsolved problem concerns the combination of single 
impurity and collective effects
at intermediate scales: depending on whether 
collective effects become important before the individual impurities 
renormalize to high barriers, a different characteristic behavior is 
expected.

In the following we consider the case of a single or double impurity,
where the effects of coherent scattering 
from many impurities are absent.
The asymptotic low-energy properties of Luttinger liquids with 
a single impurity have been investigated by 
mapping the problem onto an effective field theory,
where terms which are expected to be
irrelevant in the low-energy limit are neglected. 
For attractive interactions the impurity is
irrelevant in the renormalization-group sense and scales to zero
at low energies.  
For repulsive electron systems with $K_{\rho}<1$ the essential 
properties from the perturbative bosonic
renormalization-group calculation and the boundary
conformal field-theory analysis can be summarized as follows.
The backscattering amplitude generated by a weak impurity is a 
relevant perturbation which grows as $\Lam^{(K_{\rho}-1)/z}$,
for a decreasing energy scale $\Lambda$, where
$z$ is the number of spin components.
This behavior can be traced back
to the power-law singularity of the $2k_F$ density response
function in a Luttinger liquid. 
On the other hand, the tunneling amplitude through a weak link 
between two otherwise separate wires is irrelevant and scales to zero as 
$\Lambda^{\alpha_B}$, with the boundary exponent 
\begin{equation}
  \alpha_B = \frac{1}{z}(K_{\rho}^{-1}-1)
\end{equation}
depending only on the interaction strength and band
filling, but not on the impurity parameters.
At low energy scales any impurity thus effectively ``cuts'' 
the system into two parts with open boundary conditions at 
the end points, and physical observables are controlled by the 
open chain fixed point.

In particular,
the local density of states near the impurity is suppressed as 
\begin{equation}
  \rho (\omega) \sim |\omega|^{\alpha_B}
\end{equation}
for $|\omega| \to 0$.

Long-range Friedel oscillations in the density profile induced by 
boundaries or impurities decay 
with a power law at long distances \cite{EG95} as
\begin{equation}
  \label{eq:dens}
  n(x) \sim x^{-K_{\rho}} 
\end{equation}
for spinless fermions, 
where $x$ measures the distance from the impurity or boundary. For
spin-$\frac{1}{2}$ fermions $K_{\rho}$ is replaced by $(K_{\rho}+1)/2$ 
in Eq.~(\ref{eq:dens}).

The conductance through an infinite Luttinger liquid with a single impurity
vanishes at low temperatures as
\begin{equation}
  G(T) \sim T^{2\alpha_B} \; .
\end{equation}
The conductance through a single impurity of variable strength
can be collapsed onto a single curve by a one-parameter
scaling ansatz.  
For resonant scattering at double barriers 
the distance between the two barriers and the detuning from resonance
introduce additional scales and a more complex behavior is observed.
The Lorentzian resonance line shape for
noninteracting electrons is modified by the interaction, and
for appropriate parameters the 
conductance exhibits distinctive power-law scaling 
as a function of temperature \cite{KF92c,FN93,Fur98,NG03,PG03,YGM94}.

Note that the above power laws are strictly valid only in the
absence of two-particle backscattering. For spin-$\frac{1}{2}$ 
fermions they are in general modified by logarithmic corrections. 
The asymptotic behavior is universal in the sense that the exponents
depend only on the properties of the bulk system, via $K_{\rho}$,
while they do not depend on the impurity strength or shape, except
in special cases such as resonant scattering at double barriers, 
which require fine-tuning of parameters.

The asymptotic low-energy properties of Luttinger liquids with a
single impurity are rather well understood. Universal power laws
and scaling functions have been obtained by bosonization, conformal
field theory, and exact solutions for the low-energy asymptotics in 
special integrable cases \cite{Gia03}.
Numerical results from exact diagonalization and
DMRG applied to the lattice model of spinless fermions
with nearest-neighbor interaction confirmed 
the field theoretical scenario and the validity
of the underlying assumptions \cite{EA92,Med98}.
These methods are however limited to lattice systems with about 
$1000$ sites and do not allow for
a systematic analysis of the crossover between the weak
and strong-impurity limit.
Moreover, only a restricted set of observables can be evaluated with 
affordable computational effort.
In this context the fRG provides a complementary technique 
for microscopic models of interacting fermions with impurities,
which does not only capture correctly the universal low-energy
asymptotics, but allows one to compute observables on all energy
scales, providing thus also nonuniversal properties, and in
particular an answer to the important question at what scale the
ultimate asymptotics sets in. That scale can indeed be surprisingly
low, and the properties above it very different from the asymptotic
behavior.

%%%%%%%%%%%%%%%%%%%%%%%%%%%%%%%%%%%%%%%%%%%%%%%%%%%%%%%%%%%%%%%%%%%%%%%%

\section{Experimental realization}
\label{sec:imp:exp}

The progress in the fabrication of artificial low-dimensional structures
led to advanced experimental verification of the theoretical 
predictions. 
%Following the recent overview on the experimental verification 
%of Luttinger-liquid behavior in Ref.~\cite{Sch05}, 
We present a short list of the most
promising systems and of the employed experimental techniques.
For a detailed discussion and references to the most recent publications
and review articles on the subject we refer to Ref.~\cite{Sch05}.

Strictly one-dimensional systems are a theoretical idealization, the 
coupling to an experimental probe as well as the coupling between 
several Luttinger liquids is not completely understood \cite{Gia03}. 
The coupling between the chains in a strongly
anisotropic three-dimensional compound leads to the development of 
long-range order at very low temperatures in the phase for which the 
algebraic decay of the corresponding
correlation function of the single-chain Luttinger liquid is the slowest.
In appropriate temperature and energy regimes Luttinger-liquid behavior 
can be expected in 
several systems with a predominantly one-dimensional character, as 
highly anisotropic quasi one-dimensional conductors, organic conductors 
like the Bechgaard salts, as well as inorganic materials, artificial 
quantum wires in semiconductor
heterostructures or on surface substrates, carbon nanotubes, and 
fractional quantum Hall fluids \cite{Sch05}.
In particular, single-wall carbon nanotubes
are expected to show Luttinger-liquid behavior
with $K_{\rho} \sim 0.2 -0.3 $ down to very
low temperatures, despite the presence of two
low-energy channels \cite{EG97,KBF97}.

Experimental techniques used to verify Luttinger-liquid behavior involve 
mainly high resolution photoemission and transport measurements, in 
addition to optical properties \cite{Sch05}.
A careful analysis of experimental data 
indicating power-law behavior and signatures of spin-charge separation
reveals partly inconsistent interpretations.
The discussion on the modification of the 
quantized value $e^2/h$ for noninteracting electrons
in a single channel by the interaction to $K_{\rho} \,(e^2/h)$
indicates a sensitive dependence on the schematization of the contacts, 
a challenging theoretical as well as experimental problem \cite{Sch05}. 
Experimental results for cleaved-edge overgrowth quantum wires and
carbon nanotubes indicate power laws
of the conductance consistent with Luttinger-liquid behavior. 
In the last few years, ultracold gases in optical lattices 
have opened up an entirely new area of physics,
where strong correlations can be studied
with unprecedented flexibility and control of the parameters. 
Further work is necessary
for clear experimental evidence of Luttinger-liquid behavior.

%%% Local Variables: 
%%% mode: latex
%%% TeX-master: "thesis"
%%% End: 

\begin{fmffile}{diag} % 
\fmfcmd{%
  vardef slashbaro (expr p, len, ang) =
    ((-len/2,0)--(len/2,0))
    rotated (ang + angle direction length(p)/2 of p)
    shifted point length(p)*0.3 of p
  enddef;
  vardef slashbarm (expr p, len, ang) =
    ((-len/2,0)--(len/2,0))
    rotated (ang + angle direction length(p)/2 of p)
    shifted point length(p)*0.5 of p
  enddef;
  vardef slashbar (expr p, len, ang) =
    ((-len/2,0)--(len/2,0))
    rotated (ang + angle direction length(p)/2 of p)
    shifted point length(p)*0.7 of p
  enddef;
  vardef blob (expr z_arg, diameter) =
    save p,currentpen; path p; pen currentpen;
    pickup pencircle scaled thick;
    p = fullcircle scaled diameter shifted z_arg;
    cfill p;
  enddef;
  style_def slplaino expr p =
    cdraw p;
    ccutdraw slashbaro (p, 5mm, 45)
  enddef;
  style_def slplainm expr p =
    cdraw p;
    ccutdraw slashbarm (p, 5mm, 45)
  enddef;
  style_def slarrowm expr p =
    cdraw p;
    cfill (arrow p);
    ccutdraw slashbaro (p, 5mm, 45)
  enddef;
  style_def slplain expr p =
    cdraw p;
    ccutdraw slashbar (p, 5mm, 45)
  enddef;
  style_def sldasheso expr p =
    draw_dashes p;
    ccutdraw slashbaro (p, 5mm, 45)
  enddef;
  style_def sldashesm expr p =
    draw_dashes p;
    ccutdraw slashbarm (p, 5mm, 45)
  enddef;
  style_def dashesp expr p =
    draw_dashes p;
    blob (point length(p)*0.5 of p, 1.6mm)
  enddef;
}

% Thesis chapter on functional Renormalization Group

\chapter{Functional RG technique: a short overview}
\label{sec:rg}
\vspace{2cm}
\textit{We review the functional renormalization-group approach for
  interacting Fermi systems in the 1PI version.
  Introducing an infrared cutoff $\Lam$ in the free propagator and 
  differentiating the effective action with respect to $\Lam$,
  an exact hierarchy of differential flow equations for
  the 1PI vertex functions is derived, describing
  the gradual evolution from the microscopic model Hamiltonian 
  to the effective action as a function 
  of the continuously decreasing energy cutoff.
  We briefly discuss the relation to alternative formulations of the fRG 
  approach.}
\vspace{.75cm}

%%%%%%%%%%%%%%%%%%%%%%%%%%%%%%%%%%%%%%%%%%%%%%%%%%%%%%%%%%%%%%%%%%%%%%%%

\section{Introduction}
\label{sec:rg:intro}

The renormalization-group is a powerful method
in the study of low-dimensional Fermi systems, providing in particular 
a systematic and unbiased method to study 
competing instabilities and entangled infrared 
singularities at weak coupling.
Early renormalization-group approaches for one-dimensional systems, 
combined with exact solutions of fixed-point
models, have been a major source of physical insight \cite{Sol79,Gia03}.
From the renormalization-group point of view, the existence of the 
Luttinger liquid requires the cancellation
of contributions to the flow of the two-particle vertex
to all orders \cite{Metzner:9701012}.
This is a one-dimensional phenomenon, in
higher dimensions the interactions in general diverge and 
the flow in the fermionic variables breaks down indicating a possible
opening of a gap in the fermionic excitation spectrum.

Wilson's renormalization-group approach \cite{Wil71,WK74} of successive 
integration of degrees of freedom with different energy scales 
determines the evolution of the bare action of the system, given by the
microscopic Hamiltonian, to the final effective action, 
from which all physical quantities can be extracted.
The hierarchy of coupled differential flow equations 
for the Green or vertex functions describing the 
full functional evolution of the effective 
action has been first implemented for bosonic field theories in the context 
of critical phenomena \cite{WH73,Pol84,Wet93}. 
The intuition of the relevance of fRG methods 
for interacting Fermi systems followed in the 1990s
\cite{BG90,FT90,Sh91,Sh94}, together with important rigorous work
\cite{Salmhofer:1999}.
The infinite hierarchy of flow equations
can be solved exactly only in special cases, for  
instance the Luttinger model \cite{Schuetz04}. 
Truncations however preserve the 
successive handling of energy scales and the consequent treatment of 
infrared singularities,
characteristic of a renormalization-group treatment.

There are several variants of the fRG flow equations. The 
flow equations for the connected amputated Green functions 
correspond to the Polchinski scheme, first derived in 
Ref.~\cite{Pol84,KKS92}.
The expansion of the connected amputated Green functions in 
1PI vertex functions led to the respective flow equations 
\cite{WH73,Wei76}, 
subsequently derived from the Legendre transform of the generating 
functional \cite{Wet93,Mor94,SH01}. 
The Wick-ordered scheme is obtained from the Polchinski scheme by 
expanding the generating functional of the connected amputated Green 
functions in Wick-ordered polynomials 
\cite{Wie88,Salmhofer:9706188,Salmhofer:1999}.
Important applications of the fRG in condensed-matter physics include the
two-dimensional Hubbard model using the Polchinski scheme
\cite{Zanchi:9703189,Zanchi:9812303}, the Wick-ordered scheme
\cite{HM00} and also the 1PI scheme \cite{HSFR01}. 
In the context of classical disordered systems a fRG approach 
is necessary to overcome the problem of dimensional reduction 
\cite{Wie03}.
One-dimensional impurity problems and Luttinger-liquid physics are 
most conveniently investigated in the 1PI scheme, as self-energy
contributions are included to all orders.

In the following the hierarchy of differential flow equations
for the 1PI vertex functions is derived,
which is obtained by differentiating
the corresponding generating functional with respect to an infrared 
cutoff introduced in the free propagator 
\cite{Salmhofer:9706188}.

%%%%%%%%%%%%%%%%%%%%%%%%%%%%%%%%%%%%%%%%%%%%%%%%%%%%%%%%%%%%%%%%%%%%%%%%

\section{Generating functional}
\label{sec:rg:form}

We consider a system of interacting fermions
with single-particle propagator of the noninteracting system 
$G_0$.
The properties of the system are determined by the action
\begin{equation}
  \label{eq:act}
  S[\psi,\psib] = (\psib, G_0^{-1} \psi) - V[\psi,\psib] \; ,
\end{equation}
where $\psib$ and $\psi$ are Grassmann variables associated with 
creation and annihilation operators,
and $V[\psi,\psib]$ is an arbitrary many-body interaction.
Here we introduced the short-hand notation
$(\psib,G_0^{-1} \psi) = \sum_{K,K'} \psib_K [G_0^{-1}]_{K,K'}\psi_{K'}$, where 
$K$ contains the Matsubara
frequency in addition to the single-particle quantum numbers and
$\sum_K$ stands for summation over the discrete indices and integrals 
over the continuous ones.

All connected Green functions are obtained from 
the generating functional \cite{NO87}
defined by
\begin{align}
  \label{eq:gen}
  e^{-\cG[\eta,\etab]} & = \frac{1}{\cZ_0}
  \int\! d\psi d\psib \, 
  e^{S[\psi,\psib]} \, e^{-(\psib,\eta) -(\etab,\psi)}
  \nonumber \\[2mm]
  & =\int\! d\mu_Q[\psi,\psib] \, 
  e^{-V[\psi,\psib]} \, e^{-(\psib,\eta) -(\etab,\psi)} \; ,
\end{align}
with Grassmann source terms $\eta$ and $\etab$.
The normalized Gaussian measure with covariance $Q=G_0^{-1}$ 
\begin{equation}
  \label{eq:norm}
  d\mu_Q[\psi,\psib] = \frac{1}{\cZ_0}\,
  d\psi d\psib \, e^{(\psib,Q\psi)} 
\end{equation}
includes the exponential of the 
quadratic part of the action
and the noninteracting partition function $\cZ_0$, such that 
$\int \! d\mu_Q[\psi,\psib]=1$.
The generating functional for the connected Green functions 
is related to the partition function of the physical system with 
action (\ref{eq:act}) by 
\begin{equation}
  \cG[\eta,\etab]= - {\rm ln} \, \cZ[\eta,\etab]  \; .
\end{equation}
In the noninteracting case $V[\psi,\psib] = 0$, and the Gaussian 
integral 
\begin{equation}
  \int\! d\mu_Q[\psi,\psib] \, e^{-(\psib,\eta) -(\etab,\psi)} =
  e^{-(\etab,G_0\eta)}
\end{equation}
implies that $\cG[\eta,\etab] = (\etab,G_0\eta)$.

The connected $m$-particle Green functions are given by
the derivatives of the generating functional $\cG[\eta,\etab]$
with respect to the source terms at $\eta=\etab=0$
\begin{align}
  G_m(K'_1,\dots,K'_m;K_1,\dots,K_m) &= 
  (-1)^m \bra \psi_{K'_1} \dots \psi_{K'_m} 
  \psib_{K_m} \dots \psib_{K_1} \ket_c 
  \nonumber \\[2mm]
  & =\, \left.
    \frac{\partial^m}{\partial\eta_{K'_1} \dots \partial\eta_{K'_m}}\,
    \frac{\partial^m}{\partial\etab_{K_m} \dots \partial\etab_{K_1}}\,
    \cG[\eta,\etab] \right|_{\eta = \etab = 0} \; ,
\end{align}
where $\bra \dots \ket_c$ is the connected average of the 
product of Grassmann variables between the brackets.

The connected amputated Green functions are generated by the 
{\em effective interaction} $\cV[\chi,\chib]$ defined by
\begin{equation}
  \label{eq:effint}
  e^{-\cV[\chi,\chib]} = 
  \int\! d\mu_Q[\psi,\psib] \, 
  e^{-V[\psi+\chi,\psib+\chib]}  \; .
\end{equation}
The substitution $\chi = G_0 \eta$ and $\chib = G_0^T \etab$,
where $G_0^T$ is the transposed propagator, 
relates $\cV[\chi,\chib]$ to the functional $\cG[\eta,\etab]$ by
\begin{equation}
  \cV[\chi,\chib] = \cG[\eta,\etab]-(\etab,G_0 \eta)  \; .
\end{equation}
The functional derivatives of $\cV[\chi,\chib]$ generate 
connected Green functions divided by 
$G_0 (K_1) \dots G_0 (K_m) \, G_0 (K'_1) \dots G_0 (K'_m)$,
that is, propagators amputated from external 
legs in the corresponding Feynman diagrams. 
The term $(\etab, G_0 \eta)$ cancels the noninteracting part of
$\cG[\eta,\etab]$ such that $\cV[\chi,\chib] = 0$ for 
$V[\psi,\psib] = 0$.
Hence, the noninteracting propagator is subtracted from the
one-particle Green function generated by $\cV[\chi,\chib]$.

The generating functional $\Gam[\phi,\phib]$ for the {\em 1PI} vertex 
functions $\gamma_m$ is derived from the Legendre transform of 
$\cG[\eta,\etab]$ by
\begin{equation}
  \label{eq:onepii}
  \Gam[\phi,\phib]  +(\phib,Q\phi)= 
  \cG[\eta,\etab] + (\phib,\eta) -(\etab,\phi) \;, 
\end{equation}
with 
\begin{align}
  \label{eq:help1}
  \phi &= \frac{\partial \cG }{\partial \etab} & \eta &= \frac{\partial \Gam}{\partial \phib} +Q \phi& \nonumber\\[2mm]
  \phib & = \frac{\partial \cG }{\partial \eta} &\etab &= \frac{\partial \Gam}{\partial \phi} -Q^T \phib &
\end{align}
and
\begin{equation}
  \label{eq:help2}
  \frac{\delta^2
    \cG}{\delta\eta\,\delta\bar\eta} = \Big(\frac{\delta^2
    \Gam}{\delta\phi\,\delta\bar\phi} +Q \Big)^{-1}\; . 
\end{equation}
For the special case without interaction
$\cG[\eta,\etab]=(\etab,G_0\eta)$ leads to $\Gam[\phi,\phib]=0$.

The choice of the appropriate generating functional
for a convenient formulation of a renormalization-group approach 
depends on the physical problem under investigation.
For a detailed description of the different schemes we refer to 
Ref.~\cite{EnssThesis};
here we will concentrate on the 1PI version of the fRG.

%%%%%%%%%%%%%%%%%%%%%%%%%%%%%%%%%%%%%%%%%%%%%%%%%%%%%%%%%%%%%%%%%%%%%%%%

\section[RG differential flow equation for $\Gam$]{RG differential flow 
  equation for $\mathbf{\Gam}$}
\label{sec:rg:flow}

In this section we briefly review the general renormalization-group setup, 
introduced as a transformation that leaves the generating functional for
the correlation functions invariant,
and concentrate subsequently on the derivation of a continuous 
renormalization-group equation for the 1PI functions, following the 
derivation in the context of interacting Fermi systems in Ref.~\cite{SH01}. 

The addition principle for Gaussian fields implies that for the
decomposition $G_0=G^<_0+G^>_0$ the corresponding Gaussian measure factorizes as
\begin{align}
  e^{-\cW[\Phi,\Phib]}&=\int\! d\mu_{Q} [\Psi,\Psib] e^{-\cY[\Psi+\Phi,\Psib+\Phib]}
  \nonumber \\[2mm]
  &=\int\! d\mu_{Q^<}[\Psi_<,\Psib_<] 
  \int\! d\mu_{Q^>} [\Psi_>,\Psib_>] e^{-\cY[\Psi_<+\Psi_>+\Phi,\Psib_<+\Psib_>+\Phib]} \; ,
\end{align}
with $\Psi = \Psi_<+\Psi_>$. The generating functional $\cW$
corresponds to $\cW=\cV$ for the particular choice $\cY=V$.
This leads to the semigroup law of the renormalization group
\begin{equation}
  \label{eq:law}
  e^{-\cW[\Phi,\Phib]}=\int\! d\mu_{Q^<}[\Psi',\Psib']  e^{-\cW_>[\Psi'+\Phi,\Psib'+\Phib]} \; ,
\end{equation}
where in $\cW_>=\cW(Q^>,\cY)$ the fields with propagator $Q^>$ have been 
integrated out.
The semigroup law implies that the system $(Q,\cY)$ under analysis is 
exactly equivalent to the system $(Q^<,\cW(Q^>,\cY))$.
In the present case $Q^>$ is a covariance with infrared 
cutoff $\Lam$, and $Q^<$ has support only for fields with energies smaller 
than $\Lam$.

Set up in this way, the renormalization group is simply a symmetry of 
the generating functional $\cW(Q,\cY)$. 
In differential form $\cW(Q,\cY)$ is independent of $\Lam$, that is,
\begin{equation}
  \frac{\partial}{\partial \Lam} \cW(Q,\cY) = 0 \; .
\end{equation}
Inserting the right-hand side of Eq.~(\ref{eq:law}) leads to the 
flow equation describing
the gradual evolution from $\cY$ to the effective functional $\cW$ 
as a function of the continuously decreasing energy cutoff $\Lam$.
$\cW_>$ is an infinite power series in the fields;
the quadratic and quartic terms correspond to the self-energy and 
the effective interaction, higher order terms are however always 
present and the convergence of the infinite series is a nontrivial 
problem \cite{SH01}.

In the following the differential equation for the generating 
functional $\Gam^{\Lam}[\phi,\phib]$ of the 1PI functions,
starting point for the hierarchy of differential flow equations, 
is derived.
Introducing an infrared cutoff at an energy scale $\Lam > 0$ in the 
bare propagator leads to a $\Lam$-dependent generating functional for 
the connected Green functions defined by 
\begin{equation}
  \label{eq:glambda1}
  e^{-\G^\Lambda[\eta,\bar\eta]}
  = \int \dmu{Q^\Lambda}{\psi}\, e^{-V[\psi,\bar\psi]}\,
  e^{-\inner{\bar\psi}{\eta}-\inner{\bar\eta}{\psi}} \; .
\end{equation}
The original functional is recovered in the limit $\Lam \to 0$.
Similarly the functional $\Gam^{\Lam}[\phi,\phib]$ generating the 1PI 
vertex functions is constructed with $G_0^{\Lam}$ replacing $G_0$
in Eqs.~(\ref{eq:onepii} - \ref{eq:help2}). 
Differentiating the above Eq.~(\ref{eq:glambda1}) for 
$\cG^{\Lam}[\eta,\bar\eta]$ with respect to $\Lam$
yields 
\begin{align}
  -\frac{\partial \cG^\Lambda[\eta,\bar\eta]}{\partial \Lam}\, e^{-\G^\Lambda[\eta,\bar\eta]}
  & = - \Tr\,(G_0^\Lambda {\dot Q^\Lambda}
  ) \,e^{-\G^\Lambda[\eta,\bar\eta]} \nonumber \\[2mm]
  & \quad\, + \int \dmu{Q^\Lambda}{\psi} \,
  (\bar\psi,{\dot Q^\Lambda}\psi) \, e^{-V_0[\psi,\bar\psi]} \,
  e^{-(\bar\psi,\eta)-\inner{\bar\eta}{\psi}} \nonumber\\[2mm]
  & = -\big[ \Tr \,(G_0^\Lambda {\dot Q^\Lambda} 
  ) +\Laplace{\dot
    Q^\Lambda} \big] e^{-\G^\Lambda[\eta,\bar\eta]} \;,
\end{align}
where the first term comes from the derivative of the normalization
factor (\ref{eq:norm}), and $\Tr$ denotes the sum over all space-time
indices. The functional Laplace operator $\Laplace{Q}$ is defined as
\begin{align}
  \Laplace{Q}
  = \inner{\ddd\eta}{Q\ddd{\bar\eta}}
  = \sum_K \ddd{\eta_K}\, Q_K\, \ddd{\bar\eta_K} \;.
\end{align}
The flow of $\G^\Lambda[\eta,\bar\eta]$ is then
\begin{align}
  \label{eq:gflow}
  \frac{\partial \G^\Lambda[\eta,\bar\eta]}{\partial \Lam}
  = \Tr \,(G_0^\Lambda {\dot Q^\Lambda} )
  -\Tr\left[\dot Q^\Lambda\,
    \frac{\delta^2 \G^\Lambda[\eta,\bar\eta]}
    {\delta\eta\, \delta\bar\eta}\right]
  +\inner{\frac{\delta\G^\Lambda[\eta,\bar\eta]}{\delta\eta}} {\dot
    Q^\Lambda \frac{\delta\G^\Lambda[\eta,\bar\eta]}{\delta\bar\eta}}\;.
\end{align}
Using the Legendre transform (\ref{eq:onepii}) the derivative of 
$\Gam^{\Lam}[\phi,\phib]$ reads
\begin{align}
  \frac{\partial \Gamma^\Lambda[\phi,\bar\phi]}{\partial \Lam}
  & = \frac{\partial \G^\Lambda[\eta,\bar\eta]}{\partial \Lam}
  - (\bar\phi,\dot Q^\Lambda\phi) \nonumber\\[2mm]
  & = \Bigl(\frac{\delta\G^\Lambda[\eta,\bar\eta]}{\delta\eta},
  \dot Q^\Lambda \frac{\delta\G^\Lambda[\eta,\bar\eta]}{\delta\bar\eta}\Bigr)
  \nonumber\\[2mm]
&\quad\,-\Tr \,\Bigl[ {\dot Q^\Lambda}
  \Bigl(\frac{\delta^2 \G^\Lambda[\eta,\bar\eta]} {\delta\eta\,\delta\bar\eta}
  -G_0^\Lambda\Bigr)\Bigr] - (\bar\phi,{\dot Q^\Lambda}\phi)  \nonumber\\[2mm]
  & = 
  -\Tr \Bigl[ \dot Q^\Lambda
  \Bigl( \Bigl(\frac{\delta^2
    \Gamma^\Lambda[\phi,\bar\phi]}{\delta\phi\,\delta\bar\phi}
  + Q^\Lambda \Bigr)^{-1} - G_0^\Lambda \Bigr)\Bigr] \;,
\end{align}
leading to the exact renormalization-group equation
\begin{align}
  \label{rge}
  \frac{\partial}{\partial \Lam} \Gamma^\Lambda[\phi,\bar\phi] =
  \Tr \Bigl[G_0^\Lambda \frac{\partial  (G_0^\Lambda)^{-1}}{\partial \Lam} 
  \Bigr]-
  \Tr \Bigl[\Bigl(\frac{\delta^2
    \Gamma^\Lambda[\phi,\bar\phi]}{\delta\phi\,\delta\bar\phi}
  + (G_0^\Lambda)^{-1}\Bigr)^{-1}  \frac{\partial  (G_0^\Lambda)^{-1}}{\partial \Lam}  \Bigr] \; .
\end{align}
With the initial condition 
\begin{equation}
  \Gam^{\Lam_0}[\phi,\phib] = V[\phi,\phib]
\end{equation}
Eq.~(\ref{rge}) determines the flow of $\Gam^{\Lam}$ uniquely for 
all $\Lam < \Lam_0$.

%%%%%%%%%%%%%%%%%%%%%%%%%%%%%%%%%%%%%%%%%%%%%%%%%%%%%%%%%%%%%%%%%%%%%%%%%%%%%%

\section{Expansion in the fields and exact hierarchy of flow equations}
\label{sec:rg:exp}

The renormalization-group equations for the 1PI $m$-particle vertex functions 
$\gamma_m^{\Lam}$ are derived by expanding $\Gam^{\Lam}[\phi,\phib]$ in 
Eq.~(\ref{rge}) as a power series in the fields.
The coefficients in the expansion of $\Gam^{\Lam}[\phi,\phib]$ determine
$\gamma_m^{\Lam}$ by
\begin{align}
  \label{eq:gam}
  \Gamma^{\Lam}[\phi,\bar\phi] = \sum_{m=0}^\infty \frac{1}{(m!)^2}
  \sum_{K_1\dots K_m} \sum_{K_1'\dots K_m'} 
  \gamma_m^{\Lam}(K_1',\dots,K_m';K_1,\dots,K_m) 
  \prod_{j=1}^m \bar\phi_{K_j'} \phi_{K_j} \; .
\end{align}
Due to the antisymmetry properties of the Grassmann variables
only antisymmetric vertex functions contribute.

Similarly the second derivative on the right-hand side of Eq.~(\ref{rge})
can be expanded.
Separating the $\phi$-independent part corresponding to the self-energy 
yields
\begin{align}
  \frac{\delta^2
    \Gamma^\Lambda[\phi,\bar\phi]}{\delta\phi\,\delta\bar\phi}= 
  \frac{\delta^2
    \Gamma^\Lambda[\phi,\bar\phi]}{\delta\phi\,\delta\bar\phi} \at_{\phi=\bar\phi=0} 
  +\tilde{\Gamma}^\Lambda[\phi,\bar\phi] = -\Sg ^\Lambda
  +\tilde{\Gamma}^\Lambda[\phi,\bar\phi] \; ,
\end{align}
where the remaining functional $\tilde{\Gamma}^\Lambda[\phi,\bar\phi]$ 
is defined by Eq.~(\ref{eq:gam}) with indices $m$ and $j$ starting from 
$2$.
The second term on the right-hand side of the flow equation 
(\ref{rge}) then reads
\begin{align}
  \Bigl(\frac{\delta^2 \Gamma^\Lambda[\phi,\bar\phi]}
  {\delta\phi\,\delta\bar\phi} + (G_0^\Lambda)^{-1} \Bigr)^{-1}
  &= \bigl(  \tilde\Gamma^\Lambda + (G^\Lambda)^{-1} \bigr)^{-1} 
  \nonumber\\[2mm]
  &= G^\Lambda\, \sum_{l=0}^{\infty} (-1)^l \big(\tilde\Gamma^\Lambda \,G^\Lambda\big)^l\; ,
\end{align}
with the full propagator $G^\Lambda$ defined via the Dyson equation
$(G^\Lambda)^{-1} = (G_0^\Lambda)^{-1} - \Sigma^\Lambda$.
Introducing the {\em single-scale propagator} $S^\Lambda$ as
\begin{equation}
  \label{eq:ssp}
  S^\Lambda = G^\Lambda\,\frac{\partial (G_0^\Lambda)^{-1}}{\partial \Lam}   \, G^\Lambda 
\end{equation}
the differential equation (\ref{rge}) for $\Gam^\Lambda$ is
\begin{align}
  \label{eq:gammaflow3}
  \frac{\partial}{\partial \Lam} \Gamma^\Lambda &= 
  \Tr \Bigl[G_0^\Lambda \frac{\partial  (G_0^\Lambda)^{-1}}{\partial \Lam} 
  \Bigr]+
  \sum_{l=0}^{\infty} (-1)^{l+1} \Tr \Big[G^\Lambda \frac{\partial}{\partial \Lam}  [G_0^\Lambda]^{-1} \big(G^\Lambda\, \tilde\Gamma^\Lambda \big)^l \Big]
  \nonumber \\[2mm]
  &= 
  \Tr \Bigl[\bigl(G_0^\Lambda-G^{\Lam} \bigl)\frac{\partial  (G_0^\Lambda)^{-1}}{\partial \Lam} 
  \Bigr]+
  \sum_{l=0}^{\infty} (-1)^l \Tr \Big[S^\Lambda \tilde\Gamma^\Lambda\,\big(G^\Lambda\, \tilde\Gamma^\Lambda \big)^l \Big] \; .
\end{align}
The first term corresponds to a vacuum energy not
entering the correlation functions, while
the second one contains one-loop diagrams with $(l+1)$
vertices $\tilde\Gamma^\Lambda$ connected by one single-scale 
propagator $S^\Lambda$ and $l$ full propagators $G^\Lambda$.
The term linear in $\tilde\Gamma^\Lambda$ generates self-energy 
corrections.

Inserting the components $\gamma_m^\Lambda$ on the left-hand side and 
the components $\tilde\gamma_m^\Lambda$ on 
the right-hand side of the flow equation (\ref{eq:gammaflow3}) for
$\Gamma^\Lambda$ we obtain a system of
equations for $\gamma_m^\Lambda$. 
In a graphical representation the equations for $m \leq 3$ are
\begin{align}
  \label{eq:hiera}
  \parbox{23mm}{\unitlength=1mm\fmfframe(3,8)(2,8){
      \begin{fmfgraph*}(17,17)
        \fmfv{d.sh=circle,d.f=full,d.size=1mm}{i1}
        \fmfforce{(9mm,17mm)}{i1}
        \fmfv{d.sh=circle,d.f=30,d.si=10mm}{v}
        \fmfleft{i}
        \fmfright{o}
        \fmf{plain}{i,v,o}
      \end{fmfgraph*}}}
  &\! =\!
  \parbox{18mm}{\unitlength=1mm\fmfframe(2,10)(2,10){
      \begin{fmfgraph*}(17,17)
        \fmfsurroundn{e}{8}
        \fmfv{d.sh=circle,d.f=30,d.si=10mm}{v}
        \fmfleft{i}\fmfright{o}\fmf{phantom}{i,v,o}\fmffreeze
        \fmf{plain}{v,e6}
        \fmf{plain}{v,e8}
        \fmf{slplaino,right=100,tension=0.4,label=$S^\Lambda$}{v,v}
      \end{fmfgraph*}}}
\end{align}
\begin{align}
  \label{eq:hierb}
  \parbox{23mm}{\unitlength=1mm\fmfframe(3,8)(2,8){
      \begin{fmfgraph*}(17,17)
        \fmfv{d.sh=circle,d.f=full,d.size=1mm}{i1}
        \fmfforce{(9mm,17mm)}{i1}
        \fmfsurroundn{e}{8}
        \fmfv{d.sh=circle,d.f=30,d.si=10mm}{v}
        \fmfleft{i}\fmfright{o}\fmf{phantom}{i,v,o}\fmffreeze
        \fmf{plain}{v,e2}
        \fmf{plain}{v,e4}
        \fmf{plain}{v,e6}
        \fmf{plain}{v,e8}
      \end{fmfgraph*}}}
  & \!=\!
  \parbox{23mm}{\unitlength=1mm\fmfframe(3,10)(2,10){
      \begin{fmfgraph*}(17,17)
        \fmfsurroundn{e}{6}
        \fmfv{d.sh=circle,d.f=30,d.si=10mm}{v}
        \fmfleft{i}\fmfright{o}\fmf{phantom}{i,v,o}\fmffreeze
        \fmf{plain}{v,e1}
        \fmf{plain}{v,e4}
        \fmf{plain}{v,e5}
        \fmf{plain}{v,e6}
        \fmf{slplaino,right=100,tension=0.4,label=$S^\Lambda$}{v,v}
      \end{fmfgraph*}}}
  +
  \parbox{47mm}{\unitlength=1mm\fmfframe(-4,6)(1,6){
      \begin{fmfgraph*}(47,17)
        \fmfleftn{l}{3}
        \fmfrightn{r}{3}
        \fmfv{d.sh=circle,d.f=30,d.si=10mm}{vl}
        \fmfv{d.sh=circle,d.f=30,d.si=10mm}{vr}
        \fmffixed{(.55w,0)}{vl,vr}
        \fmfleft{i}\fmfright{o}\fmf{phantom}{i,vl,vr,o}\fmffreeze
        \fmf{plain}{vl,l1}
        \fmf{plain}{vl,l3}
        \fmf{plain}{vr,r1}
        \fmf{plain}{vr,r3}
        \fmf{slplain,left=0.5,label=$S^\Lambda$}{vl,vr}
        \fmf{plain,right=0.5,label.side=left,label=$G^\Lambda$}{vl,vr}
      \end{fmfgraph*}}}
\end{align}
\begin{align}
  \label{eq:hierc}
  \parbox{23mm}{\unitlength=1mm\fmfframe(3,8)(2,8){
      \begin{fmfgraph*}(17,17)
        \fmfv{d.sh=circle,d.f=full,d.size=1mm}{i1}
        \fmfforce{(9mm,17mm)}{i1}
        \fmfsurroundn{e}{12}
        \fmfv{d.sh=circle,d.f=30,d.si=10mm}{v}
        \fmfleft{i}\fmfright{o}\fmf{phantom}{i,v,o}\fmffreeze
        \fmf{plain}{v,e1}
        \fmf{plain}{v,e3}
        \fmf{plain}{v,e5}
        \fmf{plain}{v,e7}
        \fmf{plain}{v,e9}
        \fmf{plain}{v,e11}
      \end{fmfgraph*}}}
  &\! =\!
  \parbox{23mm}{\unitlength=1mm\fmfframe(3,10)(2,10){
      \begin{fmfgraph*}(17,17)
        \fmfsurroundn{e}{10}
        \fmfv{d.sh=circle,d.f=30,d.si=10mm}{v}
        \fmfleft{i}\fmfright{o}\fmf{phantom}{i,v,o}\fmffreeze
        \fmf{plain}{v,e1}
        \fmf{plain}{v,e6}
        \fmf{plain}{v,e7}
        \fmf{plain}{v,e8}
        \fmf{plain}{v,e9}
        \fmf{plain}{v,e10}
        \fmf{slplaino,right=100,tension=0.4,label=$S^\Lambda$}{v,v}
      \end{fmfgraph*}}}
  +
  \parbox{47mm}{\unitlength=1mm\fmfframe(-4,6)(1,6){
      \begin{fmfgraph*}(47,17)
        \fmfleftn{l}{3}
        \fmfrightn{r}{4}
        \fmfv{d.sh=circle,d.f=30,d.si=10mm}{vl}
        \fmfv{d.sh=circle,d.f=30,d.si=10mm}{vr}
        \fmffixed{(.55w,0)}{vl,vr}
        \fmfleft{i}\fmfright{o}\fmf{phantom}{i,vl,vr,o}\fmffreeze
        \fmf{plain}{vl,l1}
        \fmf{plain}{vl,l3}
        \fmf{plain}{vr,r1}
        \fmf{plain}{vr,r2}
        \fmf{plain}{vr,r3}
        \fmf{plain}{vr,r4}
        \fmf{slplain,left=0.5,label=$S^\Lambda$}{vl,vr}
        \fmf{plain,right=0.5,label.side=left,label=$G^\Lambda$}{vl,vr}
      \end{fmfgraph*}}}
\!\!\!\!\!  + \;\;\,
  \parbox{30mm}{\unitlength=1mm\fmfframe(-3,0)(-2,0){
      \begin{fmfgraph*}(33,33)
        \fmfsurroundn{e}{24}
        \fmfv{d.sh=circle,d.f=30,d.si=10mm}{vl,vr,vu}
        \fmffixed{(.5w,0)}{vl,vr}
        \fmffixed{(.25w,.43h)}{vl,vu}
        \fmf{plain}{vu,e6}
        \fmf{plain}{vu,e8}
        \fmf{plain}{vl,e14}
        \fmf{plain}{vl,e16}
        \fmf{plain}{vr,e22}
        \fmf{plain}{vr,e24}
        \fmf{slplain,left=0.5,label=$S^\Lambda$}{vl,vu}
        \fmf{plain,left=0.5,label=$G^\Lambda$}{vu,vr}
        \fmf{plain,left=0.5,label=$G^\Lambda$}{vr,vl}
      \end{fmfgraph*}}} 
\end{align}
The initial conditions for the 
vertex functions at $\Lam = \infty$ are given by the bare
interactions of the system. In particular, the flow of the two-particle 
vertex starts from the antisymmetrized bare two-particle interaction 
while $m$-particle vertices of higher order vanish at $\Lam = \infty$,
in the absence of bare $m$-body interactions with $m>2$.

Note that the right-hand side of the equation for $\gamma_m$ 
contains $\gamma_{m+1}$.
The infinite system of differential equations contains
only one-loop terms in every equation, as the differential formulation
of Eq.~(\ref{rge}) contains only a single trace,
and for the 1PI scheme no tree terms appear. 
The infinite hierarchy produces the full Green functions,
generating graphs with an arbitrary
number of loops; truncations amount to a partial inclusion of 
higher order contibutions generated during the flow,
where the internal lines contain only modes above the cutoff scale $\Lam$.
Consequences of symmetries are discussed in Ref.~\cite{SH01}.

%%%%%%%%%%%%%%%%%%%%%%%%%%%%%%%%%%%%%%%%%%%%%%%%%%%%%%%%%%%%%%%%%%%%%%%%%%%%%%

\section{Comparison to other RG schemes}
\label{sec:rg:alt}

Infrared divergencies arising in the context of perturbative expansions or
in proximity of phase transitions can alternatively be regularized by
temperature, a weak coupling strength or a finite system size. 
In the fRG approach the cutoff scale is introduced only in the quadratic 
part of the bare action, and the regularization is implemented with respect 
to energy scales.
The temperature and interaction flows are derived in Refs.~\cite{HS01,HRAE04} 
respectively, a pedagogic introduction is given in Ref.~\cite{EnssThesis}.
The renormalization-group equations describe the 
flow of the correlation functions as the cutoff scale is lowered.
The choice of the basis set for the correlation functions determines a 
particular scheme.

In addition to the 1PI scheme described previously,
the various generating functionals introduced in Sec.~\ref{sec:rg:form}
correspond to different schemes.
Starting point for the {\em Polchinski scheme} is the effective interaction,
generating functional of the connected amputated Green functions. 
The flow equation for $\cV^{\Lam}[\chi,\chib]$ is 
derived by replacing $Q$ by $Q^{\Lam}$ in Eq.~(\ref{eq:effint}) and 
taking the derivative with respect to $\Lam$.
An expansion in powers of $\chi$
and $\chib$ of the functional $\cV^{\Lam}[\chi,\chib]$ in the 
renormalization-group equation leads to Polchinski's flow 
equations for amputated connected Green functions \cite{Pol84,KKS92}, with 
a similar structure as for the connected Green functions. 
The connected amputated Green functions are the expansion coefficients
of the generating functional $\cV^{\Lam}[\chi,\chib]$ in terms of monomials
of the source fields $\chi$ and $\bar\chi$.
Alternatively, one can also expand $\cV^{\Lam}[\chi,\chib]$ with respect to 
{\em Wick-ordered} polynomials, leading to the Wick-ordered
Green functions as expansion coefficients
\cite{Wie88,Salmhofer:9706188,Salmhofer:1999}.
The flow equations are characterized by a bilinear structure in
the vertices on the right-hand side connected by bare $\Lambda$-dependent 
propagators.
The Wick ordering also implies that except for the differentiated propagator 
the internal lines are supported below scale $\Lambda$ instead of above it. 
Thus, for a momentum cutoff only momenta close to the Fermi surface 
contribute at low cutoff scale $\Lam$.
This justifies a parametrization of the coupling
functions by projecting onto the Fermi surface \cite{HM00}.
Self-energy corrections are however most conveniently taken into 
account in the 1PI formalism with full propagators on the internal lines.

In an exact treatment all schemes are equivalent,
differences arise with truncations of the infinite hierarchy of flow 
equations.
While the full hierarchy of flow equations leads to the correct solution 
to all orders in perturbation theory independently of the scheme, in the 
computation of the lowest orders a particular scheme might be more 
suitable than others, depending on the considered physical problem and
properties. 
An important point for the choice concerns the possibility of an efficient 
parametrization of the effective interactions by a manageable number of variables.

Continuous symmetries in the bare 
action lead to conservation laws and Ward identities
relating Green and response functions, as a consequence of the Noether 
theorem.
These are generally not preserved for the truncated flow equations, 
in contrast to the solution of the infinite flow-equation hierarchy, 
as shown in detail in Ref.~\cite{EnssThesis}.  
For a gauge-invariant construction however, as for 
the temperature-flow scheme, the Ward identities between Green and 
response functions are satisfied exactly despite truncations.
The related property of self-consistency is satisfied by
construction in conserving approximations
\cite{BK61}, but generally violated
in truncated fRG flows.
However, in the one-dimensional lattice models for Luttinger
liquids, the truncated fRG is nevertheless surprisingly successful and
self-consistency does not appear to play an important role.

%%% Local Variables: 
%%% mode: latex
%%% TeX-master: "thesis"
%%% End: 

% Thesis chapter on fRG for Luttinger liquids

\chapter{Functional RG for Luttinger liquids}
\label{sec:frglutt}
\vspace{2cm}
\textit{We apply the fRG in the one-particle irreducible version to
  one-dimensional Fermi systems with impurities.
  The lowest order truncation of the fRG hierarchy of flow equations, 
  where the two-particle vertex is approximated by the bare interaction, 
  considered previously for spinless fermions, is extended 
  including two-particle vertex renormalization, and generalized to 
  spin-$\frac{1}{2}$ systems.
  For spinless fermions the quantitative accuracy of the results improves 
  considerably, whereas for spin-$\frac{1}{2}$ systems vertex 
  renormalization is necessary to take into 
  account that backscattering of particles with opposite spins at 
  opposite Fermi points scales to zero in the low-energy limit.
  The underlying approximations are devised for weak 
  interactions and arbitrary impurity strengths.
  Details on the computation of the relevant observables from the 
  solution of the flow equations are presented.}
\vspace{.75cm}

%%%%%%%%%%%%%%%%%%%%%%%%%%%%%%%%%%%%%%%%%%%%%%%%%%%%%%%%%%%%%%%%%%%%%%%%%%%%%%

\section{Microscopic models}
\label{sec:frglutt:model}

We consider various lattice fermion systems with spinless and 
spin-$\frac{1}{2}$ fermions 
supplemented by different types of impurity potentials.
The Hamiltonian has the form
\begin{equation}
  H = H_0 + H_I + H_{\rm imp}
\end{equation}
where $H_0$ is the kinetic energy, $H_I$
a short-range interaction, and $H_{\rm imp}$ a static local or
nonlocal impurity potential.

We distinguish between spinless and spin-$\frac{1}{2}$ fermions.

%%%%%%%%%%%%%%%%%%%%%%%%%%%%%%%%%%%%%%%%%%%%%%%%%%%%%%%%%%%%%%%%%%%%%%%%%%%%%%

\subsection{Spinless fermions}
\label{sec:frglutt:model:spinless}

For the spinless fermion model
\begin{equation}
  \label{eq:hoppingimp}
  H_0 = -t \sum_j \bigl( \,
  c^{\dag}_{j+1} c_j^{\phantom{\dag}} + c^{\dag}_j \, c_{j+1}^{\phantom{\dag}} \, \bigr)
\end{equation}
describes nearest-neighbor hopping processes with an amplitude 
$t$ and
\begin{equation}
  H_I = U \sum_j n_j \, n_{j+1}
\end{equation}
is a nearest-neighbor interaction of strength $U$, as shown in 
Fig.~\ref{fig:spinless}.
We use standard second quantization notation, where $c^{\dag}_j$ 
and $c_j$ are creation and annihilation operators on site $j$
respectively, and $n_j = c^{\dag}_j \, c_j^{\phantom{\dag}}$ is the 
local density operator.
The impurity is represented by
\begin{equation}
  H_{\rm imp} = \sum_{j,j'} 
  V_{j'j} \; c^{\dag}_{j'} \, c_j^{\phantom{\dag}} \; ,
\end{equation}
where $V_{j'j}$ is a static potential.
For ``site impurities''
\begin{equation}
  V_{j'j} = V_j \, \delta_{jj'}
\end{equation}
this potential is local. For the special case of a
single site impurity
\begin{equation}
  V_j = V \, \delta_{jj_0}
\end{equation}
the potential acts only on one site $j_0$.
We also consider ``hopping impurities'' described by the nonlocal
potential 
\begin{equation}
  V_{j'j} = V_{jj'} = - t_{j,j+1} \, \delta_{j',j+1} \; .
\end{equation}
For a single hopping impurity
\begin{equation}
  t_{j,j+1} = (t'-t) \, \delta_{jj_0}
\end{equation}
the hopping amplitude $t$ is replaced by $t'$ on the bond linking
the sites $j_0$ and $j_0+1$.
In the following we will set the bulk hopping amplitude $t$ equal
to one, that is, all energies are expressed in units of $t$.

\fig[width=12cm]{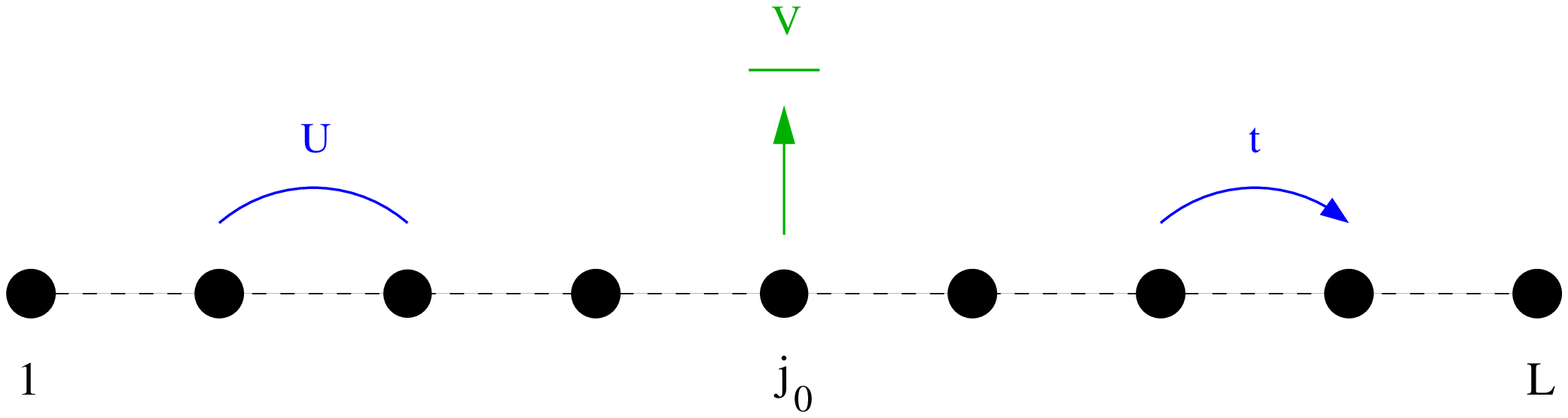}{fig:spinless}{Spinless fermion model with 
  nearest-neighbor hopping amplitude $t$, nearest-neighbor interaction $U$, 
  and a local site potential $V$ on site $j_0$.}

The clean spinless fermion model $H_0 + H_I$ is exactly soluble
via the Bethe ansatz \cite{YY66}.
The system is a Luttinger liquid for all particle densities $n$ 
and any interaction strength, except at half filling 
for $|U| > 2$. For $U>2$ a charge density wave
with wave vector $\pi$ forms; for $U<-2$ the system undergoes
phase separation.
The Luttinger-liquid parameter $K_{\rho}$, which determines all
the critical exponents of the liquid, can be computed exactly
from the Bethe ansatz solution \cite{Hal80}.
At half filling $K_{\rho}$ is related to $U$ by the simple 
explicit formula
\begin{equation}
  K_{\rho}^{-1} = \frac{2}{\pi} \, 
  \arccos \left(-\frac{U}{2} \right)
\end{equation}
for $|U| \leq 2$.

%%%%%%%%%%%%%%%%%%%%%%%%%%%%%%%%%%%%%%%%%%%%%%%%%%%%%%%%%%%%%%%%%%%%%%%%%%%%%%

\subsection[Spin-$\frac{1}{2}$ fermions]{Spin-$\mathbf{\frac{1}{2}}$ fermions}
\label{sec:frglutt:model:spin}

For spin-$\frac{1}{2}$ fermions, the kinetic energy is given by
\begin{equation}
  H_0 = -t \sum_{j,\sg} \bigl( \,
  c^{\dag}_{j+1,\sg} c_{j\sg}^{\phantom{\dag}} + c^{\dag}_{j\sg} \, c_{j+1,\sg}^{\phantom{\dag}} \,
  \bigr)\; ,
\end{equation}
where $c^{\dag}_{j\sg}$ and $c_{j\sg}^{\phantom{\dag}}$ are creation and 
annihilation operators for fermions with spin projection $\sg$ on site $j$.
The interaction term of the extended Hubbard model contains
a local interaction $U$ and a nearest-neighbor interaction $U'$
\begin{equation}
  H_I = U \sum_j n_{j\up} \, n_{j\down} + U' \sum_j n_j \, n_{j+1}\; ,
\end{equation}
with $n_{j\sg} = c^{\dag}_{j\sg} \, c_{j\sg}^{\phantom{\dag}}$ and 
$n_j = n_{j\up} + n_{j\down}$, as shown in Fig.~\ref{fig:spin}.
For the pure Hubbard model only the local 
interaction $U$ is finite, while $U'=0$.
The impurity term
\begin{equation}
  H_{\rm imp} = \sum_{j,j'} \sum_{\sg}
  V_{j'j} \; c^{\dag}_{j'\sg} \, c_{j\sg}^{\phantom{\dag}}
\end{equation}
differs from the spinless case only by the spin sum.

\fig[width=12cm]{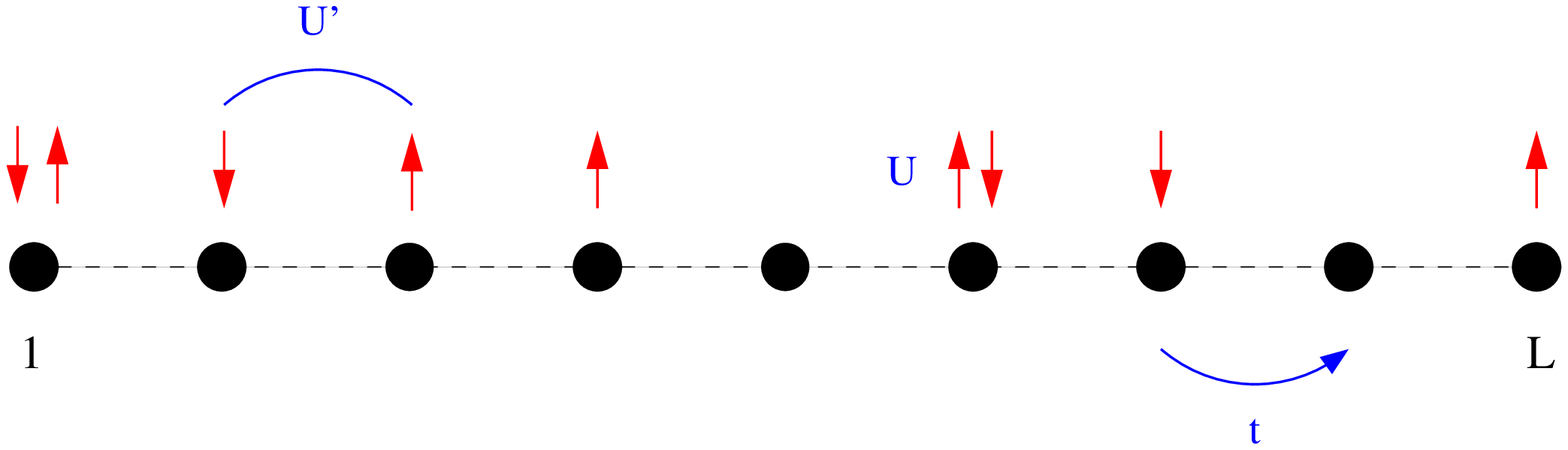}{fig:spin}{Extended Hubbard model with nearest-neighbor 
  hopping amplitude $t$, local interaction $U$ and nearest-neighbor 
  interaction $U'$.}

In the absence of impurities, the Hubbard model can be solved exactly 
using the Bethe-ansatz \cite{LW68}, while the extended 
Hubbard model is not integrable. 
The Hubbard model is a Luttinger liquid for arbitrary repulsive 
interactions at all particle densities except half filling, 
where the system becomes a Mott insulator \cite{Voi95,Gia03}.
The phase diagram of the extended Hubbard model is more complex.
Away from half filling it is a Luttinger liquid at least for 
sufficiently weak repulsive interactions \cite{Voi95}.
For the Hubbard model the Luttinger-liquid parameter $K_{\rho}$
can be computed exactly from the Bethe ansatz solution 
\cite{FK90,KY90,Schulz90b}.

%%%%%%%%%%%%%%%%%%%%%%%%%%%%%%%%%%%%%%%%%%%%%%%%%%%%%%%%%%%%%%%%%%%%%%%%%%%%%%

\section{Cutoff and flow equations}
\label{sec:frglutt:flow}

%%%%%%%%%%%%%%%%%%%%%%%%%%%%%%%%%%%%%%%%%%%%%%%%%%%%%%%%%%%%%%%%%%%%%%%%%%%%%%

\subsection{Cutoff}
\label{sec:frglutt:flow:cutoff}

The cutoff introduced in Sec.~\ref{sec:rg:flow} can be imposed in many 
different ways.
The only requirement is that the infrared singularities must be 
regularized such that the flow equations allow for a regular 
perturbation expansion in powers of the renormalized two-particle
vertex.
Since translation invariance is spoiled by the impurity, a Matsubara 
frequency cutoff is the most efficient choice,
while a momentum cutoff is less suitable.
At $T=0$ the cutoff is sharp \cite{AEMMSS04}, the extension to $T>0$ will be
addressed subsequently. 
The cutoff is imposed by excluding modes with frequencies below 
scale $\Lam$ from the functional integral representation of the
system, or equivalently, by introducing a regularized bare 
propagator
\begin{equation}
  \label{eq:cutoff}
  G_0^{\Lam}(i\om) = \Theta(|\om| - \Lam) \, G_0(i\om) \; .
\end{equation}
Here $G_0$ is the bare propagator of the pure system, involving
neither interactions nor impurities.
Instead of the sharp cutoff imposed by the step function $\Theta$
one may also choose a smooth cutoff function, but the sharp cutoff 
has the advantage that it reduces the number of integration variables
on the right-hand side of the flow equations. 
Note that we will frequently write expressions which are well defined
only if the sharp cutoff is viewed as a limit of increasingly sharp
smooth cutoff functions.
The suppression of frequencies below scale $\Lam$ affects all Green 
and vertex functions of the interacting system, which become thus 
functions of $\Lam$. The original system is recovered in the limit 
$\Lam \to 0$.

%%%%%%%%%%%%%%%%%%%%%%%%%%%%%%%%%%%%%%%%%%%%%%%%%%%%%%%%%%%%%%%%%%%%%%%%%%%%%%

\subsection{Truncation schemes}
\label{sec:frglutt:flow:trunc}

The truncation of the fRG hierarchy of differential flow 
equations for the one-particle irreducible $m$-particle vertex functions 
and their parametrization with a manageable set of variables 
or functions leads to different approximation schemes.

In the lowest order truncation of the
fRG hierarchy (cf. Sec.~\ref{sec:rg:exp}) the renormalized two-particle 
vertex is approximated by the bare interaction \cite{MMSS02a,MMSS02b}.
This truncation scheme, denoted by {\it Scheme I}, includes
only the first equation in the hierarchy for the one-particle vertex function
$\Gamma_1^{\Lam}=-\Sg^{\Lam}$, where the self-energy $\Sg^{\Lam}$ is 
related to the interacting propagator by the usual Dyson equation
\begin{equation}
  G^{\Lam} = \left[ (G_0^{\Lam})^{-1} - \Sg^{\Lam} \right]^{-1}
  \; .
\end{equation}
Here and below $G^{\Lam}$, $\Sg^{\Lam}$ etc.\ are operators, which
do not refer to any particular single-particle basis, unless we
write matrix indices explicitly. 
The right-hand side of the flow equation for $\Sg^{\Lam}$ (\ref{eq:hiera}) 
involves the two-particle vertex $\Gam^{\Lam}$
and the single-scale propagator $S^{\Lam}$ introduced in 
Eq.~(\ref{eq:ssp}) which has support only on a 
single frequency scale $|\om| = \Lam$. 
The flow equation for the self-energy reads
\begin{equation}
  \label{eq:singlesigma}
  \frac{\partial}{\partial\Lam} \Sg^{\Lam}(1',1) =
  \, - \, \frac{1}{\beta} \, \sum_{2,2'} \, e^{i\om_2 0^+} \,
  S^{\Lam}(2,2') \; \Gam^{\Lam}(1',2';1,2)\; ,
\end{equation}
where $\beta$ is the inverse temperature. 
The numbers $1$, $2$, etc.\
are a shorthand for Matsubara frequencies and labels for 
single-particle states such as site and spin indices. 
Note that $\om_1 = \om'_1$ and $\om_2 = \om'_2$ due to 
Matsubara frequency conservation.
The exponential factor in the above equation is irrelevant at
any finite $\Lam$, but is necessary to define the initial
conditions of the flow at $\Lam = \Lam_0 \to \infty$.

For a sharp frequency cutoff the frequency sum on the right-hand
side of the flow equation can be carried out analytically
in the zero temperature limit, where the Matsubara sum becomes an 
integral.
At this point one has to deal with products of delta functions 
$\delta(|\om| - \Lam)$ and expressions involving step functions 
$\Theta(|\om| - \Lam)$.
These at first sight ambiguous expressions are well defined and
unique if the sharp cutoff is implemented as a limit of increasingly 
sharp broadened cutoff functions $\Theta_{\eps}$, with the broadening 
parameter $\eps$ tending to zero. 
The expressions can then be conveniently evaluated by using the 
following relation \cite{Mor94}, valid for arbitrary continuous
functions $f$:
\begin{equation}
  \label{eq:morris}
  \delta_{\eps}(x-\Lam) \, f[\Theta_{\eps}(x-\Lam)] \to
  \delta(x-\Lam) \int_0^1 f(t) \, dt \; ,
\end{equation}
where $\delta_{\eps} = \Theta'_{\eps}$. 
Note that the functional form of $\Theta_{\eps}$ for finite $\eps$ 
does not affect the result in the limit $\eps \to 0$.

In Scheme I $\Gam^{\Lam}$ in Eq.~(\ref{eq:singlesigma}) is replaced by the 
antisymmetrized bare two-particle interaction
$\Gam^{\Lam_0}_{1',2';1,2} = I_{1',2';1,2}$,
where the lower indices $1$, $2$, etc.\ label single-particle states 
(not frequencies).
Since $I_{1',2';1,2}$ is frequency independent, 
no frequency dependence of the self-energy is generated in the flow.
Carrying out the frequency integration in the flow equation for the
self-energy (\ref{eq:singlesigma}) one obtains
\begin{equation}
  \label{eq:sigma11}
  \frac{\partial}{\partial\Lam} \Sg^{\Lam}_{1',1} =
  - \frac{1}{2\pi} \sum_{\om = \pm \Lam} \sum_{2,2'} \,
  e^{i\om 0^+} \, \tG^{\Lam}_{2,2'}(i\om) \,
  I_{1',2';1,2} \; ,
\end{equation}
where 
\begin{equation}
  \tG^{\Lam}(i\om) = 
  \left[ G_0^{-1}(i\om) - \Sg^{\Lam} \right]^{-1} \; .
\end{equation}
Note that $\tG^{\Lam}$ has no jump at $|\om| = \Lam$, in contrast
to $G^{\Lam}$.

The flow is determined uniquely by the differential flow equation
and the initial condition at $\Lam = \infty$. 
The self-energy at $\Lam = \infty$ is given by the bare impurity
(site or hopping) potential $V$.
In a numerical solution the flow starts at some large finite
initial cutoff $\Lam_0$. Here one has to take into account that, 
due to the slow decay of the right-hand side of the flow equation
for $\Sg^{\Lam}$ at large $\Lam$, the integration of the flow from 
$\Lam = \infty$ to $\Lam = \Lam_0$ yields a contribution which does
not vanish in the limit $\Lam_0 \to \infty$, but rather tends to a
finite constant.
Since $\tG^{\Lam}_{2,2'}(i\om) \to \delta_{2,2'}/(i\om)$ for
$|\om| = \Lam \to \infty$, this constant is determined as
\begin{equation}
  - \frac{1}{2\pi} \lim_{\Lam_0 \to \infty} \int_{\infty}^{\Lam_0}
  d\Lam \sum_{\om = \pm\Lam} \sum_{2,2'} e^{i\om 0^+} \,
  \frac{\delta_{2,2'}}{i\om} \, I_{1',2';1,2} =
  \frac{1}{2} \sum_{2} I_{1',2;1,2}\; .
\end{equation}
Including the bare impurity potential $V_{1,1'}$, 
the initial conditions for the self-energy at $\Lam = \Lam_0 \to \infty$ 
is
\begin{align}
  \Sg^{\Lam_0}_{1,1'} &= 
  V_{1,1'} + \frac{1}{2} \sum_{2} I_{1',2;1,2} \; .
\end{align}
For the flow at $\Lam < \Lam_0$ the factor $e^{i\om 0^+}$ in 
Eq.~\eqref{eq:sigma11} can be discarded.

A further development of the fRG approach for impurities
in Luttinger liquids includes the two-particle vertex renormalization,
denoted by {\it Scheme II} and used in the following
if not specified otherwise. 
For spinless fermions this extension does not matter qualitatively,
but the quantitative accuracy of the results improves considerably, in
particular at intermediate interaction strengths.
By contrast, for spin-$\frac{1}{2}$ systems vertex renormalization is 
necessary to take into 
account that backscattering of particles with opposite spins at 
opposite Fermi points scales to zero in the low-energy limit.
The right-hand side of the flow equation for the two-particle 
vertex $\Gam^{\Lam}$ (\ref{eq:hierb})
involves $\Gam^{\Lam}$
itself, but also the three-particle vertex $\Gam_3^{\Lam}$.
Neglecting the contribution of the three-particle vertex to the flow 
of the two-particle vertex,
the coupled system of flow equations for the two-particle vertex 
$\Gam^{\Lam}$ and the self-energy $\Sg^{\Lam}$ is closed. 
In terms of an expansion in the bare coupling function,
this truncation is exact up to second order.
However, the fRG provides more than just a second-order
calculation: the evolution of the interaction and the self-energy
is continually fed back into the fRG differential equation.
This effectively sums up contributions from arbitrarily
high orders and thus produces a scale-dependent resummation
of perturbation theory. 
We note that it does not correspond to an expansion to a fixed
loop order: the flow equations appear to be one loop, but
they also take into account two-loop effects by iteration.
The relevant question is whether higher orders significantly change the flow, 
they certainly do so if the coupling functions get too large.
The contribution of $\Gam_3^{\Lam}$ to $\Gam^{\Lam}$ is small
as long as $\Gam^{\Lam}$ is sufficiently small, because 
$\Gam_3^{\Lam}$ is initially (at $\Lam_0$) zero and is generated 
only from terms of third order in $\Gam^{\Lam}$.
A comparison of the fRG results to exact DMRG results and exact 
scaling properties shows that the truncation error is often
surprisingly small, even for rather large interactions 
\cite{MMSS02a,MMSS02b}.
The explicit form of the truncated flow equation for the two-particle 
vertex reads
\begin{align}
  \label{eq:singlegamma}
  \frac{\partial}{\partial\Lam} \,\Gam^{\Lam}(1',2';1,2)
  &= \; \frac{1}{\beta} \,
  \sum_{3,3'} \sum_{4,4'} \, G^{\Lam}(3,3') \, S^{\Lam}(4,4') \Big[
  \Gam^{\Lam}(1',2';3,4) \, \Gam^{\Lam}(3',4';1,2) 
  \nonumber \\[2mm]
  & \quad \;-  \Gam^{\Lam}(1',4';1,3) \, \Gam^{\Lam}(3',2';4,2) 
  - (3 \lra 4, 3' \lra 4') \nonumber \\[4mm]
  & \quad \;+ \Gam^{\Lam}(2',4';1,3) \, \Gam^{\Lam}(3',1';4,2)
  + (3 \lra 4, 3' \lra 4') 
  \; \Big] \; .
\end{align}
Diagrammatically, the individual contributions for the particle-particle and 
particle-hole channels written explicitly are 
\begin{align}
  \frac{\partial}{\partial \Lam} \Gam^{\Lam}\;=\;&
  \parbox{49mm}{\unitlength=1mm\fmfframe(0,2)(2,2){
      \begin{fmfgraph*}(47,17)
        \fmfleftn{l}{3}
        \fmfrightn{r}{3}
        \fmfv{d.sh=circle,d.f=30,d.si=10mm}{vl,vr}
        \fmffixed{(.55w,0)}{vl,vr}
        \fmfleft{i}\fmfright{o}\fmf{phantom}{i,vl,vr,o}\fmffreeze
        \fmflabel{$1$}{l1}
        \fmflabel{$2$}{l3}
        \fmflabel{$1'$}{r1}
        \fmflabel{$2'$}{r3}
        \fmf{plain_arrow}{l1,vl}
        \fmf{plain_arrow}{l3,vl}
        \fmf{plain_arrow}{vr,r1}
        \fmf{plain_arrow}{vr,r3}
        \fmf{slarrowm,left=0.5}{vl,vr}
        \fmf{plain_arrow,right=0.5}{vl,vr}
      \end{fmfgraph*}}} \nonumber \\[2mm]
  & - \; \;
  \parbox{19mm}{\unitlength=1mm\fmfframe(2,2)(2,0){
      \begin{fmfgraph*}(17,49)
        \fmftopn{l}{3}
        \fmfbottomn{r}{3}
        \fmfv{d.sh=circle,d.f=30,d.si=10mm}{vr,vl}
        \fmffixed{(0,1.5w)}{vr,vl}
        \fmftop{i}\fmfbottom{o}\fmf{phantom}{i,vr,vl,o}\fmffreeze
        \fmflabel{$1$}{l1}
        \fmflabel{$1'$}{l3}
        \fmflabel{$2$}{r1}
        \fmflabel{$2'$}{r3}
        \fmf{plain_arrow}{l1,vl}
        \fmf{plain_arrow}{vl,l3}
        \fmf{plain_arrow}{r1,vr}
        \fmf{plain_arrow}{vr,r3}
        \fmf{plain_arrow,right=0.5}{vr,vl}
        \fmf{slarrowm,right=0.5}{vl,vr}
      \end{fmfgraph*}}}
  \; \; \; - \; \; 
  \parbox{19mm}{\unitlength=1mm\fmfframe(2,2)(2,0){
      \begin{fmfgraph*}(17,49)
        \fmftopn{l}{3}
        \fmfbottomn{r}{3}
        \fmfv{d.sh=circle,d.f=30,d.si=10mm}{vr,vl}
        \fmffixed{(0,1.5w)}{vr,vl}
        \fmftop{i}\fmfbottom{o}\fmf{phantom}{i,vr,vl,o}\fmffreeze
        \fmflabel{$1$}{l1}
        \fmflabel{$1'$}{l3}
        \fmflabel{$2$}{r1}
        \fmflabel{$2'$}{r3}
        \fmf{plain_arrow}{l1,vl}
        \fmf{plain_arrow}{vl,l3}
        \fmf{plain_arrow}{r1,vr}
        \fmf{plain_arrow}{vr,r3}
        \fmf{slarrowm,right=0.5}{vr,vl}
        \fmf{plain_arrow,right=0.5}{vl,vr}
      \end{fmfgraph*}}}
  \; \; \; + \; \;
  \parbox{19mm}{\unitlength=1mm\fmfframe(2,2)(2,0){
      \begin{fmfgraph*}(17,49)
        \fmftopn{l}{3}
        \fmfbottomn{r}{3}
        \fmfv{d.sh=circle,d.f=30,d.si=10mm}{vr,vl}
        \fmffixed{(0,1.5w)}{vr,vl}
        \fmftop{i}\fmfbottom{o}\fmf{phantom}{i,vr,vl,o}\fmffreeze
        \fmflabel{$1$}{l1}
        \fmflabel{$2'$}{l3}
        \fmflabel{$2$}{r1}
        \fmflabel{$1'$}{r3}
        \fmf{plain_arrow}{l1,vl}
        \fmf{plain_arrow}{vl,l3}
        \fmf{plain_arrow}{r1,vr}
        \fmf{plain_arrow}{vr,r3}
        \fmf{plain_arrow,right=0.5}{vr,vl}
        \fmf{slarrowm,right=0.5}{vl,vr}
      \end{fmfgraph*}}}
  \; \; \; + \; \;
  \parbox{19mm}{\unitlength=1mm\fmfframe(2,2)(2,0){
      \begin{fmfgraph*}(17,49)
        \fmftopn{l}{3}
        \fmfbottomn{r}{3}
        \fmfv{d.sh=circle,d.f=30,d.si=10mm}{vr,vl}
        \fmffixed{(0,1.5w)}{vr,vl}
        \fmftop{i}\fmfbottom{o}\fmf{phantom}{i,vr,vl,o}\fmffreeze
        \fmflabel{$1$}{l1}
        \fmflabel{$2'$}{l3}
        \fmflabel{$2$}{r1}
        \fmflabel{$1'$}{r3}
        \fmf{plain_arrow}{l1,vl}
        \fmf{plain_arrow}{vl,l3}
        \fmf{plain_arrow}{r1,vr}
        \fmf{plain_arrow}{vr,r3}
        \fmf{slarrowm,right=0.5}{vr,vl}
        \fmf{plain_arrow,right=0.5}{vl,vr}
      \end{fmfgraph*}}}
\end{align}

Instead of solving the frequency integrated flow equation in
full generality, we implement the following approximation:
the frequency-dependent flow of the renormalized two-particle vertex 
$\Gam^{\Lam}$ is replaced by its value at vanishing (external) 
frequencies, such that $\Gam^{\Lam}$ remains frequency independent.
As a consequence, also the self-energy remains frequency independent.
Since the bare interaction is frequency independent, neglecting the
frequency dependence leads to errors only at second order (in the 
interaction strength) for the self-energy, and at third order for
the vertex function at zero frequency.
In addition to the quantitative errors we miss qualitative properties
related to the frequency dependence of the self-energy, such as the
suppression of the one-particle spectral weight in the bulk of
a pure Luttinger liquid.
On the other hand, a comparison with exact numerical results and
asymptotic analytical results shows that the impurity effects are
not qualitatively affected by the frequency dependence of $\Sg$, 
at least for weak interactions.

The frequency-integrated flow equation for the two-particle vertex, 
evaluated at vanishing external frequencies, has the form
\begin{align}
  \label{eq:flowgamma}
  \frac{\partial}{\partial\Lam} \Gam^{\Lam}_{1',2';1,2} \,& =
  \frac{1}{2\pi} \, 
  \sum_{\om = \pm\Lam} \, \sum_{3,3'} \sum_{4,4'} \!
  \Big[  \,\frac{1}{2} \,
  \tG^{\Lam}_{3,3'}(i\om) \, \tG^{\Lam}_{4,4'}(-i\om) \,
  \Gam^{\Lam}_{1',2';3,4} \, \Gam^{\Lam}_{3',4';1,2} 
  \nonumber \\[2mm]
  & \quad + \tG^{\Lam}_{3,3'}(i\om) \, \tG^{\Lam}_{4,4'}(i\om) \,
  \left( - \Gam^{\Lam}_{1',4';1,3} \, \Gam^{\Lam}_{3',2';4,2} 
    + \Gam^{\Lam}_{2',4';1,3} \, \Gam^{\Lam}_{3',1';4,2}
  \right) \Big]\; ,
\end{align}
with the initial condition $\Gam^{\Lam_0}_{1',2';1,2} = I_{1',2';1,2}$.

A crucial point is to devise an efficient parametrization of the 
vertex by a manageable number of variables.
For a finite lattice system with $L$ sites the flow of the two-particle
vertex $\Gam^{\Lam}_{1',2';1,2}$ involves $\cO(L^3)$ independent 
flowing variables, if translation invariance is assumed, and 
$\cO(L^4)$ variables, if the influence of the impurity on the flow 
of the two-particle vertex is taken into account.
For a treatment of large systems it is therefore necessary to reduce
the number of variables by a suitable approximate parametrization of
the vertex. 
In the low-energy limit (small $\Lam$) the flow is dominated by a
very small number of variables, the others being irrelevant
according to standard renormalization-group arguments \cite{Voi95}.
In particular, the frequency dependence of the vertex, discarded 
already above, is irrelevant for the flow of $\Gam^{\Lam}$ at 
small $\Lam$.
For larger $\Lam$ one can use perturbation theory as a guide for a
simple but efficient parametrization of $\Gam^{\Lam}$.

We neglect the influence of the impurity on the flow of the
two-particle vertex, such that $\Gam^{\Lam}$ remains translation
invariant. While this is sufficient for capturing the effects
of isolated impurities in otherwise pure systems, it is known
that impurity contributions to vertex renormalization become
important in macroscopically disordered systems \cite{Gia03}.
We also neglect the feedback of the bulk self-energy into the
flow of $\Gam^{\Lam}$, which yields higher order contributions
in the renormalized interaction.
The two-particle vertex is parametrized approximately by a
renormalized static short-range interaction,
which allows us to capture various features:
the low-energy flow of the vertex at $k_F$ in the pure system is 
obtained correctly to second order in the renormalized couplings;
the nonuniversal contributions at finite energy scales are correct
to second order in the bare interaction;
the algorithm for the flow of the self-energy remains as fast as
in the absence of vertex renormalization, such that one can easily 
deal with up to $10^7$ lattice sites!

For a more concrete treatment of the vertex renormalization, we now 
focus on a specific model.

%%%%%%%%%%%%%%%%%%%%%%%%%%%%%%%%%%%%%%%%%%%%%%%%%%%%%%%%%%%%%%%%%%%%%%%%%%%%%%

\subsection{Spinless fermions}
\label{sec:frglutt:flow:spinless}

For spinless fermions the two-particle vertex and the self-energy are
fully characterized by either site or momentum variables.
In the low-energy limit, the flow of the vertex is dominated by
contributions with momenta close to the Fermi points, such that the
right-hand side of the flow equation is determined by momentum 
components of the vertex 
$\Gam^{\Lam}_{k'_1,k'_2;k_1^{\phantom '},k_2^{\phantom '}}$ with 
$k_1^{\phantom '},k_2^{\phantom '},k'_1,k'_2 = \pm k_F$. 
Due to the antisymmetry of the vertex, there is only one such 
component which is nonzero:
\begin{equation}
  g^{\Lam} = \Gam^{\Lam}_{k_F,-k_F;k_F,-k_F} \; .
\end{equation}
In the low-energy limit the momentum dependence of the vertex away
from $\pm k_F$ is irrelevant. There are therefore many possible 
choices for the functional form of 
$\Gam^{\Lam}_{k'_1,k'_2;k_1^{\phantom '},k_2^{\phantom '}}$,
which all lead to the correct low-energy asymptotics. 
For a model with a bare nearest-neighbor interaction $U$, a natural
and efficient choice is to parametrize the flowing vertex simply by
a renormalized nearest-neighbor interaction $U^{\Lam}$, which leads
to a real space vertex of the form 
\begin{equation}
  \label{eq:gammaij}
  \Gam^{\Lam}_{j'_1,j'_2;j_1^{\phantom '},j_2^{\phantom '}} =
  U^{\Lam}_{j_1,j_2} \, 
  ( \delta_{j_1^{\phantom '},j'_1} \delta_{j_2^{\phantom '},j'_2} - 
  \delta_{j_1^{\phantom '},j'_2} \delta_{j_2^{\phantom '},j'_1} )\; ,
\end{equation}
with $U^{\Lam}_{j_1,j_2} = U^{\Lam} \, 
(\delta_{j_1,j_2-1} + \delta_{j_1,j_2+1}) \,$.
This yields the following structure in momentum space:
\begin{equation}
  \label{eq:momentumspacestructure}
  \Gam^{\Lam}_{k'_1,k'_2;k_1^{\phantom '},k_2^{\phantom '}} =
  2 U^{\Lam} \, [ \cos(k'_1 - k_1^{\phantom '}) - \cos(k'_2 - k_1^{\phantom '}) ] \,
  \delta^{(2\pi)}_{k_1^{\phantom '}+k_2^{\phantom '},k'_1+k'_2}\; ,
\end{equation}
where the Kronecker $\delta$ implements momentum conservation 
(modulo $2\pi$).
The flowing coupling constant $U^{\Lam}$ is linked to the value of
the vertex at the Fermi points by the relation
\begin{equation}
  \label{eq:glam}
  g^{\Lam} = 2 U^{\Lam} \, [ 1 - \cos(2k_F) ] \; .
\end{equation}
The flow equation for $g^{\Lam}$ becomes
\begin{equation}
  \label{eq:dglam}
  \frac{\partial g^{\Lam}}{\partial\Lam} =
  \frac{1}{2\pi} \sum_{\om = \pm\Lam} \int \frac{dp}{2\pi} \,
  (\, { PP + PH + PH'} \,)\; ,
\end{equation}
with the particle-particle and particle-hole contributions
\begin{align}
  { PP} =& \frac{1}{2} \, G^0_p(i\om) \, G^0_{-p}(-i\om) \,
  \Gam^{\Lam}_{k_F,-k_F;p,-p} \, \Gam^{\Lam}_{p,-p;k_F,-k_F} 
  \nonumber \\[2mm]
  { PH} =& - [G^0_p(i\om)]^2 \,
  \Gam^{\Lam}_{k_F,p;k_F,p} \, \Gam^{\Lam}_{p,-k_F;p,-k_F} 
  \nonumber \\[3mm]
  { PH'} =& G^0_{p-k_F}(i\om) \, G^0_{p+k_F}(i\om) \,
  \Gam^{\Lam}_{-k_F,p+k_F;k_F,p-k_F} \, 
  \Gam^{\Lam}_{p-k_F,k_F;p+k_F,-k_F}\; ,
\end{align}
where $\Gam^{\Lam}$ on the right-hand side of the flow equation
is given by Eq.~\eqref{eq:momentumspacestructure}. 
Using Eq.~\eqref{eq:glam} to replace $\partial_{\Lam} g^{\Lam}$ by 
$\partial_{\Lam} U^{\Lam}$ on the left-hand side of Eq.~\eqref{eq:dglam},
one obtains a flow equation for $U^{\Lam}$ of the simple form
\begin{equation}
  \label{eq:dulam}
  \partial_{\Lam} U^{\Lam} = h(\Lam) \, (U^{\Lam})^2 \; .
\end{equation}
The function $h(\Lam)$ depends only on the cutoff $\Lam$ and the
Fermi momentum $k_F$. An explicit formula for $h(\Lam)$ can be 
obtained by carrying out the momentum integral in Eq.~\eqref{eq:dglam} using 
the residue theorem.
For finite systems the momentum integral should be replaced by a
discrete momentum sum; however, this leads to sizable corrections
only for very small systems.
Inserting the momentum structure of $\Gam^{\Lam}$ 
\eqref{eq:momentumspacestructure} into the flow equation \eqref{eq:dglam} 
and replacing $g^{\Lam}$ by $U^{\Lam}$ on the left-hand side yields
\begin{equation}
  \frac{\partial U^{\Lam}}{\partial\Lam} = 
  \frac{(U^{\Lam})^2}{2\pi \sin^2 k_F} 
  \sum_{\om = \pm\Lam} \int_0^{2\pi} \frac{dp}{2\pi} \, f(p,\om) \; ,
\end{equation}
where
\begin{equation}
  f(p,\om) = 
  \frac{2 \sin^2 k_F \, \sin^2 p}{(i\om - \xi^0_p)(-i\om - \xi^0_{-p})} -
  \frac{(\cos k_F - \cos p)^2}{(i\om - \xi^0_p)^2} +
  \frac{[\cos(2k_F) - \cos p]^2}
  {(i\om - \xi^0_{p-k_F})(i\om - \xi^0_{p+k_F})} \; .
\end{equation}
Here $\xi^0_k = -2 \cos k - \mu_0$, with $\mu_0 = - 2 \cos k_F$, is
the bare dispersion relation relative to the bare Fermi level.
Since $f(p,\om)$ can be written as a rational function of $\cos p$
and $\sin p$, the $p\,$-integral can be carried out analytically using
the substitution $z = e^{ip}$ and the residue theorem.
After a lengthy but straightforward calculation one obtains the
following result for the coefficient $h(\Lam)$ in \eqref{eq:dulam}:
\begin{align}
  h(\Lam) &= - \frac{1}{2\pi} -
  \, \Re \bigg[ \frac{i}{2} \, (\mu_0 + i\Lam) 
  \, \sqrt{1 - \frac{4}{(\mu_0 + i\Lam)^2}} \nonumber \\[2mm]
  &  \quad \times \; \frac
  {3i\mu_0^4 - 10\mu_0^3\Lam - 12i\mu_0^2(\Lam^2 + 1) + 6\Lam^3\mu_0
    + 18\Lam\mu_0 + 6i\Lam^2 + i\Lam^4}
  {\pi (2\mu_0 + i\Lam)(4 - \mu_0^2 + \Lam^2 - 2i\Lam\mu_0)^2} \,
  \bigg] \; .
\end{align}
The flow equation \eqref{eq:dulam} can be integrated to
\begin{equation}
  U^{\Lam} = \frac{U}{1 - U \, H(\Lam)} \; ,
\end{equation}
where $H(\Lam)$ is the primitive function of $h(\Lam)$ with
$H(\Lam) \to 0$ for $\Lam \to \infty$.
Integrating $h(\Lam)$ one obtains
\begin{align}
  H(\Lam) &= - \frac{\Lam}{2\pi} + \frac{1}{\pi} \, \Re \Big[ \,
  \frac{(4-\mu_0^2)\Lam^2 - 2i\mu_0(2-\mu_0^2)\Lam + \mu_0^4 - 6\mu_0^2 
    + 8}{2 \, (4-\mu_0^2)\sqrt{\Lam^2 - 2i\mu_0\Lam + 4 - \mu_0^2}} \nonumber \\[2mm]
  & \quad + \, \frac{\mu_0^4}{2(4-\mu_0^2)^{3/2}} \, \tanh^{-1}
  \frac{4 + \mu_0^2 + i\mu_0\Lam}
  {\sqrt{(4 + \mu_0^2 + i\mu_0\Lam)^2 + 4(\Lam-2i\mu_0)^2}}\,\nonumber \\[2mm]
  & \quad - \, \frac{i\mu_0}{2} \, \sinh^{-1} \frac{\Lam-i\mu_0}{2} \, \Big] \; ,
\end{align}
where $\sinh^{-1}$ and $\tanh^{-1}$
denote the main branch of the inverse of the complex functions 
$\sinh$ and $\tanh$ respectively.

At half filling, corresponding to $k_F = \pi/2$, the function 
$h(\Lam)$ is particularly simple 
\begin{equation}
  h(\Lam) = - \frac{1}{2\pi} \, \left[ 1 -
    \Lam \, \frac{6 + \Lam^2}{(4 + \Lam^2)^{3/2}} \right]
\end{equation}
such that $U^{\Lam}$ reduces to
\begin{equation}
  U^{\Lam} = \frac{U}
  {1 + 
    \left(\Lam - \frac{2 + \Lam^2}{\sqrt{4 + \Lam^2}} \right) \, 
    U/(2\pi)} \; .
\end{equation}
In Fig.~\ref{fig:vert}  we show results for the renormalized nearest-neighbor
interaction $U^{\Lam}$ as obtained from the flow equation at various
densities $n$, for a bare interaction $U = 1$. 
While the renormalization does not follow any simple rule
at intermediate scales $\Lam$, all curves saturate at a finite value 
$U^*$ in the limit $\Lam \to 0$, corresponding to a finite $g^*$,
as expected for a Luttinger-liquid fixed point \cite{Voi95}.

\fig[width=10cm]{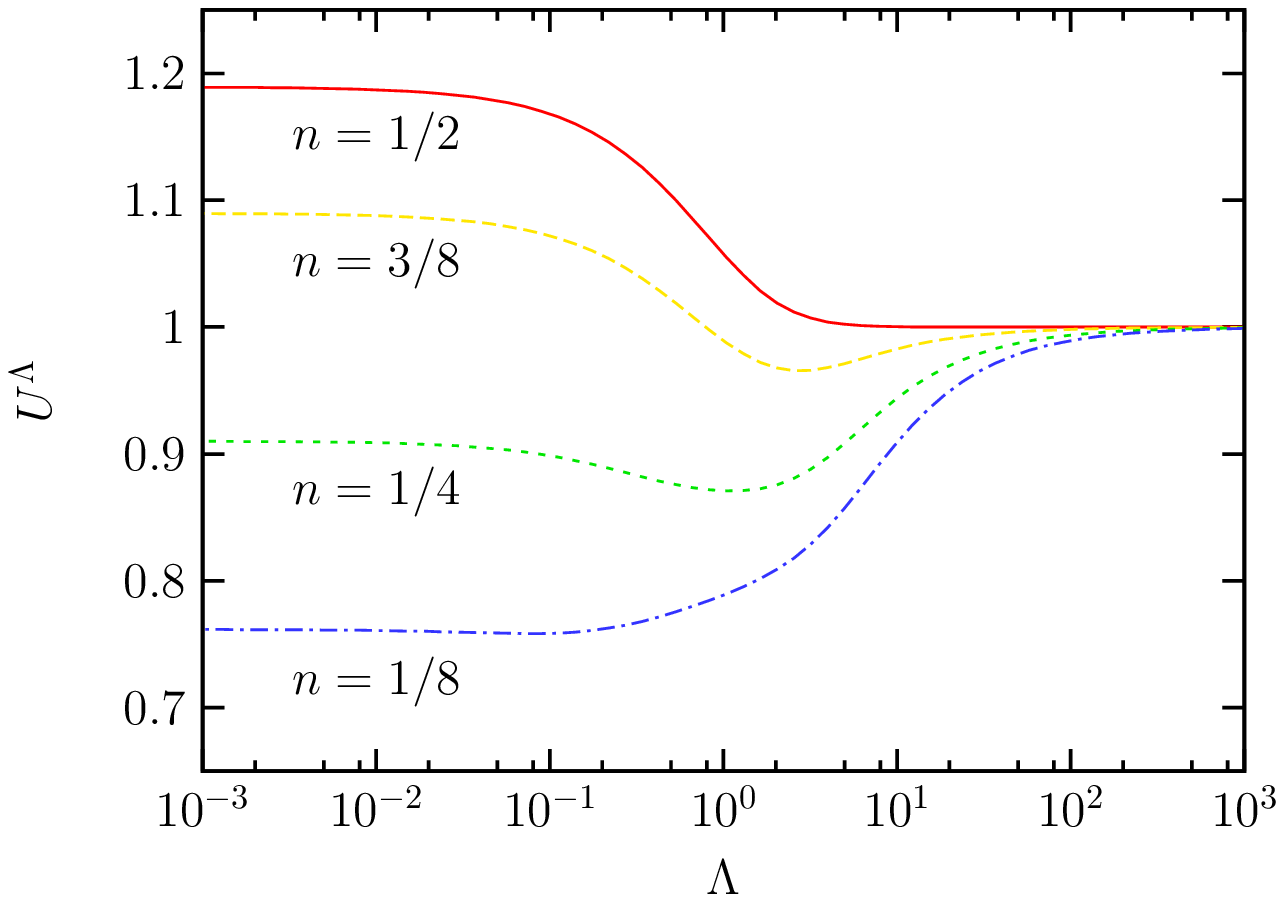}{fig:vert}{Flow of the renormalized nearest-neighbor 
  interaction $U^{\Lam}$ for the spinless fermion model, for $U=1$ and 
  various densities $n$.}

Parametrizing $\Gam^{\Lam}$ by a renormalized nearest-neighbor 
interaction has the enormous advantage that the self-energy, 
as determined by the flow equation \eqref{eq:sigma11}, is a tridiagonal matrix 
in real space, that is, only the matrix elements $\Sg^{\Lam}_{j,j}$
and $\Sg^{\Lam}_{j,j \pm 1}$ are nonzero.
Inserting $\Gam^{\Lam}$ from Eq.~\eqref{eq:gammaij} into \eqref{eq:sigma11}, 
one obtains the following simple coupled flow equations for the diagonal and 
off-diagonal matrix elements: 
\begin{align}
  \label{eq:dsigmalam}
  \frac{\partial}{\partial\Lam} \, \Sg^{\Lam}_{j,j} \; =&
  -  \frac{U^{\Lam}}{2\pi} \sum_{\om = \pm\Lam} \sum_{r = \pm 1}
  \, \tG^{\Lam}_{j+r,j+r}(i\om) \nonumber \\[2mm]
  \frac{\partial}{\partial\Lam} \, \Sg^{\Lam}_{j,j \pm 1} \; =& \quad \;
  \frac{U^{\Lam}}{2\pi} \sum_{\om = \pm\Lam}
  \tG^{\Lam}_{j,j \pm 1}(i\om) \; .
\end{align}
Note that the self-energy enters also the right-hand side of these
equations, via $\tG^{\Lam} = (G_0^{-1} - \Sg^{\Lam})^{-1}$.
Since $\Sg^{\Lam}$ and $G_0^{-1}$ are both tridiagonal in real
space, the matrix inversion required to compute the diagonal and
first off-diagonal elements of $\tG^{\Lam}$ from $\Sg^{\Lam}$
can be carried out very efficiently.
An algorithm for the numerical 
solution of the flow equation for $\Sg^{\Lam}$ scaling linearly 
with the system size is described in Ref.~\cite{AEMMSS04,EnssThesis}.
Very large systems with up to $10^7$ sites can be treated
without extensive numerical effort.

%%%%%%%%%%%%%%%%%%%%%%%%%%%%%%%%%%%%%%%%%%%%%%%%%%%%%%%%%%%%%%%%%%%%%%%%%%%%%%

\subsection[Spin-$\frac{1}{2}$ fermions]{Spin-$\mathbf{\frac{1}{2}}$ fermions}
\label{sec:frglutt:flow:spin}

We now describe the parametrization of the spatial (or momentum) dependences
of the two-particle vertex $\Gam^{\Lam}$ for spin-$\frac{1}{2}$ 
fermions \cite{AEMMSS05}, employing a natural extension of our previous 
parametrization for the
spinless case in Sec.~\ref{sec:frglutt:flow:spinless}.
We consider spin-rotation invariant lattice systems with local and 
nearest-neighbor interactions. This includes the extended Hubbard model.

For a spin-rotation invariant system the spin structure of the
two-particle vertex can be decomposed in a singlet and a triplet part: 
\begin{equation}
  \label{eq:st}
  \Gam^{\Lam} = 
  \Gam_s^{\Lam} \, S_{\sg'_1,\sg'_2;\sg_1^{\phantom '},\sg_2^{\phantom '}} +
  \Gam_t^{\Lam} \, T_{\sg'_1,\sg'_2;\sg_1^{\phantom '},\sg_2^{\phantom '}}\; ,
\end{equation}
with
\begin{align}
  S_{\sg'_1,\sg'_2;\sg_1^{\phantom '},\sg_2^{\phantom '}} &= \frac{1}{2} \,
  \left( \delta_{\sg_1^{\phantom '}\sg'_1} \delta_{\sg_2^{\phantom '}\sg'_2} -
    \delta_{\sg_1^{\phantom '}\sg'_2} \delta_{\sg_2^{\phantom '}\sg'_1} \right)
  \nonumber \\
  T_{\sg'_1,\sg'_2;\sg_1^{\phantom '},\sg_2^{\phantom '}} &= \frac{1}{2} \,
  \left( \delta_{\sg_1^{\phantom '}\sg'_1} \delta_{\sg_2^{\phantom '}\sg'_2} +
    \delta_{\sg_1^{\phantom '}\sg'_2} \delta_{\sg_2^{\phantom '}\sg'_1} \right) \; .
\end{align}
Since the total vertex is antisymmetric in the incoming and outgoing
particles, the singlet part $\Gam_s^{\Lam}$ has to be symmetric and
the triplet part $\Gam_t^{\Lam}$ antisymmetric.

Proceeding in analogy to the case of spinless fermions 
in Sec.~\ref{sec:frglutt:flow:spinless}, we first list
momentum components of the vertex with all momenta at $\pm k_F$.
For the triplet vertex the antisymmetry allows once again only one 
such component
\begin{equation}
  \label{eq:gt}
  g^{\Lam}_t = \Gam^{\Lam}_{t|\, k_F,-k_F;k_F,-k_F} \; .
\end{equation}
For the singlet vertex there are several distinct components at
$\pm k_F$.
Since we will neglect the influence of the impurity on the
vertex renormalization, the renormalized vertex remains translation
invariant. Hence the momentum components are restricted by momentum 
conservation: $k'_1 + k'_2 = k_1^{\phantom '}+ k_2^{\phantom '}$, 
modulo integer multiples of $2\pi$.
The remaining independent (not related by obvious symmetries) 
components are
\begin{align}
  \label{eq:gs1}
  g^{\Lam}_{s2} &= \Gam^{\Lam}_{s|\, k_F,-k_F;k_F,-k_F} 
  \nonumber \\[2mm]
  g^{\Lam}_{s4} &= \Gam^{\Lam}_{s|\, k_F,k_F;k_F,k_F}
\end{align}
and in the case of half filling, for which $k_F = \pi/2$, also
\begin{equation}
  \label{eq:gs2}
  g^{\Lam}_{s3} = \Gam^{\Lam}_{s|\, \pi/2,\pi/2;-\pi/2,-\pi/2} \; .
\end{equation}
The labels $2,3,4$ are chosen in analogy to the conventional 
$g$-ology notation for one-dimensional Fermi systems \cite{Sol79}.
In order to parametrize the vertex in a uniform way in all cases,
we will include the umklapp component $g^{\Lam}_{s3}$ not only
at half filling, but at any density. The effect on the other components 
is negligible for the range of interactions and fillings considered.

Extending our treatment of the spinless case 
in Sec.~\ref{sec:frglutt:flow:spinless}, we now parametrize
the vertex by renormalized local and nearest-neighbor interactions
in real space. For the triplet part, there is no local component,
and only one nearest-neighbor component compatible with the
antisymmetry, namely
\begin{equation}
  {U'_t}^{\Lam} = \Gam^{\Lam}_{t|\, j,j+1;j,j+1} \; ,
\end{equation}
which has the same form as the nearest-neighbor interaction in the 
spinless case. Note that $\Gam^{\Lam}_{t|\, j,j+1;j,j+1}$ does not
depend on $j$ and is equal to $\Gam^{\Lam}_{t|\, j,j-1;j,j-1}$.
For the symmetric singlet part, there is one local component
\begin{equation}
  U_s^{\Lam} = \Gam^{\Lam}_{s|\, j,j;j,j} 
\end{equation}
and three different components involving nearest neighbors:
\begin{align}
  {U'_s}^{\Lam} &= \Gam^{\Lam}_{s|\, j,j+1;j,j+1} \nonumber \\[2mm]
  P^{\Lam}_s    &= \Gam^{\Lam}_{s|\, j+1,j+1;j,j} \nonumber \\[2mm]
  W^{\Lam}_s    &= \Gam^{\Lam}_{s|\, j+1,j;j,j} \; .
\end{align}
For the Hubbard model, the bare vertex is purely local and the
initial condition for the vertex is given by $U_s^{\Lam_0} = 2U$,
while all the other components vanish. For the extended Hubbard model, 
${U'_s}^{\Lam_0}={U'_t}^{\Lam_0}=U'$ are nonzero.

The triplet vertex is parametrized by only one renormalized real 
space coupling, which leads to a momentum representation of the form
\begin{equation}
  \label{eq:paramt}
  \Gam^{\Lam}_{t|\, k'_1,k'_2;k_1^{\phantom '},k_2^{\phantom '}} =
  2 {U'_t}^{\Lam} \, [ \cos(k'_1 - k_1^{\phantom '}) - \cos(k'_2 - k_1^{\phantom '}) ] \,
  \delta^{(2\pi)}_{k_1^{\phantom '}+k_2^{\phantom '},k'_1+k'_2} \; ,
\end{equation}
where the Kronecker $\delta$ implements momentum conservation 
(modulo $2\pi$).
The flowing coupling ${U'_t}^{\Lam}$ is thus linked in a one-to-one 
correspondence to the Fermi momentum coupling $g^{\Lam}_t$ by
\begin{equation}
  \label{eq:paramgt}
  g^{\Lam}_t = 2 {U'_t}^{\Lam} \, [ 1 - \cos(2k_F) ] 
\end{equation}
as in the spinless case in Sec.~\ref{sec:frglutt:flow:spinless}.
In the singlet channel we have found four real space couplings,
that is, one more than necessary to match the three singlet couplings 
in momentum space, $g^{\Lam}_{s2}$, $g^{\Lam}_{s3}$, $g^{\Lam}_{s4}$. 
We discard the interaction $W^{\Lam}_s$,
because it does not appear in the bare Hubbard model,
where it is generated only at third order in $U$,
while the pair hopping $P^{\Lam}_s$ appears already in second-order
perturbation theory.
Fourier transforming the remaining interactions yields the singlet 
vertex in $k$-space
\begin{align}
  \label{eq:params}
  \Gam^{\Lam}_{s|\, k'_1,k'_2;k_1^{\phantom '},k_2^{\phantom '}} &=
  \left[ U_s^{\Lam} + 
    2 {U'_s}^{\Lam} \, [ \cos(k'_1 - k_1^{\phantom '}) + \cos(k'_2 - k_1^{\phantom '}) ] \right. \nonumber \\[2mm]
  & \left. \quad + P^{\Lam}_s \, \cos(k_1 + k_2) \right] \, \delta^{(2\pi)}_{k_1^{\phantom '}+k_2^{\phantom '},k'_1+k'_2}
\end{align}
from which we obtain a linear relation between the momentum space
couplings $g^{\Lam}_{s2}$, $g^{\Lam}_{s3}$, $g^{\Lam}_{s4}$ and the 
renormalized interaction parameters $U_s^{\Lam}$, ${U'_s}^{\Lam}$, 
$P^{\Lam}_s \,$:
\begin{align}
  \label{eq:paramgs}
  g^{\Lam}_{s2} &= U_s^{\Lam} + 
  2 {U'_s}^{\Lam} \, [1 + \cos(2k_F)] + 2 P^{\Lam}_s 
  \nonumber \\[2mm]
  g^{\Lam}_{s3} &= U_s^{\Lam} - 4 {U'_s}^{\Lam} - 2 P^{\Lam}_s
  \nonumber \\[2mm]
  g^{\Lam}_{s4} &= U_s^{\Lam} + 4 {U'_s}^{\Lam} +
  2 P^{\Lam}_s \, \cos(2k_F) \; .
\end{align} 
The determinant of this linear system is positive for all $k_F$,
except for $k_F = 0$ and $\pi$.
Hence the equations can be inverted for all densities except the 
trivial cases of an empty or completely filled band.

We can now set up the flow equations for the four independent 
couplings ${U'_t}^{\Lam}$, $U_s^{\Lam}$, ${U'_s}^{\Lam}$, and
$P^{\Lam}_s$ which parametrize the vertex. Consider the case $T=0$ first.
Inserting the spin structure (\ref{eq:st}) into the general flow 
equation for the two-particle vertex (\ref{eq:flowgamma}),
and using the momentum representation for a translation invariant 
vertex, the flow equation for the singlet and triplet vertices
$\Gam^{\Lam}_a$, for $a = s,t$, can be written as
\begin{equation}
  \label{eq:flow}
  \frac{\partial}{\partial\Lam} \, 
  \Gam^{\Lam}_{a|\, k'_1,k'_2;k_1,k_2} =
  - \frac{1}{2\pi} \sum_{\om = \pm\Lam} \, \sum_{b,b' = s,t} 
  \int \frac{dp}{2\pi} \, (\, { PP + PH + PH'} \,)\; ,
\end{equation}
with the particle-particle and particle-hole contributions
\begin{align}
  { PP} &= C^{ PP}_{a,bb'} \, 
  G^0_p(i\om) \, G^0_{k_1+k_2-p}(-i\om) \,
  \Gam^{\Lam}_{b|\, k'_1,k'_2;p,k_1+k_2-p} \, 
  \Gam^{\Lam}_{b'|\, p,k_1+k_2-p;k_1,k_2} 
  \nonumber \\[2mm]
  { PH} &= C^{ PH}_{a,bb'} \, 
  G^0_p(i\om) \, G^0_{p+k_1-k'_1}(i\om) \,
  \Gam^{\Lam}_{b|\, k'_1,p+k_1-k'_1;k_1,p} \, 
  \Gam^{\Lam}_{b'|\, p,k'_2;p+k_1-k'_1,k_2} 
  \nonumber \\[2mm]
  { PH'} &= C^{ PH'}_{a,bb'} \, 
  G^0_p(i\om) \, G^0_{p+k_1-k'_2}(i\om) \,
  \Gam^{\Lam}_{b|\, k'_2,p+k_1-k'_2;k_1,p} \, 
  \Gam^{\Lam}_{b'|\, p,k'_1;p+k_1-k'_2,k_2} \; .
\end{align}
The coefficients $C_{a,bb'}$ are obtained from the spin sums as
\begin{align}
  C^{PP}_{s,ss} &= 1 \,  \; &
  C^{ PP}_{s,st} &= C^{ PP}_{s,ts} = C^{ PP}_{s,tt} = 0&
  \nonumber \\[2mm]
  C^{ PP}_{t,tt} &= 1 \,  \; &
  C^{ PP}_{t,ss} &= C^{ PP}_{t,st} = C^{ PP}_{t,ts} = 0
  \nonumber \\[2mm]
  C^{ PH}_{s,ss} &= -1/4 \,  \; &
  C^{ PH}_{s,st}& = C^{ PH}_{s,ts} = C^{ PH}_{s,tt} = 3/4
  \nonumber \\[2mm]
  C^{ PH}_{t,tt} &= 5/4 \,  \; &
  C^{ PH}_{t,ss} &= C^{ PH}_{t,st} = C^{ PH}_{t,ts} = 1/4
  \nonumber \\[2mm]
  C^{ PH'}_{s,bb'} &= - \, C^{ PH}_{s,bb'} \,  \; &
  C^{ PH'}_{t,bb'} &= C^{ PH}_{t,bb'} \; .&
\end{align}
Note that we have neglected the self-energy feedback in the flow of 
$\Gam^{\Lam}$, such that only bare propagators $G_0$ enter.
On the right-hand side of the flow equation we insert the 
parametrization (\ref{eq:paramt}) for $\Gam^{\Lam}_t$ 
and (\ref{eq:params}) for $\Gam^{\Lam}_s$.
The flow of the triplet vertex 
$\Gam^{\Lam}_{t|\, k'_1,k'_2;k_1^{\phantom '},k_2^{\phantom '}}$
is evaluated only for 
$(k'_1,k'_2,k_1^{\phantom '},k_2^{\phantom '}) = (k_F,-k_F,k_F,-k_F)$ as 
in Eq.~(\ref{eq:gt}), which yields the flow of $g^{\Lam}_t$, while 
the flow of the singlet vertex 
$\Gam^{\Lam}_{s|\, k'_1,k'_2;k_1^{\phantom '},k_2^{\phantom '}}$
is computed for the three choices of 
$(k'_1,k'_2,k_1^{\phantom '},k_2^{\phantom '})$ which yield
the flow of $g^{\Lam}_{s2}$, $g^{\Lam}_{s3}$, $g^{\Lam}_{s4}$ corresponding
to Eqs.~(\ref{eq:gs1}) and (\ref{eq:gs2}).
Using the linear equations (\ref{eq:paramgt}) and (\ref{eq:paramgs}) 
to replace the couplings
$g^{\Lam}$ by the renormalized real space interactions on the left-hand 
side of the flow equations, we obtain a complete set of
flow equations for the four renormalized interactions ${U'_t}^{\Lam}$, 
$U_s^{\Lam}$, ${U'_s}^{\Lam}$, and $P^{\Lam}_s$ of the form
\begin{equation}
  \label{eq:flowu}
  \partial_{\Lam} U^{\Lam}_{\alf} = \sum_{\alf', \alf''}
  h_{\alf' \alf''}^{\alf}(\Lam) \, U^{\Lam}_{\alf'} \, U^{\Lam}_{\alf''}  \; ,
\end{equation}
where $\alf = 1,2,3,4$ labels the four different interactions.
The functions $h_{\alf'\alf''}^{\alf}(\Lam)$ can be computed analytically by
carrying out the momentum integrals in Eq.~(\ref{eq:flow}) via the residue 
theorem, details are reported in App.~\ref{sec:app:vertspin:frg}.
The flow equations can then be solved numerically very easily. 
For finite systems the momentum integral should be replaced by a discrete 
momentum sum; however, this leads only to negligible corrections for the 
physical observables presented in Sec.~\ref{sec:frglutt:obs}.

After computing the flow of the real space interactions, one can
also calculate the flow of the momentum space couplings $g^{\Lam}$
by using the linear relation between the two.
In the low-energy limit (small $\Lam$) one recovers the one-loop
flow of the $g$-ology model, the general effective low-energy model for 
one-dimensional fermions \cite{Sol79}, for details see 
Sec.~\ref{sec:app:vertspin:gology}.
In addition, our vertex renormalization captures also all nonuniversal
second-order contributions to the vertex at $\pm k_F$ from higher energy
scales.

In Fig.~\ref{fig:g} we show results for the renormalized real space 
interactions together with the corresponding momentum space couplings, 
as obtained by integrating the flow equations for the Hubbard model at 
quarter filling and $T=0$. Note that the couplings converge to finite 
fixed-point values in the limit $\Lam \to 0$, but the convergence is 
very slow, except for the momentum space couplings $g^{\Lam}_{s3}$ and 
$g^{\Lam}_{s4}$.
This can be traced back to the familiar behavior of the so-called
backscattering coupling 
$g^{\Lam}_{1\perp} = \frac{1}{2} \, (g^{\Lam}_{s2} - g^{\Lam}_t)$,
that is, the amplitude for the exchange of two particles with opposite 
spin at opposite Fermi points.
Backscattering is known to vanish logarithmically in the low-energy 
limit for spin-rotation invariant spin-$\frac{1}{2}$ Luttinger 
liquids \cite{Voi95}.
We emphasize that this logarithmic behavior is not promoted to a 
power law by higher order terms beyond our approximation.
By contrast, the linear combination of couplings which determines the
Luttinger-liquid parameter $K_{\rho}$ converges
very quickly to a finite fixed-point value.

\fig[width=10cm]{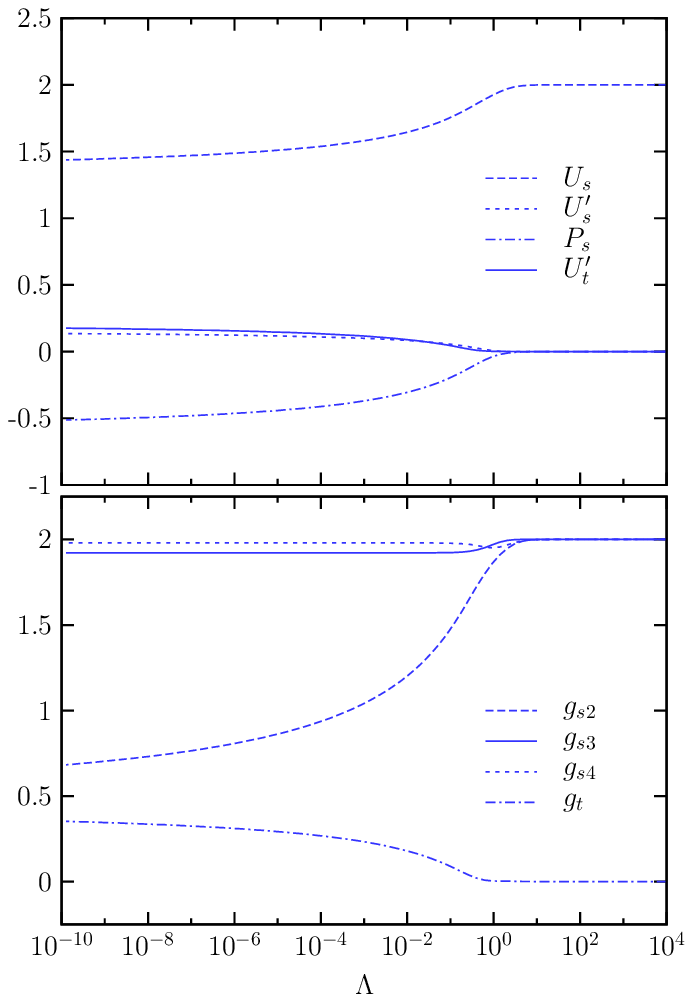}{fig:g}{Vertex flow for the
  Hubbard model at quarter filling ($n=1/2$) and $U = 1$; 
  \emph{upper panel}: 
  flow of the renormalized real space interactions, 
  \emph{lower panel}: 
  flow of the momentum space couplings.}

Due to the above parametrization of the vertex by real space interactions
which do not extend beyond nearest neighbors on the lattice, the
self-energy generated by the flow equations is frequency independent and 
tridiagonal in real space.
Inserting the spin and real space structure of $\Gamma^{\Lambda}$ 
into the general flow equation for the self-energy (\ref{eq:sigma11}), 
one obtains
\begin{align}
  \label{eq:flows}
  \frac{\partial}{\partial\Lam} \, \Sg^{\Lam}_{j,j} &=
  - \frac{1}{4\pi} \sum_{\om = \pm\Lam} \Big[ \,
  U_s^{\Lam} \, \tG^{\Lam}_{j,j}(i\om) +
  ({U'_s}^{\Lam} + 3 {U'_t}^{\Lam}) 
  \sum_{r = \pm 1} \, \tG^{\Lam}_{j+r,j+r}(i\om) \, \Big]
  \nonumber \\[2mm]
  \frac{\partial}{\partial\Lam} \, \Sg^{\Lam}_{j,j \pm 1} &=
  - \frac{1}{4\pi} \sum_{\om = \pm\Lam} \left[ \,
    ({U'_s}^{\Lam} - 3 {U'_t}^{\Lam}) \, \tG^{\Lam}_{j,j \pm 1}(i\om) +
    P^{\Lam}_s \, \tG^{\Lam}_{j \pm 1,j}(i\om) \, \right] \;  ,
\end{align}
where $\tG^{\Lam} = (G_0^{-1} - \Sg^{\Lam})^{-1}$.

Due to the slow decay of $G$ at large frequencies, the integration 
of the flow equation for $\Sg$ from $\Lam = \infty$ to 
$\Lam = \Lam_0$ yields a contribution which remains finite even in 
the limit $\Lam_0 \to \infty$, as described in 
Sec.~\ref{sec:frglutt:flow:trunc}.
For the extended Hubbard model this contribution is given by
$\Sg^{\Lam_0}_{j,j} = U/2 + 2U'$ for $j = 2, \dots, L \!-\! 1$
and $\Sg^{\Lam_0}_{1,1} = \Sg^{\Lam_0}_{L,L} = U/2 + U'$.
The numerical integration of the flow is started at a sufficiently 
large $\Lam_0$ with $\Sg^{\Lam_0}$ as initial condition.

%%%%%%%%%%%%%%%%%%%%%%%%%%%%%%%%%%%%%%%%%%%%%%%%%%%%%%%%%%%%%%%%%%%%%%%%%%%%%%

\subsection{Extension to finite temperature}
\label{sec:frglutt:flow:t}

At finite temperatures the Matsubara frequencies $\om_n$ are discrete.
The sum over $\om_n$ of a function $f$ can be written as an integral 
over a continuous variable $\om$ by introducing the
distribution function $P$ with a normalization
$\int_{|\om-\om_n|\leq\,\pi T} d\om \,P(\om)=1$ for all $n$,
\begin{equation}
  \sum_{\om_n}\, f(\om_n) =  \sum_{\om_n}\, \int_{|\om-\om_n|\leq\,\pi T} d\om\,P(\om)\,f(\om_n)=\int d\om\,P(\om) f(\om_{n, \,\om}) \;,
\end{equation}
where $\om_{n, \,\om}$ denotes the discrete Matsubara frequency 
closest to $\om$.
Introducing a sharp frequency cutoff in the continuous variable $\om$, 
the extension of the flow equations to finite temperatures is 
fairly simple, as pointed out by T. Enss \cite{AEMMSS05}.
The general form of the flow equation for the generating functional 
for the 1PI vertex functions $\Upsilon$ at $T=0$,
\begin{equation}
  \frac{\partial }{\partial \Lam}\Upsilon^{\Lam}=\int d\om\,\delta(|\om|-\Lam)\,{\cal F}\big[\Theta(|\om|-\Lam),\Upsilon^{\Lam}(\om)\big] \; ,
\end{equation}
is modified to 
\begin{equation}
  \frac{\partial }{\partial \Lam}\Upsilon^{\Lam}=T \int d\om\,P(\om) \,\delta(|\om|-\Lam)\,{\cal F}\big[\Theta(|\om|-\Lam),\Upsilon^{\Lam}(\om_{n, \,\om})\big]\,
\end{equation}
at $T>0$. 
Applying the lemma (\ref{eq:morris}) the integral over $\om$ can be 
carried out analytically 
\begin{align}
  \frac{\partial }{\partial \Lam}\Upsilon^{\Lam}&=T \int d\om \,P(\om) \,\delta(|\om|-\Lam)\,\int_0^1 dt\,{\cal F}\big[t,\Upsilon^{\Lam}(\om_{n, \,\om})\big] \nonumber \\[2mm]
  &=T \sum_{\om=\pm \Lam}P(\om) \, F\big[\Upsilon^{\Lam}(\om_{n, \,\om})\big] \;,
\end{align}
where $F(\cdot)=\int_0^1 dt\,{\cal F}(t,\cdot)$. 
The contribution to the flow on the interval
$\om_{n,\,\Lam}-\pi T\leq \Lam< \om_{n,\,\Lam}+\pi T$ is described by an
autonomous differential equation,
as the only explicit $\Lam$ dependence appears in $P$.
As a consequence the result is independent of the particular choice 
of the distribution function $P$; for simplicity we choose the constant
\begin{equation}
  P(\om) =  \frac{1}{2\pi T} \; .
\end{equation}
This leads to the final form of the flow equation
\begin{equation}
  \frac{\partial }{\partial \Lam}\Upsilon^{\Lam}=\frac{1}{2\pi}\,\sum_{\om=\pm \om_{n,\,\Lam}}  F\big[\Upsilon^{\Lam}(\om)\big]\; 
\end{equation}
for $\Upsilon$. Hence, in the flow equations for the self-energy and the 
two-particle vertex, Eqs.~(\ref{eq:sigma11}) and (\ref{eq:flowgamma}) 
respectively, the expression $\omega = \pm \Lam$ at $T=0$ is replaced by
$\omega = \pm \omega_{n,\,\Lam}$ at finite temperature,
the functional dependence on $\omega$ remains the same.

Note that for $P(\om)=\delta(\om-\om_{n,\,\om})$ the flow equation cannot be
simplified by (\ref{eq:morris}) and a smooth frequency cutoff has
to be chosen \cite{EMABMS04,EnssThesis}.

%%%%%%%%%%%%%%%%%%%%%%%%%%%%%%%%%%%%%%%%%%%%%%%%%%%%%%%%%%%%%%%%%%%%%%%%%%%%%%

\section[Calculation of $K_{\rho}$]{Calculation of $\mathbf{K_{\rho}}$}
\label{sec:frglutt:kr}

The Luttinger-liquid parameter $K_{\rho}$,
which determines the critical exponents of Luttinger liquids,
can be computed from the fixed-point couplings as obtained from the fRG. 
A relation between the fixed-point couplings and $K_{\rho}$ can be
established via the exact solution of the fixed-point Hamiltonian 
of Luttinger liquids, the Luttinger model.
A comparison of the fRG result for $K_{\rho}$ with the
exact Bethe-ansatz result for the bulk model (without impurity)
serves also as a check for the accuracy of our vertex renormalization. 
Since the above simplified flow equations yield not only the correct 
low-energy asymptotics to second order in the renormalized interaction,
but contain also all {\em nonuniversal}\/ second-order corrections 
at $\pm k_F$ from higher energy scales, the resulting $K_{\rho}$ 
is obtained correctly to second order in the interaction.

%%%%%%%%%%%%%%%%%%%%%%%%%%%%%%%%%%%%%%%%%%%%%%%%%%%%%%%%%%%%%%%%%%%%%%%%%%%%%%

\subsection{Spinless fermions}
\label{sec:frglutt:kr:spinless}

For spinless fermions, $K_{\rho}$ is determined by the Luttinger model 
parameters $g$ and $v_F$ as
\begin{equation}
  \label{eq:krspinless}
  K_{\rho} = \sqrt{\frac{1 - g/(2\pi v_F)}{1 + g/(2\pi v_F)}} \; ,
\end{equation}
where $g$ is the interaction between left and right movers and
$v_F$ the effective Fermi velocity of the model, that is, the
slope of the (linear) dispersion relation, with a possible
shift due to interactions between particles moving in the same
direction ($g_4$-coupling) already included \cite{Voi95}.
We therefore need to extract $g$ and $v_F$ from the fRG flow in
the limit $\Lam \to 0$. In order to obtain $K_{\rho}$ correctly to
order $U^2$, it is sufficient to obtain $v_F$ correctly to linear
order in $U$.

The Luttinger model interaction $g$ and the fixed-point coupling
$g^* = \Gam^{\Lam \to 0}_{k_F,-k_F;k_F,-k_F}$ from the fRG 
are not simply identical, in contrast to what one might naively 
expect.
To find the true relation between $g$ and $g^*$, one has to take
into account that the forward scattering limit of the dynamical
two-particle vertex is generally not unique (in the absence of cutoffs),
and depends on whether momentum or frequency transfers tend to
zero first. This ambiguity is well-known in Fermi-liquid theory,
where it leads to the distinction between quasi-particle interactions
and scattering amplitudes \cite{NO87}, but is equally present in 
Luttinger liquids, for the same reason in all cases: the ambiguity
of the small momentum, small frequency limit of particle-hole
propagators contributing to the vertex function.
In the {\em dynamical limit}, where the momentum transfer 
$q$ vanishes first, the singular particle-hole propagators do not 
contribute. In Fermi liquids this limit yields the quasi-particle
interaction. 
In the opposite {\em static limit} the frequency transfer 
$\nu$ vanishes first and particle-hole propagators yield a finite 
contribution.   
In the presence of an infrared cutoff $\Lam > 0$ the forward
scattering limit of the vertex function is unique, since the
ambiguity in the particle-hole propagator is due to the infrared
pole of the single-particle propagator. 
Hence $\Gam^{\Lam}_{k_F,-k_F;k_F,-k_F}$ is well defined.
However, $\Gam^{\Lam}_{k_F,-k_F;k_F,-k_F}$ and also its limit 
for $\Lam \to 0$ depend on the choice for the cutoff function.
For a momentum cutoff, which excludes states with excitation energies
below $\Lam$ around the Fermi points, particle-hole excitations with
small momentum transfers $q$ are impossible. Hence particle-hole 
propagators with infinitesimal $q$ do not contribute to the vertex
at any $\Lam > 0$, such that $\Gam^{\Lam}_{k_F,-k_F;k_F,-k_F}$
converges to the dynamical forward scattering limit, which is simply
given by the bare coupling constant $g$ in the Luttinger model.
For a frequency cutoff the particle-hole propagators with vanishing
momentum and frequency transfer yield a finite contribution at
$\Lam > 0$, which tends to the static limit for $\Lam \to 0$. 
This can be seen directly by integrating 
${\sum_{\om = \pm\Lam} \int dp} \, [G^0_p(i\om)]^2$ 
over $\Lam$ from infinity to zero. 
Hence the vertex $\Gam^{\Lam}_{k_F,-k_F;k_F,-k_F}$
obtained from our frequency cutoff fRG tends to the static
forward scattering limit.

For the Luttinger model the static forward scattering limit of the
vertex can be obtained from the dynamical effective interaction between
left and right movers $D(q,i\nu)$, which is defined as the sum of 
particle-hole chains
\begin{equation}
  D(q,i\nu) = 
  g + g \, \Pi^0_-(q,i\nu) \, g \, \Pi^0_+(q,i\nu) \, g + \dots =
  \frac{g}{1 - g^2 \, \Pi^0_-(q,i\nu) \, \Pi^0_+(q,i\nu)}\; ,
\end{equation}
where
\begin{equation}
  \Pi^0_{\pm}(q,i\nu) = 
  \pm \frac{1}{2\pi} \, \frac{q}{i\nu \mp v_F q}
\end{equation}
is the bare particle-hole bubble for right (+) and left ($-$) movers.
Note that only odd powers of $g$ contribute to the effective 
interaction between left and right movers.
This effective interaction appears naturally in the exact
solution of the Luttinger model via Ward identities \cite{DL73,MC93}.
For the static limit one obtains
\begin{equation}
  \lim_{q \to 0} D(q,0) = \frac{g}{1 - [g/(2\pi v_F)]^2}
\end{equation}
which we identify with our fixed-point coupling $g^*$ as obtained
from the fRG with frequency cutoff. 
Inverting this relation between $g$ and $g^*$ we obtain
\begin{equation}
  g = \frac{2\pi v_F}{g^*} \, \left[ - \pi v_F +
    \sqrt{(\pi v_F)^2 + (g^*)^2} \right] \; .
\end{equation}
For spinless fermions the difference between $g$ and $g^*$ appears 
only at third order in the coupling, but for models with spin the 
distinction becomes important already at second order.

\fig[width=10cm]{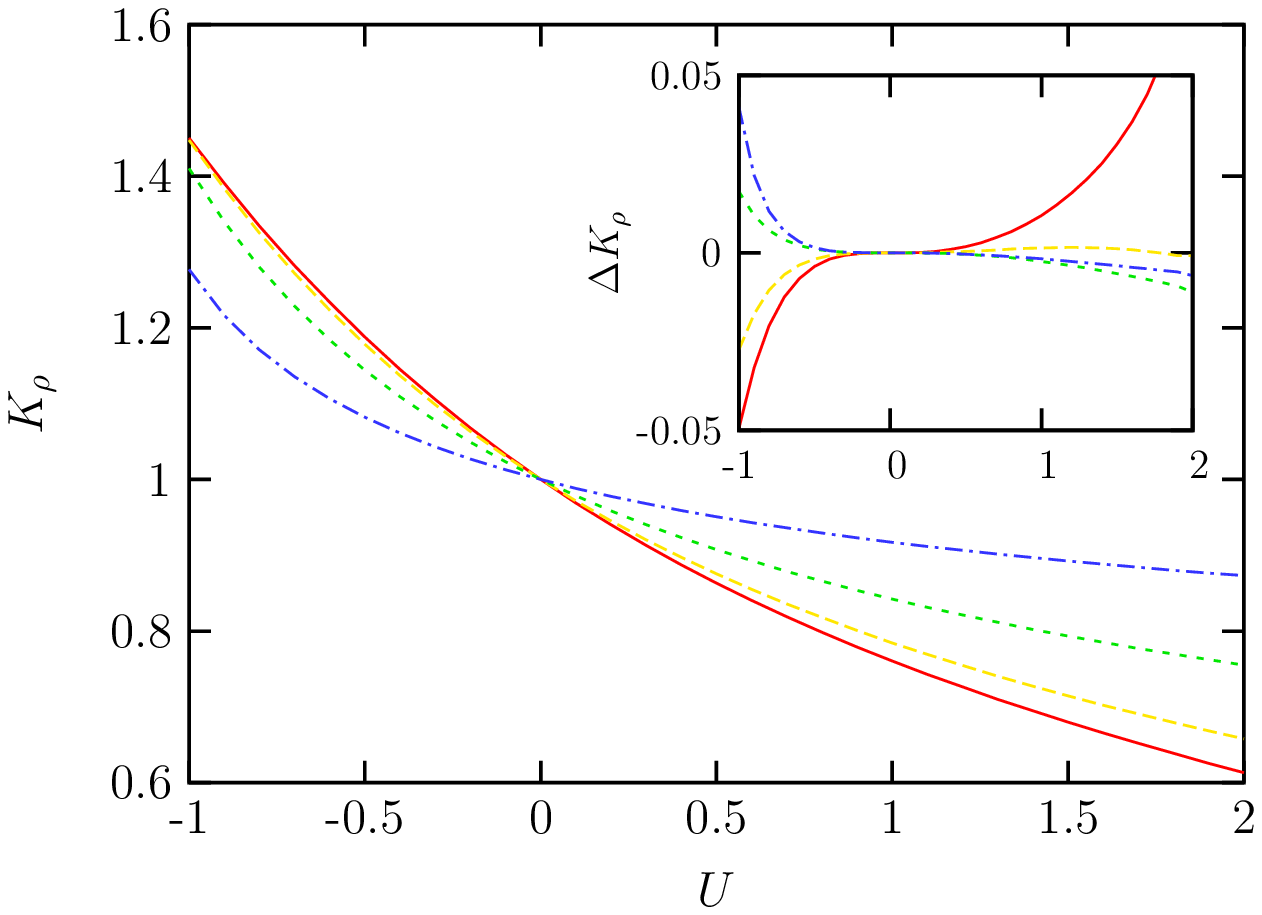}{fig:kr}{Luttinger-liquid parameter $K_{\rho}$ as a 
  function of $U$ at various densities (as in Fig.~\ref{fig:vert}) for the 
  spinless fermion model; 
  the inset shows the difference between the fRG result and the 
  exact Bethe ansatz result for $K_{\rho}$.}

The Fermi velocity $v_F$ can be computed from the (frequency-independent) 
self-energy in momentum space as
\begin{equation}
  v_F = v^0_F + \left. \partial_k \Sg_k \right|_{k_F}\; ,
\end{equation}
where $v^0_F = \partial_k \eps_k |_{k_F}$ is the bare Fermi
velocity. 
The self-energy is computed from the flow equation \eqref{eq:dsigmalam}, 
which can be rewritten in momentum space as
\begin{equation}
  \frac{\partial}{\partial\Lam} \, \Sg^{\Lam}_k =
  - \frac{U^{\Lam}}{\pi} \sum_{\om = \pm\Lam} \int \frac{dp}{2\pi} \, 
  \frac{1 - \cos(k-p)}{i\om - \xi_p - \Sg^{\Lam}_p}\; ,
\end{equation}
where $\xi_p = \eps_p - \mu$. The chemical potential $\mu$ has to be 
fixed by the final condition $\xi_{k_F} + \Sg_{k_F} = 0$, where 
$k_F = \pi n$ depends only on the density, not the interaction.
From the tridiagonal structure of $\Sg$ in real space, but also from 
the above expression it follows that $\Sg^{\Lam}_k$ has the form 
$\Sg^{\Lam}_k = a^{\Lam} + b^{\Lam} \cos k$.
The functional flow equation for $\Sg^{\Lam}_k$ yields a coupled 
set of ordinary flow equations for the coefficients $a^{\Lam}$ and 
$b^{\Lam}$, with initial conditions $a^{\Lam_0} = U$ and 
$b^{\Lam_0} = 0$. 
The momentum integrals can be evaluated analytically
via the residue theorem, such that the remaining set of two coupled
differential equations (with $U^{\Lam}$ as input) can be easily 
solved numerically.
The result for $v_F$ is correct at least to first order in $U$, but
not necessarily to second order, since our simplified parametrization
captures the two-particle vertex correctly to second order only at the 
Fermi points.

Inserting $g$ and $v_F$ into the Luttinger model formula \eqref{eq:krspinless} 
we can now compute $K_{\rho}$ as a function of $U$ and density for the 
microscopic spinless fermion model. 
In Fig.~\ref{fig:kr} we show results for $K_{\rho}(U)$ for various fixed 
densities. A comparison with exact results from the Bethe 
ansatz solution of the spinless fermion model \cite{Hal80} in the inset shows
that the fRG results are correct to second order in $U$ 
and the vertex renormalization scheme described above is very accurate.

%%%%%%%%%%%%%%%%%%%%%%%%%%%%%%%%%%%%%%%%%%%%%%%%%%%%%%%%%%%%%%%%%%%%%%%%%%%%%%

\subsection[Spin-$\frac{1}{2}$ fermions]{Spin-$\mathbf{\frac{1}{2}}$ fermions}
\label{sec:frglutt:kr:spin}

The Luttinger-liquid parameter $K_{\rho}$ for spin-$\frac{1}{2}$ 
fermions is given by 
\begin{equation}
 K_{\rho} = \sqrt{\frac{1 + (g_{\rho 4} - g_{\rho 2})/(\pi v_F)}
 {1 + (g_{\rho 4} + g_{\rho 2})/(\pi v_F)}} \; .
\end{equation}
The coupling constants $g_{\rho 2}$ and $g_{\rho 4}$ parametrize forward 
scattering interactions in the charge channel (that is, spin symmetrized) between 
opposite and equal Fermi points respectively. 
They are related to the bare singlet and triplet vertices of the Luttinger model 
by
\begin{align}
  g_{\rho 2} &= \frac{1}{4} \, 
  \bigl(\, \gam_{s|\, k_F,-k_F;k_F,-k_F} + 
  3 \gam_{t|\, k_F,-k_F;k_F,-k_F} \, \bigr)
  \nonumber \\[2mm]
  g_{\rho 4} &= \frac{1}{4} \, \gam_{s|\, k_F,k_F;k_F,k_F} \; .
\end{align}
These bare vertices are identical to the \textit{dynamical} forward scattering 
limits of the full vertex $\Gam^{\Lam}$.
On the other hand, the vertex $\Gam^{\Lam}$ obtained from the fRG with a frequency
cutoff yields the \textit{static} forward scattering limit for $\Lam \to 0$
(cf. Sec.~\ref{sec:frglutt:kr:spinless}).
For the Luttinger model, the static forward scattering limit for the vertex
can be computed from the effective interactions $D_{\rho 2}(q,i\nu)$ 
and $D_{\rho 4}(q,i\nu)$, which are defined as the sum over all particle-hole 
chains with the bare interactions $g_{\rho 2}$ and $g_{\rho 4}$ \cite{Sol79}.
The summation becomes a simple geometric series if one introduces symmetric and 
antisymmetric combinations $g_{\rho\pm} = g_{\rho 4} \pm g_{\rho 2}$ and 
$D_{\rho\pm}(q,i\nu) = D_{\rho 4}(q,i\nu) \pm D_{\rho 2}(q,i\nu)$.
The static limit of the effective interaction $D_{\rho\pm}(q,i\nu)$ yields the 
relation
\begin{equation}
  \label{eq:gpm}
  g^*_{\rho\pm} = \frac{g_{\rho\pm}}{1 - g_{\rho\pm}/(\pi v_F)}
\end{equation}
between the Luttinger model couplings $g_{\rho\pm}$ and the fixed-point couplings
\begin{equation}
  g^*_{\rho\pm} = \frac{1}{4} \, \bigl[ \,
  g^*_{s4} \pm \big( \, g^*_{s2} + 3 g^*_t \, \big)  \bigr]
\end{equation}
from the fRG with frequency cutoff. Inverting (\ref{eq:gpm}) one obtains
\begin{equation}
  K_{\rho} = \sqrt{\frac{1 - g^*_{\rho +}/(\pi v_F)}
    {1 - g^*_{\rho -}/(\pi v_F)}} \; .
\end{equation}
The Fermi velocity $v_F$ can be computed from the self-energy for the 
translation-invariant pure system as in the spinless case, 
using the momentum representation of the flow equations (\ref{eq:flows}).

The results for $K_{\rho}$ from the above procedure are correct to second 
order in the bare interaction
for the Hubbard model and also for the extended Hubbard model.
While the flowing couplings $g^{\Lam}_{s2}$ and $g^{\Lam}_t$
converge only logarithmically to their fixed-point values for
$\Lam \to 0$, the linear combination $g^{\Lam}_{s2} + 3 g^{\Lam}_t$
which enters $K_{\rho}$ converges much faster.

\fig[width=10cm]{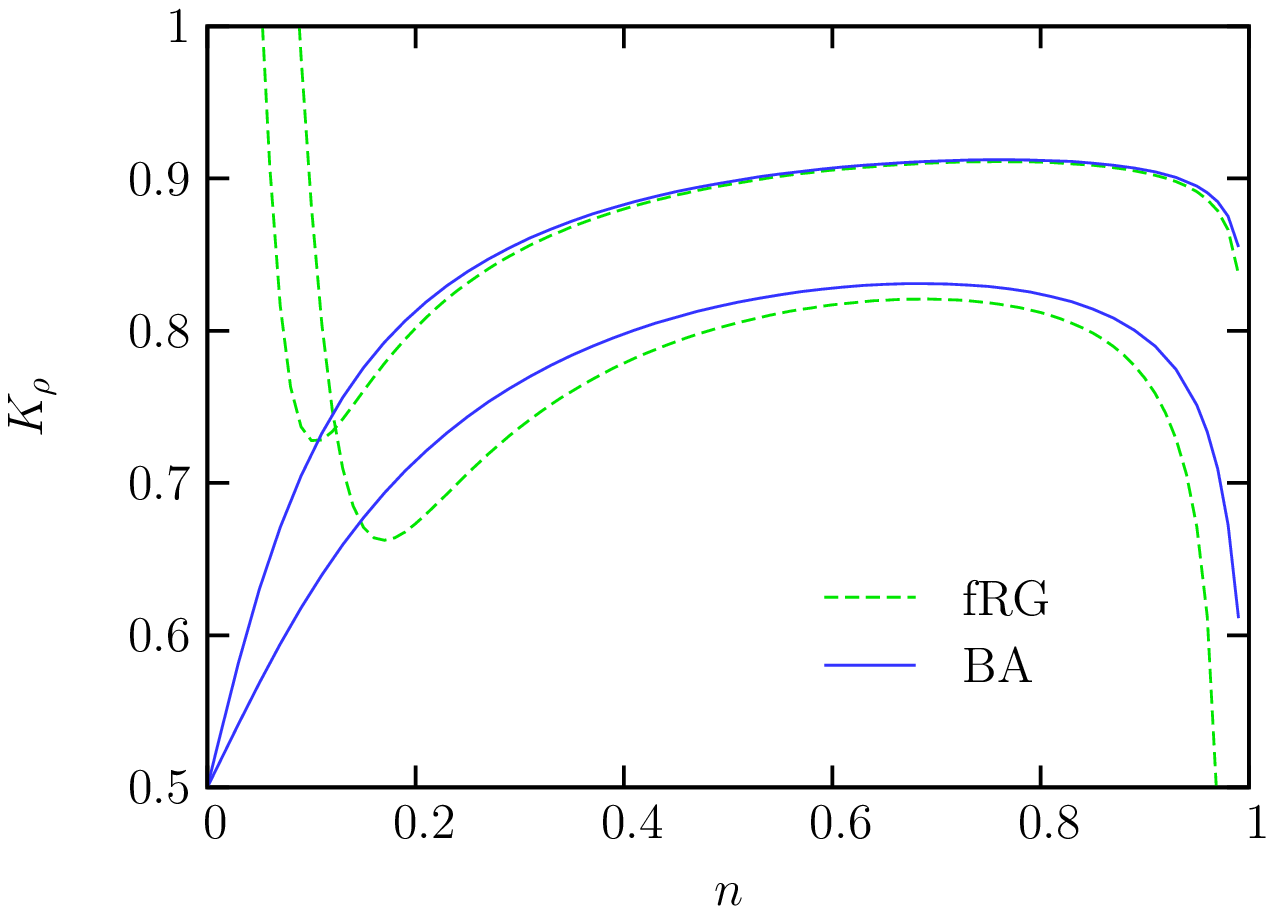}{fig:kn}{Luttinger-liquid parameter $K_{\rho}$ 
  for the Hubbard model as a function of electron
  density; results from the fRG are compared to exact results from the Bethe 
  ansatz; the upper curves are for $U = 1$ and the lower ones for $U = 2$.}

In Fig.~\ref{fig:kn} we show results for $K_{\rho}$ for the Hubbard model
as obtained from the fRG and, for comparison, from the exact Bethe ansatz 
solution \cite{FK90,KY90,Schulz90b}. Details on the solution of the 
corresponding integral equations are reported in App.~\ref{sec:app:ba}.
The truncated fRG yields accurate
results at weak coupling except for low densities and close to half filling. 
In the latter case this failure is expected since umklapp 
scattering interactions renormalize toward strong coupling,
even if the bare coupling is weak. At low densities already 
the bare dimensionless coupling $U/v_F$ is large for fixed finite
$U$, simply because $v_F$ is proportional to $n$ for small $n$, 
such that neglected higher order terms become important. 
Note, for comparison, that for spinless fermions with a fixed
nearest-neighbor interaction 
the bare dimensionless coupling at the
Fermi level vanishes in the low-density limit.

\fig[width=10cm]{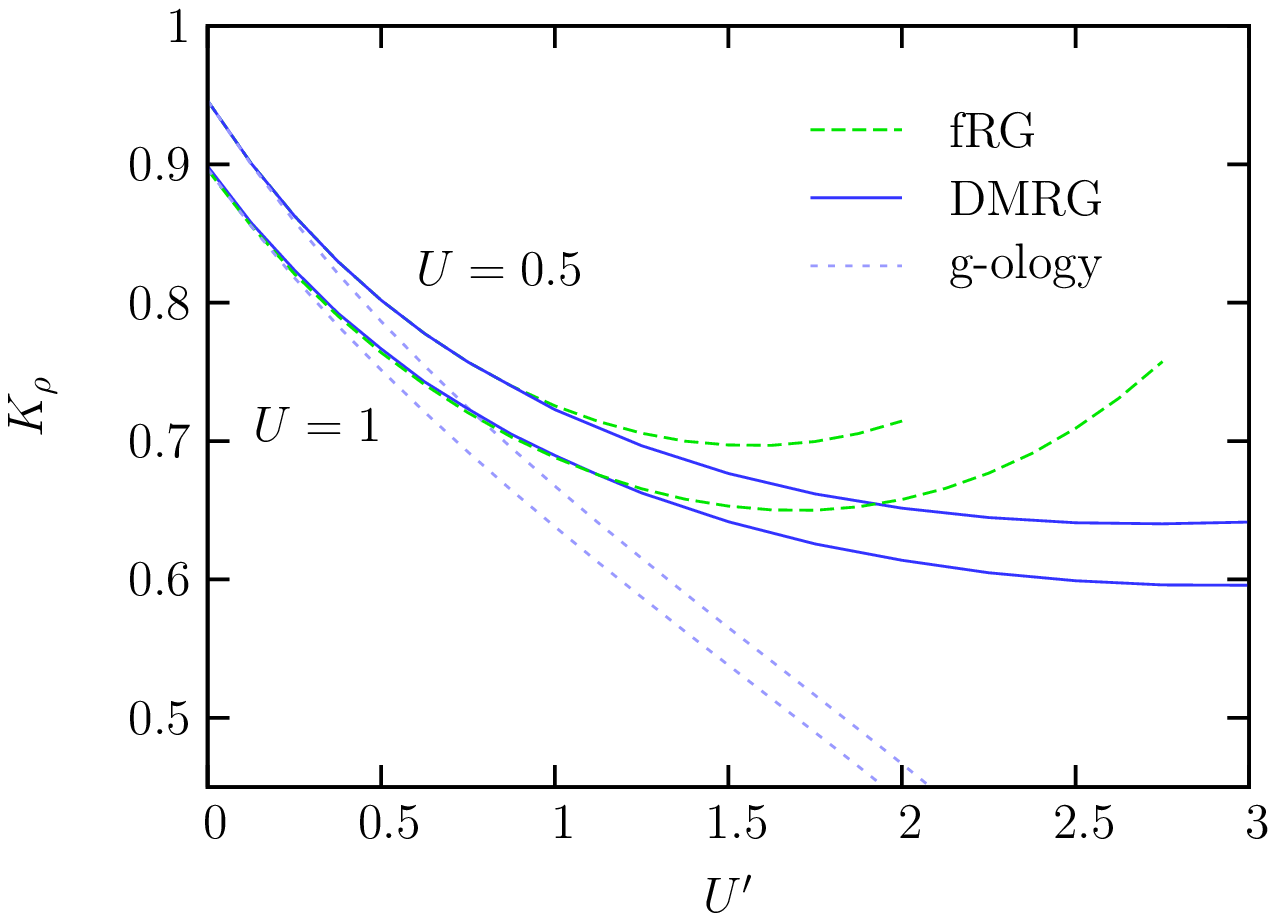}{fig:krehm}{Luttinger-liquid parameter $K_{\rho}$ for 
  the quarter-filled extended Hubbard model
  as a function of $U'$ for $U = 0.5$ and $1$; results from the fRG are 
  compared to DMRG data and to results from a one-loop $g$-ology calculation.}

For the extended Hubbard model Fig.~\ref{fig:krehm} shows 
a comparison of fRG results for $K_{\rho}$ to DMRG data \cite{EGN05}.
The fRG results are exact to second order in the interaction and are thus
very accurate for weak $U$ and $U'$. 
Results from a standard one-loop $g$-ology calculation as described 
in Sec.~\ref{sec:app:vertspin:gology} deviate
quite strongly already for $U'>0.5$. In the $g$-ology approach
interaction processes are classified 
into backward scattering $(g_{1 \perp})$, forward scattering involving 
electrons from opposite Fermi points $(g_{2 \perp})$, from the same Fermi 
points $(g_{4 \perp})$, and umklapp scattering 
$(g_{3 \perp})$. All further momentum dependences of the vertex are discarded.
This is justified by the irrelevance of these momentum dependences in the
low-energy limit, but leads to deviations from the exact flow at finite scales, 
and therefore to less accurate results for the fixed-point couplings.

The flow of $g_{i\perp}^{\Lambda}$, $i=1,...,4$, is plotted 
in Fig.~\ref{fig:gology}, in the upper panel for the quarter-filled Hubbard 
model with bare interaction $U=1$, 
and for the extended Hubbard model with $U'=U/\sqrt{2}$ in the lower.
The fRG result is compared to the result from a one-loop $g$-ology calculation.
The backscattering coupling $g_{1 \perp}$ vanishes 
logarithmically in both cases, as expected for the Luttinger-liquid fixed point
\cite{Voi95}.
For the pure Hubbard model the good agreement with $g$-ology results stems from 
the purely local interaction in 
real space,
since in that case pronounced momentum dependences of the vertex develop
only in the low-energy regime where the $g$-ology parametrization is a good
approximation.
By contrast, for the extended Hubbard model momentum dependences of the vertex 
which are not captured by the $g$-ology classification (except for small $\Lam$) 
are obviously more important. A generalization of the $g$-ology parametrization
of the vertex to higher dimensions, which amounts to neglecting the
momentum dependence normal to the Fermi surface, is frequently used in one-loop
fRG calculations in two dimensions \cite{HM00,Zanchi:9812303,HSFR01,KK03}.
The above comparison indicates that this parametrization works well for
the pure Hubbard model, but could be improved for models with nonlocal 
interactions. The parametrization of the vertex by an effective short-range 
interaction used here could be easily extended to higher dimensions,
where it will probably yield more accurate results, too.
The relevance of an improved parametrization of the vertex beyond
the conventional $g$-ology classification has also been demonstrated in a recent
fRG analysis of the phase diagram of the 
half-filled extended Hubbard model \cite{TTC05}.

\fig[width=10cm]{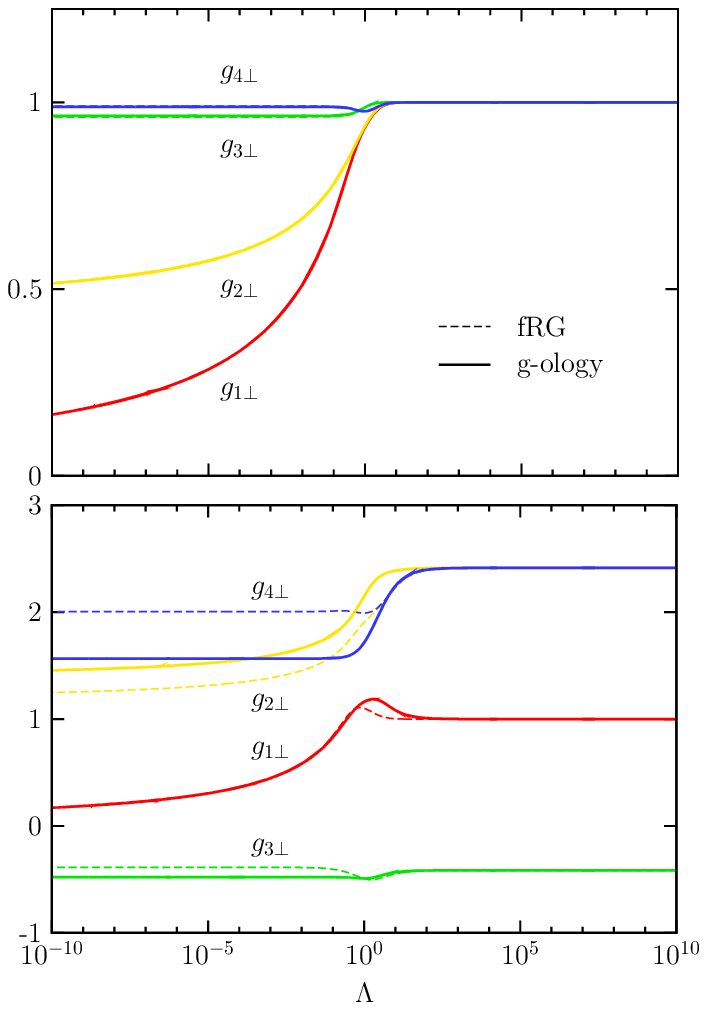}{fig:gology}{Flow of vertex on the Fermi points
  (in $g$-ology notation) at quarter filling and $U = 1$;
  \emph{upper panel}: Hubbard model,  
  \emph{lower panel}: extended Hubbard model with $U' = U/\sqrt{2}$;
  the fRG flow is compared to the one-loop $g$-ology flow;
  note that in the upper panel fRG and $g$-ology results almost
  coincide.}

The inclusion of the momentum dependence 
due to the nearest-neighbor interaction component in real space on the
right-hand side of the flow equation for the two-particle vertex modifies 
the flow of the couplings
at intermediate scales,
before reaching the regime where a $g$-ology description  at
weak coupling applies.
A small repulsive initial backscattering amplitude may renormalize to 
an effective \emph{attractive} one.
For negative $g_{1\perp}$ the renormalization
group scales to strong coupling, indicating  
an instability of the model towards a different ground state
characterized by a gap in the spin excitation spectrum \cite{Voi95}.
In Fig.~\ref{fig:phase} the phase boundary for the Luttinger liquid
and spin gap is shown as a function of $n$ and $U'$, as 
obtained from the fRG together with the result 
from a one-loop $g$-ology calculation. 
The fRG results confirm the spin gap phase at low densities found with
numerical Quantum Monte Carlo methods \cite{CSC99}, where the 
spin gap develops with increasing $U'$ from $U'=0$.
At low densities and close to half filling the truncated fRG results 
are not meaningful, since renormalization towards
strong coupling occurs in these limits and neglected higher 
order terms become important, see also Fig.~\ref{fig:kn}.
Close to half filling the spin gap opens with increasing $U'$ 
from $U'=U/2$, as for the one-loop $g$-ology calculation.
A full functional implementation of the momentum dependence of the 
two-particle vertex would allow a more detailed analysis of the phase diagram.

\fig[width=10cm]{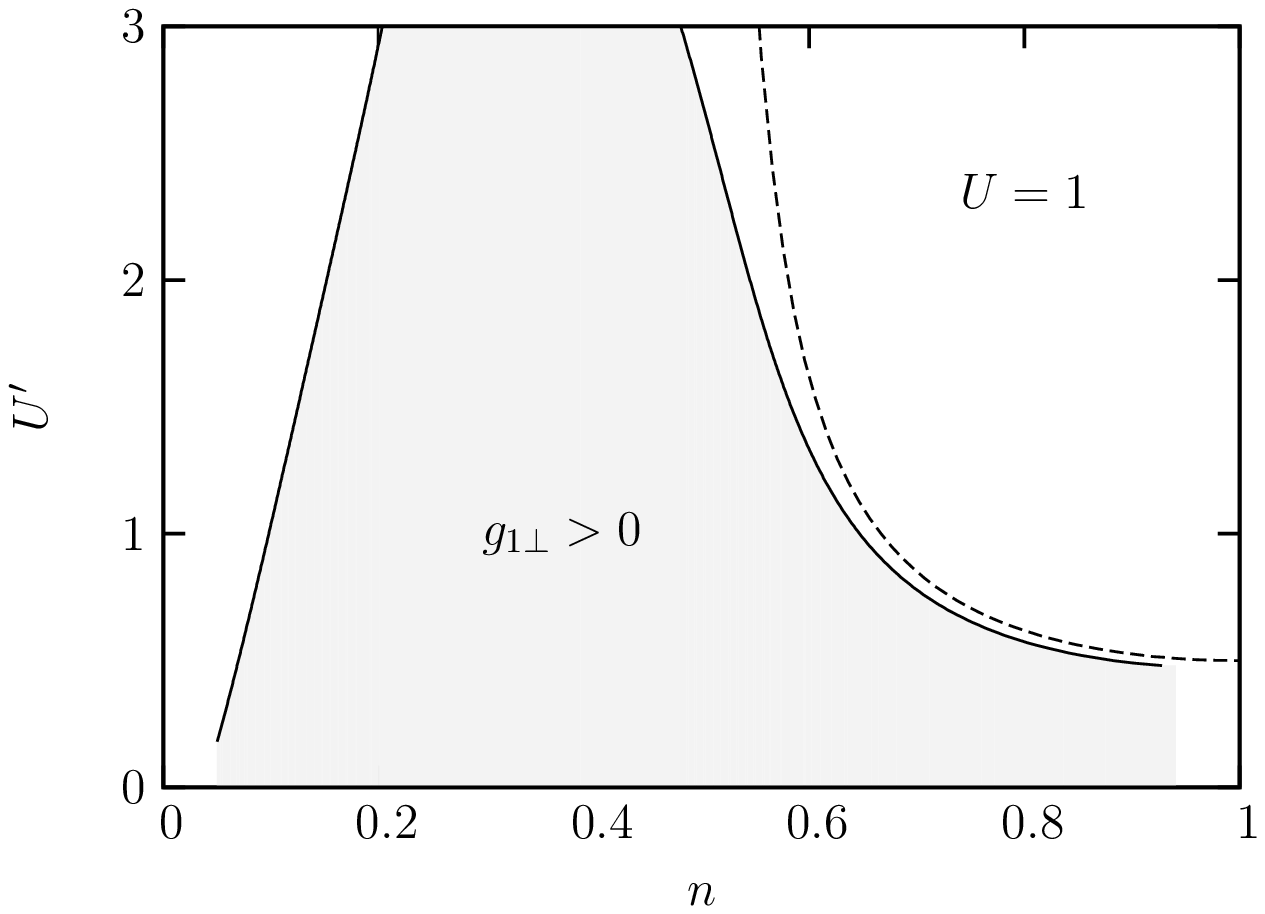}{fig:phase}{Phase boundary between the Luttinger liquid 
  ($g_{1\perp}>0$) and spin gap phase ($g_{1\perp}<0$) 
  for the extended Hubbard model as a 
  function of $n$ and $U'$ for $U=1$;
  results from the fRG (solid lines) are compared to results from a 
  one-loop $g$-ology calculation (dashed line).}

%%%%%%%%%%%%%%%%%%%%%%%%%%%%%%%%%%%%%%%%%%%%%%%%%%%%%%%%%%%%%%%%%%%%%%%%%%%%%%

\section{Observables}
\label{sec:frglutt:obs}

In the next section we will present results for spectral 
properties of single-particle excitations near an impurity
or boundary, the density profile and the linear conductance. 
Here we describe how the relevant observables are computed from 
the solution of the flow equations.

%%%%%%%%%%%%%%%%%%%%%%%%%%%%%%%%%%%%%%%%%%%%%%%%%%%%%%%%%%%%%%%%%%%%%%%%%%%%%%

\subsection{Single-particle excitations}
\label{sec:frglutt:obs:dos}

Integrating the flow equation for the self-energy $\Sg^{\Lam}$
down to $\Lam = 0$ yields the physical (cutoff-free) self-energy
$\Sg$ and the single-particle propagator 
$G = (G_0^{-1} - \Sg)^{-1}$.
From the latter spectral properties of single-particle 
excitations can be extracted.
We focus on local spectral properties, which are described by 
the \emph{local spectral function}
\begin{equation}
  \rho_j(\om) = - \frac{1}{\pi} \, 
  \Im \, G_{jj}(\om + i0^+) \; ,
\end{equation}
where $G_{jj}(\om + i0^+)$ is the local propagator, analytically
continued to the real frequency axis from above.

In our approximation the self-energy is frequency independent
and can therefore be viewed as an effective single-particle
potential. The propagator $G$ is thus the Green function of an 
effective single-particle Hamiltonian. 
In real space representation this Hamiltonian is given by the
tridiagonal matrix $h_{\rm eff} = h_0 + \Sg$, where the matrix 
elements of $h_0$ are the hopping amplitudes in $H_0$, 
Eq.~\eqref{eq:hoppingimp}.
For a lattice with $L$ sites this matrix has $L$ (including
possible multiplicities) eigenvalues $\eps_{\lam}$ and an 
orthonormal set of corresponding eigenvectors $\psi_{\lam}$.
For the spectral function $\rho_j(\om)$ one thus obtains a 
sum of $\delta$ peaks
\begin{equation}
  \rho_j(\om) = \sum_{\lam} w_{\lam j} \, 
  \delta(\om - \xi_{\lam}) \; ,
\end{equation}
where $\xi_{\lam} = \eps_{\lam} - \mu$, and the \emph{spectral weight} 
$w_{\lam j}$ is the squared modulus of the amplitude of $\psi_{\lam}$ 
on site $j$. For large $L$ the level spacing between neighboring
eigenvalues is usually of order $L^{-1}$, except for one or a few 
levels outside the band edges which correspond to bound states.

Due to even-odd effects etc.\ the spectral weight $w_{\lam j}$ 
generally varies quickly from one eigenvalue to the next one. 
A smooth function of $\om$ which suppresses these usually
irrelevant finite-size details can be obtained by averaging
over neighboring eigenvalues. 
In addition, dividing the spectral weight $w_{\lam j}$ by the level 
spacing between eigenvalues yields 
the local \emph{density of states}, which we denote by 
$D_j(\om)$.

%%%%%%%%%%%%%%%%%%%%%%%%%%%%%%%%%%%%%%%%%%%%%%%%%%%%%%%%%%%%%%%%%%%%%%%%%%%%%%

\subsection{Density profile}
\label{sec:frglutt:obs:density}

Boundaries or impurities induce a density profile with long-range 
Friedel oscillations, which are expected to decay with a power law
at long distances \cite{EG95}.
The expectation value of the local density $n_j$ could be
computed from the local one-particle propagator $G_{jj}$, if
$G$ was known exactly. However, the approximate flow equations
for $\Sg$ can be expected to describe the asymptotic behavior
of $G$ correctly only at long distances between creation and
annihilation operator in time and/or space, while in the local 
density operator time and space variables coincide. In the
standard renormalization-group terminology $n_j$ is a \emph{composite} 
operator, which has to be renormalized separately \cite{Zin02}.

The flow equation for $n_j^{\Lam}$ can be derived by computing the 
shift of the grand canonical potential $\Omega^{\Lambda}$ generated 
by a small field $\phi_j$ coupled to the local density. 
Alternatively to the numerical differentiation one may carry out the 
$\phi_j$ derivative analytically in the flow equations, which yields a 
flow equation for the density in terms of the density response vertex.
The general structure at $T=0$ is described in 
Refs.~\cite{AEMMSS04,EnssThesis}; the final form of the flow equation for 
the density is
\begin{equation}
  \label{eq:dnlam}
  \frac{\partial}{\partial\Lam} n_j^{\Lam} = 
  - \frac{1}{2\pi} \sum_{\om = \pm\Lam} 
  {\rm tr} \left[ e^{i\om 0^+}
    \tG^{\Lam}(i\om) \, R_j^{\Lam}(i\om) \right] \; ,
\end{equation}
with the density response vertex given by  
\begin{equation}
  \frac{\partial}{\partial\Lam} R_{j;1',1}^{\Lam} = 
  - \frac{1}{2\pi} \sum_{\om = \pm\Lam} 
  \sum_{2,2'} \sum_{3,3'} 
  \tG_{2,3}^{\Lam}(i\om) \, R_{j;3,3'}^{\Lam} \, 
  \tG_{3',2'}^{\Lam}(i\om) \,
  \Gam_{1',2';1,2}^{\Lam} \; .
\end{equation}
Note that within the approximate treatment 
of the two-particle vertex described in Sec.~\ref{sec:frglutt:flow:trunc}, 
the density-response vertex is frequency independent.

For $\Gam^{\Lam}$ parametrized by local and nearest-neighbor interactions
in real space, the matrix $R_j^{\Lam}$ 
is tridiagonal, that is,
only the components $R_{j;3,3}^{\Lam}$ and 
$R_{j;3,3\pm 1}^{\Lam}$ are nonzero.
The initial condition for the density is $n_j^{\Lam_0} = \frac{1}{2}$ for 
any filling, due to the slow 
convergence of the flow equation \eqref{eq:dnlam} at large frequencies, 
which yields a finite contribution to the integrated flow from 
$\Lam = \infty$ to $\Lam_0$ for arbitrarily large finite $\Lam_0$,
as in the case of the self-energy discussed in more detail
in Sec.~\ref{sec:frglutt:flow:trunc}.
The initial condition for the response vertex 
is $R_{j;l,l'}^{\Lam_0} = \delta_{jl} \delta_{ll'}$.

To avoid the interference of Friedel oscillations emerging
from the impurity or one boundary with those coming from the 
(other) boundaries of our systems we suppress the influence 
of the latter by coupling the finite chain to semi-infinite
noninteracting leads, with a smooth decay of the interaction
at the contacts, as described in detail in the next section.

%%%%%%%%%%%%%%%%%%%%%%%%%%%%%%%%%%%%%%%%%%%%%%%%%%%%%%%%%%%%%%%%%%%%%%%%%%%%%%

\subsection{Conductance}
\label{sec:frglutt:obs:cond}

For the calculation of the conductance $G$ a finite interacting chain 
(with sites $1,\dots,L$) is coupled to noninteracting leads at both ends,
corresponding to an experimental setup where the Luttinger-liquid
wires are connected to (higher dimensional) Fermi-liquid
leads. 
Using a projection method the system with leads
can be reduced to an effective $L$-site problem \cite{Tay}. 
The leads are modeled by a one-dimensional tight-binding lattice 
with nearest-neighbor hopping amplitude $t$.
The influence of the leads 
on the interacting chain can be taken into account 
by an additional dynamical boundary potential
\begin{equation} \label{vlead}
  V_j^{\rm lead}(i\omega_n) = 
  \frac{i\omega_n+\mu_0}{2} \left( 1 - 
    \sqrt{1 - \frac{4}{(i\omega_n+\mu_0)^2}} \, \right) 
  \left( \delta_{1,j} + \delta_{L,j} \right)
\end{equation}
in the bare propagator $G_0$ of the interacting chain \cite{EMABMS04}. 
The parameter $\mu_0$ is the chemical potential, which is related 
to the density $n$ in the leads by \mbox{$\mu_0 = -2 \cos k_F$}.
Uncontrolled conductance drops due to scattering at the contacts
between leads and the interacting part of the chain can be avoided
by switching off the interaction potential smoothly near the 
contacts. 
As long as the switching on of the interaction is
smooth enough and the bulk part of the wire is large compared to the 
switching region, the results are independent of the microscopic details of 
this procedure. In addition, interaction-induced bulk shifts of the 
density have to be compensated by a suitable bulk potential \cite{EMABMS04}.

In linear response the conductance is computed via the 
Kubo formula from the current-current correlation function \cite{Mah00}.
Within our approximation scheme the self-energy has no
imaginary part, which implies that there are no vertex corrections 
\cite{Ogu01}.
In this approximation the Kubo formula reduces to the Landauer-B\"uttiker
formula \cite{Dat95}, relating the transmission probability directly to the 
linear conductance $G(T,L)$ of noninteracting fermions. 
For a detailed derivation we refer to Refs.~\cite{EMABMS04,EnssThesis}; 
the final expression for the conductance is given by 
\begin{eqnarray}
  \label{eq:conductf}
  G(T,L) = - z\,\frac{e^2}{h} \int_{-2-\mu_0}^{2-\mu_0}
  |t(\varepsilon,T,L)|^2  \; f'(\varepsilon) \; d \varepsilon \; ,
\end{eqnarray} 
with $|t(\varepsilon,T,L)|^2 = [4- (\mu_0+\varepsilon)^2]
|G_{1,L}(\varepsilon,T)|^2$, and $f$ the Fermi function; 
$z$ is the number of spin components.

%%% Local Variables: 
%%% mode: latex
%%% TeX-master: "thesis"
%%% End: 

% Thesis chapter on fRG for Luttinger liquids

\chapter{Solution of fRG equations and results}
\label{sec:results}
\vspace{2cm}
\textit{In this section we present results for the local density of states 
  near boundaries and
  impurities, the density profile, and the linear conductance 
  for the spinless fermion model and the extended Hubbard model, 
  as obtained from the solution of the fRG flow equations.
  A comparison with exact DMRG data is made for those observables 
  and system sizes for which such data could be obtained. 
  The asymptotic
  low-energy behavior for weak and intermediate impurity strengths 
  is approached only at rather low energy scales,
  accessible only for very large systems.
  For spin-$\frac{1}{2}$ systems two-particle backscattering leads to striking 
  effects, which are not captured if the bulk system is approximated by
  its low-energy fixed point, the Luttinger model.
  In particular, the expected decrease of spectral weight near the 
  impurity and of the conductance at low energy scales is often 
  preceded by a pronounced increase, and the asymptotic power 
  laws are modified by logarithmic corrections.}
\vspace{.75cm}

%%%%%%%%%%%%%%%%%%%%%%%%%%%%%%%%%%%%%%%%%%%%%%%%%%%%%%%%%%%%%%%%%%%%%%%%%%%%%%

\section{Spinless fermions}
\label{sec:results:spinless}

%%%%%%%%%%%%%%%%%%%%%%%%%%%%%%%%%%%%%%%%%%%%%%%%%%%%%%%%%%%%%%%%%%%%%%%%%%%%%%

\subsection{Effective impurity potential}
\label{sec:results:spinless:sigma}

The typical shape of the self-energy representing 
the effective impurity potential
can be seen in Fig.~\ref{fig:friedel}, where we plot the diagonal elements 
$\Sg_{j,j}$ and the off-diagonal elements $\Sg_{j,j+1}$ near a
site impurity of strength $V=1.5$ added to the spinless fermion
model with interaction strength $U=1$ at quarter filling. 
Recall that the self-energy is tridiagonal in real
space and frequency independent within our treatment.
The diagonal elements can be interpreted as a local effective 
potential, the off-diagonal elements as a nonlocal effective 
potential which renormalizes the hopping amplitudes.
At long distances from the impurity both $\Sg_{j,j}$ and
$\Sg_{j,j+1}$ tend to a constant. The former describes just a bulk
shift of the chemical potential, the latter a bulk renormalization
of the hopping amplitude toward larger values.
The oscillations around the bulk shifts are generated by the
impurity. The wave number of the oscillations is $2k_F=\pi/2$, 
where $k_F$ is the Fermi wave vector of the bulk system at 
quarter filling. 
Fig.~\ref{fig:sigmaimp} shows $\Sigma_{j_0,j_0}^{\Lambda}$ for a site impurity 
of strength $V=1.5$ as a function of $\Lam$ for different system sizes $L$. 
For finite $L$ the flow is effectively cut off on a scale $\sim 1/L$, 
a sequence of $L$ provides an extrapolation to the thermodynamic limit.
The renormalized potential at the
impurity site remains \emph{finite} in the limit $L \to \infty$, 
while the Fourier transform $\Sigma_{k,k'}^{\Lambda}$ for
momenta with $k-k'=~\pm 2 k_F$ diverges.
Similarly for a hopping impurity the effective amplitude does not scale to zero 
in the limit $L \to \infty$,
and the weak-link behavior is associated to the 
long-range oscillations in real space.
The straight line in a log-log plot of the difference between 
the asymptotic value $\Sigma_{j_0,j_0}^0(L=\infty)$ and $\Sigma_{j_0,j_0}^0(L)$
shown in the inset indicates a power law dependence.

\fig[width=10cm]{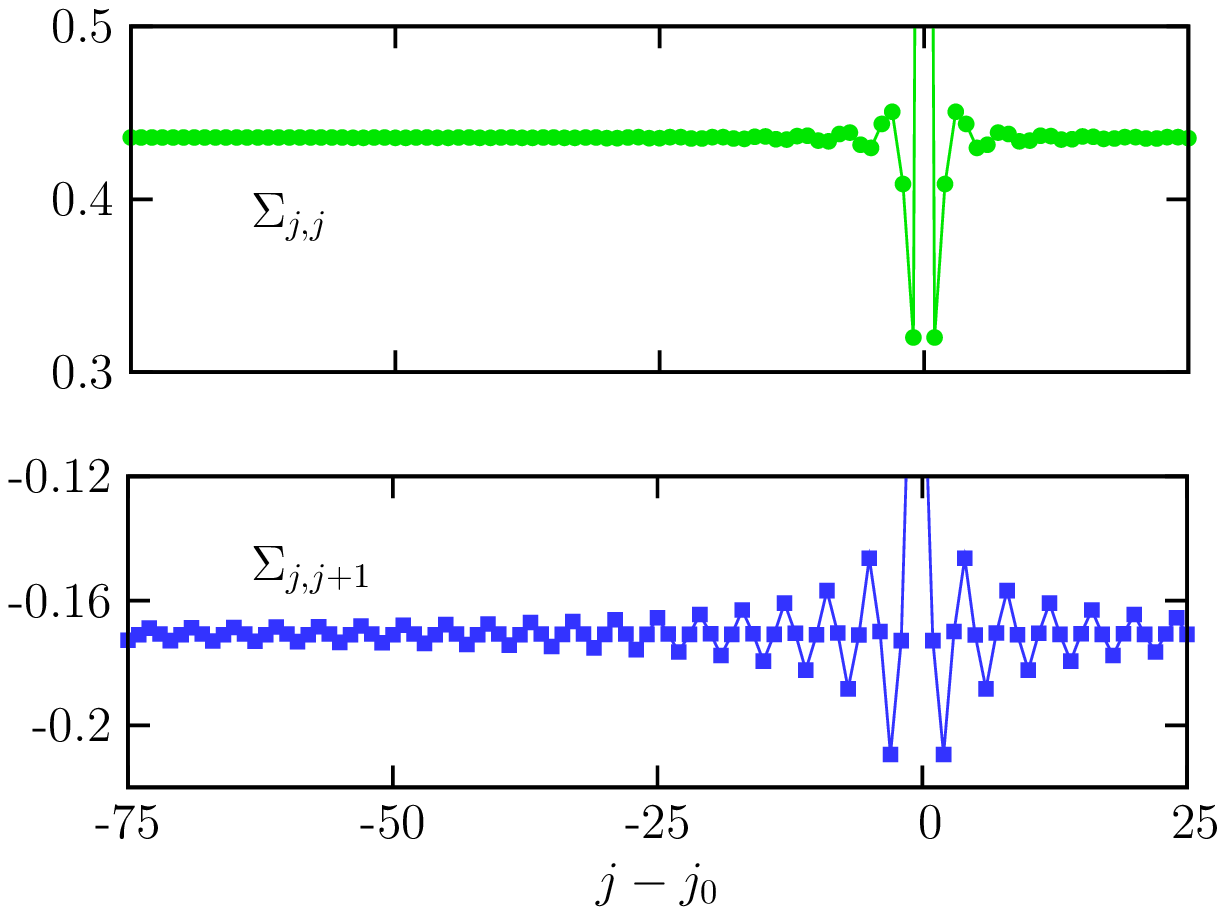}{fig:friedel}{Self-energy near a site impurity of 
  strength $V=1.5$
  for the spinless fermion model at quarter filling and 
  interaction strength $U=1$; the impurity is situated at the
  center of a chain with $L = 1025$ sites.}

\fig[width=10cm]{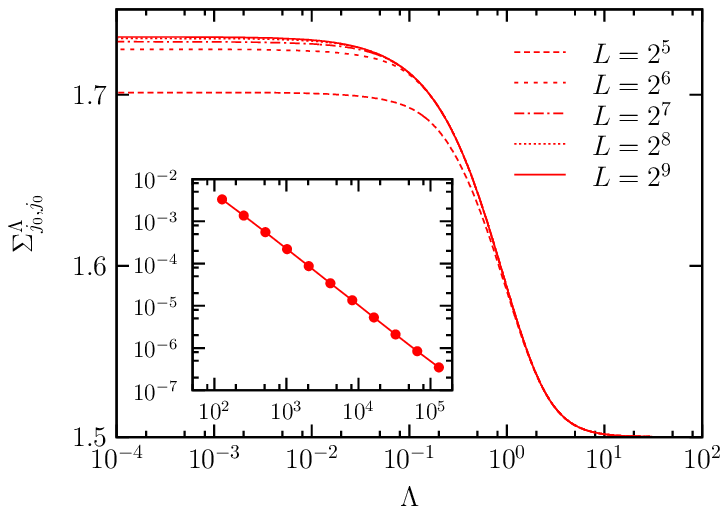}{fig:sigmaimp}{$\Sigma_{j_0,j_0}^{\Lambda}$ 
  as a function of $\Lambda$ 
  for a site impurity of strength $V=1.5$ 
  for spinless fermions at half filling and $U=1$;
  the impurity is situated at the center of a chain of length $L$;
  the inset shows the difference from the asymptotic value
  $\Sigma_{j_0,j_0}^{0}(\infty)-\Sigma_{j_0,j_0}^{0}(L)$ 
  as a function of $L$.}

The amplitude of the oscillations generated by an impurity in
$\Sg$ decays slower than
the inverse distance from the impurity at intermediate length
scales, but approaches a decay proportional to $1/|j-j_0|$ for
$|j-j_0| \to \infty$.
This can be seen most clearly by plotting an effective exponent
$\beta_j$ for the decay of the oscillations, defined as the
negative logarithmic derivative of the oscillation amplitude 
with respect to the distance $|j-j_0|$.
In Fig.~\ref{fig:self} we show the effective exponent resulting from the 
oscillations of $\Sg_{j,j}$ as a function of the distance from 
a site impurity, for $U=1$ and half filling. 
The impurity is situated at the center of a long chain with 
$L = 2^{18}+1$ sites. 
To avoid interferences with oscillations from the boundaries we 
have attached semi-infinite noninteracting leads to the ends 
of the interacting chain, as described in Sec.~\ref{sec:frglutt:obs:cond}.
Only for relatively large impurity strengths the asymptotic regime 
corresponding to $\beta_j = 1$ is reached before finite-size 
effects set in. For small $V$ one can see that $\beta_j$ increases 
from values below one, but the asymptotic long-distance behavior 
is cut off by the boundaries of the interacting region. For very small $V$
(for example $V=0.01$ in Fig.~\ref{fig:self}) we observe a plateau in $\beta_j$ 
for intermediate distances from the impurity site. In this regime $\beta_j$
is close to $K_\rho$ which can be understood by analytically solving the flow 
equations for small $V$ \cite{MMSS02a,MMSS02b}.

\fig[width=10cm]{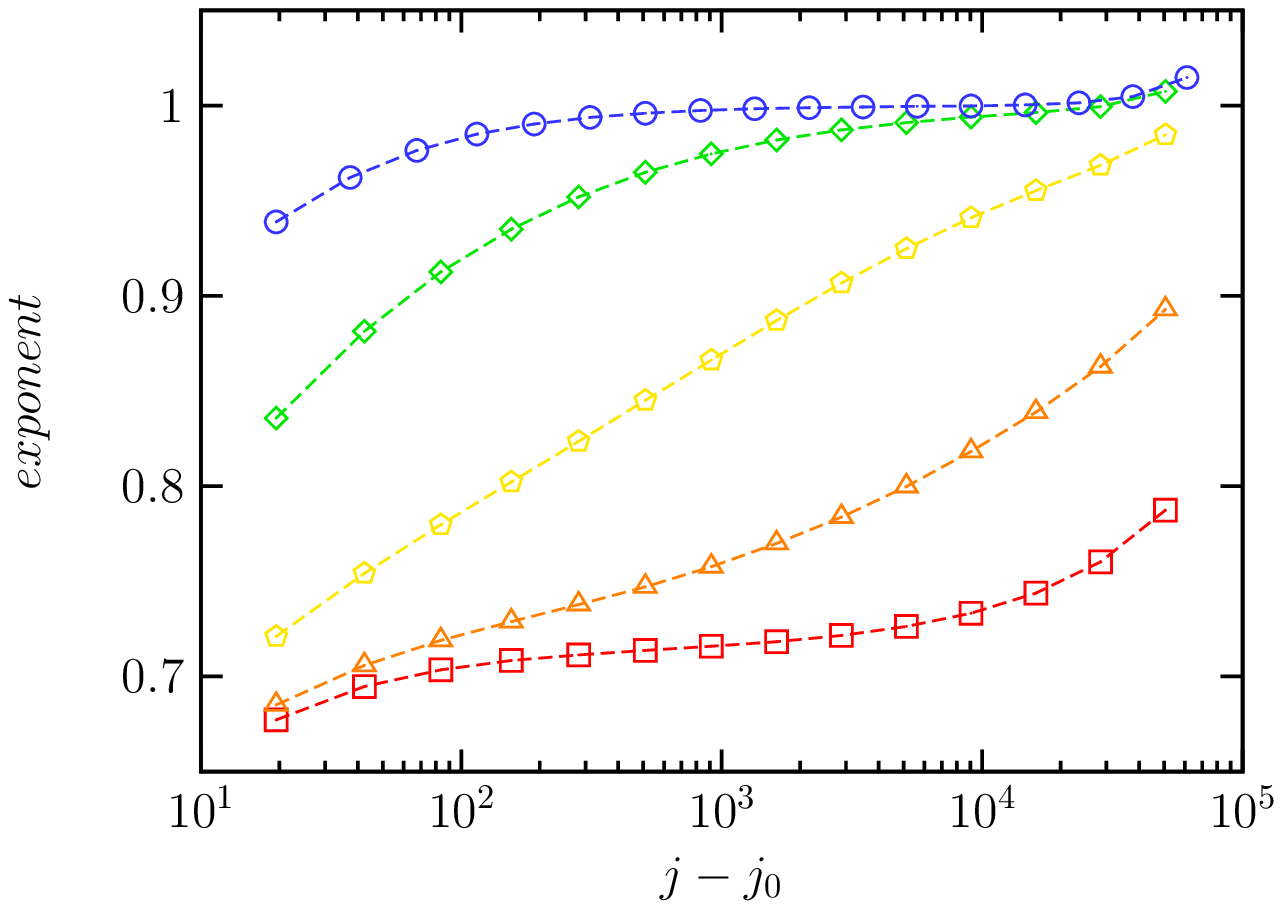}{fig:self}{Effective exponent for the decay 
  of oscillations of 
  $\Sg_{j,j}$ as a function of the distance from a site impurity 
  of strengths $V=0.01$, $0.1$, $0.3$, $1$, $10$ (from bottom 
  to top), for the spinless fermion model at half filling and 
  interaction strength $U=1$; 
  the impurity is situated at the center of a chain with 
  $L=2^{18}+1$ sites.}

%%%%%%%%%%%%%%%%%%%%%%%%%%%%%%%%%%%%%%%%%%%%%%%%%%%%%%%%%%%%%%%%%%%%%%%%%%%%%%

\subsection{Local density of states}
\label{sec:results:spinless:dos}

The long-range $2k_F$-oscillations of the self-energy lead to
a marked suppression of the spectral weight for single-particle
excitations at the Fermi level, that is, at $\om = 0$.
In Fig.~\ref{fig:dos12} we show the local density of states $D_j(\om)$ on
the site next to a site impurity of strength $V=1.5$ for the
spinless fermion model at half filling. The result for the 
interacting system at $U=1$ is compared to the noninteracting 
case. Even-odd effects have been eliminated by averaging over 
neighboring eigenvalues (cf. Sec.~\ref{sec:frglutt:obs:dos}). 
$\delta$ peaks outside the band edges corresponding to bound states 
are not plotted. 
The interaction leads to a global broadening of the band, which
is due to an enhancement of the bulk hopping amplitude, and also
to a strong suppression of $D_j(\om)$ at low frequencies which
is not present in the noninteracting system. 
For a finite system (here $L=1025$) the spectral weight at the
Fermi level remains finite, but tends to zero with increasing 
system size.
In Fig.~\ref{fig:dos14} we show results for the density of states choosing 
the same parameters as in Fig.~\ref{fig:dos12}, but now for densities away
from half filling: $n=1/4$ and $n=3/4$. 
In addition to the dip near $\om=0$ 
a second singularity appears at a finite frequency. This effect
is due to the fact that a long-range potential with a wave number 
$2k_F$ does not only strongly scatter states with momenta near 
$k_F$, but also those with momenta close to $\pi-k_F$. Indeed
the singularity is situated at $\om = \eps_{\pi-k_F} - \mu$,
where $\eps_k$ is the renormalized (bulk) dispersion.
In the half-filled case only one singularity is seen simply
because $\pi-k_F = k_F$ for $k_F = \pi/2$.

\fig[width=10cm]{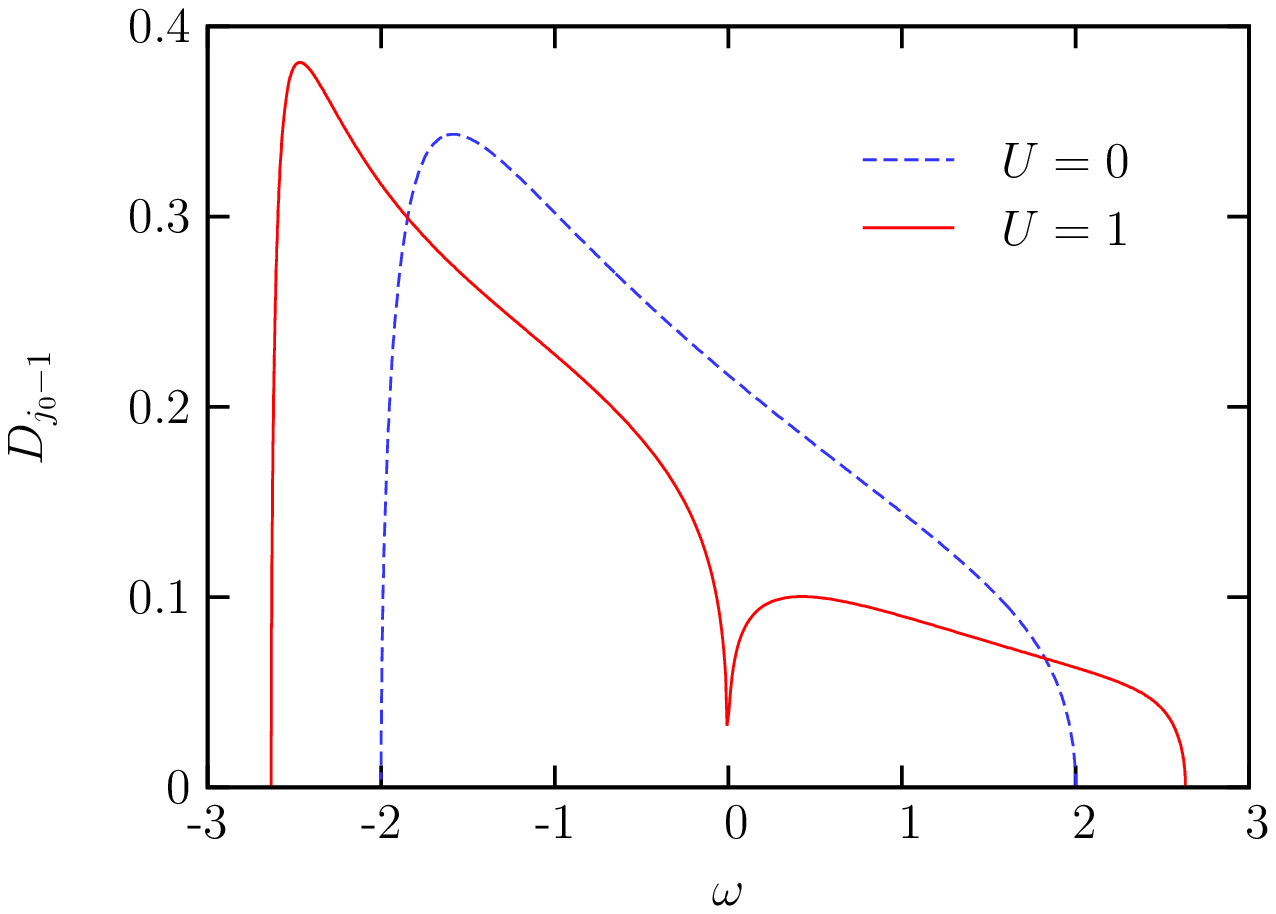}{fig:dos12}{Local density of states 
  on the site next to a site
  impurity of strength $V=1.5$ for spinless fermions at 
  half filling and $U=1$; the impurity is situated at the
  center of a chain with $1025$ sites; the noninteracting
  case $U=0$ is shown for comparison.}

\fig[width=10cm]{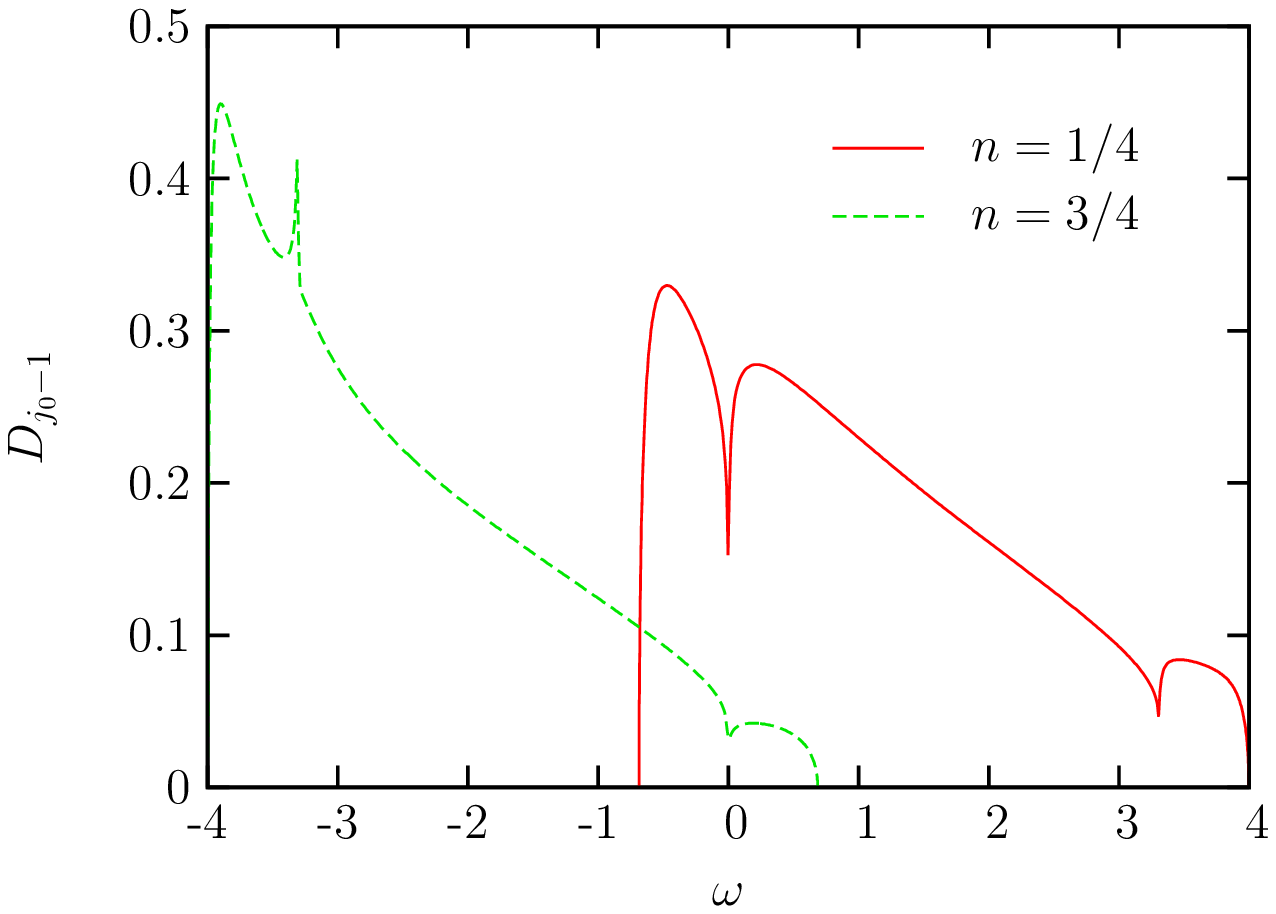}{fig:dos14}{Local density of states on the site next 
  to a site impurity as in Fig.~\ref{fig:dos12} (same parameters), but now for 
  densities $n=1/4$ and $3/4$.}

Similar results are found for a hopping impurity, shown in Fig.~\ref{fig:ldosi}.
For an attractive interaction the density of states is strongly enhanced at the 
Fermi level. 
For comparison, typical results for the local density of states at 
a boundary are presented in Fig.~\ref{fig:ldosb}.

\fig[width=10cm]{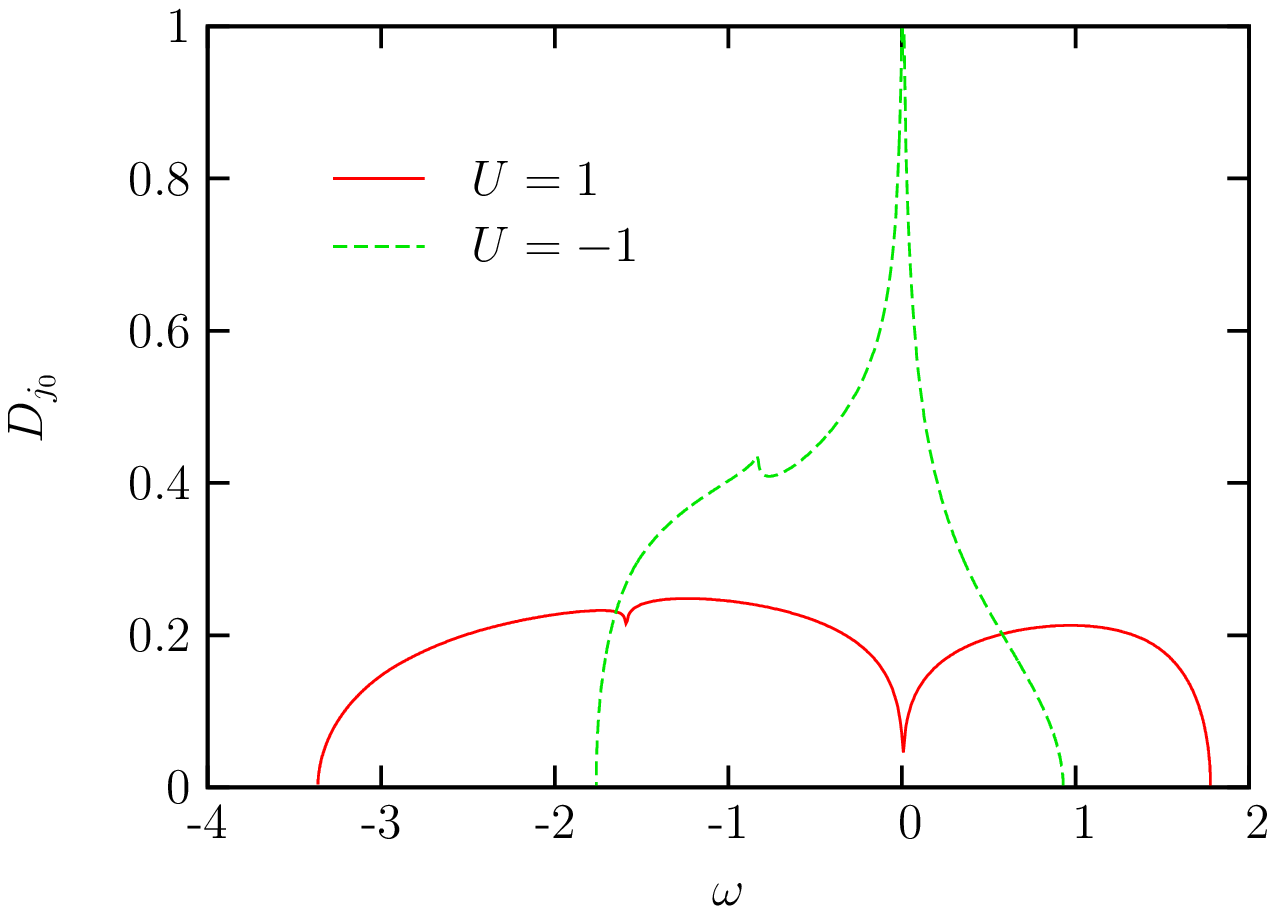}{fig:ldosi}{Local density of states on the site next 
  to a hopping impurity $t'=0.25$  
  for spinless fermions at density $n=0.6$ and $U=1$;
  the impurity is situated at the center of a chain with
  $1024$ sites; 
  the attractive case $U=-1$ is shown for comparison.}

\fig[width=10cm]{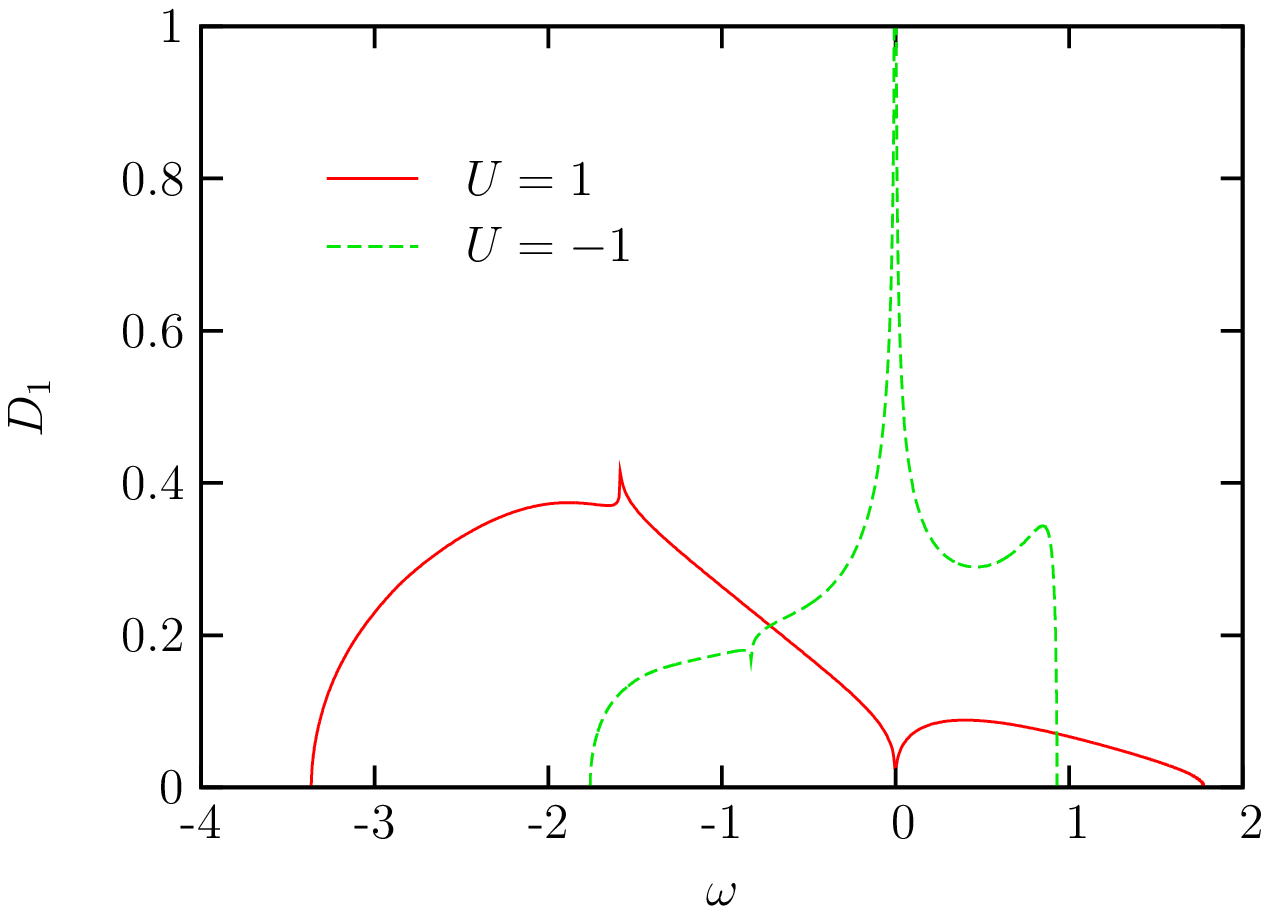}{fig:ldosb}{Local density of states at the 
  boundary,
  for the same parameters as in Fig.~\ref{fig:ldosi}.}

The spectral weight at the Fermi level is expected to vanish
asymptotically as a power law $|\om|^{\alf_B}$, where 
\begin{equation}
  \alf_B = \frac{1}{K_{\rho}} - 1
\end{equation}
is the \emph{boundary exponent} describing the
power-law suppression of the density of states at the boundary
of a semi-infinite chain with repulsive interactions \cite{KF92c}. 
That exponent depends only
on the bulk parameters of the model, not on the impurity
strength. For the spinless fermion model it can be computed 
exactly from the Bethe ansatz solution \cite{Hal80}.

In the above figures the envelope of the $\delta$ peaks of weight $w$
characterizing the spectral function of a finite system size introduced in 
Sec.~\ref{sec:frglutt:obs:dos} is shown. 
As a consequence the energy range over which a power-law suppression is observed
is cut off by the finite size of the system at low energies.
For a reliable analysis of the exponential behavior it is more convenient
to consider the finite-size scaling of the spectral weight at the
chemical potential. 
The large $L$ dependence of the spectral weight is expected to exhibit a 
power law with the same exponent, and the scale where the
power-law behavior sets in is given by $L \sim \om/ \pi v_F$.
In Fig.~\ref{fig:alpha_w} the spectral weight at a boundary of a half-filled
chain of length $L=10^6$ and different interaction strengths $U$ is shown
as a function of $\omega$. The straight line in a log-log plot corresponds to
a power law. 
Results for the same density and interaction parameters, but now 
as a function of system size $L$ are presented in Fig.~\ref{fig:wspinless}.
Within the extension to finite temperatures (cf. Sec.~\ref{sec:frglutt:flow:t}), 
the same power-law behavior is found as a function of temperature,
a detailed analysis in the context of transport phenomena follows below.

\fig[width=10cm]{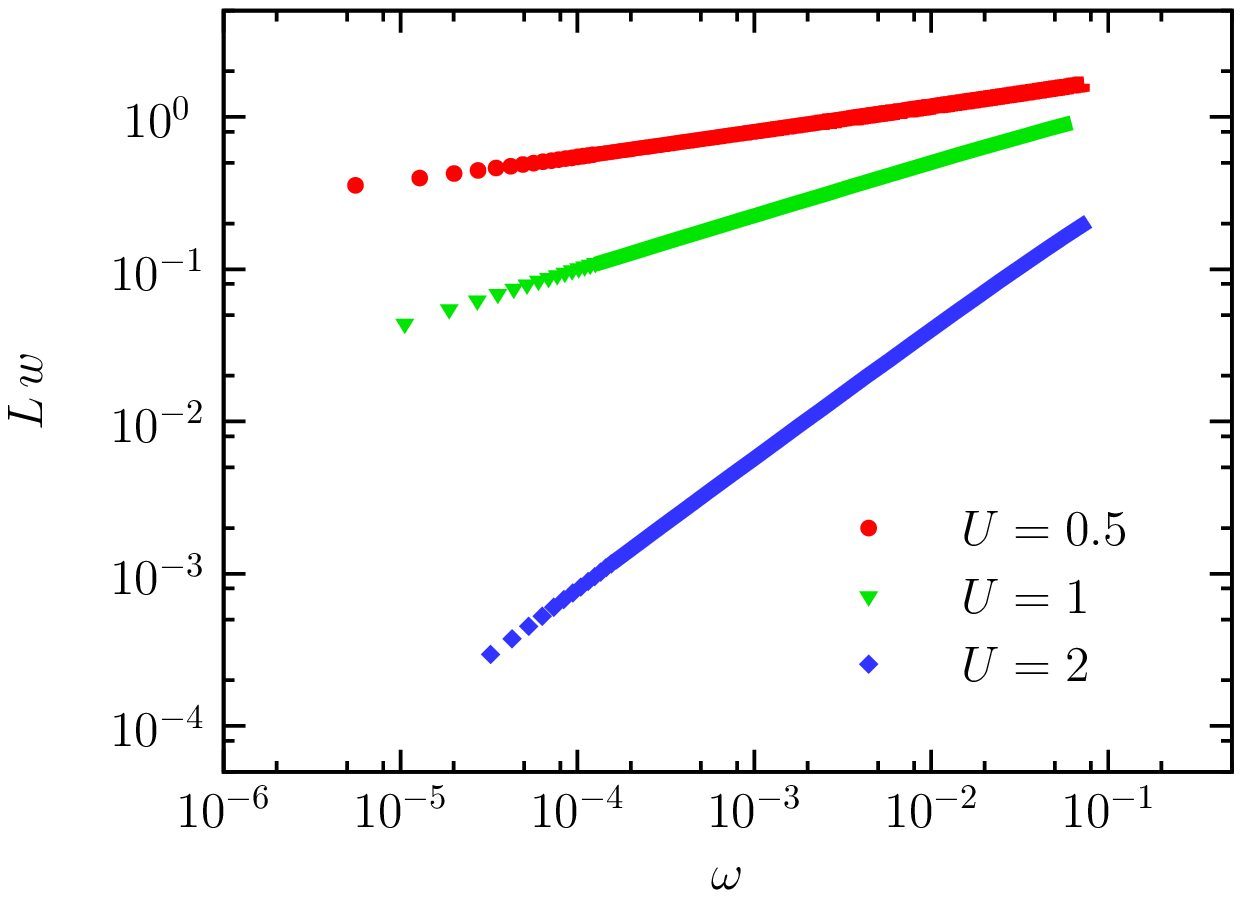}{fig:alpha_w}{Spectral weight at a boundary
  as a function of $\omega$
  for the spinless fermion model at 
  half filling, $L=10^6$ and different interaction strengths $U$.}

\fig[width=10cm]{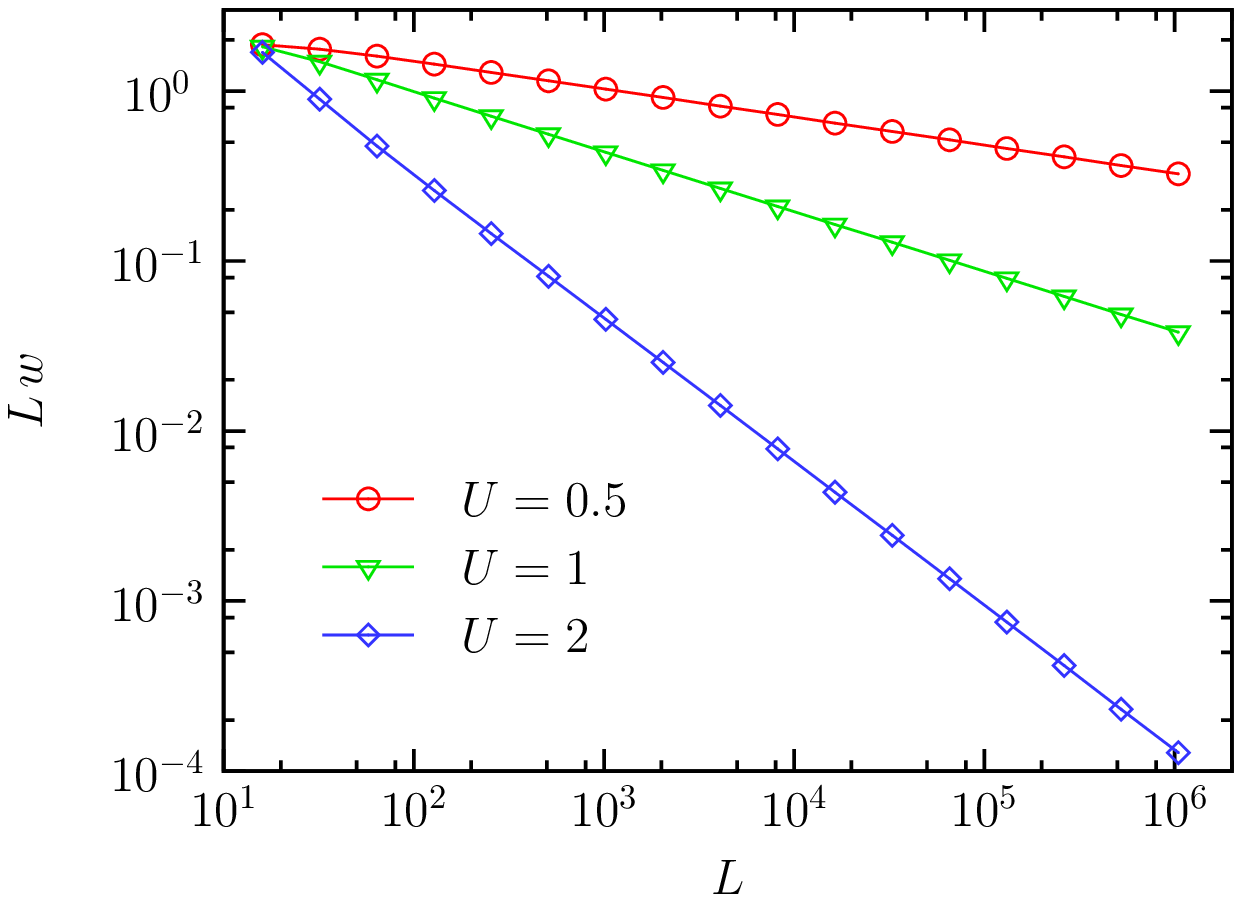}{fig:wspinless}{Spectral weight at the Fermi 
  level as a function of system size $L$, for the same parameters as in 
  Fig.~\ref{fig:alpha_w}.}

\fig[width=10cm]{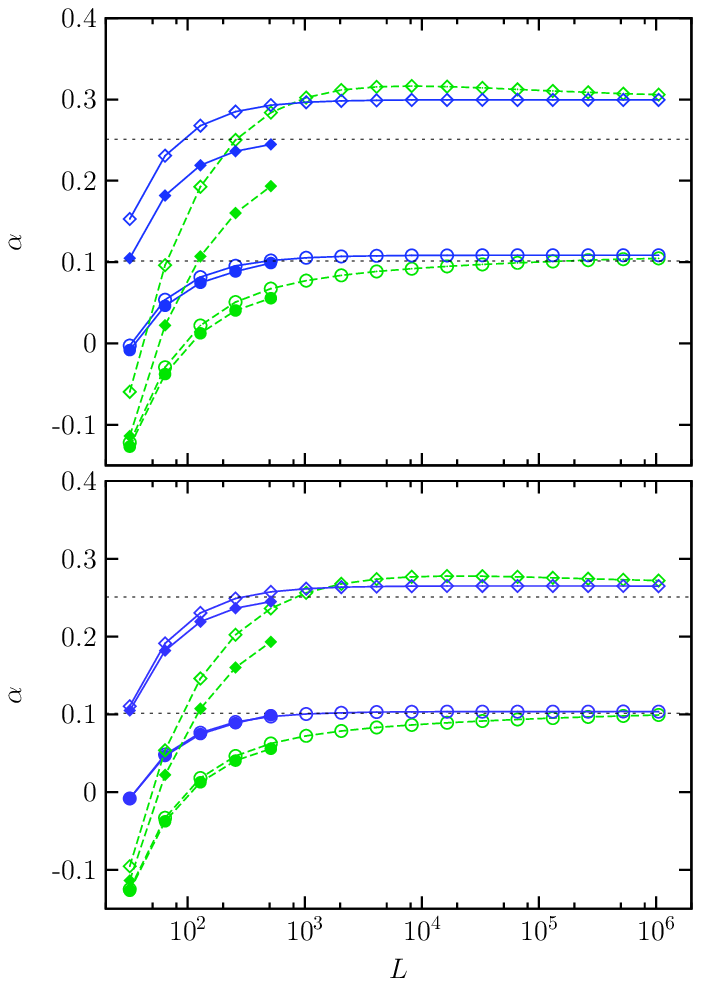}{fig:alpha025}{Logarithmic derivative of the 
  spectral weight at the Fermi 
  level near a boundary (solid lines) or hopping impurity (dashed
  lines) as a function of system size $L$, for spinless fermions at 
  quarter filling and interaction strength $U=0.5$ (circles) or 
  $U=1.5$ (squares); 
  \emph{upper panel}: without vertex renormalization,
  \emph{lower panel}: with vertex renormalization;
  the open symbols are fRG, the filled symbols DMRG results;
  the horizontal lines represent the exact boundary exponents for
  $U=0.5$ and $1.5$.;
  in the boundary case (solid lines) the spectral weight has been 
  taken on the first site of a homogeneous chain,
  in the impurity case (dashed lines) on one of the two sites 
  next to a hopping impurity $t'=0.5$ in the center of the chain.}

We now analyze the asymptotic behavior of the spectral
weight at the Fermi level by defining an effective exponent 
$\alf(L)$ as the negative logarithmic derivative of the spectral 
weight with respect to the system size, such that $\alf(L)$ 
tends to a (positive) constant in case of a power law suppression.
In Fig.~\ref{fig:alpha025} we show results for $\alf(L)$ as obtained from the
fRG for the spinless fermion model at quarter filling 
with up to about $10^6$ sites, for a weak ($U=0.5$) and an
intermediate ($U=1.5$) interaction parameter. 
The spectral weight has been computed either at a boundary, or 
near a hopping impurity of strength $t'=0.5$. Results obtained
from the fRG without (upper panel) and with (lower panel) vertex renormalization,
corresponding to Scheme I and Scheme II 
introduced in Sec.~\ref{sec:frglutt:flow:trunc},
are compared to exact numerical DMRG results (for up to 
512 sites) and the exact boundary exponents $\alf_B$, plotted as 
horizontal lines. 
The fRG results follow a power law for large
$L$, with the same asymptotic exponent for the boundary and
impurity case, confirming thus the expected universality.
However, the asymptotic regime is reached only for fairly large
systems, even for the intermediate interaction strength $U=1.5$.
For the fRG Scheme I 
developed previously for impurities in spinless Luttinger liquids 
the effects of a single static impurity in a spinless Luttinger liquid
are fully captured qualitatively, and in the weak-coupling limit also
quantitatively.
Originally developed for the analysis of spectral densities of single-particle
excitations, this scheme has been applied recently also to transport 
problems, such as persistent currents in mesoscopic rings
and the conductance of interacting wires connected to noninteracting
leads \cite{Med03,MS03b,MS03a}. 
The comparison with the exact DMRG results and exact exponents
shows that the fRG Scheme II is also quantitatively rather
accurate, and that the inclusion of vertex renormalization leads
to a substantial improvement at intermediate coupling strength.

A quantitative estimate of the accuracy of the exponents can be obtained from
a comparison to exact results from the Bethe ansatz solution of the spinless
fermion model \cite{Hal80} shown in Fig.~\ref{fig:deltaalpha}. The results
obtained without vertex renormalization (Scheme I) are represented with open 
symbols. The fRG results are correct to first order in the interaction $U$.
In the approximate treatment of the two-particle vertex 
(cf. Sec.~\ref{sec:frglutt:flow:trunc})
terms of order $U^2$ are only partially included, 
and an agreement to higher order can not be expected. 
Nevertheless the quantitative accuracy is improved considerably.
The $n$ dependence of the accuracy is related to the different importance
of the vertex renormalization
for different densities, as can be seen in Fig.~\ref{fig:vert}.
For half filling the vertex renormalization leads to a pronounced increase of 
the renormalized interaction, whereas for smaller fillings the effect is 
smaller, eventually leading to a decrease for $n\lesssim 1/3$.

\fig[width=10cm]{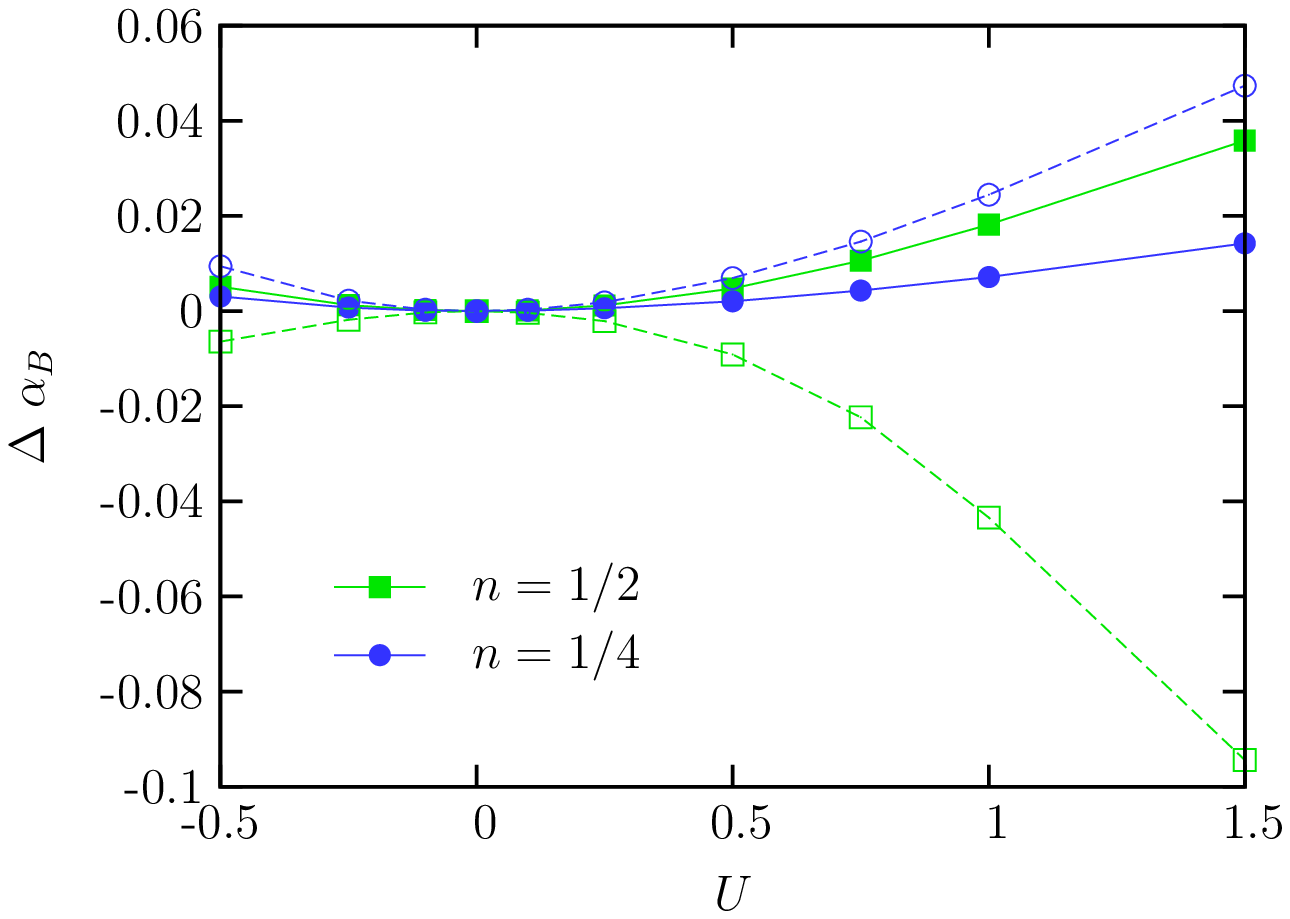}{fig:deltaalpha}{Difference between the fRG result 
  and the exact Bethe ansatz result for the boundary exponent $\alpha_B$
  as a function of $U$, at densities $n=1/2$ 
  and $1/4$ for the spinless fermion model
  as obtained from the power-law
  suppression of the spectral weight at the boundary;
  the open symbols are results without vertex renormalization,
  the filled symbols results with vertex renormalization.}

Results for the effective exponent $\alf$ in the case of a site
impurity are shown in Fig.~\ref{fig:imp}, at quarter filling and for an
interaction strength $U=1$. The comparison of the different
curves obtained for different impurity strengths confirms once 
again the expected asymptotic universality, and also how the 
asymptotic regime shifts rapidly toward larger systems as the 
bare impurity strength decreases.
The crossover scale depends on the bare impurity strength $V$ as 
\cite{KF92b,KF92a,FLS95} 
\begin{equation}
  \label{eq:imps}
  \frac{1}{L_c} \sim \Big(\frac{V}{\pi v_F}\Big)^{\frac{1}{1-K_{\rho}}} \;,
\end{equation}
in agreement with our findings.
The scale on which the impurity flows to strong coupling 
depends on the initial strength
$V$ and the ``flow velocity'' given by $1-K_{\rho}$. 
The effective flowing impurity strength can be estimated by
\begin{equation}
  V_{\rm eff} \sim V L^{1-K_{\rho}} \; ,
\end{equation}
where the crossover scale $L_c$ in Eq.~(\ref{eq:imps}) 
corresponds to $V_{\rm eff} \sim \pi v_F$. 
Note that for weak impurities and intermediate system sizes the 
spectral weight follows a power law corresponding to the bulk behavior, 
and approaches the boundary exponent only at large distances.
The bulk suppression of the spectral weight described by the anomalous 
dimension $\alpha=(K_{\rho}^{\phantom '}+K_{\rho}^{-1}-2)/2$ is not 
captured within the present scheme, since the self-energy is frequency 
independent.
In the 1PI version of the fRG the frequency 
dependence is generated by the two-particle vertex.
The anomalous dimension could be included in an improved scheme by 
an iterative solution of the fRG flow inserting 
the two-particle vertex into the flow equation for the self-energy 
without neglecting its frequency dependence.
This gives a two-loop diagram including the full
two-particle vertex functions at scale $\Lam$. The
right-hand side of the differential equation for the self-energy
is then nonlocal in $\Lam$: the change of the two-particle vertex
and self-energy at scale $\Lam$ involves two-particle vertices and
self-energies at scales $\Lam \geq \Lam'$. 
In presence of sufficiently strong effective impurity potentials however, the
boundary behavior prevails over the bulk suppression of the 
spectral weight, as $\alpha_B \sim U$ and $\alpha \sim U^2$.

\fig[width=10cm]{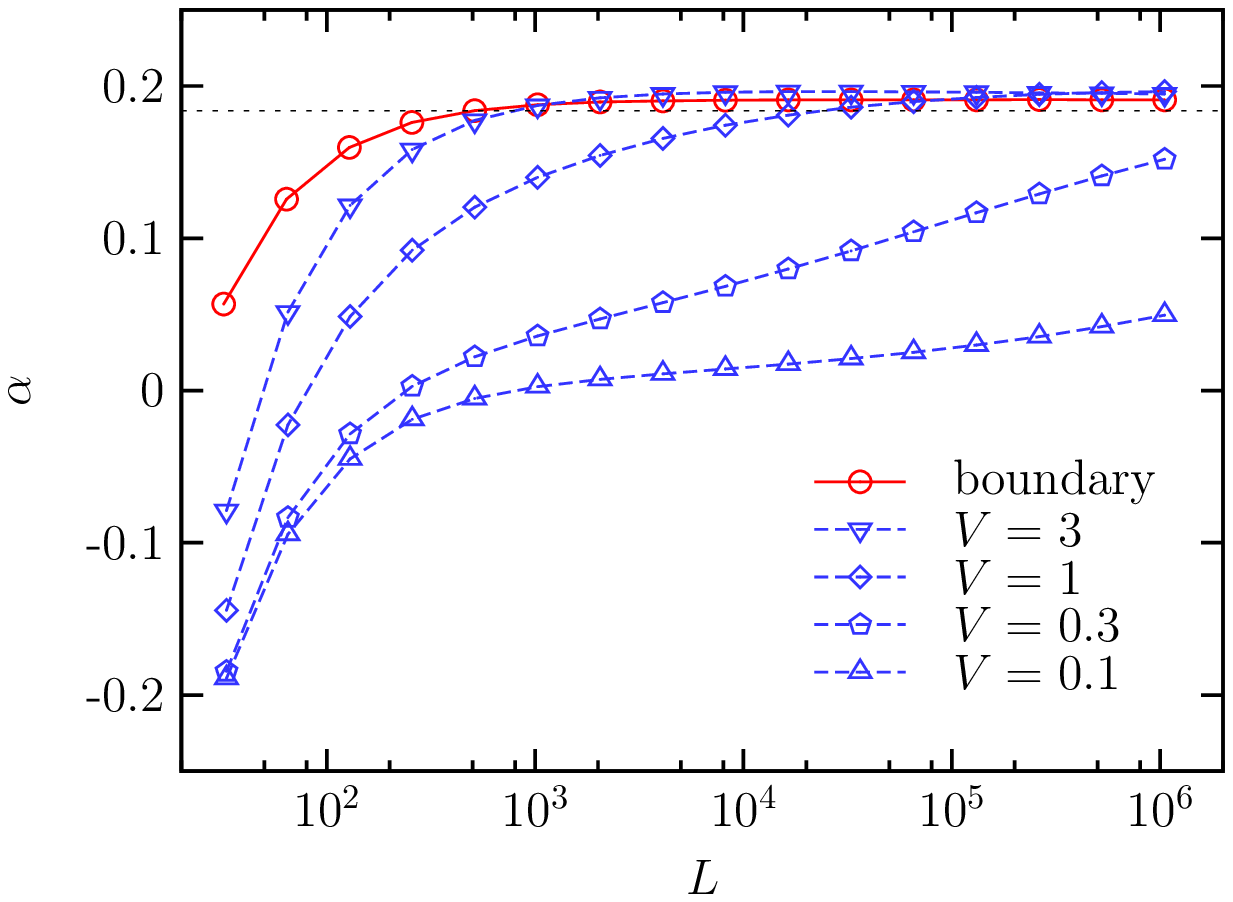}{fig:imp}{Logarithmic derivative of the spectral weight 
  at the Fermi level near a boundary (solid line) or site impurity 
  (dashed lines) as a function of system size $L$, for spinless 
  fermions at quarter filling and interaction strength $U=1$;
  in the boundary case the spectral weight has been 
  taken on the first site of a homogeneous chain, in the impurity 
  case on the site next to a site impurity of strength $V$ in the 
  center of the chain; 
  the horizontal line represents the exact boundary exponent for
  $U=1$.}

%%%%%%%%%%%%%%%%%%%%%%%%%%%%%%%%%%%%%%%%%%%%%%%%%%%%%%%%%%%%%%%%%%%%%%%%%%%%%%

\subsection{Friedel oscillations}
\label{sec:results:spinless:density}

We now discuss results for the density profile $n_j$.
Boundaries and impurities induce Friedel oscillations of the
local density with a wave vector $2k_F$.
In a noninteracting system these oscillations decay 
proportionally to the inverse distance from the boundary or
impurity. 
In an interacting Luttinger liquid the Friedel oscillations are 
expected to decay as $|j-j_0|^{-K_{\rho}}$ at long distances 
$|j-j_0|$.
For a very weak impurity one expects a slower decay proportional
to $|j-j_0|^{1-2K_{\rho}}$ at intermediate distances, and a
crossover to the asymptotic power law with exponent $K_{\rho}$
at very long distances \cite{EG95}. At intermediate distances the
response of the density to a weak impurity can be treated in
linear response theory, such that the density modulation is
determined by the density-density response function at $2k_F$, 
which leads to the power-law decay with exponent $2K_{\rho}-1$.

\fig[width=10cm]{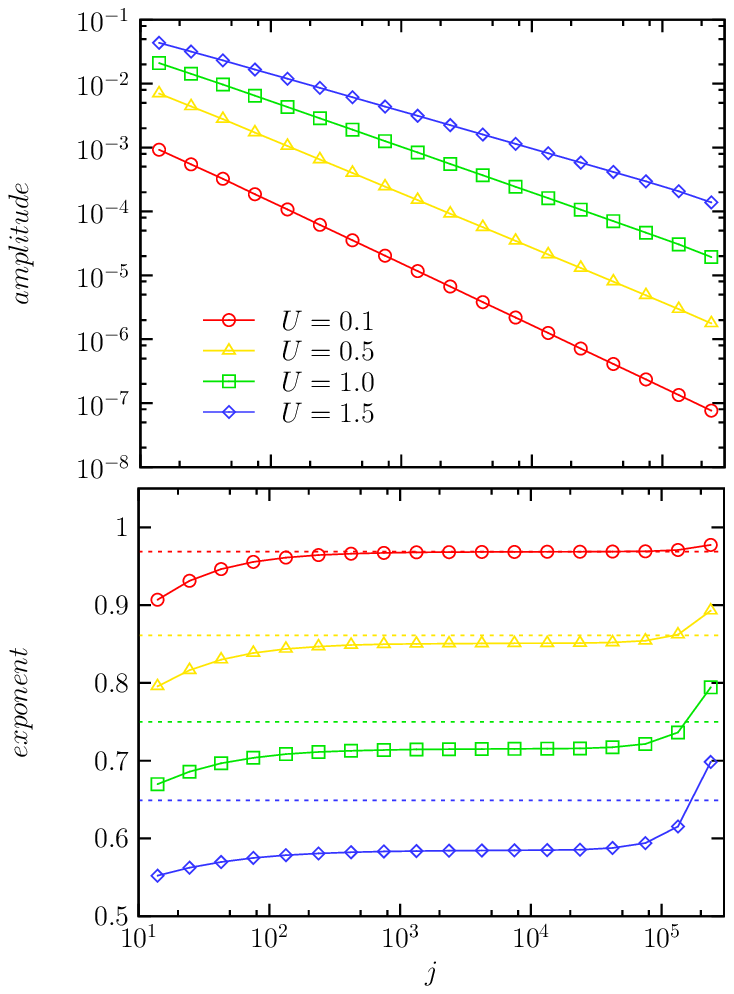}{fig:friedelexpU}{Amplitude (envelope) 
  of oscillations of the density 
  profile $n_j$ induced by a boundary as a function of the distance 
  from the boundary, for spinless fermions with various interaction
  strengths $U$ at half filling; the interacting chain with
  $2^{19}+1$ sites is coupled to a semi-infinite noninteracting
  lead at the end opposite to the boundary; 
  \emph{upper panel}: log-log plot of the amplitude,
  \emph{lower panel}: effective exponents for the decay, and the 
  exact asymptotic exponents as horizontal lines.}

We analyze the long-distance behavior of the amplitudes
more closely for the half-filled case, and compare to exact
results for the asymptotic exponents.
For incommensurate filling factors the density profile looks
more complicated, but at long distances from the boundary the 
oscillation amplitude has a well-defined envelope which exhibits
a power law as a function of $j$.
In Fig.~\ref{fig:friedelexpU} we show fRG results for the amplitude of
density oscillations emerging from an open boundary, for a
very long spinless fermion chain with $2^{19}+1$ sites and
various interaction strengths $U$.
The end opposite to the open boundary is
smoothly connected to a noninteracting lead. 
In a log-log plot (upper panel of Fig.~\ref{fig:friedelexpU}) the amplitude 
follows a straight line for almost all $j$, corresponding to a power-law 
dependence. 
Deviations from a perfect power law can be seen more neatly by
plotting the effective exponent $\alf_j$, defined as the
negative logarithmic derivative of the amplitude with respect
to $j$ (see the lower panel of Fig.~\ref{fig:friedelexpU}). The effective 
exponent is almost constant except at very short distances or
when $j$ approaches the opposite end of the interacting chain,
which is not surprising. From a comparison with the exact 
exponent (horizontal lines in the figure) one can assess the 
quantitative accuracy of the fRG results.

\fig[width=10cm]{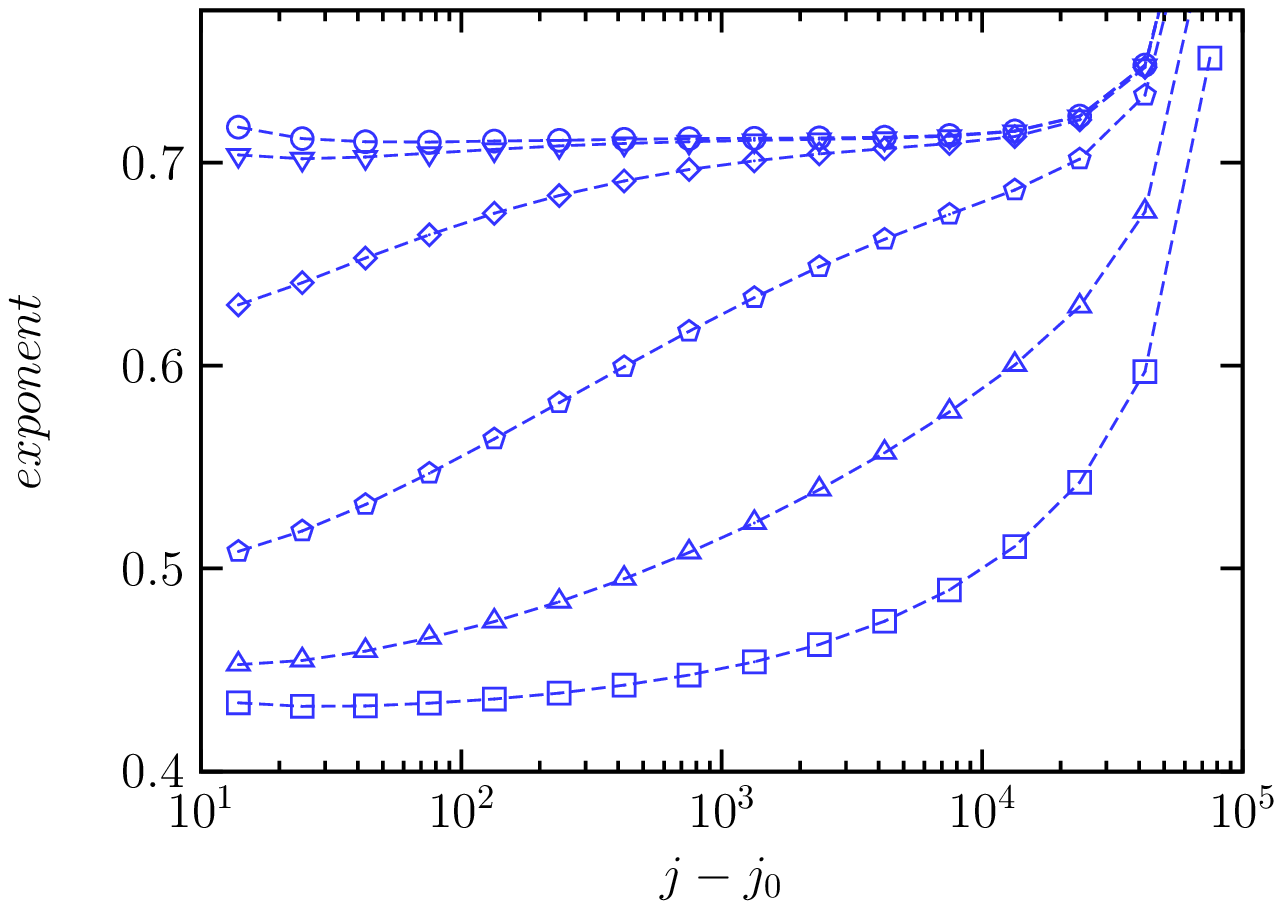}{fig:friedelexpV}{Effective exponent 
  for the decay of density oscillations
  as a function of the distance from a site impurity of strengths
  $V=0.01$, $0.1$, $0.3$, $1$, $3$, $10$ (from bottom to top); 
  the impurity is situated at the center of a spinless 
  fermion chain with $2^{18}+1$ sites and interaction strength 
  $U=1$ at half filling; the interacting chain is coupled to
  semi-infinite noninteracting leads at both ends.}

Effective exponents describing the decay of Friedel oscillations 
generated by site impurities of various strengths are shown in 
Fig.~\ref{fig:friedelexpV}, for a half-filled spinless fermion chain with 
$2^{18}+1$ sites and interaction $U=1$. Both ends of the interacting chains 
are coupled to noninteracting leads to suppress oscillations 
coming from the boundaries.
For strong impurities the results are close to the boundary result
(cf.\ Fig.~\ref{fig:friedelexpU}), as expected. 
For weaker impurities the oscillations decay more slowly, that is,
with a smaller exponent, and approach the boundary behavior only 
asymptotically at large distances (beyond the range of our chain
for $V<1$). For very weak impurities ($V=0.01$ in Fig.~\ref{fig:friedelexpV}) 
the oscillation amplitude follows a power law corresponding to the 
linear response behavior with exponent $2K_{\rho} - 1$ at 
intermediate distances.
Similar results are obtained for attractive interactions \cite{AEMMSS04}.

A comparison with the exact exponents from the Bethe ansatz solution
of the spinless fermion model \cite{Hal80} is shown in
Fig.~\ref{fig:deltak}, where the difference $\Delta K_{\rho}$ is 
reported as a function of the interaction $U$ at densities $n=1/2$ and $n=1/4$.
The open symbols represent the accuracy for the
weak-impurity behavior in the linear response regime and the filled ones for
the asymptotic exponent for strong impurities at long distances.
Deviations from the exact result are quadratic in the interaction $U$.

\fig[width=10cm]{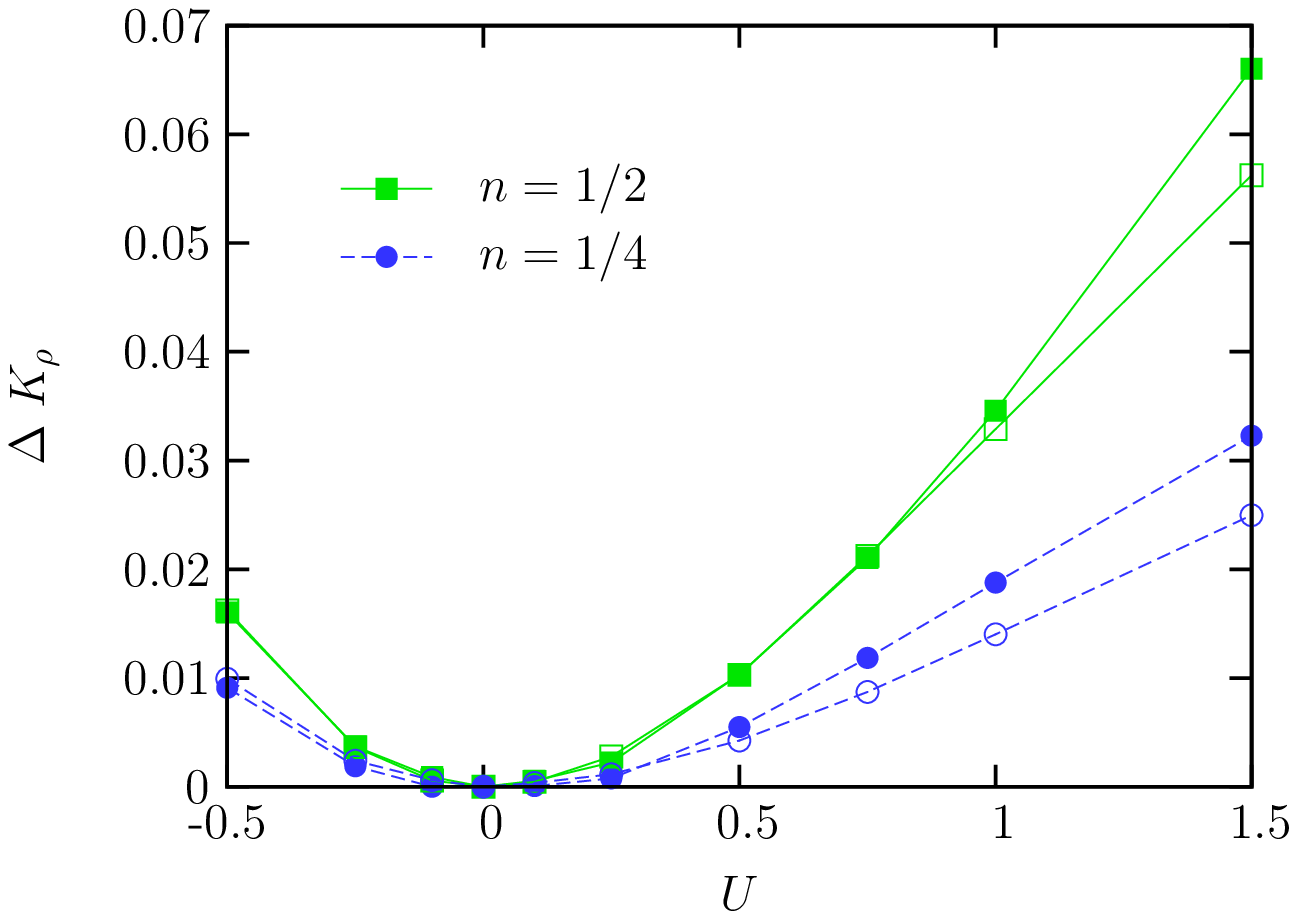}{fig:deltak}{Difference between the fRG result and
  the exact Bethe ansatz result   for the Luttinger-liquid parameter $K_{\rho}$ 
  as a function of $U$ at densities $n=1/2$ and $1/4$ for the spinless 
  fermion model, as obtained 
  from the power-law decay of Friedel oscillations generated by a strong impurity 
  at long distances (filled symbols), and by a weak impurity 
  at intermediate distances (open symbols).}

%%%%%%%%%%%%%%%%%%%%%%%%%%%%%%%%%%%%%%%%%%%%%%%%%%%%%%%%%%%%%%%%%%%%%%%%%%%%%%

\subsection{Scaling of the conductance}
\label{sec:results:spinless:oneparam}

In this section we study the transport through an interacting wire with a single
impurity connected to semi-infinite noninteracting leads. 
The conductance exhibits the same asymptotic scaling behavior
as a function of temperature $T$ for an infinite wire 
as at $T=0$ as a function of $L$.
For more than one impurity the $T$ dependence is richer, showing 
nonmonotonic behavior and distinctive power-laws with different universal
exponents in various regimes;
an extensive analysis is reported in Ref.~\cite{EMABMS04,EnssThesis}. 
Although this difference does not appear for a single impurity, we will mainly 
focus on the more physical temperature dependence of the conductance.

Before analyzing the scaling behavior we present in Fig.~\ref{fig:g_u}
a comparison of fRG results 
to numerical DMRG data for the conductance for a short wire 
at $T=0$, determined from the persistent current observed in the presence of 
a magnetic flux piercing a noninteracting ring in which the interacting wire 
is embedded \cite{Med03,MS03b,MS03a}. 
The excellent agreement proves the reliability of the approximate fRG scheme 
for interactions in the range $1/2 \leq  K_{\rho} \leq 1$.

\fig[width=10cm]{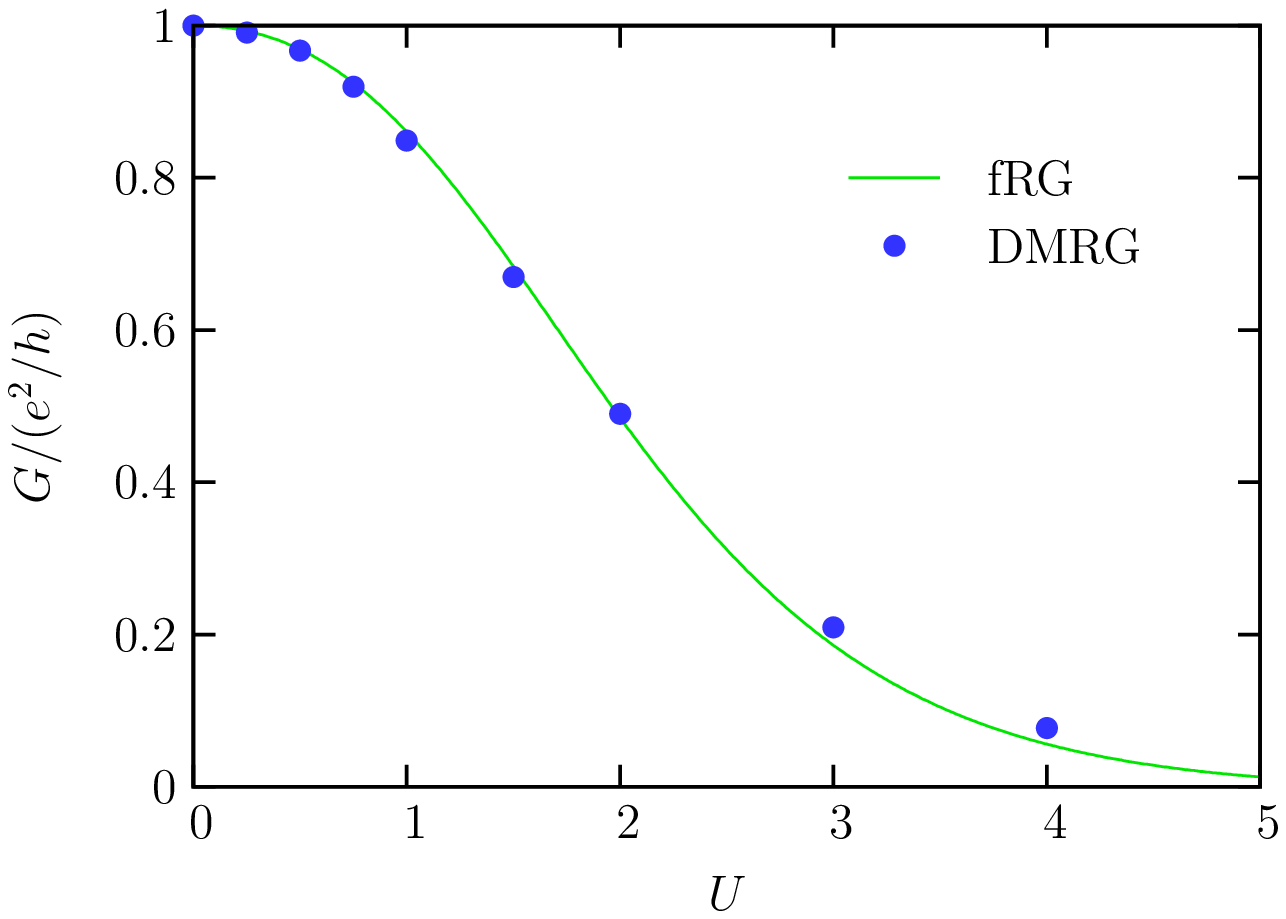}{fig:g_u}{Conductance as a function of the 
  interaction $U$ for a homogeneous spinless
  fermion chain at half filling, with $L=12$ sites; 
  the interaction is turned on sharply at the contacts.}

Fig.~\ref{fig:g_Tv} shows typical fRG results for the $T$ dependence of the 
conductance through a single site impurity of strength $V$. 
The $1/T$ scaling observed for high temperatures (of the order of the bandwidth)
results from the $1/T$ behavior of the derivative of the Fermi function in 
Eq.~(\ref{eq:conductf}), together with the
weak temperature dependence of $|t(\varepsilon,T,L)|^2$ at high $T$. 
For a strong impurity $V=10$, $G(T)$ follows a
power law with exponent $2\alpha_B$ as indicated by the dashed line in
Fig.~\ref{fig:g_Tv}, until saturation sets in for
$T \sim \pi v_F/L$.
For an intermediate impurity
the slope of the data tends towards the asymptotic exponent, 
but is still significantly away from it when finite-size saturation 
sets in.
This slow change of the slope is a general feature of intermediate $V$. 
For a weak impurity $G(T)$ approaches $e^2/h$. 
Similar behavior is found for 
the scaling of  $1-G(T)/(e^2/h)$ in the limit of a weak impurity  
predicted to follow $T^{2(K_{\rho}-1)}$, which holds as long as the
correction to perfect conductance stays small \cite{KF92b,KF92a,FLS95}.

\fig[width=10cm]{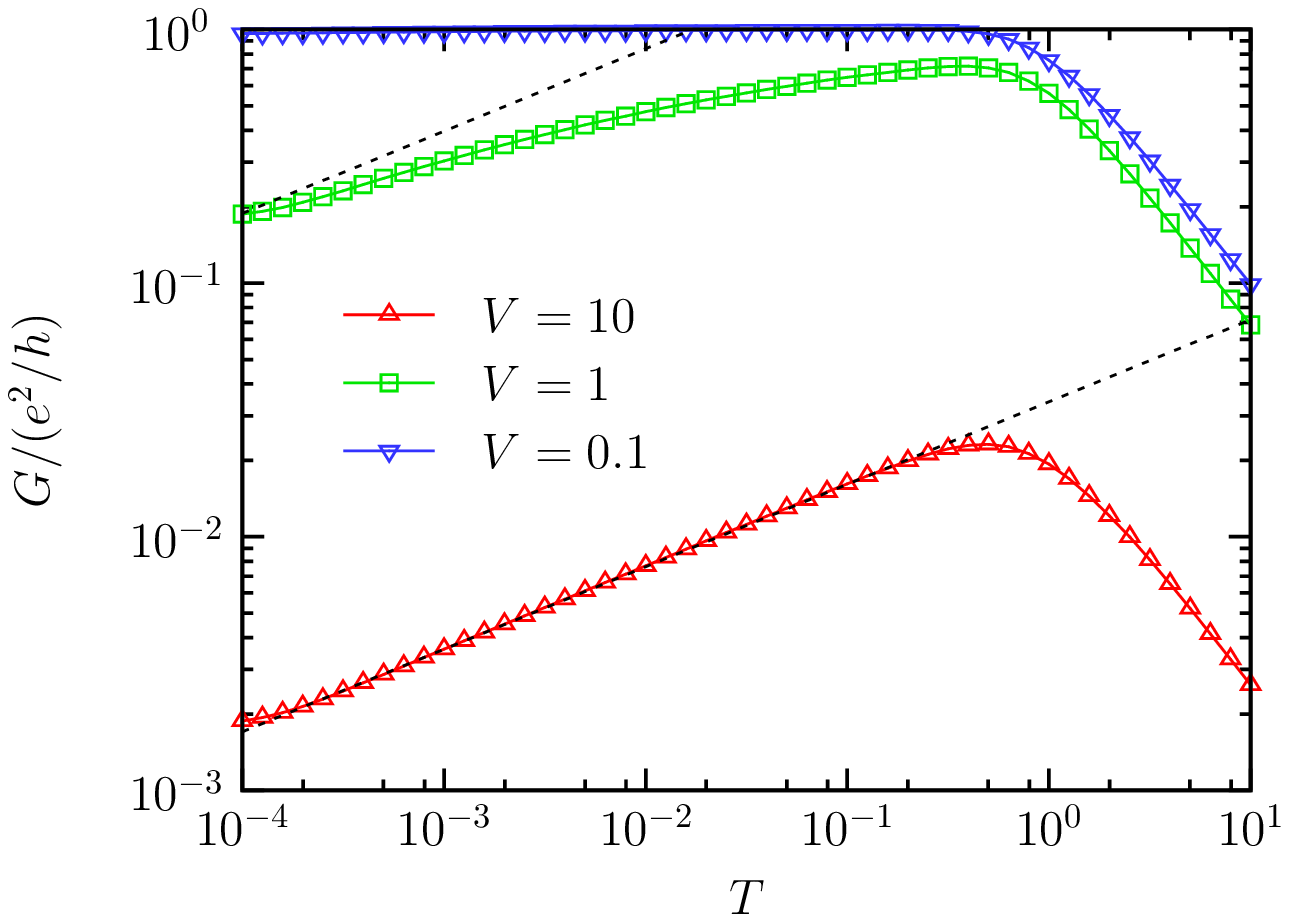}{fig:g_Tv}{Temperature dependence of the conductance 
  for a half-filled spinless fermion wire 
  of length $L=10^4$, interaction $U=0.5$ and a single site
  impurity of strengths $V$ at the center of the wire;
  the dotted lines highlight power-law behavior.}

The above results are generic as long as the impurity is placed
sufficiently away from the contact regions.
The scale
$\delta_{j_0} = \pi v_F/j_0$, where $j_0$ is the impurity position,
sets a lower bound for the
power-law scaling with the exponents discussed above \cite{FN96}. 
For $T \simeq \delta_{j_0}$ a crossover to a power-law scaling with 
different exponents is found. For impurity positions in the contact region the 
exponent $\alpha_B$ describes the tunneling between 
the noninteracting and the interacting Luttinger liquids \cite{EMABMS04}. 
Restrictions on the temperature range where universal scaling behavior
might be detected arise from the bandwidth from above and the finite wire length 
from below. For an interacting wire of $L$ lattice sites the energy 
scale $\delta_L=\pi v_F/L$ represents a lower bound for any temperature 
scaling.
Depending on the impurity and interaction strength an asymptotic
low-energy regime might not be reachable in experiments on finite wires.
For finite temperatures systems of $10^4$ lattice sites are 
considered, comparable for typical lattice constants 
to quantum wires in the micrometer range accessible to transport experiments.

Considering the conductance as a function of temperature and
impurity strength, for a fixed 
$K_{\rho}$ the renormalization-group flow from weak to strong impurity strength 
determines a scaling function $\tilde G_{K_{\rho}}(x)$ on which the data for 
different $T$ and $V$ collapse \cite{KF92b,KF92a,Moo93,FLS95}.
Using a one-parameter scaling ansatz 
\begin{eqnarray}
  \label{oneparascaling}
  G = \frac{e^2}{h} \tilde G_{K_{\rho}}(x) \; , \qquad {\rm with}\qquad 
  x=\left[ T/T_0(U,n,V)\right]^{K_{\rho}-1} \, ,
\end{eqnarray}
the curves for $G(T)$ and different $V$
can be collapsed onto the $K_{\rho}$-dependent scaling function $\tilde
G_{K_{\rho}}(x)$ for an appropriate nonuniversal scale 
$T_0(U,n,V)$. 
It has the limiting behavior $\tilde G_{K_{\rho}}(x) \sim 1-x^2$ for $x
\to 0$, and $\tilde G_{K_{\rho}}(x) \sim x^{-2/K_{\rho}}$ for $x \to \infty$;
for $K_{\rho}=1/2$ and $K_{\rho}=1/3$ 
the functional form of $\tilde G_{K_{\rho}}$ 
was determined explicitly \cite{KF92a,Moo93,FLS95}.
An example is shown in Fig.~\ref{fig:singleparam} for $U=0.5$,
the different colors
stand for different impurity strengths $V$. 
As a consequence of the extended crossover region
between weak and strong-impurity behavior, 
even for the fairly large system size of 
$L=10^4$ sites and the large range of temperatures we can treat,
it is impossible to directly demonstrate the full 
crossover for a single set of parameters. 
A power-law behavior in both limits is found, with exponents which can be 
expressed consistently in terms of a single 
approximate Luttinger-liquid parameter $K_{\rho}$. 
Data for the same $K_{\rho}$ but different interaction and filling 
parameters collapse on the results for half filling, since the scaling 
function depends on $U$ and $n$ only 
via the Luttinger-liquid parameter $K_{\rho}$.   
Considering different types of impurity potentials extending over more
than one site or bond does not modify $\tilde G_{K_{\rho}}$.

\fig[width=10cm]{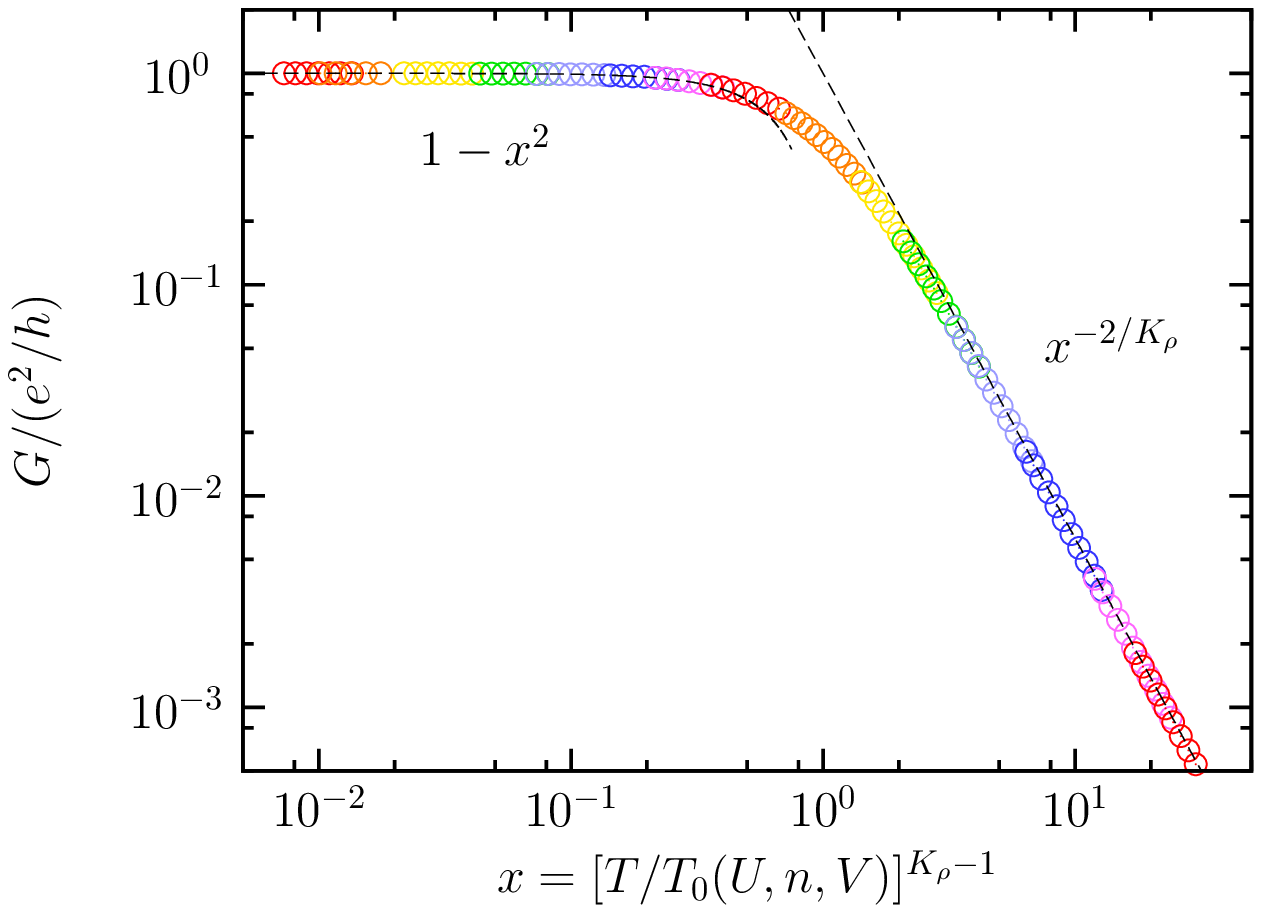}{fig:singleparam}{One-parameter scaling plot 
  of the conductance for the spinless fermion model at half filling with 
  $U=0.5$; the colors represent results obtained for different impurity 
  strengths; the dashed lines indicate the asymptotic behavior for 
  small and large $x$.}

A previous analysis of the zero temperature scaling behavior in the 
wire length $L$, replacing $T$ in the above ansatz by $\pi v_F /L$, 
showed that one-parameter scaling is not affected 
by the presence of leads if the interaction is 
turned on very smoothly at the contacts 
and no one-particle scattering terms at the contacts are considered 
\cite{Med03}. 
In addition, fRG data are found to collapse onto the $K=1/2$ local 
sine-Gordon scaling function known analytically \cite{Med03,EMABMS04}.
We finally remark that one-parameter scaling 
represents an excellent example for the power of the fRG technique 
capturing complex crossover phenomena at intermediate scales.

There is an interesting 
correspondence between the transport through impurities in a Luttinger 
liquid and the quantum Brownian motion in a cosine potential examined in 
Ref.~\cite{Wei99}. The duality symmetry of the latter
maps a weak impurity (small $V$) exactly to a strong impurity (large $V$) 
under the substitution $K_{\rho} \to 1/K_{\rho}$.
For $K_{\rho}<1$ the system becomes localized, whereas the effective
barrier height vanishes for $K_{\rho}>1$.
This symmetry can be extended also to finite temperatures
by a frequency-dependent transformation \cite{Wei99}.
The equivalence between the mobility of the Brownian particle and the
conductance through an impurity in a Luttinger liquid relates
the behavior for a strong impurity to the one for an appropriate weak
impurity by
\begin{equation}
  G_{\rm strong}(T,K_{\rho})/(e^2/h)=1-G_{\rm weak}(T,1/K_{\rho})/(e^2/h) \; .
\end{equation}
In particular, the above relation holds for expansions around $V\ll 1$ and 
$V\gg 1$. The respective convergence radius defines
the crossover scale between the two dual descriptions by 
$T_c^{\rm strong}(K_{\rho})=T_c^{\rm weak}(1/K_{\rho})$. For $K_{\rho}>1$ 
high and low temperatures 
are exchanged. Similar scaling behavior is found for the nonlinear conductance,
where the bias assumes the role of the temperature \cite{Wei99}.

As a consequence of this symmetry 
and the knowledge of the analytic form for a particular value of 
$K_{\rho}\neq 1$ the explicit scaling function can be derived,
as well as an expression for the crossover scale \cite{Wei99}.
The solution reproduces the result from the thermodynamic Bethe ansatz 
\cite{FLS95}.

%%%%%%%%%%%%%%%%%%%%%%%%%%%%%%%%%%%%%%%%%%%%%%%%%%%%%%%%%%%%%%%%%%%%%%%%%%%%%%

\section[Spin-$\frac{1}{2}$ fermions]{Spin-$\mathbf{\frac{1}{2}}$ fermions}
\label{sec:results:spin}

%%%%%%%%%%%%%%%%%%%%%%%%%%%%%%%%%%%%%%%%%%%%%%%%%%%%%%%%%%%%%%%%%%%%%%%%%%%%%%

\subsection{Single-particle excitations}
\label{sec:results:spin:dos}

For $\omega \to 0$ the spectral weights and the local density of states near 
a boundary or impurity are ultimately suppressed according to a power law with 
the boundary exponent
\begin{equation}
  \label{eq:ab}
  \alpha_B =\frac{1}{2K_{\rho}}+\frac{1}{2K_{\sigma}}-1 \; ,
\end{equation}
with $K_{\sigma}=1$ for spin-rotation invariant systems \cite{Gia03}.
However, due to the slow logarithmic decrease of the two-particle 
backscattering amplitude, the fixed point value of $K_{\sigma}$ is reached
only logarithmically from above. Hence, we can expect that the asymptotic value 
of $\alpha_B$ is usually reached only very slowly from below.

The local density of states at the boundary of a quarter-filled Hubbard chain,
computed by the fRG, is shown in Fig.~\ref{fig:dosehm} for various values of the
local interaction $U$. 
Contrary to the expected asymptotic power-law suppression the spectral weight 
near the chemical potential is strongly \emph{enhanced}. 
The predicted suppression occurs only at very small energies for 
sufficiently large systems. 
In the main panel of Fig.~\ref{fig:dosehm} the crossover to the asymptotic 
behavior cannot be observed, as the finite-size cutoff $\sim \pi v_F/L$ is too 
large. Results for a larger system with $L = 10^6$ sites at $U=2$ in the inset 
show the crossover to the asymptotic suppression, albeit only at very small 
energies.
The dependence of the boundary spectral weight at the Fermi level on the system
size $L$ is plotted in Fig.~\ref{fig:wspin}. 
The $L$ dependence of the spectral weight at zero energy is expected to display 
the same asymptotic power-law behavior for large $L$ as the
$\omega$ dependence discussed above.
Instead of decreasing with increasing $L$, the spectral weight increases even
for rather large systems for small and moderate values of $U$. For $U>2$ the 
crossover to a suppression is visible in Fig.~\ref{fig:wspin}.
For $U=0.5$ only an increase is obtained up to the largest systems studied.
The crossover depends sensitively on the interaction strength $U$, for small $U$ 
it is exponentially large in $\pi v_F/U$.

\fig[width=10cm]{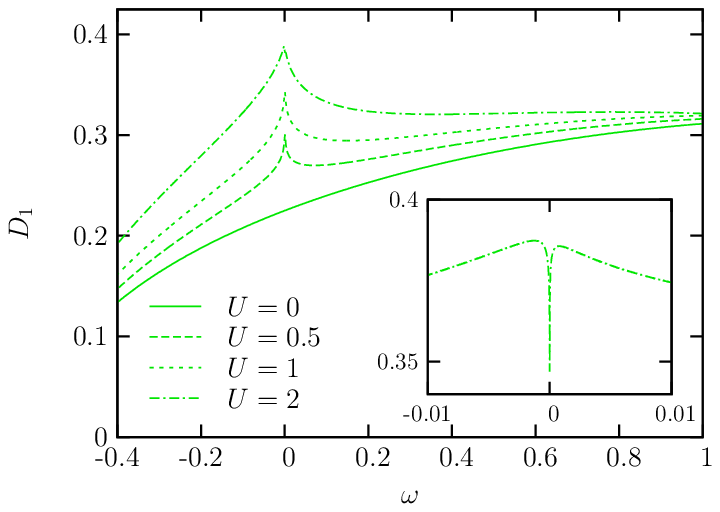}{fig:dosehm}{Local density of states 
  at the boundary of a Hubbard chain of length 
  $L = 4096$ at quarter filling and various interaction strengths $U$; 
  the inset shows results for $U=2$ and $L = 10^6$ at very low
  $\om$.}

\fig[width=10cm]{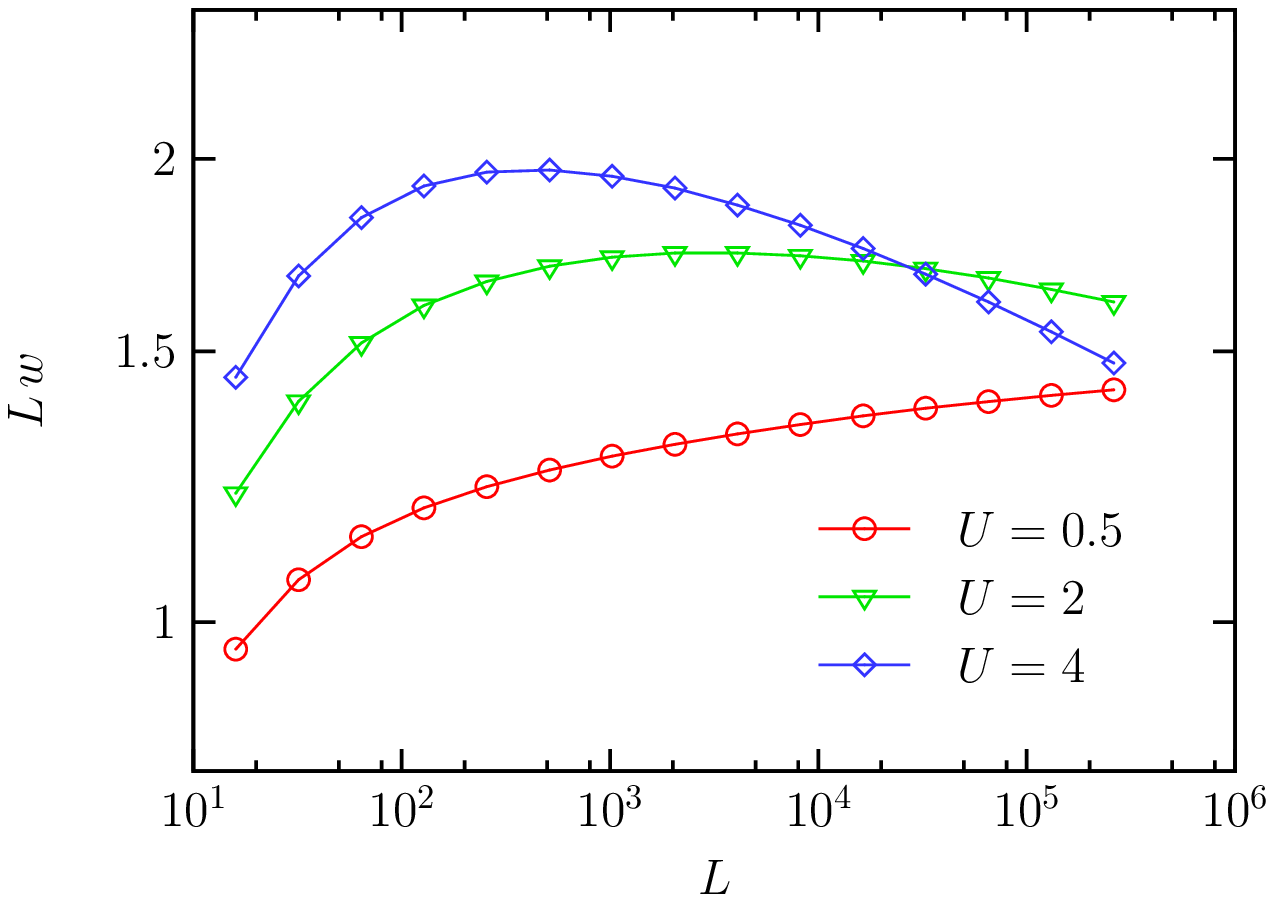}{fig:wspin}{Spectral weight at the Fermi level 
  at the boundary of a quarter-filled Hubbard chain as a function of 
  system size $L$, for various different interaction strengths.}

The above behavior of the spectral weight and the density of states near a 
boundary of the Hubbard chain, that is, a pronounced increase preceding the
asymptotic power-law suppression, is captured qualitatively already by the
Hartree-Fock approximation \cite{Med00,Gun00}. This is at first sight
surprising, as the Hartree-Fock theory does not capture any 
Luttinger-liquid features in the bulk of a translational invariant system.
In particular, a self-consistent Hartree-Fock calculation leads to the 
unphysical result of a charge-density-wave ground state for all $U>0$ 
\cite{CRB98}, since a single impurity can not modify bulk properties 
of the system.
The initial increase of $D_j (\omega)$ near a boundary is actually obtained 
already within perturbation theory at first order in the interaction 
\cite{Med00},
\begin{equation}
  D_j(\om) = D_j^0(\om) \, \left[ \, 1 +
    \frac{\tilde V(0) - z \tilde V(2k_F)}{2\pi v_F} \, 
     {\rm ln}\, |\om/\eps_F| + {\cal O}(\tilde V^2) \, \right]
\end{equation}
where $D_j(\omega)^0$ is the noninteracting density of states, $\tilde{V}(q)$ 
the Fourier transform of the real space interaction, and $z$ the number of spin
components. For spinless fermions $(z=1)$ with repulsive interactions the 
coefficient in front of the logarithm is always positive such that the 
first-order term leads to a suppression of $D_j(\omega)$. For the Hubbard model, 
one has $z=2$ and $\tilde{V}(0)-2\tilde{V}(2k_F)=-U$ is negative for repulsive 
$U$. Hence, at least for weak $U$ the density of states increases for decreasing 
$\omega$ until terms beyond first oder become important. For the extended 
Hubbard model, $\tilde{V}(0)-2\tilde{V}(2k_F)=2U'[1-2\cos(2k_F)]-U$, 
which can be positive or negative for $U,U'>0$, depending on the density and 
the relative strength of the two interaction parameters. At quarter filling 
$\tilde{V}(0)-2\tilde{V}(2k_F)$ is negative and therefore leads to an enhanced 
density of states for $U'<U/2$.

Using $g$-ology notation (cf. Sec.~\ref{sec:app:vertspin:gology}), 
one can write $\tilde{V}(0)-2\tilde{V}(2k_F)=g_{2\perp}-2g_{1\perp}$, which 
reveals that substantial two-particle backscattering
$(g_{1\perp}>g_{2\perp}/2)$ is necessary to obtain an enhancement of 
$D_j(\omega)$ for repulsive interactions. Backscattering vanishes at the 
Luttinger-liquid fixed point, but only very slowly. In case of a negative 
$\tilde{V}(0)-2\tilde{V}(2k_F)$ the crossover to a suppression of $D_j(\omega)$
is due to higher order terms, which are expected to become important when
the first-order correction is of order one, that is, for energies below the
scale
\begin{equation}
  \omega_c= \epsilon_F \exp \Bigl( \frac{2\pi v_F}{\tilde{V}(0)-2\tilde{V}(2k_F)} \Bigr)
\end{equation}
corresponding to a system size $L_c=\pi v_F/\omega_c$. The scale $\omega_c$ is 
exponentially small for weak interactions. A more accurate analytical estimate 
of the crossover scale from enhancement to suppression has been derived for the
Hubbard model within Hartree-Fock approximation in Ref.~\cite{Med00}.
In a renormalization-group treatment $\omega_c$ is somewhat enhanced by the
downward renormalization of backscattering.

A comparison of fRG results with DMRG data for 
the spectral weight at the Fermi level is shown in Fig.~\ref{fig:whm}, for 
a boundary site in the upper panel, and
near a hopping impurity of strength $t'=0.5$ in the lower. 
The agreement improves at weaker coupling, as expected, and is generally better 
for the impurity case, compared to the boundary case. 
The deviations in the boundary case are probably due to our approximate
translation-invariant parametrization of the two-particle vertex.
Boundaries and to a minor extent impurities spoil the translation invariance
of the two-particle vertex. Although the deviations from translation invariance 
of the vertex become irrelevant in the low-energy or long-distance limit, and 
therefore do not affect the asymptotic behavior, they are nevertheless present 
at intermediate scales. This feedback of impurities into the vertex increases of 
course with the impurity strength and is thus particularly important near a 
boundary.
The scale for the crossover from enhancement to suppression of spectral weight
discussed above depends sensitively on effective interactions at intermediate 
scales and can therefore be shifted considerably even by relatively small errors 
in that regime.

\fig[width=10cm]{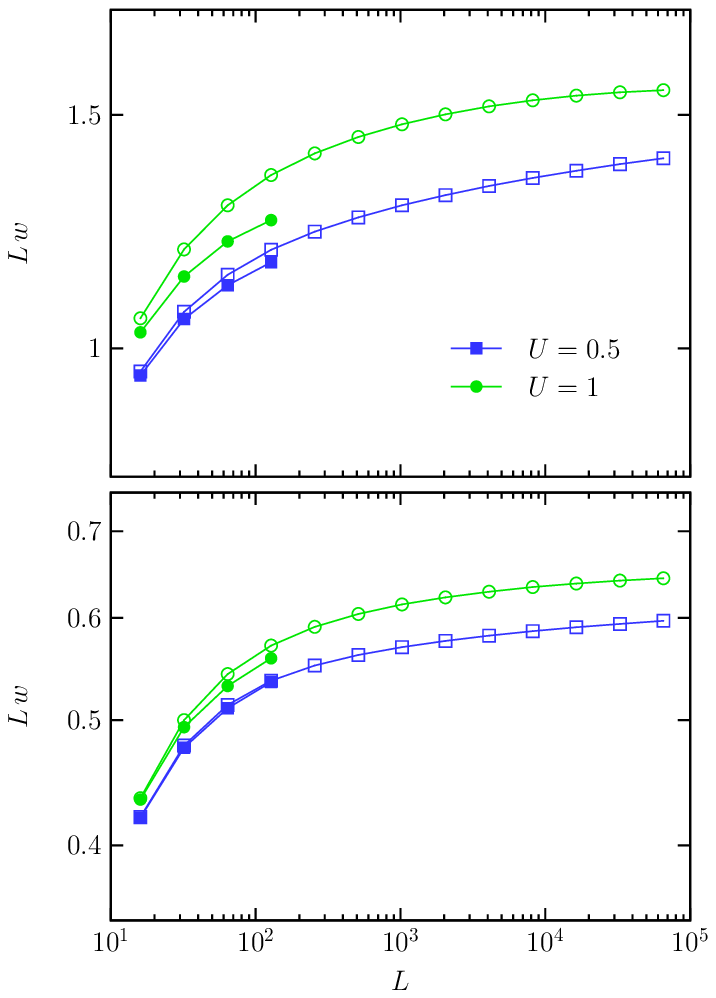}{fig:whm}{Spectral weight at the Fermi level near a 
  boundary (\emph{upper panel}) and a hopping impurity $t'=0.5$ 
  (\emph{lower panel}) as a function of system size $L$ for the Hubbard model at 
  quarter filling and different interaction strengths $U$; 
  results from the fRG (open symbols) are compared to DMRG data (filled symbols).}

\fig[width=10cm]{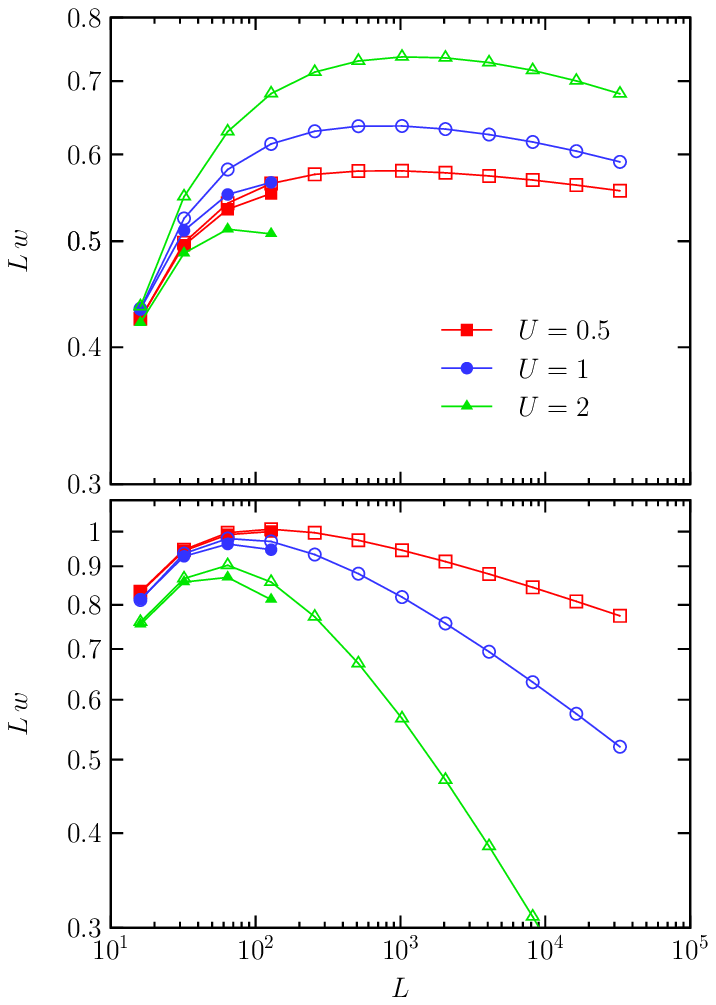}{fig:w_gology}{Spectral weight at the Fermi level near 
  a hopping impurity $t'=0.5$ as a function of system size $L$ for 
  the extended Hubbard model with $U' = U/\sqrt{2}$, for various choices of $U$;
  \emph{upper panel}:
  $n = 1/2$ (leading to sizeable backscattering),
  \emph{lower panel}:
  \mbox{$n = 3/4$} (leading to small backscattering); 
  results from the fRG (open symbols) are compared to DMRG data (filled symbols).}

With the additional nearest-neighbor interaction in the extended Hubbard model 
it is possible to tune parameters such that the two-particle backscattering 
amplitude becomes negligible.
In that case the asymptotic power-law suppression of spectral weight
should be free from logarithmic corrections and accessible already for smaller 
systems and at higher energy scales.
The bare backscattering interaction in the extended Hubbard model 
is given by $g_{1\perp} = U + 2U'\cos(2k_F)$ and therefore vanishes
for $U' = - U/[2\cos(2k_F)]$, which is repulsive for $U > 0$ if 
$n > 1/2$.
In a one-loop calculation a slightly different value of $U'$ has to 
be chosen to obtain a negligible renormalized $g_{1\perp}^{\Lam}$
for small finite $\Lam$, since the flow generates backscattering
terms at intermediate scales even if the bare $g_{1\perp}$ 
vanishes.
In Fig.~\ref{fig:w_gology} we show fRG and DMRG results

for the spectral weight 
of the extended Hubbard model at the Fermi level near a hopping impurity.
In the upper panel a generic case with sizeable backscattering is shown, while 
the parameters leading to the curves in the lower panel have been chosen such 
that the two-particle backscattering amplitude is negligible at low
energy.
Only in the latter case a pronounced suppression of spectral weight is reached 
already for intermediate system size, similar to the behavior obtained
previously for spinless fermions with nearest-neighbor interaction 
\cite{AEMMSS04,MMSS02a,MMSS02b}.
This is also reflected in the energy dependence of the local 
density of states near the impurity. 
For parameters leading to negligible two-particle backscattering 
as in Fig.~\ref{fig:dosehmb} the suppression of the density of 
states sets in already at relatively high energies and is not 
preceded by any interaction-induced increase;
the slight increase for small system sizes is a finite-size effect 
present also in the noninteracting case.
Note also that the fRG results are much more accurate for small 
backscattering, as can be seen by comparing the agreement with
DMRG data in the upper and lower panel of Fig.~\ref{fig:w_gology}
especially for larger $U$.
This indicates that the influence of the impurity on the vertex 
flow, which we have neglected, is more important in the presence 
of a sizable backscattering interaction.

\fig[width=10cm]{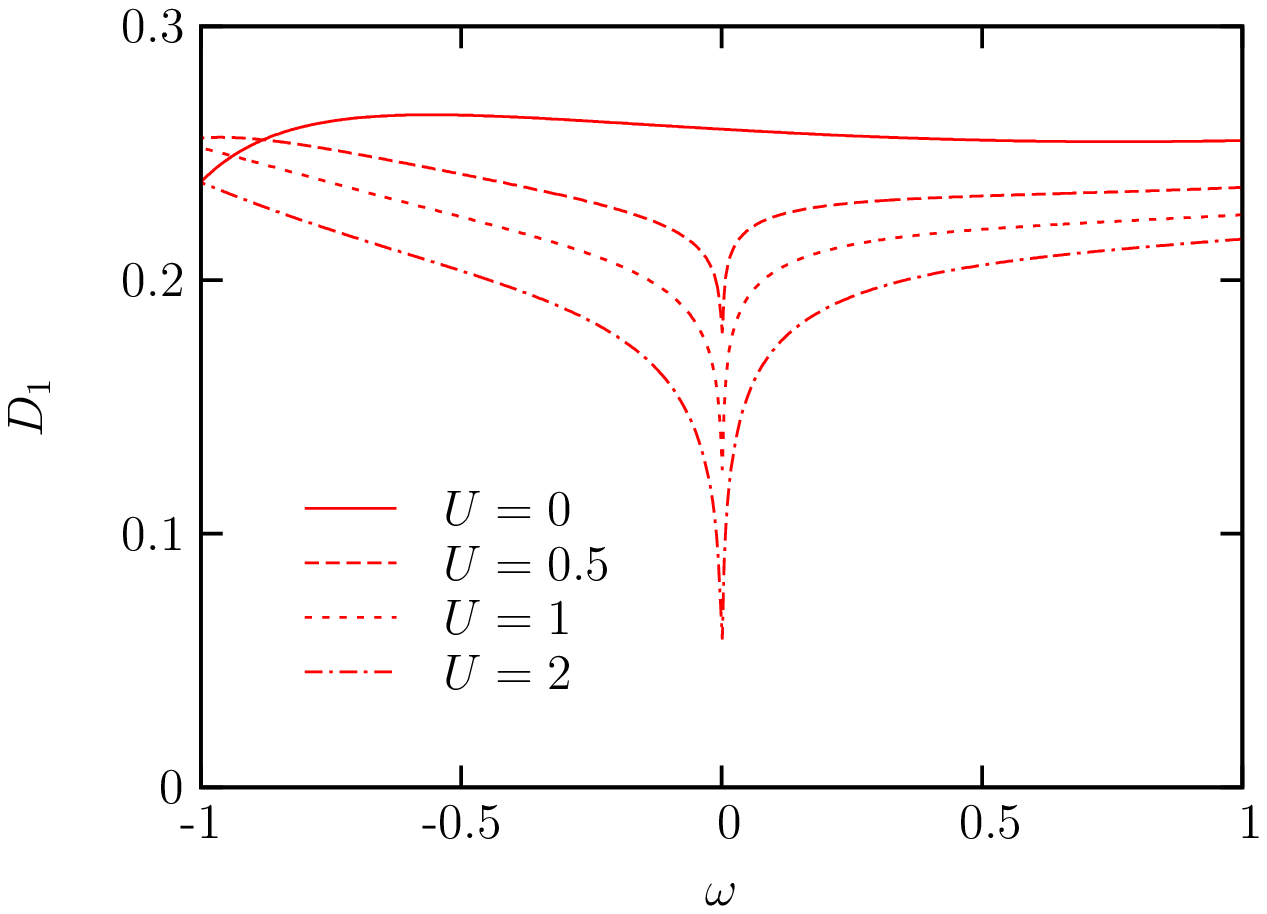}{fig:dosehmb}{Local density of states near a hopping
  impurity $t' = 0.5$ in an extended Hubbard model with density 
  $n=3/4$ and interaction $U' = U/\sqrt{2}$ 
  (leading to a small backscattering interaction) 
  for various choices of $U$;
  the size of the chain is $L = 4096$.}

In the case of a negligible backscattering amplitude, the 
spectral weight at the Fermi level approaches a power law without logarithmic
corrections for accessible system sizes if the impurity is
sufficiently strong.
The power law is seen most clearly by plotting 
the effective exponent $\alpha(L)$, that is, the negative logarithmic derivative 
of the spectral weight with respect to the system size.
Fig.~\ref{fig:impehm} shows $\alpha(L)$ on the site next to a site impurity 
of strength $V$ for the extended Hubbard model 
with  $U = 1$, $U' = 0.65$ and $n = 3/4$. The backscattering amplitude is very 
small for these parameters.
The fRG results approach the expected universal $V$-independent power law for 
large $L$, but only very slowly for small $V$.
For weak bare impurity potential $V$, the crossover to a strong effective
impurity occurs only on a large length scale of order $V^{2/(K_{\rho}-1)}$ 
\cite{KF92b,KF92a}.
For $V=0.1$ this scale is obviously well above the largest system
size reached in Fig.~\ref{fig:impehm}.
The Hartree-Fock approximation also yields power laws for large $L$, but
the exponents depend on the impurity parameters.
This failure of Hartree-Fock theory was already observed earlier for spinless
fermions \cite{MMSS02a,MMSS02b}.

\fig[width=10cm]{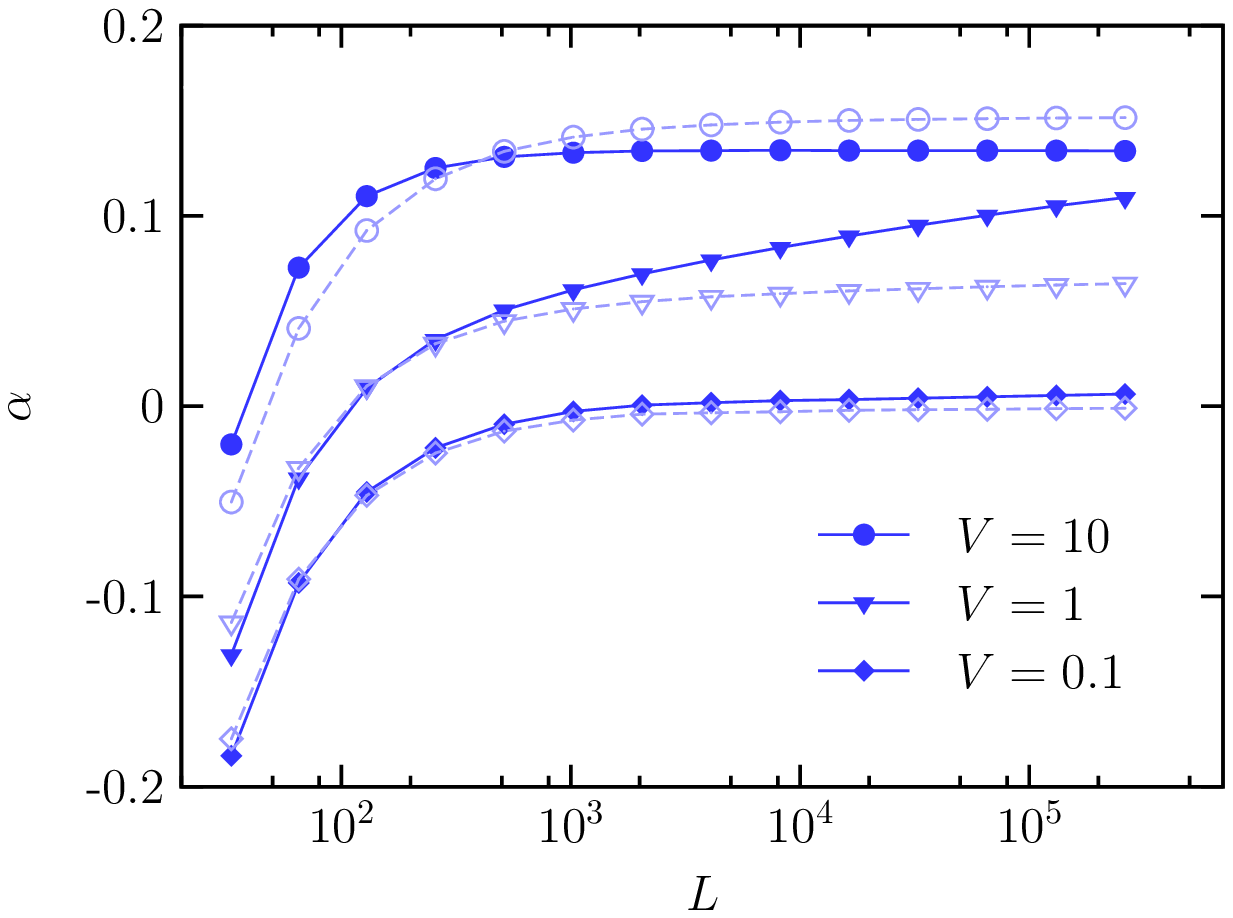}{fig:impehm}{Logarithmic derivative of the spectral 
  weight at
  the Fermi level on the site next to a site impurity of strength $V$ in the 
  center of the chain as a function of system size $L$, for the extended Hubbard 
  model with interaction parameters $U = 1$, $U' = 0.65$ and density $n = 3/4$; 
  here the filled symbols are fRG, the open symbols Hartree-Fock results.}

The effective exponent obtained from the fRG calculation agrees
with the exact boundary exponent to linear order in the bare
interaction, but not to quadratic order.
A quantitative estimate of the accuracy is obtained from
a comparison to exact DMRG results \cite{EGN05} 
shown in Fig.~\ref{fig:alphaehm}, for the extended Hubbard model at 
$n=3/4$ and with $U'=U/\sqrt{2}$ leading to a negligible backscattering amplitude. 
To improve this, the frequency dependence of the two-particle 
vertex, has to be taken into account. This is also necessary to describe
inelastic processes and to capture the anomalous dimension of the 
bulk system (see also Sec.~\ref{sec:results:spinless:dos}).

\fig[width=10cm]{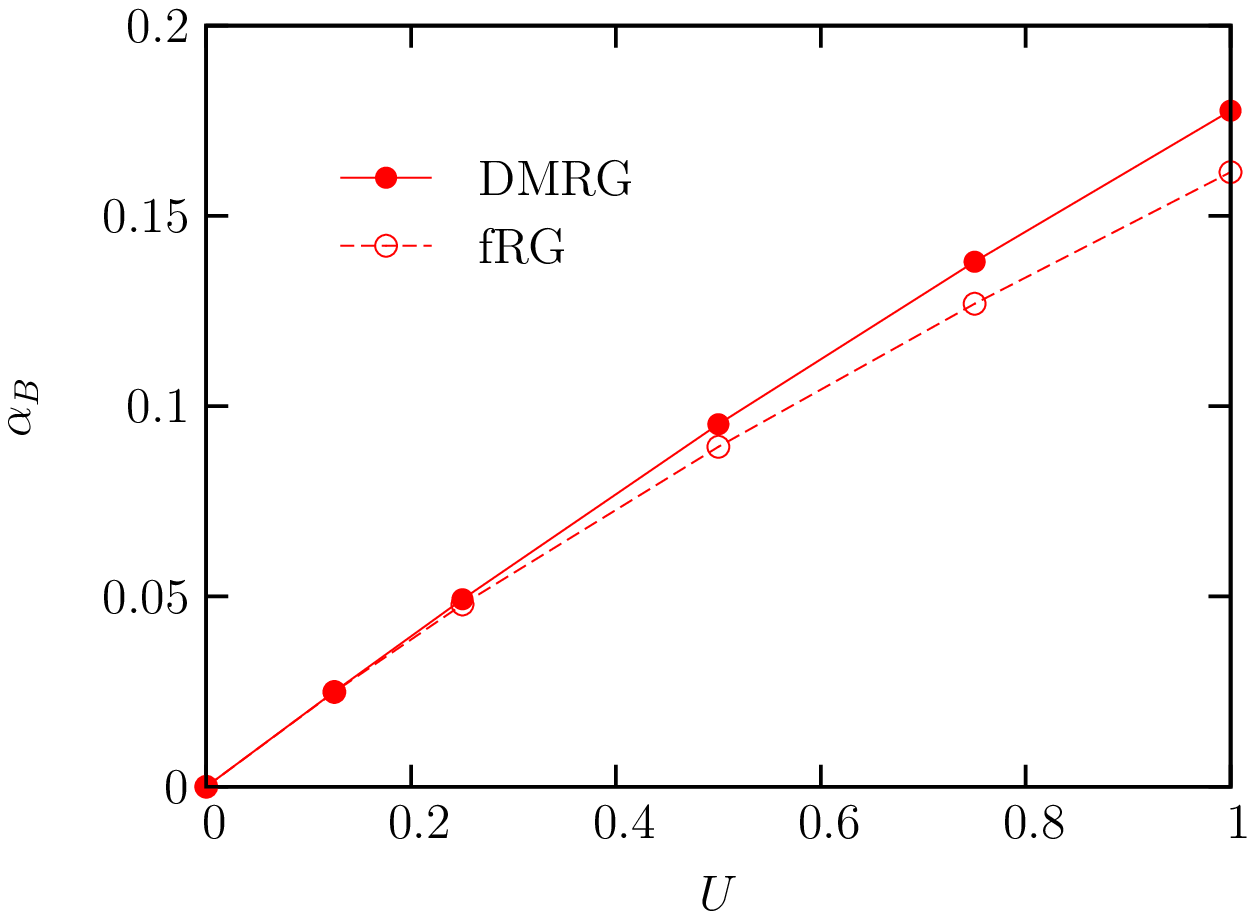}{fig:alphaehm}{Boundary exponent $\alpha_B$ as a 
  function of $U$ at densities $n=3/4$, with 
  \mbox{$U'=U/\sqrt{2}$} (small backscattering interaction) as obtained from the 
  power-law suppression of the spectral weight at the boundary;
  fRG results are compared to DMRG data.}

%%%%%%%%%%%%%%%%%%%%%%%%%%%%%%%%%%%%%%%%%%%%%%%%%%%%%%%%%%%%%%%%%%%%%%%%%%%%%%

\subsection{Density profile}
\label{sec:results:spin:density}

Boundaries and impurities induce a density profile with long-range 
Friedel oscillations, which are expected to decay with a power law
with exponent $(K_{\rho} + K_{\sigma})/2$ at long distances,
where $K_{\sg} = 1$ for spin-rotation invariant systems \cite{EG95}.
For weak impurities linear response theory predicts a decay as 
$|j-j_0|^{1 - K_{\rho} - K_{\sg}}$ at intermediate distances.

As an additional benchmark for the fRG technique, we compare
in Fig.~\ref{fig:densityhm} fRG and DMRG results for the density profile $n_j$ 
for a quarter-filled Hubbard chain with 
$L=128$ lattice sites and open boundaries. Friedel oscillations emerge from 
both boundaries and interfere in the center of the chain. 
The fRG results have been shifted by a small constant amount to 
allow for a better comparison of the oscillations. Note that the
mean value of $n_j$ in the tails of the oscillations deviates from
the average density by a finite-size correction of order $1/L$, 
which is related to the asymmetry of the oscillations near the 
boundaries.

\fig[width=10cm]{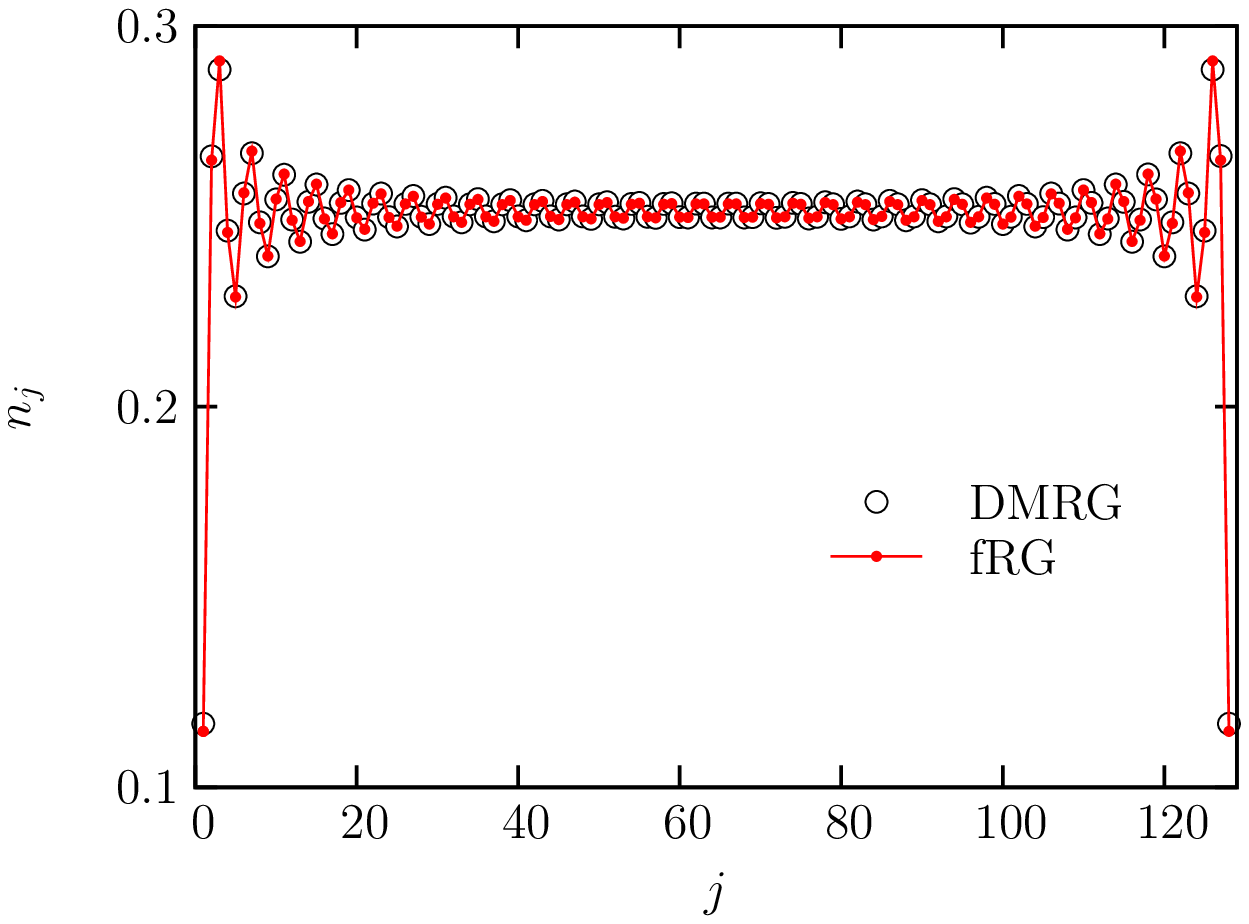}{fig:densityhm}{Density profile $n_j$ for the 
  Hubbard model with $128$ sites and interaction strength $U = 1$ at 
  quarter filling; fRG results are compared to DMRG data.}

The long-distance behavior of the density oscillations as obtained
within the fRG scheme has been analyzed in detail for spinless
fermions in Sec.~\ref{sec:results:spinless:density}.
For fermions with spin, asymptotic power laws can be identified 
only for special parameters leading to negligible two-particle 
backscattering. In general, the asymptotic behavior of Friedel 
oscillations is realized only at very long distances, and the
power laws are modified by logarithmic corrections.
We finally remark the presence of a $4k_F$-component of the Friedel 
oscillations for spin-$\frac{1}{2}$ fermions, which decays as 
$|j-j_0|^{-2K_{\rho}}$.
In the present weak-coupling treatment this contribution is negligible, 
since the $2k_F$ component dominates for $K_{\rho} > 1/3$ \cite{EG95}.

%%%%%%%%%%%%%%%%%%%%%%%%%%%%%%%%%%%%%%%%%%%%%%%%%%%%%%%%%%%%%%%%%%%%%%%%%%%%%%

\subsection{Conductance}
\label{sec:results:spin:cond}

%%%%%%%%%%%%%%%%%%%%%%%%%%%%%%%%%%%%%%%%%%%%%%%%%%%%%%%%%%%%%%%%%%%%%%%%%%%%%%

\subsubsection{Single impurity}
\label{sec:results:spin:cond:single}

For a system of spinless fermions with a single impurity it was 
already shown that the conductance obtained from the truncated fRG 
obeys the expected power laws, in particular $G(T) \propto 
T^{2\alf_B}$ at low $T$, and one-parameter scaling behavior
\cite{Med03,EMABMS04}.
The corresponding scaling function agrees remarkably well with
an exact result for $K_{\rho} = 1/2$, although the interaction
required to obtain such a small $K_{\rho}$ is quite strong.
The more complex temperature dependence of the conductance in
the case of a double barrier at or near a resonance is also
fully captured by the fRG \cite{EMABMS04,Med04}.

Fig.~\ref{fig:condsingleehm} shows typical fRG results for the 
temperature dependence of the conductance for the extended Hubbard 
model with a single strong site impurity ($V=10$). 
Similar results were obtained for a hopping impurity.
The considered size $L=10^4$ corresponds to interacting wires in 
the micrometer range, which is the typical size of quantum wires 
available for transport experiments. 
For $U'=0$ the conductance \emph{increases} as a function of 
decreasing $T$ down to the lowest temperatures in the plot. 
For increasing nearest-neighbor interactions $U'$ a suppression
of $G(T)$ at low $T$ becomes visible, but in all the data obtained
at quarter-filling the suppression is much less pronounced than 
what one expects from the asymptotic power law with exponent 
$2\alpha_B$. 
By contrast, the suppression is much stronger and follows the
expected power law more closely if parameters are chosen such
that two-particle backscattering becomes negligible at low $T$,
as can be seen from the conductance curve for $n=3/4$ and 
$U'=0.65$ in Fig.~\ref{fig:condsingleehm}. The value of $K_{\rho}$
for these parameters almost coincides with the one for another
parameter set in the plot, $n=1/2$ and $U'=0.75$, but the
behavior of $G(T)$ is completely different.
Note that at $T \sim \pi v_F/L$ finite-size effects set in, 
as can be seen at the low $T$ end of some of the curves in the
figure.
An enhancement of the conductance due to backscattering has
been found already earlier in a renormalization-group study
of impurity scattering in the $g$-ology model \cite{MYG93,YGM94}.

\fig[width=10cm]{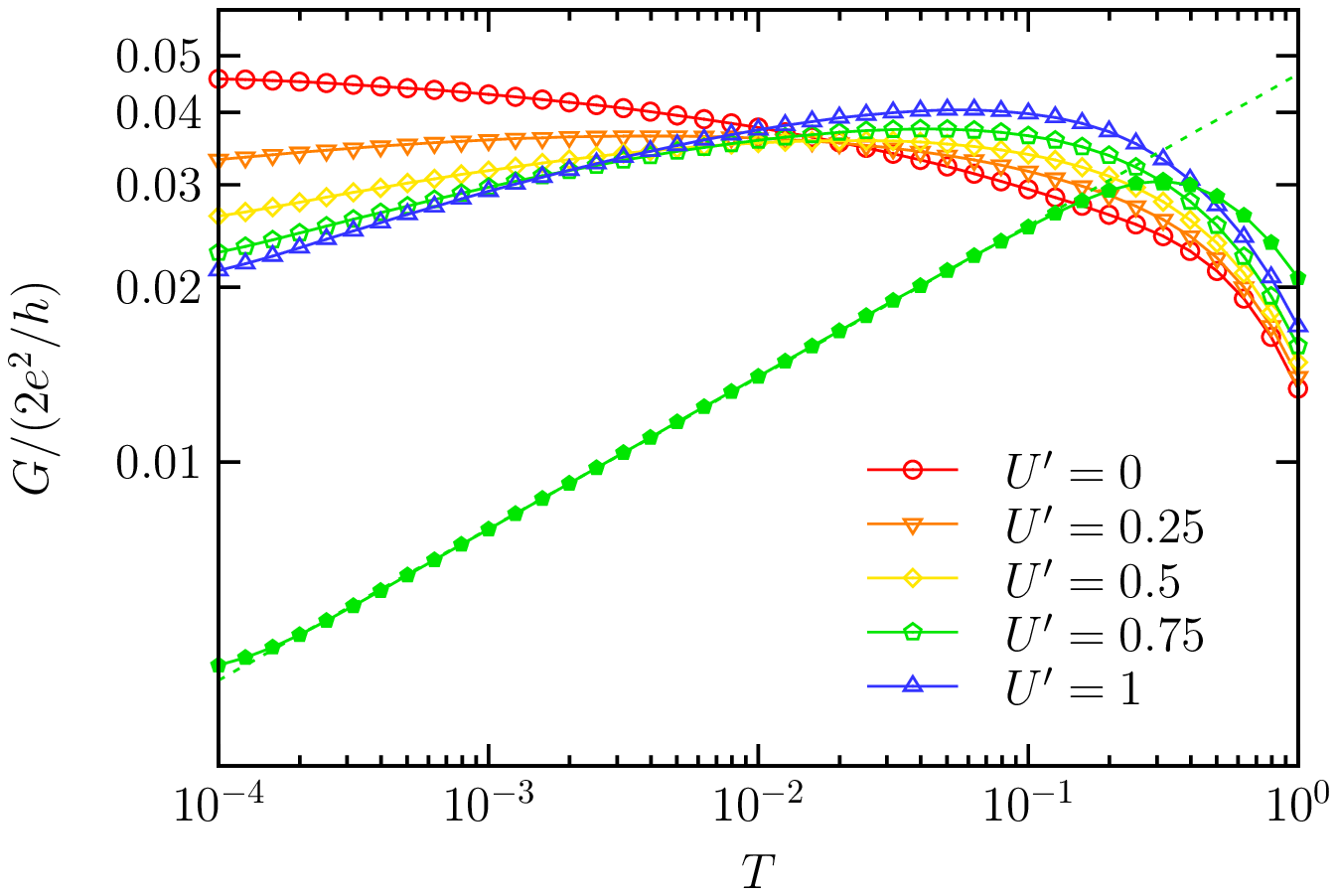}{fig:condsingleehm}{Temperature dependence 
  of the conductance for the extended Hubbard
  model with $L = 10^4$ sites and a single site impurity of strength 
  $V = 10$, for a Hubbard interaction $U = 1$ and various choices 
  of $U'$; 
  the density is $n=1/2$, except for the lowest curve,
  which has been obtained for $n=3/4$ and $U' = 0.65$ (leading to a 
  very small backscattering interaction);
  the dashed line is a power-law fit for the latter parameter set.}

\fig[width=10cm]{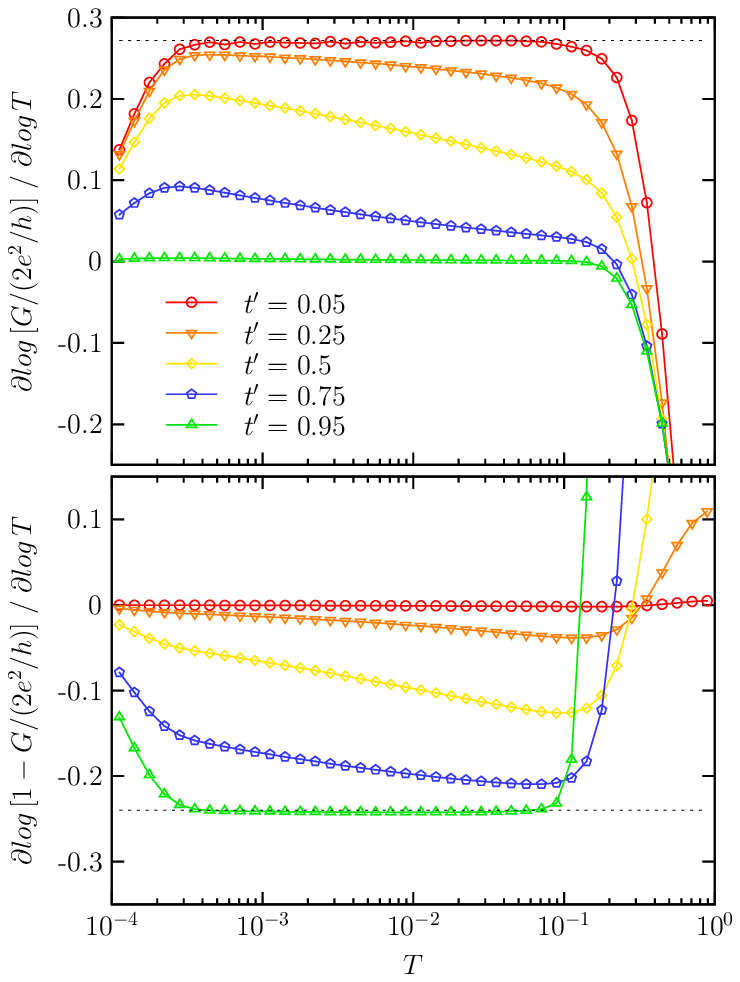}{fig:condsingletprime}{Logarithmic 
  temperature derivative of the conductance (upper panel)
  and of its deviation from the unitarity limit (lower panel) for the
  extended Hubbard model with $L = 10^4$ sites and various hopping
  impurities. The density is $n=3/4$, interaction parameters are
  $U=1$ and $U'=0.65$. The dashed horizontal lines highlight
  power-law behavior.}

Results for the conductance of the extended Hubbard model with 
a hopping impurity with various amplitudes $t'$ are shown in 
Fig.~\ref{fig:condsingletprime}. The bulk parameters have been chosen 
such that the two-particle backscattering is practically 
zero at low $T$. From the plot of the logarithmic derivative 
of $G(T)$ in the upper panel one can see that for a strong
impurity (small $t'$) the conductance 
follows a well-defined power law $G(T) \propto T^{2\alf_B}$
over a large temperature range. For intermediate $t'$ the
curves approach the asymptotic exponent at low $T$ from below, 
but do not reach it before finite-size effects lead to a 
saturation of $G(T)$ for $T < \pi v_F/L$.
For the weakest impurity in the plot, $t'=0.95$, the conductance
remains very close to the unitarity limit. However, the plot
of the logarithmic derivative of $1 - G/(2e^2/h)$ in the lower
panel of Fig.~\ref{fig:condsingletprime} shows that $1 - G/(2e^2/h)$
increases as $T^{K_{\rho} - 1}$ for decreasing $T$, as expected
for a weak impurity in the perturbative regime \cite{KF92b,KF92a}.
The effective exponents indicated by the two horizontal lines 
in the figure deviate from the exact values (determined from
the DMRG result \cite{EGN05} for $K_{\rho}$) by about $20\%$ 
in the case of $2\alf_B$ and only by $5\%$ for $K_{\rho} - 1$ (cf. 
Sec.~\ref{sec:results:spinless:oneparam}).

Depending on the bare impurity and interaction parameters,
nonuniversal behavior dominates at intermediate energy and length scales.
Moreover, in the presence of backscattering the asymptotic power laws
are modified by logarithmic corrections.

%%%%%%%%%%%%%%%%%%%%%%%%%%%%%%%%%%%%%%%%%%%%%%%%%%%%%%%%%%%%%%%%%%%%%%%%%%%%%%

\subsubsection{Double impurity}
\label{sec:results:spin:cond:double}

We finally present results for the conductance of a wire with a double-barrier
impurity \cite{AEM05}. The setup modeling a quantum dot is shown in 
Fig.~\ref{fig:model}.
Applying a gate voltage $V_g$ on the dot sites $j \in [j_l,j_r]$ by
\begin{equation}
  H_{\rm gate} = V_g \sum_{j, \sigma} n_{j,\sigma}
\end{equation}
the conductance can be tuned to resonance.
$L_D=j_r-j_l+1$ is sufficiently far away 
from the contacts at sites $1$ and $L$ the position of the dot does
not play a role.

\fig[width=12cm]{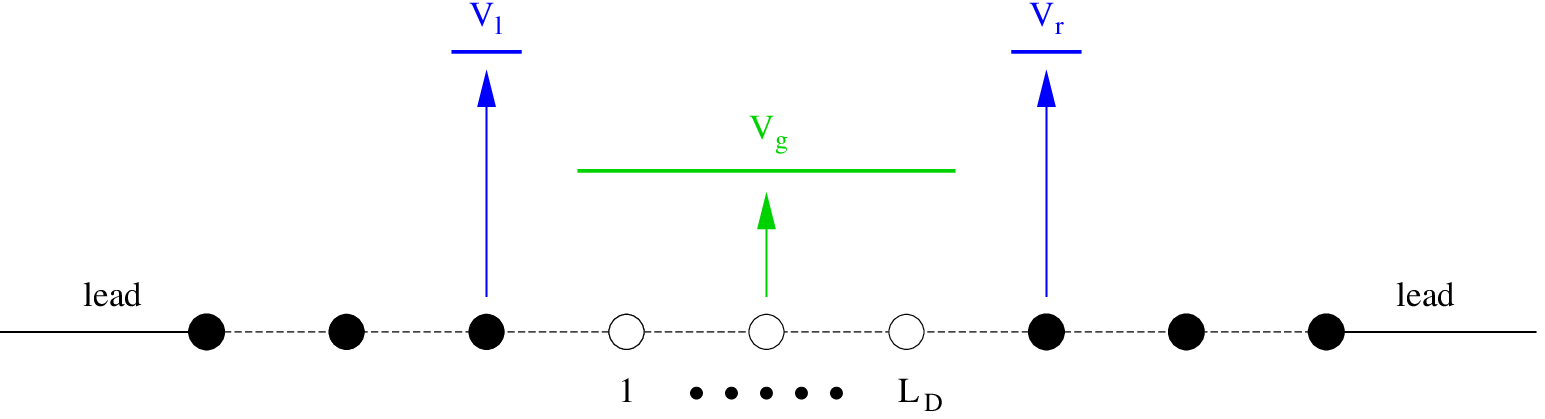}{fig:model}{Quantum dot schematization.}

Earlier studies of tunneling through a quantum dot embedded in a spinless
Luttinger liquid showed that at $T = 0$ and for finite $L$ 
the resonances in the linear conductance
$G(V_g)$ are characterized by an almost Lorentzian shape with 
unitary height for symmetric barriers and a width 
$w$ vanishing as a power law with a $K_{\rho}$-dependent
exponent in the limit $L \to \infty$. 
For asymmetric barriers the resonances disappear for increasing $L$ 
\cite{KF92b,KF92a}. At $T>0$ the peak value of the conductance 
shows distinctive power-law behavior as a function of temperature 
\cite{EMABMS04,Med04}. Including the spin degree of freedom the
physics becomes more complex due to the appearance of the Kondo effect.
For noninteracting leads ($L=1$) the Kondo physics was investigated 
theoretically for the single-impurity Anderson model \cite{GR88,NL88}. 
At low temperatures and for sufficiently high tunnel barriers
the Kondo
effect leads to a {\it broad plateau-like line} shape of the
resonance replacing the Lorentzian.
On resonance the number of electrons on the dot is odd implying a 
local spin-$\frac{1}{2}$ degree of
freedom responsible for the Kondo effect \cite{Hew93}.
For the single-impurity Anderson model the conductance is proportional 
to the one-particle
spectral weight of the dot at the chemical potential \cite{MW92}.
Varying $V_g$ within an energy range of order $U$
the Kondo resonance of the spectral function is pinned at $\mu_0$
at height $2e^2/h$ explaining the
broad plateau-like resonance in $G(V_g)$ \cite{Hew93,GEX00}.  
The problem of a single spin-$\frac{1}{2}$ coupled to a Luttinger liquid 
was investigated generalizing the Kondo model \cite{Fur05}.
The fRG approach allows for a direct computation of the
electron transport through a quantum dot embedded in a Luttinger liquid
in the presence of the Kondo effect.
Here we address the question of the resonance line shape and the 
power-law scaling of $G(T)$ resulting from the 
\emph{competition} between the two correlation effects.

For this purpose we first consider 
the situation $L=L_D=1$ at $T=0$ corresponding to the single-impurity
Anderson model. Unless otherwise stated we
consider symmetric dot-lead couplings.
In Fig.~\ref{fig:g_vgkondo} the conductance $G$ as
a function of gate voltage $V_g$ for the single-impurity
Anderson model is shown for different tunnel barriers $t'$ in the upper panel, 
together with the occupation of the dot in the lower.
For $t'\ll U$  the resonance has a plateau-like shape \cite{GEX00}. 
In this region
the occupation is close to $1$ while it sharply
raises/drops to $2$/$0$ to the left/right of the plateau. 
Also for asymmetric barriers
we reproduce the exact resonance height 
$4\Delta_L \Delta_R(\Delta_L+\Delta_R)^2\,(2e^2/h)$,
where $\Delta=\Delta_L+\Delta_R$ measures the hybridization of the dot and the
(left and right) lead states \cite{Hew93,GEX00}.
Here we are interested in the interplay of Kondo
and Luttinger-liquid physics and thus focus on small $t'$, that is, on
tunnel barriers with small transmission.

\fig[width=10cm]{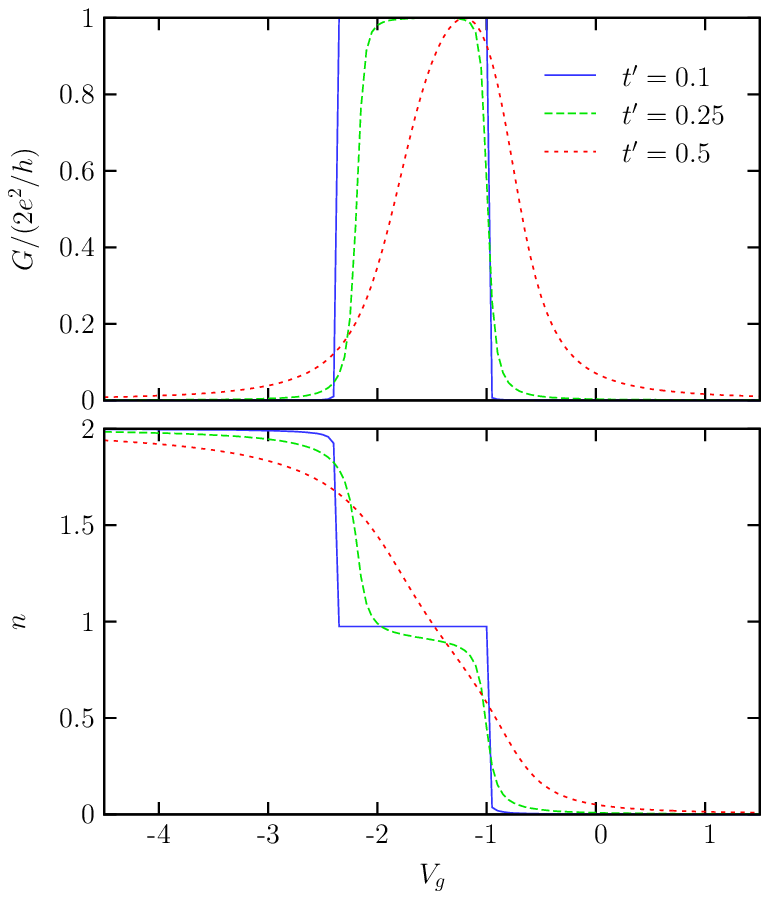}{fig:g_vgkondo}{\emph{Upper panel:} conductance as 
  a function of gate voltage for the Hubbard model at quarter filling, with
  $U = 1$, for $L=L_D=1$ and different $t'$ ; \emph{lower panel:}
  average number of electrons on the dot.}

The pinning of the spectral function and the subsequent
plateau-like resonance can be derived within the fRG Scheme I 
(cf. Sec.~\ref{sec:frglutt:flow:trunc}). For $L=1$ the chemical potential
only shifts the position of the resonance. For half filling $G(V_g)$ 
is symmetric around $0$ and the flow equation for the effective onsite
energy $V  = V_g +\Sg^{\Lam}_{j_D,j_D}$  on the dot site $j_D$ reduces
to
\begin{equation}
  \label{eq:diffkondo}
  \frac{\partial}{\partial \Lam} V^{\Lam}= -\frac{U}{\pi}\,\Re \, {\cal G}^{\Lam}_{j_D,j_D} (i\Lam)= \frac{UV^{\Lam}/\pi}{(\Lam+\Delta)^2+(V^{\Lam})^2}
\end{equation}
in the limit of $\Delta \ll U$. Here $\Delta= 2\pi {t'}^2 \rho$ is the 
hybridization,
and $\rho$ denotes the spectral weight at the end of the leads in the 
infinite band width limit \cite{Hew93}. The initial condition is 
$V^{\Lam_0} =V_g$. 
In this scheme the self-energy is frequency independent leading to a
Lorentzian spectral function of width $2\Delta$ and height $1/(\pi \Delta)$
centered around $V=V^{\Lam=0}$. 
This implies that
the spectral weight at $\mu_0$ and thus $G(V_g)$ is determined
by $V$ \cite{MW92}. The solution of the differential equation 
(\ref{eq:diffkondo}) at $\Lam=0$ is obtained in implicit form 
\begin{equation}
  \label{eq:kondosol}
  \frac{vJ_1(v)-\delta J_0(v)}{vY_1(v)-\delta Y_0(v)}=\frac{J_0(v_g)}{Y_0(v_g)}\; ,
\end{equation}
with $v = V\pi/U$, $v_g = V_g\pi/U$, $\delta = \Delta \pi/U$, and Bessel
functions $J_n$, $Y_n$. For $|V_g| < V_c$ this equation has a
solution with a small $|V|$, where $v_c = V_c \pi /U$ is
the first zero of $J_0$ corresponding to $V_c \simeq 0.77 U$.
For $U\gg \Delta$ the crossover to a
solution with $|V|$ being of order $U$ (for $|V_g|> V_c$) is fairly
sharp. Expanding both sides of Eq.~(\ref{eq:kondosol}) for small $|v|$ and
$|v_g|$ gives 
\begin{equation}
  V = V_g \,{\rm exp}\,\Big( \frac{U}{\pi \Delta}\Big) \; .
\end{equation}
The exponential pinning
of the spectral weight at $\mu_0=0$ for small $|V_g|$ and the
sharp crossover to a $V$ of order $U$ when $|V_g| > V_c$ leads to
the observed resonance line shape. For $U\gg \Delta$ the width
of the plateau is $2V_c \simeq 1.5 U$, which is
larger than the width $U$ found with the numerical renormalization-group 
method \cite{GEX00}. Our approximation
furthermore slightly overestimates the sharpness
of the box-shaped resonance. We expect the agreement
is improved with the more accurate fRG Scheme II 
(cf. Sec.~\ref{sec:frglutt:flow:trunc}).

\fig[width=10cm]{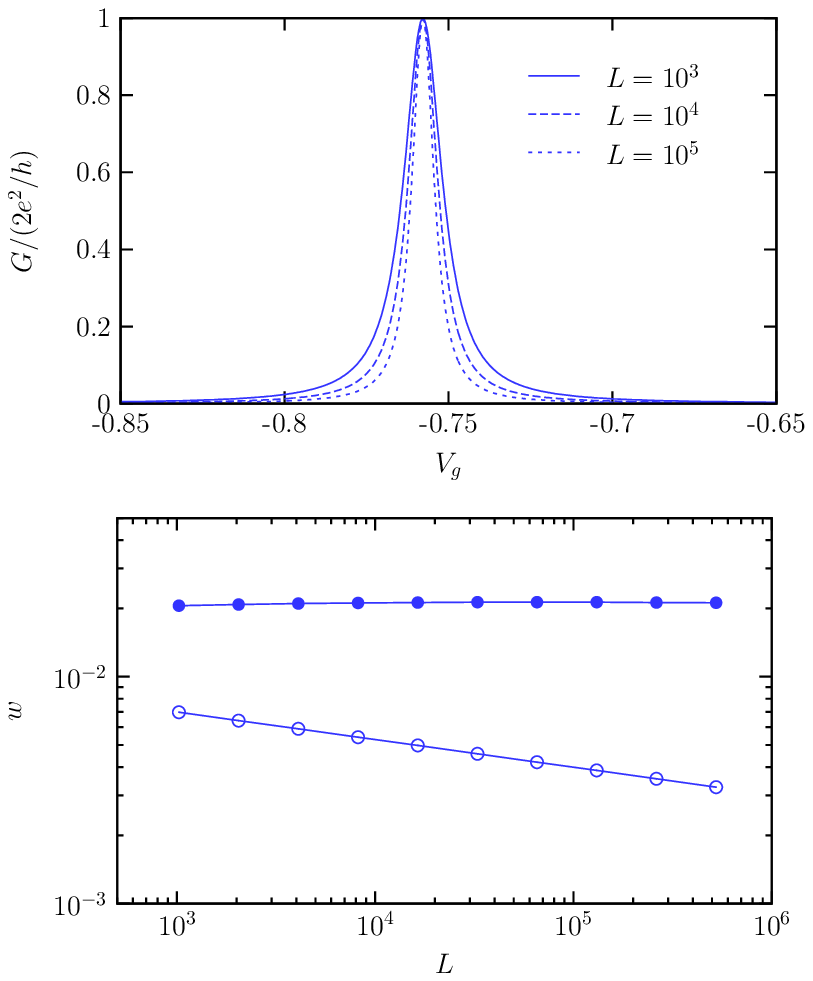}{fig:width}{\emph{Upper panel:} conductance 
  as a function of gate voltage for the extended Hubbard model at $n=3/4$, 
  with $U=1$, and $U'=0.65$, for a noninteracting dot with $L_D=1$, $t'=0.1$  
  and different $L$;
  \emph{lower panel:} scaling of the resonance width for the same 
  parameters as in the 
  upper panel corresponding to a small backscattering interaction (open symbols),
  and for $n=1/4$, $U=1$, $U'=0$ corresponding to a large
  backscattering interaction (filled symbols).}

In the complementary case of Luttinger-liquid leads 
and in the absence of Kondo effect, suppressed by
turning off the interaction on the dot,
similar results as for spinless fermions are obtained.
Luttinger-liquid behavior leads to 
{\it infinitely sharp} resonances in the limit $L \to \infty$, in strong 
contrast to the broad resonances induced by the Kondo effect \cite{EMABMS04}. 
To clearly observe Luttinger-liquid behavior for spin-$\frac{1}{2}$ fermions 
at experimentally accessible scales one has to consider a situation
in which the backscattering process yielding 
logarithmic corrections to the power-laws is small
by tuning the nearest-neighbor interaction $U'$, see 
Sec.~\ref{sec:results:spin:dos}. In Fig.~\ref{fig:width} the
$L$ dependence of $G(V_g)$ for a single-site dot computed for
a small backscattering amplitude and lead length $L=10^4$ typical 
for experiments is shown.
At $T=0$ the width of the resonance in $G(V_g)$ tends to
zero for $L\to \infty$. The extracted width $w$ as a function of $L$ reported 
in the lower panel with open symbols follows a power law $L^{(K_{\rho}-1)/2}$ 
with an fRG approximation to the Luttinger-liquid parameter that is, correct
to leading order in the interaction.
Off resonance $G$ asymptotically
vanishes as $L^{-2\alpha_B}$ \cite{KF92b,KF92a,EMABMS04}.
For $V_g$ close to resonance and
$1-G/(2e^2/h)\ll 1$,  the deviation from the unitary limit 
increases as $L^{1-K_{\rho}}$ characteristic
of the scaling of a weak single impurity. Further increasing $L$ the behavior
eventually crosses over to the off-resonance power-law
suppression of $G$ mentioned above. 
Due to an exponentially
large crossover scale, even for the very large system sizes accessible
with our method the complete crossover from
one to the other power law can not be seen for a single
fixed $V_g$ but follows from one-parameter scaling \cite{KF92b,KF92a,EMABMS04}.
For a sizeable backscattering amplitude the off-resonance conductance and
thus the width first slightly increase for increasing $L$
- becoming larger than for the noninteracting dot - 
followed by a crossover to a decrease for exponentially large $L$, as indicated
by the filled symbols in the lower panel.
Due to the logarithmic vanishing of the backscattering process
this behavior can in general not be observed on experimentally accessible 
scales, confirming the important role of two-particle backscattering on 
intermediate length scales.
An upper bound of the length of quasi one-dimensional wires realized in 
experiments is of the order of $\mu$m, roughly corresponding to $10^4$ 
lattice sites \cite{BEX99,YEX99,Aus00,Pic01}.

\fig[width=10cm]{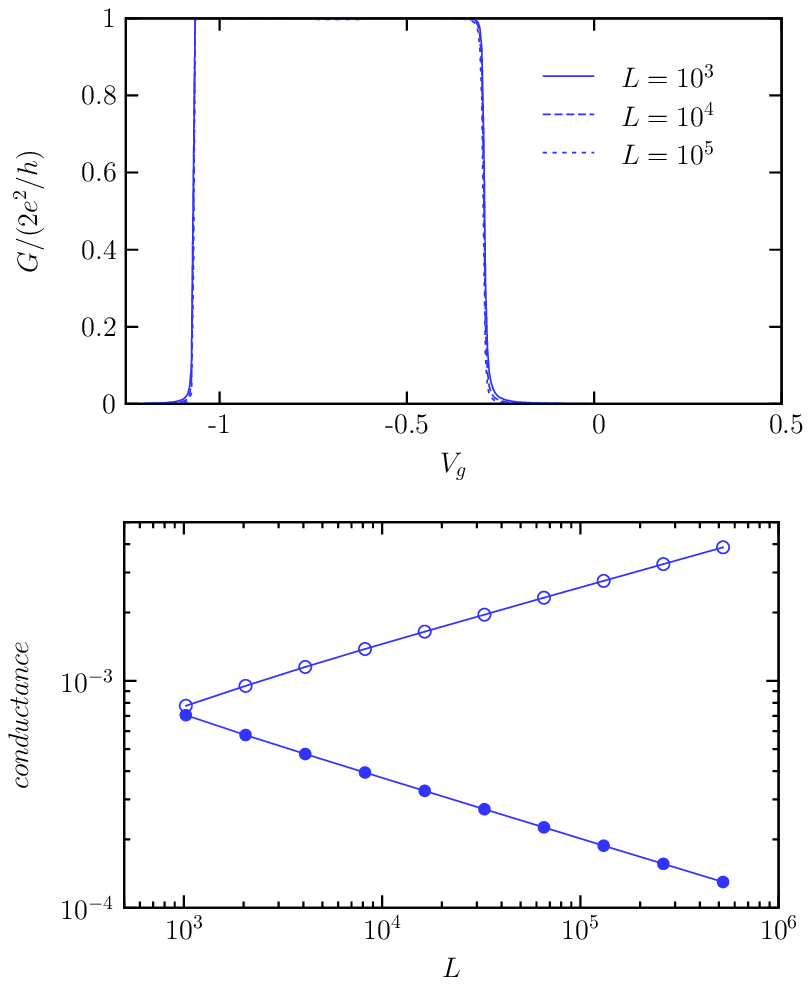}{fig:widthg}{\emph{Upper panel:}
  conductance 
  $G(V_g)$ as in Fig.~\ref{fig:width} (same parameters), but now with 
  interaction on the dot;
  \emph{lower panel:} scaling of $G/(2e^2/h)$ at $V_g=0$ outside the plateau 
  (filled symbols),
  and of $\,1-G/(2e^2/h)$ on the plateau at $V_g=-0.685$ (open symbols).}

We now analyze the linear conductance through a quantum dot in the presence of 
both Kondo effect as well as Luttinger-liquid leads.
The upper panel of Fig.~\ref{fig:widthg} shows the $L$ dependence
of $G(V_g)$ for the same parameters as in Fig.~\ref{fig:width}, but now
including the interaction on the dot. Note the different scales on the 
$x$-axis.
For interactions large compared to the hybridization the broad plateau-like 
resonance induced by the Kondo effect is also present at
least for {\it finite} Luttinger-liquid leads. The same holds for $L_D>1$.
The width of the plateaus is proportional to the ratio of the local
component of the effective interaction at the end of the fRG flow and $L_D$.
The differences between the
curves for different $L$ are barely visible, in particular the
changes of the resonance width are marginal. 
For generic parameters with sizeable backscattering, the
difference between curves computed for different $L$ are
even smaller. 
We note that the plateaus vanish
if $U$, $U'$, and $n$ are chosen such that at the end of the fRG
flow the local part of the interaction is small, and the resonance peaks are 
sharp.
To analyze the $L$ dependence at small backscattering in more detail
in the lower panel of Fig.~\ref{fig:widthg} the scaling of $G/(2e^2/h)$ for a
gate voltage outside the plateau (filled symbols) and of $1-G/(2e^2/h)$
for a gate voltage on the resonance plateau (open symbols) are shown.
Off resonance $G$ follows a power-law with the exponent $2\alpha_B$
and $G$ vanishes for $L\to\infty$. Within
every plateau there a value of $V_g=V_g^r$ where $G=2e^2/h$ independently of
$L$. For $V_g \neq V_g^r$ still within the plateau the
deviation of $G$ from the unitary limit scales as $L^{1-K_{\rho}}$,
that is, with the weak single-impurity exponent. This shows
that any deviation from $V_g^r$ acts as an impurity. By analogy
with the single-impurity behavior discussed in the previous sections
we conclude
that in the asymptotic low-energy limit the impurity will
effectively grow and in the limit $L\to\infty$ the plateaus will vanish.
For infinitely long Luttinger-liquid leads the resonances are infinitely 
sharp even in the presence of Kondo physics. However, for
tunnel barriers with small transmission the
plateaus at finite $L$ are well developed and the length scale
on which the plateaus start to deteriorate is extremely
large. For sizeable two-particle backscattering this 
scale is enhanced and the plateaus are more pronounced.

Also for asymmetric barriers we find (almost)
plateau-like resonances. To discuss this in more detail
we focus on typical parameters with $N=10^4$ and an
asymmetry $\Delta_l/\Delta_R\sim 2$. Then the width is
almost unaffected by the asymmetry. For the interaction and filling as
in Fig.~\ref{fig:widthg} the height within
the plateaus varies by a few percent (with maxima at the
left and right boundaries) and
has average value  $\sim 0.85 \,(2e^2/h)$. With increasing $L$
the variation of the conductance on the
plateaus increases while the average value decreases. We
expect that for $L\to \infty$ the resonance disappears.

\fig[width=10cm]{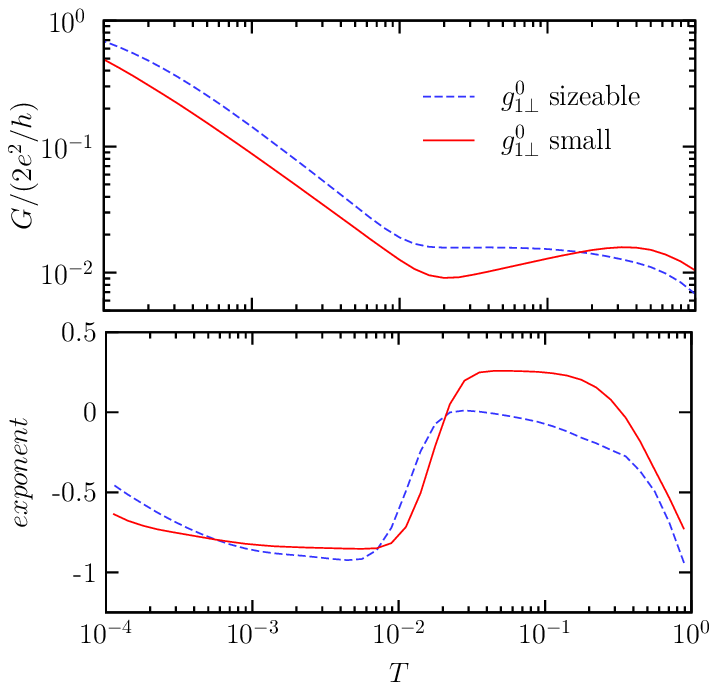}{fig:g_Tdot}{Temperature dependence of 
  the conductance for the extended Hubbard model with $L=10^4$ sites,
  Hubbard interaction $U = 1$, for $U'=0.65$ and $n=3/4$ (leading to a small
  backscattering interaction), and for $U'=0.75$ and $n=1/4$ (leading to a 
  sizable backscattering interaction); dot parameters: $t' = 0.1$, $L_D=100$;
  \emph{upper panel}: log-log plot of the conductance,
  \emph{lower panel}: effective exponents;
  the gate potential is chosen at the center of the resonance plateau closest
  to $V_g=0$.}

Concerning the power-law scaling of the conductance as a function of temperature, 
$G(T)$, 
the resonant tunneling behavior in general extends over a range of gate 
voltages defined by the width of the resonance plateau,
since for all experimentally
accessible length scales and for typical 
asymmetries of the dot-lead hybridizations the plateau-like
resonances characteristic for Kondo physics will also
be present if the leads are Luttinger liquids.
The vanishing of the resonance plateau 
in the limit of infinite system size is beyond the
infrared cutoff scale set by the system size $T\sim \pi v_F/L$ for
the appearance of
power-law scaling with interaction-dependent
exponents characteristic for Luttinger-liquid behavior. 
Hence the behavior of $G(T)$ is similar to the one found for spinless 
fermions \cite{EMABMS04,Med04}.
In Fig.~\ref{fig:g_Tdot} the temperature dependence of the conductance 
for a large dot with $L_D=100$ and for $V_g$ placed at the center of the 
resonance plateau closest to $V_g = 0$ is shown. 
We distinguish the case of sizable and small backscattering processes 
$g_{1\perp}^0$.
For a small backscattering amplitude we 
identify temperature regimes in 
which $G(T)$ follows distinctive power-law behavior 
with universal exponents.
For $T$ larger than the level spacing of the dot given by $\pi v_F/L_D$  
a power law with the single-impurity exponent $2\alpha_B$ is found, arising 
from two independent single barriers acting as resistors in series. It follows a
sequential tunneling regime characterized by the power-law exponent 
$\alpha_B-1$, until saturation sets in at $T\sim \pi v_F/L$. 
Small dots with $L_D$ of order $1$ exhibit only the latter.
Similar results are obtained for other $V_g$ within the resonance plateau, 
outside the plateaus the conductance follows the on-resonance
behavior down to a scale set by the deviation from resonance.
For smaller $T$ we find a crossover to $T^{2\alpha_B}$.
By contrast, for a sizable backscattering amplitude no clear power law can 
be distinguished, as can be seen from the conductance curve for $n=1/4$ and 
$U'=0.75$ in Fig.~\ref{fig:g_Tdot}. The value of $K_{\rho}$
for these parameters almost coincides with the one for another
parameter set in the plot, $n=3/4$ and $U'=0.65$, but the
behavior of $G(T)$ is completely different. 
For weak barriers analogous power-laws are detected in $1-G/(2e^2/h)$, 
described by the single-impurity exponent $K_{\rho}-1$ at high temperatures 
and by $K_{\rho}+1$ in the sequential tunneling regime, similarly to the 
spinless case investigated in detailed in Ref.~\cite{EMABMS04}.

For systems with long-range interactions backscattering is 
strongly reduced compared to forward scattering. 
This seems to be the case in carbon nanotubes \cite{EG97,KBF97}.
Hence, the conductance can be expected to follow the asymptotic 
power law at accessible temperature scales for sufficiently 
strong impurities in these systems, as is indicated also by 
experiments \cite{YEX99}.
However, the effects due to two-particle backscattering should be 
observable in systems with a screened Coulomb interaction.
Whether Luttinger-liquid behavior has convincingly been
demonstrated experimentally is still a matter
of debate. Nonetheless, the above scenario has to be taken
into consideration in the discussion of the influence of 
boundaries or impurities in quasi one-dimensional conductors.

%%% Local Variables: 
%%% mode: latex
%%% TeX-master: "thesis"
%%% End: 

\end{fmffile}
% Thesis conclusions+outlook

\chapter{Conclusions and outlook}
\label{sec:concl}

The fRG provides a powerful tool to compute 
the intriguing properties of Luttinger liquids 
with static impurities. 
It captures the physics  
at all energy scales from the
Fermi energy to the ultimate low-energy limit.
The presented computation scheme extends
previous work for spinless fermions 
\cite{MMSS02a,MMSS02b} to spin-$\frac{1}{2}$ fermions,
including the vertex renormalization in addition to 
the renormalization of the effective impurity potential.
The underlying approximations are devised for weak short-range 
interactions and arbitrary impurity potentials.
The results agree remarkably well with exact asymptotic results 
up to intermediate interaction strength,
and cover the universal low-energy asymptotics, as well as 
nonuniversal behavior and crossover phenomena at higher scales.

Various observables have been computed for different fermion lattice models: 
spectral properties of single-particle excitations, 
the oscillations in the density profile 
induced by impurities or boundaries,
and the temperature dependence of the linear conductance.
The comparison to DMRG results,
for those observables and system sizes for which such data could
be obtained, yields a good agreement at weak coupling.
For intermediate interaction strengths with sizable 
two-particle backscattering and strong impurities the deviations 
are significantly larger for spin-$\frac{1}{2}$ than for spinless fermions.
For the computation of the nonuniversal behavior 
at intermediate scales
the neglected influence of impurities on vertex
renormalization, in particular 
the interplay of impurities and the two-particle backscattering amplitude, 
is probably important for fermions with spin.

We confirm the universality of the open chain 
fixed point, but it turns out
that very large systems are required to reach the fixed point 
for realistic choices of the impurity and interaction parameters.
The spectral properties of single-particle excitations and the 
Friedel oscillations in the density profile induced by impurities or 
boundaries present
the characteristic asymptotic power laws at low energy or large distance.
For the linear conductance through a single impurity in 
Luttinger liquids connected to noninteracting leads the
fRG captures the expected power-law scaling, as well as 
the complete crossover from the weak to the strong-impurity limit
determined by a one-parameter scaling function. 
For resonant tunneling in a Luttinger liquid with a double 
barrier enclosing a dot region, depending on the dot parameters
several temperature regimes with distinctive power-law behavior 
of the resonance conductance as well as
regimes characterized by non-universal behavior are identified
\cite{EMABMS04,Med04}.

Including the spin degree of freedom, 
two-particle backscattering of particles with opposite spin at
opposite Fermi points leads to two important effects, not
present in the case of spinless fermions. First, the expected
decrease of spectral weight and of the conductance at low
energy scales is often preceded by an increase, which 
can be particularly pronounced for the density of states near 
an impurity or boundary as a function of $\om$.
For the density of states near a boundary this effect has been
found already earlier within a Hartree-Fock and DMRG study of 
the Hubbard model \cite{Med00,Gun00}, and for the conductance by a 
renormalization-group analysis of the $g$-ology model \cite{MYG93,YGM94}. 
Second, the asymptotic low-energy power laws are usually 
modified by logarithmic corrections.
In the extended Hubbard model the backscattering can be 
eliminated for a special fine-tuned choice of parameters.
Then the results are very similar to those for spinless fermions.
For weak and intermediate impurity strengths the asymptotic
low-energy behavior is approached only at rather low scales,
which are accessible only for very large systems.
This slow convergence observed already for spinless fermions
holds also in the absence of two-particle backscattering.

Interesting further extensions of the fRG for 
impurities in Luttinger liquids include 
the investigation of non-equilibrium phenomena and
the analysis of disorder.
For isolated impurities the
influence of impurities on the vertex renormalization is 
irrelevant for the asymptotic low-energy or long-distance 
behavior, although it may contribute quantitatively at intermediate
scales.
For disordered systems with a finite impurity
density the influence of the latter on the two-particle 
vertex is crucial and must be taken into account \cite{Gia03}.
In principle this is possible by computing the vertex flow
with full propagators, which contain the renormalized impurity 
potential via the self-energy.
A further challenging extension concerns the 
inclusion of inelastic processes. They appear
at second order in the interaction and
can be included in the flow 
equations by inserting the two-particle vertex into the flow
equation for the self-energy without neglecting its frequency
dependence. 
Finally, a flexible microscopic modeling feasible with the 
fRG approach allows for a more realistic description of
contacts and leads in experimental systems.

%%% Local Variables: 
%%% mode: latex
%%% TeX-master: "thesis"
%%% End: 

%%%%%%%%%%%%%%%%%%%%%%%%%%%%%%%%%%%%%%%%%%%%%%%%%%%%%%%%%%%%%%%%%%%%%%%%%%%%%%

\appendix
% Thesis appendices

\chapter{Evaluation of vertex flow for spin-$\mathbf{\frac{1}{2}}$ fermions}
\label{sec:app:vertspin}

%%%%%%%%%%%%%%%%%%%%%%%%%%%%%%%%%%%%%%%%%%%%%%%%%%%%%%%%%%%%%%%%%%%%%%%%%%%%%%

\section{Functional RG}
\label{sec:app:vertspin:frg}

Here we present a detailed derivation of the flow equations for the two-particle 
vertex $\Gam^{\Lam}$. 
Starting from the flow equation (\ref{eq:flow}) we insert on the right-hand side 
the parametrization (\ref{eq:paramt}) for $\Gam^{\Lam}_t$ 
and (\ref{eq:params}) for $\Gam^{\Lam}_s$.
The flow of the singlet vertex 
$\Gam^{\Lam}_{s|\, k'_1,k'_2;k_1^{\phantom '},k_2^{\phantom '}}$
is computed for the three choices of 
$(k'_1,k'_2,k_1^{\phantom '},k_2^{\phantom '})$ which yield
the flow of $g^{\Lam}_{s2}$, $g^{\Lam}_{s3}$, $g^{\Lam}_{s4}$ corresponding to 
Eqs.~(\ref{eq:gs1}) and (\ref{eq:gs2}), while the flow of the triplet vertex 
$\Gam^{\Lam}_{t|\, k'_1,k'_2;k_1^{\phantom '},k_2^{\phantom '}}$
is evaluated only for 
$(k'_1,k'_2,k_1^{\phantom '},k_2^{\phantom '}) = (k_F,-k_F,k_F,-k_F)$ as 
in (\ref{eq:gt}), which yields the flow of $g^{\Lam}_t$. For $\alpha=s2,s3,s4,t$
we obtain
\begin{equation} 
  \label{eq:pint}
  \frac{\partial g_{\alpha}}{\partial\Lam} =
  \frac{1}{4\pi^2}\sum_{\om = \pm\Lam}  \int_0^{2\pi} \frac{dp}{2\pi} \,  
  f_{\alpha}(p,\om) \; , 
\end{equation} 
with
\begin{align}
  f_{s2}(p,\om) =&
  \frac{(2P^{\phantom '}_s+U^{\phantom '}_s-2\mu_0U'_s\cos p)^2}{2(i\om - \xi^0_p)(-i\om - \xi^0_{-p})}
  + 
  \frac{(2U'_s+U^{\phantom '}_s-\mu_0(U'_s+P^{\phantom '}_s)\cos p)^2}{4(i\om - \xi^0_p)^2}
  \nonumber\\[2mm]&- 
  \frac{(4-\mu_0^2)(U'_s-P^{\phantom '}_s)^2\sin^2 p+6(2+\mu_0\cos p)U'_t(2U'_s+U^{\phantom '}_s)}{4(i\om - \xi^0_p)^2}
  \nonumber\\[2mm]&+\frac{6(2+\mu_0\cos p)\mu_0U'_t(U'_s+P^{\phantom '}_s)\cos p-(4-\mu_0^2)(U'_s-P^{\phantom '}_s)\sin^2 p}{4(i\om - \xi^0_p)^2}
  \nonumber\\[2mm]&-\frac{3(2\cos p+\mu_0)^2{U'}^2_t}{4(i\om - \xi^0_p)^2} 
  +\frac{((\mu_0^2-2)U'_s+U^{\phantom '}_s+2(U'_s+P^{\phantom '}_s)\cos p)^2}{4(i\om - \xi^0_{p-k_F})(i\om - \xi^0_{p+k_F})}
  \nonumber\\[2mm]& -\frac{6(\mu_0^2-2-2\cos p)U'_t((\mu_0^2-2)U'_s+U^{\phantom '}_s+2(U'_s+P^{\phantom '}_s)\cos p)}{4(i\om - \xi^0_{p-k_F})(i\om - \xi^0_{p+k_F})}
  \nonumber\\[2mm]& -\frac{3(\mu_0^2-2-2\cos p)^2 {U'}^2_t}
  {4(i\om - \xi^0_{p-k_F})(i\om - \xi^0_{p+k_F})}
  \\[9mm]
  f_{s3}(p,\om) =&
  \frac{(2U'_s-U^{\phantom '}_s)^2-16{U'}^2_s\sin^2 p}{2(i\om - \xi^0_p)(-i\om - \xi^0_{\pi-p})} 
  -
  \frac{4(U'_s+P^{\phantom '}_s)^2\sin^2 p -(2U'_s-U^{\phantom '}_s)^2}{2(i\om - \xi^0_{p})(i\om - \xi^0_{\pi+p})}
  \nonumber\\[2mm]&-\frac{6U'_t(2(U'_s+P^{\phantom '}_s)\sin^2 p-2U'_s+U^{\phantom '}_s)+6{U'}^2_t\cos^2 p}
  {(i\om - \xi^0_{p})(i\om - \xi^0_{\pi+p})} 
  \\[9mm]
  f_{s4}(p,\om) =&
  \frac{(4U'_s\cos p+(\mu_0^2-2)P^{\phantom '}_s+U^{\phantom '}_s)^2}{2(i\om - \xi^0_{p-k_F})(-i\om - \xi^0_{p+k_F})} 
  \nonumber\\[2mm]&+
  \frac{(2U'_s+U^{\phantom '}_s-\mu_0(U'_s+P^{\phantom '}_s)\cos p+2(U'_s-P^{\phantom '}_s)\sin p \sin k_F)^2}{2(i\om - \xi^0_p)^2}
  \nonumber\\[2mm]&-
  \frac{6(2+\mu_0\cos p-2\sin p\sin k_F)(2U'_s+U^{\phantom '}_s-\mu_0(U'_s+P^{\phantom '}_s)\cos p)U'_t}{2(i\om - \xi^0_p)^2}
  \nonumber\\[2mm]&-
  \frac{
    12(2+\mu_0\cos p-2\sin p\sin k_F)U'_t(U'_s-P^{\phantom '}_s)\sin p\sin k_F
  }{2(i\om - \xi^0_p)^2}
  \nonumber\\[2mm]&
  -\frac{3(2+\mu_0\cos p-2\sin p\sin k_F)^2{U'}^2_t}{2(i\om - \xi^0_p)^2}
  \\[9mm]
  f_t (p,\om)=&
 \frac{2(4-\mu_0^2){U'}^2_t\sin^2 p}{(i\om - \xi^0_p)(-i\om - \xi^0_{-p})} -
 \frac{(2U'_s+U^{\phantom '}_s-\mu_0(U'_s+P^{\phantom '}_s)\cos p)^2}{4(i\om - \xi^0_p)^2}
 \nonumber\\[2mm]&+\frac{
   (4-\mu_0)(U'_s-P^{\phantom '}_s)^2\sin^2 p-2(2+\mu_0\cos p)(2U'_s+U^{\phantom '}_s)U'_t}{4(i\om - \xi^0_p)^2}
 \nonumber\\[2mm]&+\frac{
   2(\mu_0(2+\mu_0\cos p)(U'_s+P^{\phantom '}_s)\cos p-(4-\mu_0^2)(U'_s-P^{\phantom '}_s)\sin^2 p)U'_t}{4(i\om - \xi^0_p)^2}
 \nonumber\\[2mm]&-\frac{
   5(\mu_0+2\cos p)^2}{4(i\om - \xi^0_p)^2{U'}^2_t} +
 \frac{((\mu_0^2-2)U'_s+U^{\phantom '}_s+2(U'_s+P^{\phantom '}_s)\cos p)^2}{4(i\om - \xi^0_{p-k_F})(i\om - \xi^0_{p+k_F})} 
 \nonumber\\[2mm]&+\frac{
   2(\mu_0^2-2-2\cos p)((\mu_0^2-2)U'_s+U^{\phantom '}_s+2(U'_s+P^{\phantom '}_s)\cos p)U'_t}{4(i\om - \xi^0_{p-k_F})(i\om - \xi^0_{p+k_F})} 
 \nonumber\\[2mm]&+\frac{
   5(\mu_0^2-2-2\cos p)^2{U'}^2_t}
  {4(i\om - \xi^0_{p-k_F})(i\om - \xi^0_{p+k_F})} 
  \;.\\[-5mm]\nonumber
\end{align}
Here $\xi^0_k = -2 \cos k - \mu_0$ with $\mu_0 = - 2 \cos k_F$ is
the bare dispersion relation relative to the bare Fermi level.
Since the functions $f_{\alpha}(p,\om)$ can be written as rational functions 
of $\cos p$ and $\sin p$, the $p\,$-integral of Eq.~(\ref{eq:pint}) can be 
carried out analytically using
the substitution $z = e^{ip}$ and the residue theorem, in analogy to the
spinless case described in Sec.~\ref{sec:frglutt:flow:spinless}.
The resulting differential equations for the momentum space couplings 
$g^{\Lam}_{\alpha}$ read
\\[-5mm]  
\begin{align}    
  \frac{\partial g_{s2}}{\partial\Lam} =&\frac{2({U'}^2_s+P^{2}_s-3{U'}^2_t+6U'_sU'_t) 
    +\mu_0^2(2{U'}^2_s+U'_sP^{\phantom '}_s+3P^{\phantom '}_sU'_t)}{4\pi}
  \nonumber\\[2mm] &
  +\Re\, \Big[\,\frac{\gam(\Lam) }{\pi(4-(\mu_0 + i\Lam)^2)}
  \,
  \Big(\,
  \frac{(2P^{\phantom '}_s+U^{\phantom '}_s+\mu_0(\mu_0+i\Lam)U'_s)^2}{ i\Lam}
  \nonumber\\[2mm] +&\frac{
    (\mu_0 + i\Lam)((2U'_s+U^{\phantom '}_s)^2+2\mu_0^2(U'_s+P^{ \phantom '}_s)^2) 
    +4\mu_0(2U'_s+U^{\phantom '}_s)(U'_s+P^{\phantom '}_s)}{2(4-(\mu_0 + i\Lam)^2)}
  \nonumber\\[2mm] -&\frac{
    \mu_0^2(\mu_0 + i\Lam)^3(U'_s+P^{\phantom '}_s)^2
    +24(\mu_0 + i\Lam)(4U'_s+2U^{\phantom '}_s-\mu_0^2(U'_s+P^{\phantom '}_s))U'_t}{8(4-(\mu_0 + i\Lam)^2)}
  \nonumber\\[2mm] +&\frac{
    12\mu_0(U^{\phantom '}_s-2P^{\phantom '}_s)U'_t-3((\mu_0 + i\Lam)(\mu_0^2+4)-8\mu_0){U'}^2_t}{2(4-(\mu_0 + i\Lam)^2)}
  \nonumber\\[2mm] -&\frac{
    (\mu_0 + i\Lam)(((\mu_0^2-2)U'_s+U^{\phantom '}_s)^2+4(U'_s+P^{\phantom '}_s)^2)
  }{2((\mu_0 + i\Lam)^2-\mu_0^2)}
  \nonumber \\[2mm] -&\frac{
    2\mu_0(U'_s+P^{\phantom '}_s)((\mu_0^2-2)U'_s+U^{\phantom '}_s)
    -6\mu_0((\mu_0^2-2)P^{\phantom '}_s-U^{\phantom '}_s)U'_t }{(\mu_0 + i\Lam)^2-\mu_0^2}
  \nonumber \\[2mm] +&\frac{
    3(\mu_0 + i\Lam)((\mu_0^2-2)((\mu_0^2-2)U'_s+U^{\phantom '}_s) 
    -4(U'_s+P^{\phantom '}_s))U'_t }{(\mu_0 + i\Lam)^2-\mu_0^2}
  \nonumber \\[2mm] +&\frac{
    3((\mu_0 + i\Lam)((\mu_0^2-2)^2+4)-4\mu_0(\mu_0^2-2))  
    {U'}^2_t}{2((\mu_0 + i\Lam)^2-\mu_0^2)}+
  (\mu_0 + i\Lam)({U'}^2_s+P^{2 \phantom '}_s)
  \nonumber \\[2mm] -&\frac{
    (\mu_0 + i\Lam)(\mu_0^2(U'_s-P^{\phantom '}_s)^2-12\mu_0^2P^{\phantom '}_sU'_t 
    -48U'_sU'_t+24{U'}^2_t) }{8}
  \, 
  \Big)\,\Big]
  \\[9mm]
  \frac{\partial g_{s3}}{\partial\Lam}=&-\frac{4{U'}^2_s+(U'_s+P^{\phantom '}_s)(U'_s+P^{\phantom '}_s+6U'_t)-3{U'}^2_t }{2\pi}
  \nonumber\\[2mm] &
  +\Re \,\Big[\,\frac{\gam(\Lam) }{\pi}
  \,\Big(\,\frac{4{U'}^2_s}{\mu_0}+\frac{ (U'_s+P^{\phantom '}_s)(U'_s+P^{\phantom '}_s+6U'_t) 
    }{\mu_0 + i\Lam}
  \nonumber\\[2mm] &  
  -\frac{(2U'_s-U^{\phantom '}_s)(2U'_s-U^{\phantom '}_s-12U'_t)}{4(\mu_0 + i\Lam)}
+\frac{3(\mu_0 + i\Lam){U'}^2_t
  }{4-(\mu_0 + i\Lam)^2}
  \nonumber  \\[2mm] &
  -\frac{4(2P^{\phantom '}_s-U^{\phantom '}_s)^2
    +\mu_0 (\mu_0 + i\Lam)(2U'_s-U^{\phantom '}_s)(2U'_s-U^{\phantom '}_s-12U'_t)}{4\mu_0(4-(\mu_0 + i\Lam)^2)}
   \,
  \Big)\,\Big]
  \\[9mm]
  \frac{\partial g_{s4}}{\partial\Lam}=&\frac{8{U'}^2_s 
    +(\mu_0^2-2)({U'}^2_s+P^{2 \phantom '}_s-3{U'}^2_t+6U'_sU'_t) 
    +4U'_sP^{\phantom '}_s+12P^{\phantom '}_sU'_t }{4\pi}
  \nonumber\\[2mm] &
  +\Re \,\Big[\,\frac{\gam(\Lam)}{\pi(4-(\mu_0 + i\Lam)^2)}
  \,\Big(\,2(\mu_0 + i\Lam)P^{\phantom '}_s(U'_s+3U'_t)
  \nonumber\\[2mm] &
  +\frac{(\mu_0 + i\Lam)(\mu_0^2-2)({U'}^2_s+P^{2 \phantom '}_s
    -3{U'}^2_t+6U'_tU'_s)}{2}
  \nonumber\\[2mm] &
  -\frac{((\mu_0^2-2)P^{\phantom '}_s+U^{\phantom '}_s+((\mu_0 + i\Lam)\mu_0-4i\gam(\Lam)\sin k_F)U'_s)^2 }{\mu_0 \Lam^2}\, \big(\mu_0(\mu_0 + i\Lam)
  \nonumber\\[2mm] &
  +4i \gam(\Lam) \sin k_F -4 \big)+\frac{
    (\mu_0 + i\Lam)((2U'_s+U^{\phantom '}_s)^2+\mu_0^2(U'_s+P^{\phantom '}_s)^2)}{4-(\mu_0 + i\Lam)^2}              
  \nonumber\\[2mm] &+\frac{
    4\mu_0(2U'_s+U^{\phantom '}_s)(U'_s+P^{\phantom '}_s) -6(\mu_0 + i\Lam)(4U'_s+2U^{\phantom '}_s-\mu_0^2(U'_s+P^{\phantom '}_s)) U'_t  }{4-(\mu_0 + i\Lam)^2}              
  \nonumber\\[2mm] &
  +\frac{12\mu_0(U^{\phantom '}_s-2P^{\phantom '}_s)U'_t  -3((\mu_0 + i\Lam)(4+\mu_0^2)-8\mu_0) 
    {U'}^2_t }{4-(\mu_0 + i\Lam)^2} 
  \,\Big)\,\Big]
  \\[9mm]
  \frac{\partial g_{t}}{\partial\Lam}=&\frac{(4-\mu_0^2)(U'_sP^{\phantom '}_s-P^{\phantom '}_sU'_t-2{U'}^2_t) }{4\pi}
  \nonumber\\[2mm] &
  +\Re \,\Big[\,\frac{\gam(\Lam) }{\pi(4-(\mu_0 + i\Lam)^2)}
  \,\Big(\,\frac{(\mu_0 + i\Lam)(4-\mu_0^2)P^{\phantom '}_s(U'_s-U'_t)}{2}
  \nonumber\\[2mm] &
  +\frac{(4-\mu_0^2)(4-(\mu_0 + i\Lam)^2){U'}^2_t}{i\Lam}
- \frac{
     2\mu_0(2U'_s+U^{\phantom '}_s)(U'_s+P^{\phantom '}_s) }{(4-(\mu_0 + i\Lam)^2)}
  \nonumber\\[2mm] &- \frac{
    (\mu_0 + i\Lam)((2U'_s+U^{\phantom '}_s)^2+\mu_0^2(U'_s+P^{\phantom '}_s)^2) 
     }{2(4-(\mu_0 + i\Lam)^2)}
  \nonumber\\[2mm] &-\frac{
    ((\mu_0 + i\Lam)(4U'_s+2U^{\phantom '}_s-\mu_0^2(U'_s+P^{\phantom '}_s))-2\mu_0(U^{\phantom '}_s-2P^{\phantom '}_s))U'_t  }{4-(\mu_0 + i\Lam)^2}
  \nonumber       \\[2mm] &-\frac{
    5((\mu_0 + i\Lam)(\mu_0^2+4)-8\mu_0){U'}^2_t}{2(4-(\mu_0 + i\Lam)^2)}
  -\frac{(\mu_0 + i\Lam)((\mu_0^2-2)U'_s+U^{\phantom '}_s)^2 }{2((\mu_0 + i\Lam)^2-\mu_0^2)}
  \nonumber\\[2mm] 
  &-\frac{2(\mu_0 + i\Lam)
    (U'_s+P^{\phantom '}_s)^2 
    +2\mu_0(U'_s+P^{\phantom '}_s)((\mu_0^2-2)U'_s+U^{\phantom '}_s)}{(\mu_0 + i\Lam)^2-\mu_0^2}
  \nonumber\\[2mm] &
  -\frac{(\mu_0 + i\Lam)((\mu_0^2-2)((\mu_0^2-2)U'_s+U^{\phantom '}_s) 
    -4(U'_s+P^{\phantom '}_s))U'_t}{(\mu_0 + i\Lam)^2-\mu_0^2}
  \nonumber\\[2mm] &
  -\frac{2\mu_0((\mu_0^2-2)P^{\phantom '}_s-U^{\phantom '}_s)U'_t}{(\mu_0 + i\Lam)^2-\mu_0^2}
  -\frac{5(\mu_0 + i\Lam)((\mu_0^2-2)^2+4){U'}^2_t}{2((\mu_0 + i\Lam)^2-\mu_0^2)}
  \nonumber\\[2mm] &
  +\frac{10\mu_0(\mu_0^2-2){U'}^2_t 
  }{(\mu_0 + i\Lam)^2-\mu_0^2}
  \,\Big)\,\Big] \;,\\[-5mm]\nonumber
\end{align}
with
\begin{align}
  \label{eq:helpf}
  \gam (\Lam)= \frac{(\mu_0 + i\Lam)}{2}\, \sqrt{1 - \frac{4}{(\mu_0 + i\Lam)^2}} \; .
\end{align}
Using the linear equations (\ref{eq:paramgt}) and (\ref{eq:paramgs}) 
to replace $g_{\alpha}$ by the renormalized real space interactions 
on the left-hand 
side of the flow equations, we obtain a complete set of
flow equations for the four renormalized interactions ${U'_t}$, 
$U^{\phantom '}_s$, ${U'_s}$, and $P^{\phantom '}_s$ of the form (\ref{eq:flowu}).

%%%%%%%%%%%%%%%%%%%%%%%%%%%%%%%%%%%%%%%%%%%%%%%%%%%%%%%%%%%%%%%%%%%%%%%%%%%%%%

\section{One-loop $\mathbf{g}$-ology calculation}
\label{sec:app:vertspin:gology}

In the low-energy limit the flow of the two-particle vertex $\Gam ^{\Lam}$, as
described in Sec.~\ref{sec:app:vertspin:frg},
reduces to the one-loop flow of the $g$-ology model, 
the general effective low-energy model for 
one-dimensional fermions \cite{Sol79}. 
In the $g$-ology approach
interaction processes are classified 
into backward scattering $(g_{1 \perp})$, forward scattering involving electrons 
from opposite 
Fermi points $(g_{2 \perp})$, from the same Fermi points $(g_{4 \perp})$, and 
umklapp scattering 
$(g_{3 \perp})$. All further momentum dependences of the vertex are discarded.

The $g$-ology couplings are related to the momentum space couplings 
$g_{s2},g_{s3},g_{s4}$ and $g_t$ by
\begin{align}
  g_{1\perp} &= \frac{1}{2}(g_{s2}-g_t) \nonumber\\[2mm]
  g_{2\perp} &= \frac{1}{2}(g_{s2}+g_t) \nonumber\\[2mm]
  g_{3\perp} &= \frac{g_{s3}}{2} \nonumber\\[2mm]
  g_{4\perp} &= \frac{g_{s4}}{2}
\end{align}
and to the real space couplings by
\begin{align}
  g_{1\perp} &= \frac{1}{2}(\mu_0^2 \Gam_{s2}+2 \Gam_{s3}+\Gam_{s4}-(4-\mu_0^2)\Gam_t) \nonumber\\[2mm]
  g_{2\perp} &= \frac{1}{2}(\mu_0^2 \Gam_{s2}+2 \Gam_{s3}+\Gam_{s4}+(4-\mu_0^2)\Gam_t) \nonumber\\[2mm]
  g_{3\perp} &= \frac{1}{2}(-4 \Gam_{s2}-2 \Gam_{s3}+\Gam_{s4}) \nonumber\\[2mm]
  g_{4\perp} &= \frac{1}{2}(4 \Gam_{s2}-(2-\mu_0^2) \Gam_{s3}+\Gam_{s4})
\end{align}
respectively.

The flow equation for $\Gam ^{\Lam}$ (\ref{eq:flow}) reduces to the standard
one-loop $g$-ology calculation \cite{Sol79}, 
once the dependence on the internal momentum $p$ on the right-hand side
of the flow equation is neglected.
Applying the above parametrization, we obtain a complete set of flow equations
for $g_{i\perp}$ for $i=1,\dots,4$ of the form
\begin{align}
  \label{eq:flowgg}
  \frac{\partial g_{1\perp}}{\partial \Lam} &= \frac{1}{2\pi}\,\big[\,
  g_{1\perp}g_{2\perp}{PP}(0)+g_{1\perp}g_{4\perp}{ PH}(0)+2g_{1\perp}(g_{2\perp}-g_{1\perp}){ PH}(2k_F)
  \,\big] \nonumber\\[2mm]
  \frac{\partial g_{2\perp}}{\partial \Lam} &= \frac{1}{2\pi}\,\big[\,
  \frac{1}{2}(g_{1\perp}^2+g_{2\perp}^2){ PP}(0)+g_{4\perp}(g_{1\perp}-g_{2\perp}){ PH(0)}+g_{2\perp}^2{ PH}(2k_F)
  \,\big] \nonumber\\[2mm]
  \frac{\partial g_{3\perp}}{\partial \Lam} &= \frac{1}{2\pi}\,\big[\,
  g_{3\perp}g_{4\perp}{PP}(\pi)+g_{3\perp}(2g_{2\perp}-g_{1\perp}){ PH}(\pi)
  \,\big] \nonumber\\[2mm]
  \frac{\partial g_{4\perp}}{\partial \Lam} &= \frac{1}{2\pi}\,\big[\,
  g_{4\perp}^2{PP}(2k_F)+\frac{1}{2}(g_{1\perp}^2+2g_{1\perp}g_{2\perp}-2g_{2\perp}^2+g_{4\perp}^2){ PH}(0)
  \,\big] \; ,
\end{align}
where 
\begin{align}
  \label{eq:parthole}
  {PP}(q)&= \frac{1}{2\pi}\sum_{\om = \pm\Lam}  \int_0^{2\pi} \frac{dp}{2\pi} \,
  \frac{1}{i\om-\xi_{p+q/2}}\frac{1}{-i\om-\xi_{-p+q/2}} \nonumber\\[2mm]
  { PH}(q)&= \frac{1}{2\pi}\sum_{\om = \pm\Lam}  \int_0^{2\pi} \frac{dp}{2\pi} \,
  \frac{1}{i\om-\xi_{p+q/2}}\frac{1}{i\om-\xi_{p-q/2}} \; .
\end{align}
The initial conditions are $g_{1\perp}^{\Lam_0}=g_{3\perp}^{\Lam_0}=U+(\mu_0^2-2)U'$ 
and $g_{2\perp}^{\Lam_0}=g_{4\perp}^{\Lam_0}=U+2U'$.
The integrals in Eq.~(\ref{eq:parthole}) can be computed analytically using the
residue theorem, as discussed in Sec.~\ref{sec:app:vertspin:frg}; for $q=0$ and 
$q=2k_F$ we obtain
\begin{align}
  \label{eq:exp}
  {PP}(0)&=-\frac{4}{\Lam}\, \Re \,\Big( \frac{i\gam (\Lam)  }{4-(\mu_0 + i\Lam)^2} \Big) \nonumber\\[2mm]
  {PH}(0)&=4\, \Re \,\Big( \frac{(\mu_0 + i\Lam)\gam (\Lam) }{(4-(\mu_0 + i\Lam)^2)^2}\Big) \nonumber\\[2mm]
  {PP}(2k_F)&=-\frac{4}{\mu_0}\, \Re \,\Big( \frac{\gam (\Lam)}{(4-(\mu_0 + i\Lam)^2)(4-\mu_0(\mu_0 + i\Lam)+4i\gam (\Lam)\sin k_F )}\Big) \nonumber\\[2mm]
  {PH}(2k_F)&=4\, \Re \,\Big( \frac{ (\mu_0 + i\Lam)\gam (\Lam) }{(4-(\mu_0 + i\Lam)^2)(\mu_0^2-(\mu_0 + i\Lam)^2)}\Big) \; ,
\end{align}
where $\gam(\Lam)$ is defined by Eq.~(\ref{eq:helpf}).

For $\mu\neq 0$ the above equations reduce to
\begin{align}
  { PP}(0)&=\frac{2}{\Lam \sqrt{4-\mu_0^2}} \nonumber\\[2mm]
  { PH}(0)&= -\frac{4(2+\mu_0^2)\Lam }{(\sqrt{4-\mu_0^2})^5}\nonumber\\[2mm]
  { PP}(2k_F)&= \frac{8\Lam }{(\sqrt{4-\mu_0^2})^5}\nonumber\\[2mm]
  { PH}(2k_F)&=- \frac{1}{\Lam  \sqrt{4-\mu_0^2}} \; 
\end{align}
in the limit $\Lam \to 0$.
The resulting flow equations in the low-energy limit are
\begin{align}
  \frac{\partial g_{1\perp}}{\partial \Lam} &= \frac{ g_{1\perp}^2}{\Lam\pi\sqrt{4-\mu_0^2}} \nonumber\\[2mm]
  \frac{\partial g_{2\perp}}{\partial \Lam} &= \frac{ g_{1\perp}^2}{2\Lam\pi\sqrt{4-\mu_0^2}} 
\end{align}
for $g_{1\perp}$ and  $g_{2\perp}$,  whereas the flow for $g_{3\perp}$ and 
$g_{4\perp}$ vanishes.
The solution reads
\begin{align}
   g_{1\perp}^{ \Lam}&= \frac{U+(\mu_0^2-2)U'}{1+\frac{U+(\mu_0^2-2)U'}{\pi\sqrt{4-\mu_0^2}}\ln \frac{\Lam}{\Lam_0} }  \nonumber\\[2mm]
   g_{2\perp}^{ \Lam}&= \frac{1}{2}\Big(\frac{U+(\mu_0^2-2)U'}{1+\frac{U+(\mu_0^2-2)U'}{\pi\sqrt{4-\mu_0^2}}\ln \frac{\Lam}{\Lam_0}}+U+(6-\mu_0^2)U' \Big)\; ,
\end{align}
yielding the fixed-point couplings $g_{1\perp}=0$, 
$g_{2\perp}=\frac{1}{2}(U+(6-\mu_0^2)U')$, and
$g_{3\perp}=U+(\mu_0^2-2)U'$,
$g_{4\perp}=U+2U'$.

%%%%%%%%%%%%%%%%%%%%%%%%%%%%%%%%%%%%%%%%%%%%%%%%%%%%%%%%%%%%%%%%%%%%%%%%%%%%%%

\chapter{Bethe-ansatz calculation of $\mathbf{K_{\rho}}$ for the Hubbard model}
\label{sec:app:ba}

The Luttinger-liquid parameter $K_{\rho}$ can be evaluated exactly from
the Bethe ansatz solution for the one-dimensional Hubbard model \cite{LW68}.
The Bethe ansatz provides the energies of the ground and excited states as
solution of specific integral equations.

In the description of critical properties the ``dressed charge matrix''
is introduced \cite{FK90}. 
This $2 \times 2$ matrix
contains the effective renormalized coupling constants within and between
the low-energy charge and spin sectors of the Hilbert space of the 
Hubbard model, and therefore directly determines all critical exponents. 
It is defined as
\begin{equation}
  Z = \left( 
    \begin{array}{cc}
      Z_{cc} & Z_{cs} \\
      Z_{sc} & Z_{ss} 
    \end{array}
  \right) 
  \; .
\end{equation}
In the absence of a magnetic field, $Z$ is completely
determined by its first element  $Z_{cc}$ as
\begin{equation}
  Z = \left( 
    \begin{array}{cc}
      \xi(k_0) & 0 \\
      \frac{\xi(k_0)}{2} & \frac{1}{\sqrt{2}}
    \end{array}
  \right) \; ,
\end{equation}
where $k_0$ is a cutoff determined by the particle 
density $n$  \cite{FK90}.

$\xi(k)$ obeys the integral equation
\begin{equation}
  \xi (k) = 1+\frac{4}{U} \int_{-k_0}^{k_0}\, \cos k' \,R\big(\,\frac{4}{U}\,(\sin k-\sin k')\,\big)\,\xi(k') \,dk' \; ,
\end{equation}
with the kernel 
\begin{equation}
  \label{eq:csi}
  R(x)=\frac{1}{2\pi} \int_0^{\infty}\frac{\cos(\frac{xy}{2})}{1+e^y}\, dy \; .
\end{equation}
The cutoff momentum $k_0$ is defined by
\begin{equation}
  \label{eq:n}
  \int_{-k_0}^{k_0}\rho(k') \,dk' = n \; ,
\end{equation}
where the integral equation for the 
ground-state charge distribution function $\rho (k)$ 
reads
\begin{equation}
  \label{eq:rho}
  \rho (k) = \frac{1}{2\pi} +\frac{4}{U}\cos k \int_{-k_0}^{k_0}\, R\,\big(\,\frac{4}{U}\,(\sin k-\sin k')\,\big)\,\rho(k') \,dk' 
\end{equation}
in the limit $L \to \infty$ .

The compressibility $\kappa$ is related to the dressed charge matrix element  
$\xi(k_0)$ by $\xi^2(k_0)=\pi v_{\rho} \kappa$, where $v_{\rho}$ is the charge 
velocity \cite{FK90}. On the other hand, the Luttinger-liquid parameter $K_{\rho}$ 
is given by $K_{\rho}=\pi v_{\rho} \kappa /2$. 
Hence $K_{\rho}$ is determined by
\begin{equation}
  K_{\rho}=\frac{\xi^2(k_0)}{2} \; .
\end{equation}
Solving the integral equation for $\xi(k_0)$ with  
$k_0$ determined from equations (\ref{eq:n}) and (\ref{eq:rho}),
yields the exact Luttinger-liquid parameter.
For the numerical computation of the kernel $R(x)$ defined in Eq.~(\ref{eq:csi}), 
the following expression is more convenient
\begin{equation}
  R(x)=\frac{1}{\pi} \,\sum_{l=1}^{\infty} (-1)^{l+1}\frac{2l}{x^2+(2l)^2} \; .
\end{equation}

Similarly the chemical potential $\mu$ is determined.
A comparison of the fRG results with the exact results allows a quantitative 
estimate of the effect due to the neglected 
frequency dependence of the two-particle vertex.
The chemical potential can be derived from the ground-state energy via
\begin{equation}
  \mu=\frac{\partial \eps}{\partial n}=\frac{\partial \eps}{\partial k_0}\,\Big( \frac{\partial n}{\partial k_0}\Big)^{-1} \; ,
\end{equation}
where the energy per lattice site is given by
\begin{equation}
  \eps (k) = -2 \int_{-k_0}^{k_0}\, \cos k \,\rho(k) \,dk \; .
\end{equation}
Using equations (\ref{eq:n}) and (\ref{eq:rho}), the solution of the 
following integral equations
\begin{align}
  \frac{\partial \eps}{\partial k_0}&=-4\rho(k_0)\cos k_0 -2 \int_{-k_0}^{k_0}\,\cos k \,\frac{\partial \rho(k)}{\partial k_0}\, dk\nonumber\\[2mm]
  \frac{\partial n}{\partial k_0}&=2\rho(k_0)+\int_{-k_0}^{k_0}\,\frac{\partial \rho(k)}{\partial k_0}\, dk \; ,
\end{align}
with 
\begin{align}
  \frac{\partial \rho(k) }{\partial k_0}=&\,\frac{4}{U}\,\rho(k_0)\cos k \,\big[\,R\,\big(\,\frac{4}{U}\,(\sin k-\sin k_0)\,\big)\,+\,
  R\,\big(\,\frac{4}{U}\,(\sin k+\sin k_0)\,\big)\,\big]\,\nonumber\\[2mm]
  &+\frac{4}{U}\cos k \int_{-k_0}^{k_0}\, R\,\big(\,\frac{4}{U}\,(\sin k-\sin k')\,\big)\,\frac{\partial \rho(k')}{\partial k_0} \,dk'
\end{align}
provides the exact chemical potential.

Analogous expressions can be derived for the spinless fermion model \cite{Hal80}.

%%% Local Variables: 
%%% mode: latex
%%% TeX-master: "thesis"
%%% End: 

%%%%%%%%%%%%%%%%%%%%%%%%%%%%%%%%%%%%%%%%%%%%%%%%%%%%%%%%%%%%%%%%%%%%%%%%%%%%%%

\backmatter
\setsize{1.1}
\setbibpreamble{
The electronic \abb{PDF} file contains links to the online 
\texttt{arXiv} and journal references.
  \bigskip}
\bibliographystyle{thesisnat}
%{\raggedright
\bibliography{thesis}

\providecommand{\href}[2]{#2}
\begin{thebibliography}{102}
\expandafter\ifx\csname natexlab\endcsname\relax\def\natexlab#1{#1}\fi
\expandafter\ifx\csname url\endcsname\relax
  \def\url#1{{\tt #1}}\fi

\bibitem[Andergassen et~al.(2005{\natexlab{a}})Andergassen, Enss, and
  Meden]{AEM05}
S.~Andergassen, T.~Enss, and V.~Meden, {\em Kondo physics in transport through
  a quantum dot with Luttinger liquid leads},
  \href{http://arXiv.org/abs/cond-mat/0509576}{\texttt{cond-mat/0509576}}.

\bibitem[Andergassen et~al.(2004)Andergassen, Enss, Meden, Metzner,
  Schollw{\"o}ck, and Sch{\"o}n-hammer]{AEMMSS04}
S.~Andergassen, T.~Enss, V.~Meden, W.~Metzner, U.~Schollw{\"o}ck, and
  K.~Sch{\"o}n-hammer, {\em Functional renormalization group for Luttinger
  liquids with impurities},
  \href{http://dx.doi.org/10.1103/PhysRevB.70.075102}{Phys.\ Rev.~B
  \textbf{70}, 075102 (2004)},
  \href{http://arXiv.org/abs/cond-mat/0403517}{\texttt{cond-mat/0403517}}.

\bibitem[Andergassen et~al.(2005{\natexlab{b}})Andergassen, Enss, Meden,
  Metzner, Schollw{\"o}ck, and Sch{\"o}n-hammer]{AEMMSS05}
S.~Andergassen, T.~Enss, V.~Meden, W.~Metzner, U.~Schollw{\"o}ck, and
  K.~Sch{\"o}n-hammer, {\em Renormalization group analysis of the
  one-dimensional extended Hubbard model with a single impurity},
  \href{http://arXiv.org/abs/cond-mat/0509021}{\texttt{cond-mat/0509021}}.

\bibitem[Apel and Rice(1982)]{AR82}
W.~Apel and T.~M. Rice, {\em Combined effect of disorder and interaction on the
  conductance of a one-dimensional fermion system},
  \href{http://dx.doi.org/10.1103/PhysRevB.26.7063}{Phys.\ Rev.~B \textbf{26},
  R7063 (1982)}.

\bibitem[Auslaender and Fishman(2000)]{Aus00}
O.~M. Auslaender and S.~Fishman, {\em Correlations in the Adiabatic Response of
  Chaotic Systems}, \href{http://dx.doi.org/10.1103/PhysRevLett.84.1886}{Phys.\
  Rev.\ Lett. \textbf{84}, 1886 (2000)}.

\bibitem[{Barnab{\'e}-Th{\'e}riault}
  et~al.(2005{\natexlab{a}}){Barnab{\'e}-Th{\'e}riault}, Sedeki, Meden, and
  Sch{\"o}nhammer]{BSMS04}
X.~{Barnab{\'e}-Th{\'e}riault}, A.~Sedeki, V.~Meden, and K.~Sch{\"o}nhammer,
  {\em A junction of three quantum wires: restoring time-reversal symmetry by
  interaction}, \href{http://dx.doi.org/10.1103/PhysRevLett.94.136405}{Phys.\
  Rev.\ Lett. \textbf{94}, 136405 (2005{\natexlab{a}})},
  \href{http://arXiv.org/abs/cond-mat/0411612}{\texttt{cond-mat/0411612}}.

\bibitem[{Barnab{\'e}-Th{\'e}riault}
  et~al.(2005{\natexlab{b}}){Barnab{\'e}-Th{\'e}riault}, Sedeki, Meden, and
  Sch{\"o}nhammer]{BSMS05}
X.~{Barnab{\'e}-Th{\'e}riault}, A.~Sedeki, V.~Meden, and K.~Sch{\"o}nhammer,
  {\em Junctions of one-dimensional quantum wires - correlation effects in
  transport}, \href{http://dx.doi.org/10.1103/PhysRev.71.205327}{Phys.\ Rev.~B
  \textbf{71}, 205327 (2005{\natexlab{b}})},
  \href{http://arXiv.org/abs/cond-mat/0501742}{\texttt{cond-mat/0501742}}.

\bibitem[Baym and Kadanoff(1961)]{BK61}
G.~Baym and L.~P. Kadanoff, {\em Conservation Laws and Correlation Functions},
  \href{http://dx.doi.org/10.1103/PhysRev.124.287}{Phys.\ Rev. \textbf{124},
  287 (1961)}.

\bibitem[Benfatto and Gallavotti(1990)]{BG90}
G.~Benfatto and G.~Gallavotti, {\em Perturbation Theory of the Fermi Surface in
  a Quantum Liquid. A General Quasiparticle Formalism and One-Dimensional
  Systems}, \href{http://dx.doi.org/10.1038/17569}{J. Stat. Phys. \textbf{59},
  541 (1990)}.

\bibitem[Bockrath et~al.(1999)Bockrath, Cobden, Lu, Rinzler, Smalley, Balents,
  and Mceuen]{BEX99}
M.~Bockrath, D.~H. Cobden, J.~Lu, A.~G. Rinzler, R.~Smalley, L.~Balents, and
  P.~L. Mceuen, {\em Luttinger-liquid behaviour in carbon nanotubes},
  \href{http://dx.doi.org/10.1038/17569}{Nature \textbf{397}, 598 (1999)}.

\bibitem[Clay et~al.(1999)Clay, Sandvik, and Campbell]{CSC99}
R.~T. Clay, A.~W. Sandvik, and D.~K. Campbell, {\em Possible exotic phases in
  the one-dimensional extended Hubbard model},
  \href{http://dx.doi.org/10.1103/PhysRevB.59.4665}{Phys.\ Rev.~B \textbf{59},
  4665 (1999)}.

\bibitem[Cohen et~al.(1998)Cohen, Richter, and Berkovits]{CRB98}
A.~Cohen, K.~Richter, and R.~Berkovits, {\em Spin and interaction effects on
  charge distribution and currents in one-dimensional conductors and rings
  within the Hartree-Fock approximation},
  \href{http://dx.doi.org/10.1103/PhysRevB.57.6223}{Phys.\ Rev.~B \textbf{57},
  6223 (1998)},
  \href{http://arXiv.org/abs/cond-mat/9804018}{\texttt{cond-mat/9804018}}.

\bibitem[Datta(1995)]{Dat95}
S.~Datta, {\em Electronic Transport in Mesoscopic Systems}, Cambridge
  University Press, Cambrigde, 1995.

\bibitem[Dzyaloshinskii and Larkin(1973)]{DL73}
I.~E. Dzyaloshinskii and A.~I. Larkin, {\em Correlation functions for a
  one-dimensional fermi system with long-range interaction (Tomonaga model)},
  Zh.\ Eksp.\ Teor.\ Fiz. \textbf{65}, 411 (1973), [Sov.\ Phys.\ JETP 38, 202
  (1974)].

\bibitem[Egger and Gogolin(1997)]{EG97}
R.~Egger and A.~O. Gogolin, {\em Effective Low-Energy Theory for Correlated
  Carbon Nanotubes},
  \href{http://dx.doi.org/10.1103/PhysRevLett.79.5082}{Phys.\ Rev.\ Lett.
  \textbf{79}, 5082 (1997)},
  \href{http://arXiv.org/abs/cond-mat/9708065}{\texttt{cond-mat/9708065}}.

\bibitem[Egger and Grabert(1995)]{EG95}
R.~Egger and H.~Grabert, {\em Friedel Oscillations for Interacting Fermions in
  One Dimension}, \href{http://dx.doi.org/10.1103/PhysRevLett.75.3505}{Phys.\
  Rev.\ Lett. \textbf{75}, 3505 (1995)},
  \href{http://arXiv.org/abs/cond-mat/9509100}{\texttt{cond-mat/9509100}}.

\bibitem[Eggert and Affleck(1992)]{EA92}
S.~Eggert and I.~Affleck, {\em Magnetic impurities in half-integer-spin
  Heisenberg antiferromagnetic chains},
  \href{http://dx.doi.org/10.1103/PhysRevB.47.10866}{Phys.\ Rev.~B \textbf{46},
  10866 (1992)}.

\bibitem[Ejima et~al.(2005)Ejima, Gebhard, and Nishimoto]{EGN05}
S.~Ejima, F.~Gebhard, and S.~Nishimoto, {\em Tomonaga-Luttinger parameters for
  doped Mott insulators},
  \href{http://dx.doi.org/10.1209/epl/i2005-10020-8}{Europhys.\ Lett.
  \textbf{70}, 492 (2005)},
  \href{http://arXiv.org/abs/cond-mat/0507508}{\texttt{cond-mat/0507508}},
  private communication.

\bibitem[Enss(2005)]{EnssThesis}
T.~Enss, {\em Renormalization, Conservation Laws and Transport in Correlated
  Electron Systems}, PhD thesis, University of Stuttgart, Germany, 2005,
  \href{http://arXiv.org/abs/cond-mat/0504703}{\texttt{cond-mat/0504703}}.

\bibitem[Enss et~al.(2005)Enss, Meden, Andergassen,
  {Barnab{\'e}-Th{\'e}riault}, Metzner, and Sch{\"o}nhammer]{EMABMS04}
T.~Enss, V.~Meden, S.~Andergassen, X.~{Barnab{\'e}-Th{\'e}riault}, W.~Metzner,
  and K.~Sch{\"o}nhammer, {\em Impurity and correlation effects on transport in
  one-dimensional quantum wires},
  \href{http://dx.doi.org/10.1103/PhysRevB.71.155401}{Phys.\ Rev.~B
  \textbf{71}, 155401 (2005)},
  \href{http://arXiv.org/abs/cond-mat/0411310}{\texttt{cond-mat/0411310}}.

\bibitem[Feldman and Trubowitz(1990)]{FT90}
J.~Feldman and E.~Trubowitz, {\em Perturbation Theory for Many Fermion
  Systems}, Helv. Phys. Acta \textbf{63}, 156 (1990).

\bibitem[Fendley et~al.(1995)Fendley, Ludwig, and Saleur]{FLS95}
P.~Fendley, A.~W.~W. Ludwig, and H.~Saleur, {\em Exact Conductance through
  Point Contacts in the {$\nu$} = 1/3 Fractional Quantum Hall Effect},
  \href{http://dx.doi.org/10.1103/PhysRevLett.74.3005}{Phys.\ Rev.\ Lett.
  \textbf{74}, 3005 (1995)},
  \href{http://arXiv.org/abs/cond-mat/9408068}{\texttt{cond-mat/9408068}}.

\bibitem[Frahm and Korepin(1990)]{FK90}
H.~Frahm and V.~E. Korepin, {\em Critical exponents for the one-dimensional
  Hubbard model}, \href{http://dx.doi.org/10.1103/PhysRevB.42.10553}{Phys.\
  Rev.~B \textbf{42}, 10553 (1990)}.

\bibitem[Furusaki(1998)]{Fur98}
A.~Furusaki, {\em Resonant tunneling through a quantum dot weakly coupled to
  quantum wires or quantum Hall edge states},
  \href{http://dx.doi.org/10.1103/PhysRevB.57.7141}{Phys.\ Rev.~B \textbf{57},
  7141 (1998)},
  \href{http://arXiv.org/abs/cond-mat/9712054}{\texttt{cond-mat/9712054}}.

\bibitem[Furusaki(2005)]{Fur05}
A.~Furusaki, {\em Kondo Problems in Tomonaga-Luttinger liquids},
  \href{http://dx.doi.org/10.1143/JPSJ.74.73}{J. Phys. Soc. Jpn. \textbf{74},
  73 (2005)},
  \href{http://arXiv.org/abs/cond-mat/0409016}{\texttt{cond-mat/0409016}}.

\bibitem[Furusaki and Nagaosa(1993{\natexlab{a}})]{FN93}
A.~Furusaki and N.~Nagaosa, {\em Resonant tunneling in a Luttinger liquid},
  \href{http://dx.doi.org/10.1103/PhysRevB.47.3827}{Phys.\ Rev.~B \textbf{47},
  3827 (1993{\natexlab{a}})}.

\bibitem[Furusaki and Nagaosa(1993{\natexlab{b}})]{FN93s}
A.~Furusaki and N.~Nagaosa, {\em Single-barrier problem and Anderson
  localization in a one-dimensional interacting electron system},
  \href{http://dx.doi.org/10.1103/PhysRevB.47.4631}{Phys.\ Rev.~B \textbf{47},
  4631 (1993{\natexlab{b}})}.

\bibitem[Furusaki and Nagaosa(1996)]{FN96}
A.~Furusaki and N.~Nagaosa, {\em Tunneling through a barrier in a
  Tomonaga-Luttinger liquid connected to reservoirs},
  \href{http://dx.doi.org/10.1103/PhysRevB.54.R5239}{Phys.\ Rev.~B \textbf{54},
  R5239 (1996)},
  \href{http://arXiv.org/abs/cond-mat/9604193}{\texttt{cond-mat/9604193}}.

\bibitem[Gerland et~al.(2000)Gerland, von Delft, Costi, and Oreg]{GEX00}
U.~Gerland, J.~von Delft, T.~A. Costi, and Y.~Oreg, {\em Transmission Phase
  Shift of a Quantum Dot with Kondo Correlations},
  \href{http://dx.doi.org/10.1103/PhysRevLett.84.3710}{Phys.\ Rev.\ Lett.
  \textbf{84}, 3710 (2000)},
  \href{http://arXiv.org/abs/cond-mat/9909401}{\texttt{cond-mat/9909401}}.

\bibitem[Giamarchi(2004)]{Gia03}
T.~Giamarchi, {\em Quantum Physics in One Dimension}, Oxford University Press,
  New York, 2004.

\bibitem[Giamarchi and Schulz(1988)]{GS88}
T.~Giamarchi and H.~J. Schulz, {\em Anderson localization and interactions in
  one-dimensional metals},
  \href{http://dx.doi.org/10.1103/PhysRevB.37.325}{Phys.\ Rev.~B \textbf{37},
  325 (1988)}.

\bibitem[Glazman and Raikh(1988)]{GR88}
L.~I. Glazman and M.~{\'E}. Raikh, {\em Resonant Kondo transparency of a
  barrier with quasilocal impurity states}, JETP Lett. \textbf{47}, 452 (1988).

\bibitem[Halboth and Metzner(2000)]{HM00}
C.~J. Halboth and W.~Metzner, {\em Renormalization group analysis of the
  {\twoD} Hubbard model},
  \href{http://dx.doi.org/10.1103/PhysRevB.61.7364}{Phys.\ Rev.~B \textbf{61},
  7364 (2000)},
  \href{http://arXiv.org/abs/cond-mat/9908471}{\texttt{cond-mat/9908471}}.

\bibitem[Haldane(1980)]{Hal80}
F.~D.~M. Haldane, {\em General Relation of Correlation Exponents and Spectral
  Properties of One-Dimensional Fermi Systems: Application to the Anisotropic
  $S=1/2$ Heisenberg Chain},
  \href{http://dx.doi.org/10.1103/PhysRevLett.45.1358}{Phys.\ Rev.\ Lett.
  \textbf{45}, 1358 (1980)}.

\bibitem[Haldane(1981{\natexlab{a}})]{Hal81b}
F.~D.~M. Haldane, {\em Effective Harmonic-Fluid Approach to Low-Energy
  Properties of One-Dimensional Quantum Fluids},
  \href{http://dx.doi.org/10.1103/PhysRevLett.47.1840}{Phys.\ Rev.\ Lett.
  \textbf{47}, 1840 (1981{\natexlab{a}})}.

\bibitem[Haldane(1981{\natexlab{b}})]{Hal81a}
F.~D.~M. Haldane, {\em `Luttinger liquid theory' of one-dimensional quantum
  fluids. I. Properties of the Luttinger model and their extension to the
  general {\oneD} interacting spinless Fermi gas},
  \href{http://dx.doi.org/10.1088/0022-3719/14/19/010}{J. Phys.~C \textbf{14},
  2585 (1981{\natexlab{b}})}.

\bibitem[Hewson(1993)]{Hew93}
A.~Hewson, {\em The Kondo Problem to Heavy Fermions}, Cambridge University
  Press, Cambridge, 1993.

\bibitem[Honerkamp et~al.(2004)Honerkamp, Rohe, Andergassen, and Enss]{HRAE04}
C.~Honerkamp, D.~Rohe, S.~Andergassen, and T.~Enss, {\em Interaction flow
  method for many-fermion systems},
  \href{http://dx.doi.org/10.1103/PhysRevB.70.235115}{Phys.\ Rev.~B
  \textbf{70}, 235115 (2004)},
  \href{http://arXiv.org/abs/cond-mat/0403633}{\texttt{cond-mat/0403633}}.

\bibitem[Honerkamp and Salmhofer(2001)]{HS01}
C.~Honerkamp and M.~Salmhofer, {\em The temperature-flow renormalization group
  and the competition between superconductivity and ferromagnetism},
  \href{http://dx.doi.org/10.1103/PhysRevB.64.184516}{Phys.\ Rev.~B
  \textbf{64}, 184516 (2001)},
  \href{http://arXiv.org/abs/cond-mat/0105218}{\texttt{cond-mat/0105218}}.

\bibitem[Honerkamp et~al.(2001)Honerkamp, Salmhofer, Furukawa, and
  Rice]{HSFR01}
C.~Honerkamp, M.~Salmhofer, N.~Furukawa, and T.~M. Rice, {\em Breakdown of the
  Landau-Fermi liquid in Two Dimensions due to Umklapp Scattering},
  \href{http://dx.doi.org/10.1103/PhysRevB.63.035109}{Phys.\ Rev.~B
  \textbf{63}, 035109 (2001)},
  \href{http://arXiv.org/abs/cond-mat/9912358}{\texttt{cond-mat/9912358}}.

\bibitem[Kampf and Katanin(2003)]{KK03}
A.~P. Kampf and A.~A. Katanin, {\em Competing phases in the extended U-V-J
  Hubbard model near the Van Hove fillings},
  \href{http://dx.doi.org/10.1103/PhysRevB.67.125104}{Phys.\ Rev.~B
  \textbf{67}, 125104 (2003)},
  \href{http://arXiv.org/abs/cond-mat/0212190}{\texttt{cond-mat/0212190}}.

\bibitem[Kane et~al.(1997)Kane, Balents, and Fisher]{KBF97}
C.~Kane, L.~Balents, and M.~P.~A. Fisher, {\em Coulomb Interactions and
  Mesoscopic Effects in Carbon Nanotubes},
  \href{http://dx.doi.org/10.1103/PhysRevLett.79.5086}{Phys.\ Rev.\ Lett.
  \textbf{79}, 5086 (1997)},
  \href{http://arXiv.org/abs/cond-mat/9708054}{\texttt{cond-mat/9708054}}.

\bibitem[Kane and Fisher(1992{\natexlab{a}})]{KF92b}
C.~L. Kane and M.~P.~A. Fisher, {\em Resonant tunneling in an interacting
  one-dimensional electron gas},
  \href{http://dx.doi.org/10.1103/PhysRevB.46.7268}{Phys.\ Rev.~B \textbf{46},
  7268 (1992{\natexlab{a}})}.

\bibitem[Kane and Fisher(1992{\natexlab{b}})]{KF92c}
C.~L. Kane and M.~P.~A. Fisher, {\em Transmission through barriers and resonant
  tunneling in an interacting one-dimensional electron gas},
  \href{http://dx.doi.org/10.1103/PhysRevB.46.15233}{Phys.\ Rev.~B \textbf{46},
  15233 (1992{\natexlab{b}})}.

\bibitem[Kane and Fisher(1992{\natexlab{c}})]{KF92a}
C.~L. Kane and M.~P.~A. Fisher, {\em Transport in a one-channel Luttinger
  liquid}, \href{http://dx.doi.org/10.1103/PhysRevLett.68.1220}{Phys.\ Rev.\
  Lett. \textbf{68}, 1220 (1992{\natexlab{c}})}.

\bibitem[Kawakami and Yang(1990)]{KY90}
N.~Kawakami and S.-K. Yang, {\em Luttinger anomaly exponent of momentum
  distribution in the Hubbard chain},
  \href{http://dx.doi.org/10.1016/0375-9601(90)90818-9}{Phys.\ Lett.~A
  \textbf{148}, 359 (1990)}.

\bibitem[Keller et~al.(1992)Keller, Kopper, and Salmhofer]{KKS92}
G.~Keller, C.~Kopper, and M.~Salmhofer, {\em Perturbative renormalization and
  effective Lagrangians in $\varphi^4$ in four dimensions}, Helv.\ Phys.\ Acta
  \textbf{65}, 32 (1992).

\bibitem[Lieb and Wu(1968)]{LW68}
E.~H. Lieb and F.~Y. Wu, {\em Absence of Mott Transition in an Exact Solution
  of the Short-Range, One-Band Model in One Dimension},
  \href{http://dx.doi.org/10.1103/PhysRevLett.20.1445}{Phys.\ Rev.\ Lett.
  \textbf{135}, A1505 (1968)}.

\bibitem[Luther and Peschel(1974)]{LP74}
A.~Luther and I.~Peschel, {\em Single-particle states, Kohn anomaly, and
  pairing fluctuations in one dimension},
  \href{http://dx.doi.org/10.1103/PhysRevB.9.2911}{Phys.\ Rev.~B \textbf{9},
  2911 (1974)}.

\bibitem[Luttinger(1963)]{Lut63}
J.~M. Luttinger, {\em An exactly soluble model of a many-fermion system}, J.
  Math.\ Phys. \textbf{4}, 1154 (1963).

\bibitem[Mahan(2000)]{Mah00}
G.~D. Mahan, {\em Many-particle physics}, Kluwer Academic Publishers, New York,
  3. edition, 2000.

\bibitem[Mattis(1974)]{Mat74}
D.~C. Mattis, {\em New wave-operator identity applied to the study of
  persistent currents in {\oneD}},
  \href{http://dx.doi.org/10.1063/1.1666693}{J. Math.\ Phys. \textbf{15}, 609
  (1974)}.

\bibitem[Mattis and Lieb(1965)]{ML65}
D.~C. Mattis and E.~H. Lieb, {\em Exact solution of a many-fermion system and
  its associated boson field}, J. Math.\ Phys. \textbf{6}, 304 (1965).

\bibitem[Matveev et~al.(1993)Matveev, Yue, and Glazman]{MYG93}
K.~A. Matveev, D.~Yue, and L.~I. Glazman, {\em Tunneling in one-dimensional
  non-Luttinger electron liquid},
  \href{http://dx.doi.org/10.1103/PhysRevLett.71.3351}{Phys.\ Rev.\ Lett.
  \textbf{71}, 3351 (1993)},
  \href{http://arXiv.org/abs/cond-mat/9306041}{\texttt{cond-mat/9306041}}.

\bibitem[Meden et~al.(2003)Meden, Andergassen, Metzner, Schollw{\"o}ck, and
  Sch{\"o}nhammer]{Med03}
V.~Meden, S.~Andergassen, W.~Metzner, U.~Schollw{\"o}ck, and
  K.~Sch{\"o}nhammer, {\em Scaling of the conductance in a quantum wire},
  \href{http://dx.doi.org/10.1209/epl/i2003-00624-x}{Europhys.\ Lett.
  \textbf{64}, 769 (2003)},
  \href{http://arXiv.org/abs/cond-mat/0303460}{\texttt{cond-mat/0303460}}.

\bibitem[Meden et~al.(2005)Meden, Enss, Andergassen, Metzner, and
  Sch{\"o}nhammer]{Med04}
V.~Meden, T.~Enss, S.~Andergassen, W.~Metzner, and K.~Sch{\"o}nhammer, {\em
  Correlation effects on resonant tunneling in one-dimensional quantum wires},
  \href{http://dx.doi.org/10.1103/PhysRevB.71.041302}{Phys.\ Rev.~B
  \textbf{71}, 041302(R) (2005)},
  \href{http://arXiv.org/abs/cond-mat/0403655}{\texttt{cond-mat/0403655}}.

\bibitem[Meden et~al.(2000)Meden, Metzner, Schollw{\"o}ck, Schneider, Stauber,
  and Sch{\"o}n-hammer]{Med00}
V.~Meden, W.~Metzner, U.~Schollw{\"o}ck, O.~Schneider, T.~Stauber, and
  K.~Sch{\"o}n-hammer, {\em Luttinger liquids with boundaries: Power-laws and
  energy scales}, Eur.\ Phys.~J.~B \textbf{16}, 631 (2000),
  \href{http://arXiv.org/abs/cond-mat/0002215}{\texttt{cond-mat/0002215}}.

\bibitem[Meden et~al.(2002{\natexlab{a}})Meden, Metzner, Schollw{\"o}ck, and
  Sch{\"o}nhammer]{MMSS02a}
V.~Meden, W.~Metzner, U.~Schollw{\"o}ck, and K.~Sch{\"o}nhammer, {\em Scaling
  behavior of impurities in mesoscopic Luttinger liquids},
  \href{http://dx.doi.org/10.1103/PhysRevB.65.045318}{Phys.\ Rev.~B
  \textbf{65}, 045318 (2002{\natexlab{a}})},
  \href{http://arXiv.org/abs/cond-mat/0104336}{\texttt{cond-mat/0104336}}.

\bibitem[Meden et~al.(2002{\natexlab{b}})Meden, Metzner, Schollw{\"o}ck, and
  Sch{\"o}nhammer]{MMSS02b}
V.~Meden, W.~Metzner, U.~Schollw{\"o}ck, and K.~Sch{\"o}nhammer, {\em A single
  impurity in a Luttinger liquid: How it ``cuts'' the chain},
  \href{http://dx.doi.org/10.1023/A:1013823514926}{J. Low Temp.\ Phys.
  \textbf{126}, 1147 (2002{\natexlab{b}})},
  \href{http://arXiv.org/abs/cond-mat/0109013}{\texttt{cond-mat/0109013}}.

\bibitem[Meden et~al.(1998)Meden, Schmitteckert, and Shannon]{Med98}
V.~Meden, P.~Schmitteckert, and N.~Shannon, {\em Orthogonality catastrophe in a
  one-dimensional system of correlated electrons},
  \href{http://dx.doi.org/10.1103/PhysRevB.57.8878}{Phys.\ Rev.~B \textbf{57},
  8878 (1998)},
  \href{http://arXiv.org/abs/cond-mat/9707082}{\texttt{cond-mat/9707082}}.

\bibitem[Meden and Schollw{\"o}ck(2003{\natexlab{a}})]{MS03b}
V.~Meden and U.~Schollw{\"o}ck, {\em The conductance of interacting
  nano-wires}, \href{http://dx.doi.org/10.1103/PhysRevB.67.193303}{Phys.\
  Rev.~B \textbf{67}, 193303 (2003{\natexlab{a}})},
  \href{http://arXiv.org/abs/cond-mat/0210515}{\texttt{cond-mat/0210515}}.

\bibitem[Meden and Schollw{\"o}ck(2003{\natexlab{b}})]{MS03a}
V.~Meden and U.~Schollw{\"o}ck, {\em Persistent currents in mesoscopic rings: A
  numerical and renormalization group study},
  \href{http://dx.doi.org/10.1103/PhysRevB.67.035106}{Phys.\ Rev.~B
  \textbf{67}, 035106 (2003{\natexlab{b}})},
  \href{http://arXiv.org/abs/cond-mat/0209588}{\texttt{cond-mat/0209588}}.

\bibitem[Meir and Wingreen(1992)]{MW92}
Y.~Meir and N.~S. Wingreen, {\em Landauer formula for the current through an
  interacting electron region},
  \href{http://dx.doi.org/10.1103/PhysRevLett.68.2512}{Phys.\ Rev.\ Lett.
  \textbf{68}, 2512 (1992)}.

\bibitem[Metzner et~al.(1998)Metzner, Castellani, and
  {Di~Castro}]{Metzner:9701012}
W.~Metzner, C.~Castellani, and C.~{Di~Castro}, {\em Fermi Systems with Strong
  Forward Scattering}, Advances in Physics \textbf{47}, 317 (1998),
  \href{http://arXiv.org/abs/cond-mat/9701012}{\texttt{cond-mat/9701012}}.

\bibitem[Metzner and {Di~Castro}(1993)]{MC93}
W.~Metzner and C.~{Di~Castro}, {\em Conservation laws and correlation functions
  in the Luttinger liquid},
  \href{http://dx.doi.org/10.1103/PhysRevB.47.16107}{Phys.\ Rev.~B \textbf{47},
  16107 (1993)}.

\bibitem[Moon et~al.(1993)Moon, Yi, Kane, Girvin, and Fisher]{Moo93}
K.~Moon, H.~Yi, C.~L. Kane, S.~M. Girvin, and M.~P.~A. Fisher, {\em Resonant
  tunneling between quantum Hall edge states},
  \href{http://dx.doi.org/10.1103/PhysRevLett.71.4381}{Phys.\ Rev.\ Lett.
  \textbf{71}, 4381 (1993)},
  \href{http://arXiv.org/abs/cond-mat/9304010}{\texttt{cond-mat/9304010}}.

\bibitem[Morris(1994)]{Mor94}
T.~R. Morris, {\em The Exact Renormalisation Group and Approximate Solutions},
  Int.\ J. Mod.\ Phys.~A \textbf{9}, 2411 (1994),
  \href{http://arXiv.org/abs/hep-ph/9308265}{\texttt{hep-ph/9308265}}.

\bibitem[Nazarov and Glazman(2003)]{NG03}
Y.~V. Nazarov and L.~I. Glazman, {\em Resonant Tunneling of Interacting
  Electrons in a One-Dimensional Wire},
  \href{http://dx.doi.org/10.1103/PhysRevLett.91.126804}{Phys.\ Rev.\ Lett.
  \textbf{91}, 126804 (2003)},
  \href{http://arXiv.org/abs/cond-mat/0209090}{\texttt{cond-mat/0209090}}.

\bibitem[Negele and Orland(1987)]{NO87}
J.~W. Negele and H.~Orland, {\em Quantum Many-Particle Systems},
  Addison-Wesley, Reading, 1987.

\bibitem[Ng and Lee(1988)]{NL88}
T.~K. Ng and P.~A. Lee, {\em On-site Coulomb repulsion and resonant tunneling},
  \href{http://dx.doi.org/10.1103/PhysRevLett.61.1768}{Phys.\ Rev.\ Lett.
  \textbf{61}, 1768 (1988)}.

\bibitem[Oguri(2001)]{Ogu01}
A.~Oguri, {\em Transmission Probability for Interacting Electrons Connected to
  Reservoirs}, \href{http://dx.doi.org/10.1143/JPSJ.70.2666}{J. Phys.\ Soc.\
  Jpn \textbf{70}, 2666 (2001)},
  \href{http://arXiv.org/abs/cond-mat/0106033}{\texttt{cond-mat/0106033}}.

\bibitem[Picciotto et~al.(2001)Picciotto, Stormer, Pfeiffer, Baldwin, and
  West]{Pic01}
R.~D. Picciotto, H.~L. Stormer, L.~N. Pfeiffer, K.~W. Baldwin, and K.~W. West,
  {\em Four-terminal resistance of a ballistic quantum wire},
  \href{http://dx.doi.org/10.1038/35075009}{Nature \textbf{411}, 51 (2001)}.

\bibitem[Polchinski(1984)]{Pol84}
J.~Polchinski, {\em Renormalization and effective lagrangians},
  \href{http://dx.doi.org/10.1016/0550-3213(84)90287-6}{Nucl.\ Phys.~B
  \textbf{231}, 269 (1984)}.

\bibitem[Polyakov and Gornyi(2003)]{PG03}
D.~G. Polyakov and I.~V. Gornyi, {\em Transport of interacting electrons
  through a double barrier in quantum wires},
  \href{http://dx.doi.org/10.1103/PhysRevB.68.035421}{Phys.\ Rev.~B
  \textbf{68}, 035421 (2003)},
  \href{http://arXiv.org/abs/cond-mat/0212355}{\texttt{cond-mat/0212355}}.

\bibitem[Salmhofer(1998)]{Salmhofer:9706188}
M.~Salmhofer, {\em Continuous Renormalization for Fermions and Fermi Liquid
  Theory}, \href{http://dx.doi.org/10.1007/s002200050358}{Commun.\ Math.\ Phys.
  \textbf{194}, 249 (1998)},
  \href{http://arXiv.org/abs/cond-mat/9706188}{\texttt{cond-mat/9706188}}.

\bibitem[Salmhofer(1999)]{Salmhofer:1999}
M.~Salmhofer, {\em Renormalization. An Introduction}, Springer, Berlin, 1999.

\bibitem[Salmhofer and Honerkamp(2001)]{SH01}
M.~Salmhofer and C.~Honerkamp, {\em Fermionic renormalization group flows:
  Technique and theory}, \href{http://ptp.ipap.jp/link?PTP/105/1}{Prog.\
  Theor.\ Phys. \textbf{105}, 1 (2001)}.

\bibitem[Sch\"onhammer(2004)]{Sch05}
K.~Sch\"onhammer, {\em Luttinger liquids: the basic concepts}, In D.~Baeriswyl
  and L.~Degiorgi, editors, {\em Strong Interactions in Low Dimensions}, Kluwer
  Academic Publishers, 2004,
  \href{http://arXiv.org/abs/cond-mat/0305035}{\texttt{cond-mat/0305035}}.

\bibitem[Sch\"onhammer et~al.(2000)Sch\"onhammer, Meden, Metzner,
  Scholl-w\"ock, and Gunnarsson]{Gun00}
K.~Sch\"onhammer, V.~Meden, W.~Metzner, U.~Scholl-w\"ock, and O.~Gunnarsson,
  {\em Boundary effects on one-particle spectra of Luttinger liquids},
  \href{http://dx.doi.org/10.1103/PhysRevB.61.4393}{Phys.\ Rev.~B \textbf{61},
  4393 (2000)},
  \href{http://arXiv.org/abs/cond-mat/9903121}{\texttt{cond-mat/9903121}}.

\bibitem[Schulz(1990)]{Schulz90b}
H.~J. Schulz, {\em Correlation exponents and the metal-insulator transition in
  the one-dimensional Hubbard model},
  \href{http://dx.doi.org/10.1103/PhysRevLett.64.2831}{Phys.\ Rev.\ Lett.
  \textbf{64}, 2831 (1990)}.

\bibitem[Sch{\"u}tz et~al.(2004)Sch{\"u}tz, Bartosch, and Kopietz]{Schuetz04}
F.~Sch{\"u}tz, L.~Bartosch, and P.~Kopietz, {\em Collective fields in the
  functional renormalization group for fermions, Ward identities, and the exact
  solution of the Tomonaga-Luttinger model},
  \href{http://dx.doi.org/10.1103/PhysRevB.72.035107}{Phys.\ Rev.~B
  \textbf{72}, 035107 (2004)},
  \href{http://arXiv.org/abs/cond-mat/0409404}{\texttt{cond-mat/0409404}}.

\bibitem[Shankar(1991)]{Sh91}
R.~Shankar, {\em Renormalization group for interacting fermions in d > 1},
  \href{http://dx.doi.org/10.1016/0378-4371(91)90197-K}{Physica A \textbf{177},
  530 (1991)}.

\bibitem[Shankar(1994)]{Sh94}
R.~Shankar, {\em Renormalization-group approach to interacting fermions},
  \href{http://dx.doi.org/10.1103/RevModPhys.66.129}{Reviews of Modern Physics
  \textbf{66}, 129 (1994)},
  \href{http://arXiv.org/abs/cond-mat/9307009}{\texttt{cond-mat/9307009}}.

\bibitem[S\'olyom(1979)]{Sol79}
J.~S\'olyom, {\em The Fermi gas model of one-dimensional conductors}, Adv.
  Phys. \textbf{28}, 201 (1979).

\bibitem[Tam et~al.(2005)Tam, Tsai, and Campbell]{TTC05}
K.~M. Tam, S.~W. Tsai, and D.~K. Campbell, {\em Functional Renormalization
  Group Analysis of the Half-filled One-dimensional Extended Hubbard Model},
  \href{http://arXiv.org/abs/cond-mat/0505396}{\texttt{cond-mat/0505396}}.

\bibitem[Taylor(1972)]{Tay}
J.~R. Taylor, {\em Scattering theory}, John Wiley and Sons, New York, 1972.

\bibitem[Tomonaga(1950)]{Tom50}
S.~Tomonaga, {\em Remarks on Blochs method of sound waves applied to
  many-fermion problems}, Prog.\ Theor.\ Phys. \textbf{5}, 544 (1950).

\bibitem[Voit(1995)]{Voi95}
J.~Voit, {\em One-Dimensional Fermi liquids},
  \href{http://dx.doi.org/10.1088/0034-4885/58/9/002}{Rep.\ Prog.\ Phys.
  \textbf{58}, 977 (1995)},
  \href{http://arXiv.org/abs/cond-mat/9510014}{\texttt{cond-mat/9510014}}.

\bibitem[Wegner and Houghton(1973)]{WH73}
F.~J. Wegner and A.~Houghton, {\em Renormalization Group Equation for Critical
  Phenomena}, \href{http://dx.doi.org/10.1103/PhysRevA.8.401}{Phys.\ Rev.~A
  \textbf{8}, 401 (1973)}.

\bibitem[Weinberg(1976)]{Wei76}
S.~Weinberg, {\em Critical phenomena for field theorists},
  \href{http://ccdb3fs.kek.jp/cgi-bin/img_index?197610218}{Erice Subnucl.\
  Phys., 1 (1976)}.

\bibitem[Weiss(1999)]{Wei99}
U.~Weiss, {\em Quantum Dissipative Systems}, World Scientific, Singapore, 2.
  edition, 1999, and references therein.

\bibitem[Wetterich(1993)]{Wet93}
C.~Wetterich, {\em Exact evolution equation for the effective potential},
  \href{http://dx.doi.org/10.1016/0370-2693(93)90726-X}{Phys.\ Lett.~B
  \textbf{301}, 90 (1993)}.

\bibitem[Wieczerkowski(1988)]{Wie88}
C.~Wieczerkowski, {\em Symanzik improved actions from the viewpoint of the
  renormalization-group}, Commun.\ Math.\ Phys. \textbf{120}, 149 (1988).

\bibitem[Wiese(2003)]{Wie03}
K.~J. Wiese, {\em The Functional Renormalization Group Treatment of Disordered
  Systems: a Review},
  \href{http://arXiv.org/abs/cond-mat/0302322}{\texttt{cond-mat/0302322}}.

\bibitem[Wilson(1971)]{Wil71}
K.~G. Wilson, {\em Renormalization Group and Critical Phenomena. II.
  Phase-Space Cell Analysis of Critical Behavior},
  \href{http://dx.doi.org/10.1103/PhysRevB.4.3184}{Phys.\ Rev.~B \textbf{4},
  3184 (1971)}.

\bibitem[Wilson and Kogut(1974)]{WK74}
K.~G. Wilson and J.~Kogut, {\em The renormalization group and the $\epsilon$
  expansion}, \href{http://dx.doi.org/10.1016/0370-1573(74)90023-4}{Phys.\ Rep.
  \textbf{12}, 75 (1974)}.

\bibitem[Yang and Yang(1966)]{YY66}
C.~N. Yang and C.~P. Yang, {\em One-Dimensional Chain of Anisotropic Spin-Spin
  Interactions. I.+II.},
  \href{http://dx.doi.org/10.1103/PhysRev.150.321}{Phys.\ Rev. \textbf{150},
  321 (1966)}.

\bibitem[Yao et~al.(1999)Yao, Postma, Balents, and Dekker]{YEX99}
Z.~Yao, H.~W.~C. Postma, L.~Balents, and C.~Dekker, {\em Carbon nanotube
  intramolecular junctions}, \href{http://dx.doi.org/10.1038/46241}{Nature
  \textbf{402}, 273 (1999)}.

\bibitem[Yue et~al.(1994)Yue, Glazman, and Matveev]{YGM94}
D.~Yue, L.~I. Glazman, and K.~A. Matveev, {\em Conduction of a weakly
  interacting one-dimensional electron gas through a single barrier},
  \href{http://dx.doi.org/10.1103/PhysRevB.49.1966}{Phys.\ Rev.~B \textbf{49},
  1966 (1994)}.

\bibitem[Zanchi and Schulz(1998)]{Zanchi:9703189}
D.~Zanchi and H.~J. Schulz, {\em Weakly correlated electrons on a square
  lattice: A renormalization group theory},
  \href{http://dx.doi.org/10.1209/epl/i1998-00462-x}{Europhys.\ Lett.
  \textbf{44}, 235 (1998)},
  \href{http://arXiv.org/abs/cond-mat/9703189}{\texttt{cond-mat/9703189}}.

\bibitem[Zanchi and Schulz(2000)]{Zanchi:9812303}
D.~Zanchi and H.~J. Schulz, {\em Weakly correlated electrons on a square
  lattice: Renormalization-group theory},
  \href{http://dx.doi.org/10.1103/PhysRevB.61.13609}{Phys.\ Rev.~B \textbf{61},
  13609 (2000)},
  \href{http://arXiv.org/abs/cond-mat/9812303}{\texttt{cond-mat/9812303}}.

\bibitem[Zinn-Justin(2002)]{Zin02}
J.~Zinn-Justin, {\em Quantum Field Theory and Critical Phenomena}, Clarendon
  Press, Oxford, 4. edition, 2002.

\end{thebibliography}
%}
\end{document}